\documentclass[11pt,twoside,british,czech,english]{article}
\usepackage{ae,aecompl}
\usepackage{helvet}
\usepackage[T1]{fontenc}
\usepackage[latin2,latin9]{inputenc}
\usepackage{geometry}
\usepackage{color}
\usepackage{float}
\usepackage{textcomp}
\usepackage{amsbsy}
\usepackage{amstext}
\usepackage{amssymb}
\usepackage{graphicx}
\usepackage{setspace}
\usepackage{esint}
\usepackage[authoryear]{natbib}
\usepackage{subscript}

\makeatletter

\providecommand{\LyX}{L\kern-.1667em\lower.25em\hbox{Y}\kern-.125emX\@}
\newcommand{\lyxmathsym}[1]{\ifmmode\begingroup\def\b@ld{bold}
  \text{\ifx\math@version\b@ld\bfseries\fi#1}\endgroup\else#1\fi}

\providecommand{\tabularnewline}{\\}
\newcommand{\lyxdot}{.}

\newenvironment{lyxlist}[1]
{\begin{list}{}
{\settowidth{\labelwidth}{#1}
 \setlength{\leftmargin}{\labelwidth}
 \addtolength{\leftmargin}{\labelsep}
 }}
{\end{list}}

\usepackage{aas_macros}
\usepackage{natbib}

\makeatother

\usepackage{babel}
\begin{document}
\selectlanguage{czech}%
\inputencoding{latin2}\thispagestyle{empty}  \ 

\selectlanguage{english}%
\inputencoding{latin9}\vspace{1cm}

\begin{center}
{\Large Charles University in Prague}\\
{\Large{} Faculty of Mathematics and Physics}
\par\end{center}{\Large \par}

\bigskip{}

\bigskip{}

\begin{center}
\textbf{\huge DOCTORAL THESIS }
\par\end{center}{\huge \par}

\begin{center}
{\large \includegraphics[width=60mm]{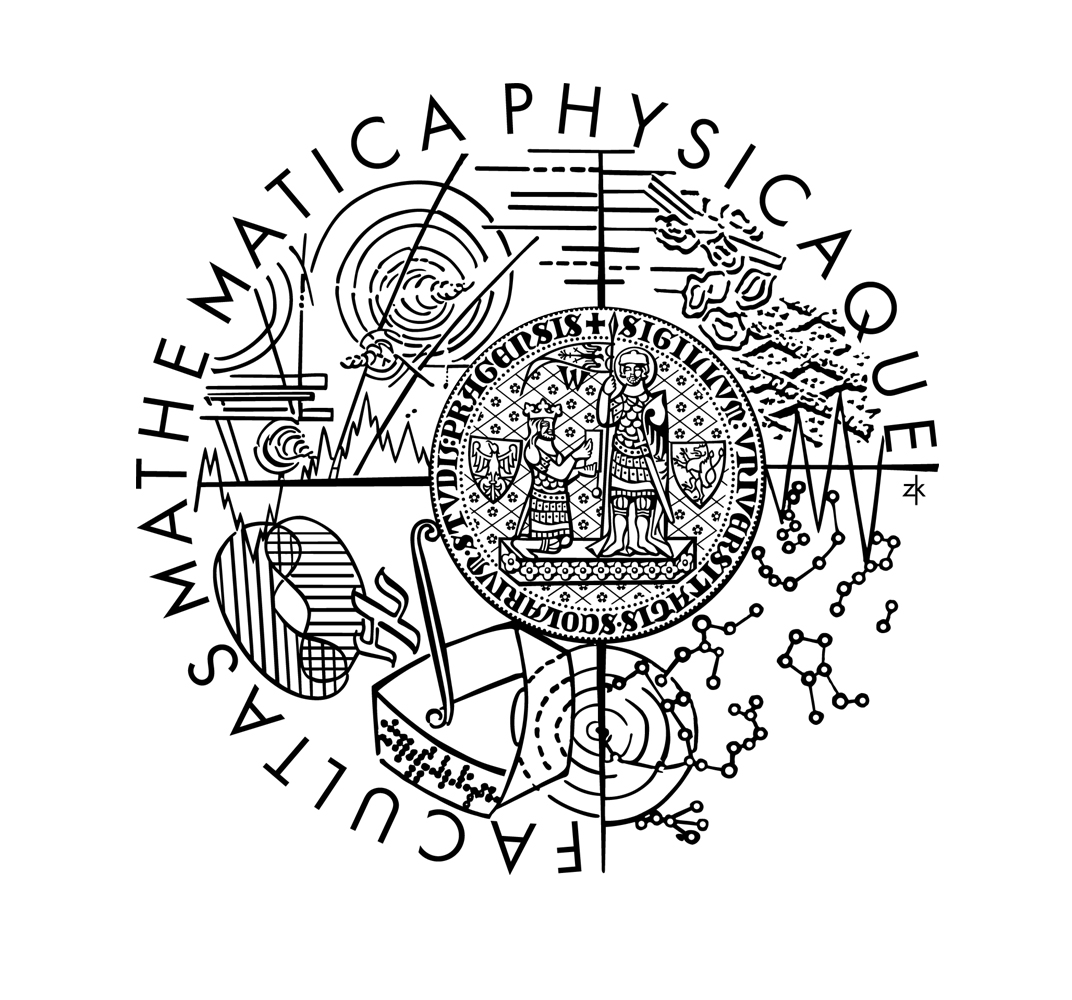}}
\par\end{center}{\large \par}

\begin{center}
{\large %\normalsize
}{\Large Ivana Ebrov{\'a}}
\par\end{center}{\Large \par}

\begin{center}
{\large \vspace{2mm}
}
\par\end{center}{\large \par}

\begin{center}
\textbf{\LARGE Shell galaxies:}
\par\end{center}{\LARGE \par}

\begin{center}
\textbf{\Large kinematical signature of shells, satellite galaxy disruption
and dynamical friction}
\par\end{center}{\Large \par}

\begin{center}
{\large \vspace{2mm}
}
\par\end{center}{\large \par}

\begin{center}
{\Large Astronomical Institute}\\
{\Large of the Academy of Sciences of the Czech Republic}{\large{} }
\par\end{center}{\large \par}

\begin{center}
{\large \vspace{2mm}
}
\par\end{center}{\large \par}

\begin{center}
{\Large Supervisor of the doctoral thesis: RNDr. Bruno Jungwiert,
Ph.D. }
\par\end{center}{\Large \par}

\begin{center}
{\large \vspace{2mm}
}{\Large Study program: Physics}\\
{\Large Specialization: Theoretical Physics, Astronomy and Astrophysics}
\par\end{center}{\Large \par}

\vspace{1.5cm}

\begin{center}
{\large Prague 2013 }
\par\end{center}{\large \par}

\selectlanguage{czech}%
\inputencoding{latin2}

\selectlanguage{english}%
\inputencoding{latin9}\newpage{}

\setcounter{page}{2} \ 

This research has made use of NASA's Astrophysics Data System, micronised
purified flavonoid fraction, and a lot of iso-butyl-propanoic-phenolic
acid. Typeset in \LyX{}, an open source document processor. For graphical
presentation, we used Gnuplot, the PGPLOT (a graphics subroutine library
written by Tim Pearson) and scripts and programs written by Miroslav
K{\v r}{\'{\i}}{\v z}ek using Python and matplotlib. Calculations and simulations
have been carried out using Maple 10, Wolfram Mathematica 7.0, and
own software written in programming language FORTRAN 77, Fortran 90
and Fortran 95. The software for simulation of shell galaxy formation
using test particles are based on the source code of the MERGE 9 (written
by Bruno Jungwiert, 2006; unpublished); kinematics of shell galaxies
in the framework of the model of radial oscillations has been studied
using the smove software (written by Lucie J{\'{\i}}lkov{\'a}, 2011; unpublished);
self-consistent simulations have been done by Kate\v{r}ina Barto{\v s}kov{\'a}
with GADGET-2 \citep{2005MNRAS.364.1105S}.

We acknowledge support from the following sources: grant No.\ 205/08/H005
by Czech Science Foundation; research plan AV0Z10030501 by Academy
of Sciences of the Czech Republic; and the project SVV-267301 by Charles
University in Prague. This work has been done with the support for
a long-term development of the research institution RVO67985815.

\newpage{}

\noindent {\large \vglue 0pt plus 1fill }\vfill{}
\noindent {\large I declare that I carried out this doctoral thesis
independently, and only with the cited sources, literature and other
professional sources.}{\large \par}

{\large \medskip{}
I understand that my work relates to the rights and obligations under
the Act No. 121/2000 Coll., the Copyright Act, as amended, in particular
the fact that the Charles University in Prague has the right to conclude
a license agreement on the use of this work as a school work pursuant
to Section 60 paragraph 1 of the Copyright Act.}{\large \par}

\noindent {\large \vspace{10mm}
}{\large \par}

\ 

\noindent \noindent In Prague, 19. 8. 2013\hspace{\fill}Ivana Ebrov{\'a}

\noindent {\large \vspace{20mm}
} 

\newpage{}

\selectlanguage{english}%
\noindent \inputencoding{latin9}\vfill{}

\noindent Title: Shell galaxies: kinematical signature of shells,
satellite galaxy disruption and dynamical friction\\
Author: Ivana Ebrov{\'a}\\
Department / Institute: Astronomical Institute of the Academy of Sciences
of the Czech Republic\\
Supervisor of the doctoral thesis: RNDr. Bruno Jungwiert, Ph.D., Astronomical
Institute of the Academy of Sciences of the Czech Republic\\

\noindent Abstract: {Stellar shells observed in many giant
elliptical and lenticular as well as a few spiral and dwarf galaxies
presumably result from radial minor mergers of galaxies. We show that
the line-of-sight velocity distribution of the shells has a quadruple-peaked
shape. We found simple analytical expressions that connect the positions
of the four peaks of the line profile with the mass distribution of
the galaxy, namely, the circular velocity at the given shell radius
and the propagation velocity of the shell. The analytical expressions
were applied to a test-particle simulation of a radial minor merger,
and the potential of the simulated host galaxy was successfully recovered.
Shell kinematics can thus become an independent tool to determine
the content and distribution of dark matter in shell galaxies up to
$\sim$$100$\,kpc from the center of the host galaxy. Moreover we
investigate the dynamical friction and gradual disruption of the cannibalized
galaxy during the shell formation in the framework of a simulation
with test particles. The coupling of both effects can considerably
redistribute positions and luminosities of shells. Neglecting them
can lead to significant errors in attempts to date the merger in observed
shell galaxies.}\\


\newpage{}

\def\as{\hbox{\,$^{\prime\prime}$}}
\def\am{\hbox{\,$^{\prime}$}}
\def\sun{$_{\odot}$}
\def\suns{\sun \space}
\sloppy
\clubpenalty=9999
\widowpenalty=9999
\defcitealias{mk98}{MK98}
\tableofcontents{}

\newpage{}

\section{Objectives and motivation \label{sec:Objectives-and-motivation}}

The most successful theory of the evolution of the Universe so far
seems to be the theory of the hierarchical formation based on the
assumption of the existence of cold dark matter, significantly dominating
the baryonic one. In such a universe, large galaxies are formed by
merging of small galaxies, protogalaxies and diffuse accretion of
surrounding matter. Galactic interaction and dark matter play thus
a crucial role in the life of every galaxy.

But the determination of both the dark matter content and the merger
history of a galaxy is difficult. Firstly, the cold dark matter interacts
only gravitationally (and possibly via the weak interaction) and thus
the mapping of its distribution in galaxies is tricky. Secondly, the
nature disallows us to see individual galaxies from different angles,
thus our knowledge of their spatial properties is degenerate. Thirdly,
it is non-trivial to determine anything about the history of a given
galaxy as the whole existence of humanity presents only a snapshot
in the evolution of the Universe. Yet this knowledge is important
to confirm or disprove theories of the creation and evolution of the
Universe, improve their accuracy and to understand how the Universe
we live in actually looks. 

The deal of the galactic astronomy is to try to circumvent these obstacles.
One of the possibilities is to use tidal features left by the galactic
interactions. They act as dynamical tracers of the potential of their
host galaxies and as hints left behind by the accreted galaxies in
the past. The special case is that of arc-like fine structures found
in shell galaxies. Their unique kinematics carries both qualitative
and quantitative information on the distribution of the dark matter,
the shape of the potential of the host galaxy and its merger history.
Moreover, shell galaxies have their own mysteries that call for an
explanation. 

\begin{figure}[!bh]
\centering{}\includegraphics[width=11cm]{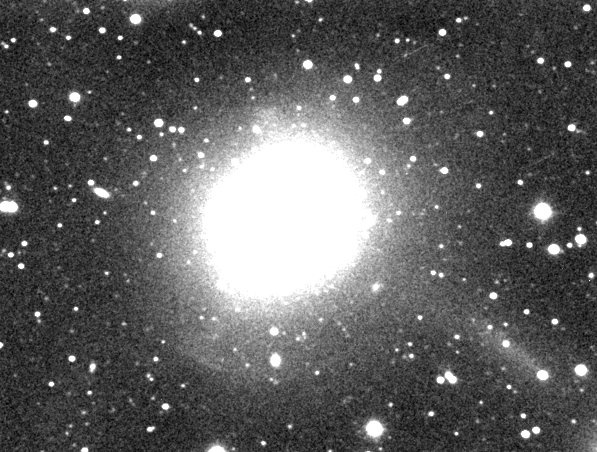}
\caption{\textsf{\small Shell galaxy M89. \label{fig:M89}}}
\end{figure}

Some shells need to be discovered using deep photometry, e.g., \citet{duc11},
whereas others can be today captured using amateur technology. The
photography of galaxy M89 in Fig.~\ref{fig:M89} was taken by a member
of our research group Michal B{\'{\i}}lek using his own amateur equipment
(taking 4.4 hours of exposure with an 8\textquotedbl{}, f/4 Schmidt-Newton
telescope equipped with a CCD at a site about 50 km from Prague).
Faint structures were first identified by \citet{1979Natur.277..279M}
and \citet{2005ApJ...631..809X} who concluded that the galaxy possibly
possesses a low-luminosity active galactic nucleus. Michal's image
shows fairly well the shell at bottom left, the jet at bottom right
and a less prominent shell at top right. 

However all the information is hidden so deep in the structure and
kinematics of shell galaxies that it is not clear that they could
be practically unraveled. Certainly, a lot of effort and invention
is required. In this work we focus mainly on the possibility to deduce
the potential of the host galaxy using shell kinematics (Part~\ref{PART II-S.kin}).
We aim at creating equations and algorithms applicable to observed
data. Now comes the era when the instrumental equipment begins to
allow us to actually obtain such kind of data and that requires deeper
theoretical understanding of the topic. Having no such data yet at
hand, we apply our methods to simulated data. This method requires
that the shell is formed by stars on mainly radial orbits. According
to present state of knowledge, shells in one galaxy are probably bound
by common origin in a radial minor merger. Reproducing their overall
structure is nevertheless complicated by physical processes such as
the dynamical friction and the gradual decay of the cannibalized galaxy.
We deal with these phenomena in Part~\ref{PART III-DF}.

Self-consistent simulations allow us to simulate many physical processes
at once. Some of them are difficult or outright impossible to reproduce
by analytical or semi-analytical methods. At the same time, the manifestation
of these processes in self-consistent simulations is difficult to
separate and sometimes they may even be confused with non-physical
outcomes of used methods. Moreover, self-consistent simulations with
high resolution necessary to analyze delicate tidal structures such
as the shells are demanding on computation time. This demand is even
larger if we want to explore a significant part of the parameter space.

Attempts to date a merger from observed positions of shells have been
made in previous works. Recently, \citet{canalizo07} presented HST/ACS
observations of spectacular shells in a quasar host galaxy (Fig.~\ref{obr.MC})
and, by simulating the position of the outermost shell by means of
restricted $N$-body simulations, attempted to put constraints on
the age of the merger. They concluded that it occurred a few hundred
Myr to $\sim2$\,Gyr ago, supporting a potential causal connection
between the merger, the post-starburst ages in nuclear stellar populations,
and the quasar. A typical delay of 1--2.5\,Gyr between a merger and
the onset of quasar activity is suggested by both $N$-body simulations
by \citet{2005MNRAS.361..776S} and observations by \citet{2008ApJ...678..751R}.
It might therefore appear reassuring to find a similar time lag between
the merger event and the quasar ignition in a study of an individual
spectacular object. In Part~\ref{PART III-DF} we explore the options
for inclusion of the dynamical friction and the gradual decay of the
cannibalized galaxy in test-particle simulations and we look at what
these simulations tell us about the potential and merger history of
shell galaxies. 

In Appendix~\ref{Apx:units}, we show the conversion of units used
in the thesis to SI units. List of abbreviations can be found in Appendix~\ref{apx:List-of-Abbreviations}.
Videos mostly illustrating the formation and evolution of shell structures
are part of the electronic attachment of the thesis. Their description
can be found in Appendix~\ref{Apx:Videos} and the videos can be
downloaded at: pc048b.fzu.cz/$\sim$ivana/shells/phd

\newpage{}

\part{Introduction\label{PART I}}

\section{Shell galaxies in brief }

\begin{figure}[H]
\centering{}\includegraphics[width=12cm]{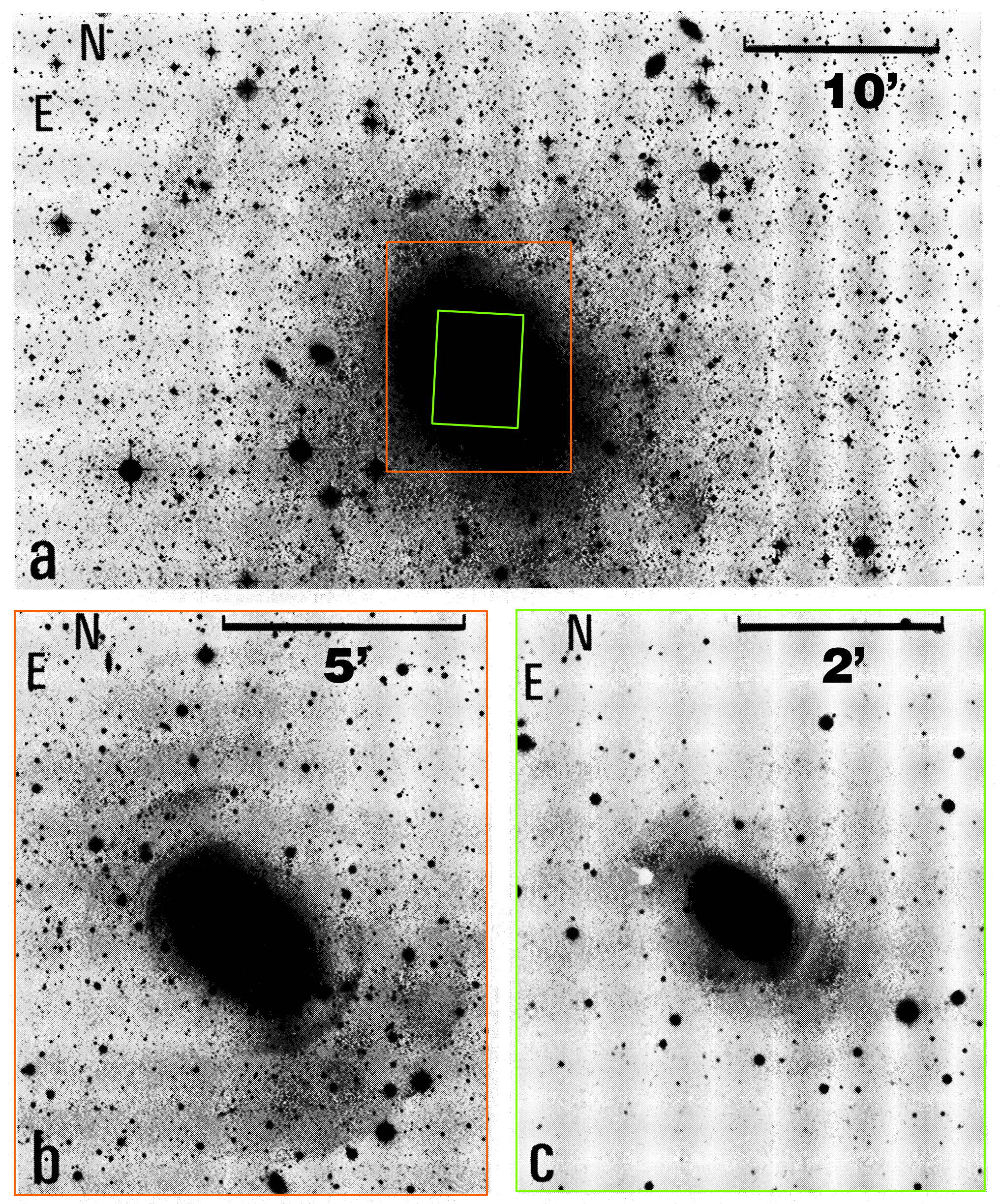}
\caption{\textsf{\small NGC~3923 from \citet{1983ApJ...274..534M} made from
UK Shmidt IIIa-J plates. The bottom row shows more central parts of
the galaxy. All images were processed (unsharp masking) to emphasize
the shell structure. 10\as\ roughly corresponds to 1}\,\textsf{\small kpc
in the galaxy. \label{fig:3923}}}
\end{figure}

Shell galaxies, like e.g. the beautiful and renowned NGC\,3923 in
Fig.~\ref{fig:3923}, are galaxies containing fine structures. These
structures are made of stars and form open, concentric arcs that do
not cross each other. The term \textit{shells} has spread throughout
the literature, gradually superseding the competing term \textit{ripples}.
According to the knowledge gained over the past more than thirty years,
their origin lies in the interactions between galaxies.

\section{Observational knowledge of shell galaxies \label{sec:Observational-knowledge}}

This section is mostly based on the review of literature presented
in \citet{EbrovaMAT}.

\subsection{Observational history \label{sub:history}}

It was Halton Arp, who first noticed the shell galaxies in his \textbf{Atlas
of Peculiar Galaxies} \citep{1966apg..book.....A} and the accompanying
article \citet{1966ApJS...14....1A}. He used the term {}``shells''
to describe the structures associated with galaxy Arp\,230. The Atlas
contains 338 objects, divided into several subgroups. Shell galaxies
are found under {}``concentric rings'' (Arp numbers 227 to 231),
but many other objects are in fact shell galaxies (Arp\,92, 103,
104, 153--155, 171, 215, 223, 226 and probably others).

To date, the only (at least partial) list of shell galaxies is {}``\textbf{A
catalogue of elliptical galaxies with shells}'' from \citet{1983ApJ...274..534M}.
The authors present a catalogue of 137 galaxies (with declination
south of $-17\lyxmathsym{\textdegree}$) that exhibit shell or ripple
features at large distances from the galaxy or in the outer envelope.
Some further work has been done on this set of galaxies: \citet*{1987MNRAS.224..895W,1987MNRAS.228..933W}
examined these shell galaxies to find radio and infrared sources,
\citet{1987IAUS..127..465W} carried out two-color CCD photometry
of 66 Malin-Carter galaxies, \citet{1988MNRAS.235..813C} obtained
nuclear spectra for 100 of the galaxies in the catalogue. In a series
of articles, \citet{1998A&AS..130..251L,1998A&AS..130..267L,1999A&A...341..357R,2000A&A...353..917L,1999A&A...345..419L}
(the fifth part surprisingly preceding the fourth) examined star formation
history in 21 catalogued shell galaxies. \citet{1994AJ....107.1713F}
were searching for secondary nuclei in 29 shell galaxies. Larger samples
of shell galaxies were studied for example by \citet{1983IAUS..100..319S},
\citet{1989AJ.....97..363T}, \citet{1992MNRAS.254..723F} or \citet{2001AJ....121..808C}.
Their results will be mentioned in the following chapters. Unsurprisingly,
many observational studies have been carried out over decades for
smaller samples or many individual shell galaxies.

\subsection{Occurrence of shell galaxies \label{sub:V=0000FDskyt-shell-galaxies}}

Originally \citep{1966ApJS...14....1A,1983ApJ...274..534M}, shells
were discovered basically in galaxies of E, E/S0 or S0 \textbf{morphological
type}. \citet{1988ApJ...328...88S} revealed that they can be found
also in S0/Sa and Sa galaxies (NGC\,3032, NGC\,3619, NGC\,4382,
NGC\,5739, and a Seyfert galaxy NGC\,5548) and even one Sbc galaxy
(NGC\,3310) was found likely to contain a shell. In fact, \citeauthor{1988ApJ...328...88S}
were against the term {}``shells'', supporting the term {}``ripples''
being more descriptive and not forcing a particular geometric interpretation.
NGC\,2782 (Arp\,215) is probably a spiral galaxy with shells which
Arp misclassified as spiral arms rather than as shells. NGC\,7531,
NGC\,3521, and NGC\,4651 \citep{2010AJ....140..962M} are examples
of some other lesser known cases of spiral galaxies with shells. The
last of them, NGC\,4651 and also M31 \citep{fardal07,fardal12} are
the only spiral galaxies where a multiple shell system has been discovered.
\citet{2004PASA...21..379C} and \citet{coleman04} reported a shell,
immediately followed by another one \citep{2005PASA...22..162C,2005AJ....129.1443C}
in Fornax dwarf spheroidal galaxy and it became the only shell galaxy
of this type.

The realistic estimate of the \textbf{relative abundance} of shell
galaxies (\citealp{1983IAUS..100..319S,1985LNP...232..145S}; cited
in \citealp{1988ApJ...331..682H}, and \citealp{1983ApJ...274..534M})
is about 10\% in early-type galaxies.%
\footnote{We use the term \textit{early-type galaxies} to denote all the Hubble
types E, E/S0, and S0 (elliptical and lenticular galaxies), because
many galaxies gradually wander between these classes according to
different classifications or simply in time (not physically, of course,
e.g. because of better or other observations).%
} \citet{1983ApJ...274..534M} state a surface brightness detection
limit $\mu_{\mathrm{max}}=26.5$\,mag$/$arcsec\textsuperscript{2}
in B filter%
\footnote{It is interesting to note that according to \citet{2005AJ....130.2647V},
galaxy surveys in blue filters would miss the majority of faint features
in their sample even if they met the same surface brightness limit.%
}. \citet{1988ApJ...328...88S} quoted similar results for their sample
of more than a hundred of galaxies, with the abundance of 6\% for
S0 and 10\% for E type galaxies, but with significantly lower number
among spirals (around 1\%). \citet{1993ASPC...48..629W} state that
\citet{1990dig..book..270S} found 56\% and 32\% of 74 E and S0 type
galaxies respectively posses ripples. 

In a complete sample of 55 elliptical galaxies at distances 15\textendash{}50
Mpc and luminosity cut of M\textsubscript{B}$<-20$ with detection
limit $\mu_{\mathrm{max}}=27.7$\,mag$/$arcsec\textsuperscript{2}
in V band, at least 22\% of galaxies have shells, making them the
most common interaction signature identified by \citet{tal09}. Shells
are also the most commonly detected feature in a sample of radio galaxies
of \citet{2011MNRAS.410.1550R} with $\mu_{\mathrm{max}}\sim26$\,mag$/$arcsec\textsuperscript{2}
in V filter.

On the contrary, in ATLAS\textsuperscript{3D} sample of 260 early-type
galaxies \citet{krajnovic11} found only 9 (3.5\%) galaxies with shells
at the limiting surface brightness $\mu_{\mathrm{max}}\sim26$\,mag$/$arcsec\textsuperscript{2}
in r band. \citet{2012ApJ...753...43K} examined a sample of 65 early
types drawn from the Spitzer Survey of Stellar Structure in Galaxies
(S\textsuperscript{4}G) and identified 4 shell galaxies (6\%). Their
detection limit was 25.2\,mag$/$arcsec\textsuperscript{2} for newly
obtained S\textsuperscript{4}G data and 26.5\,mag$/$arcsec\textsuperscript{2}
for some Spitzer archival images, both at 3.6\,$\mu$m, which correspond
to 26.9 and 28.2\,mag$/$arcsec\textsuperscript{2} in B band, respectively.
But they failed to detect some previously known shells in at least
three cases: NGC\,2974 and NGC\,5846 \citep{tal09} and NGC\,680
\citep{duc11} -- these three galaxies alone increase the percentage
of shell galaxies in their sample to 11\%. \citet{2013ApJ...765...28A}
found shells in 6\% of blue galaxies and around 14\% in red galaxies.%
\footnote{Red and blue galaxies are defined based on position in the color-magnitude
diagram in order to discriminate between systems on the red sequence
and blue cloud. It corresponds to a morphological segregation as well.
Vast majority of the red sequence galaxies are early-type galaxies,
while the blue sequence represents the late-type galaxies \citep{2009A&A...500..981C}.%
} The survey concerns 1781 luminous galaxies with the redshift range
$0.04<z<0.2$ and detection limit 27.7\,mag$/$arcsec\textsuperscript{2}
in $\textrm{g}^{\prime}$ filter. 

The occurrence of tidal features of any kind in galaxies is quite
high: 73\% in the sample of \citet{tal09}; 53\% in a sample of 126
red galaxies at a median redshift of $z=0.1$ and $\mu_{\mathrm{max}}\sim28$\,mag$/$arcsec\textsuperscript{2}
using B, V, and R fi{}lters \citep{2005AJ....130.2647V}; 71\% in
the subsample of 86 color- and morphology- selected bulge-dominated
early-type galaxies of the previous sample; about 24\% in s sample
of 474 close to edge-on early-type galaxies using the Sloan Digital
Sky Survey DR7 archive with $\mu_{\mathrm{max}}\sim26$\,mag$/$arcsec\textsuperscript{2}
using $\mathrm{\textrm{u}^{\prime}}$, $\mathrm{\textrm{g}^{\prime}}$,
$\textrm{r}^{\prime}$, $\mathrm{\textrm{i}^{\prime}}$, $\mathrm{\textrm{z}^{\prime}}$
bends \citep{2011A&A...536A..66M}; 12--26\% (according to confidence
level of a feature identification) in the sample of \citet{2013ApJ...765...28A}.
The lower detection rate in \citet{2013ApJ...765...28A} is explained
by authors by assertion that the majority of tidal features in early-type
galaxies are seen at surface brightness near (or below) 28\,mag$/$arcsec\textsuperscript{2}.
Since shells are generally low surface brightness features, the abundance
of shell galaxies will probably rise with deeper photometric observations. 

Another important piece of information from the above mentioned studies
is the \textbf{environmental dependence} of occurrence of shell structures.
They are seen about five times more often in isolated galaxies than
in galaxies in clusters. \citet{1983ApJ...274..534M} explored 137
shell galaxies -- 65 (47.5\%) are isolated, 42 (30.9\%) occur in loose
groups (of these 13\% have one or two close companions), only 5 (3.6\%)
occur in clusters or rich groups, and the remaining 25 (18\%) occur
in groups of two to five galaxies. Taking into account only isolated
galaxies, the relative abundance of shell galaxies increases to 17\%.
Similar result was reached more recently by \citet{2001AJ....121..808C}
-- they detected shell/tidal features in nine of the 22 isolated galaxies
(41\%), but only one of the twelve (8\%) group early-type galaxies
shows evidence for shells. \citet{1996MNRAS.282..149R} presented
their result that 4\% of 54 pairs of galaxies (pairs are located in
low-density environments) and 16\% of 61 isolated early-type galaxies
exhibit shells. \citet{2012AJ....144..128A} found abundance of tidal
features about 3\% in a sample of 54 galaxy clusters ($0.04<z<0.15$)
containing 3551 early-type galaxies, $\mu_{\mathrm{max}}=26.5$\,mag$/$arcsec\textsuperscript{2}
in $\textrm{r}^{\prime}$ filter.

\citet{1985LNP...232..145S} have investigated an unbiased sample
of 36 isolated giant ellipticals, in order to study their fine morphology.
They found that 16 of them (44\%) possess ripples (some of them very
weak, as \citeauthor{1985LNP...232..145S} note). In contrast to this,
\citet{2004AJ....127.3213M} did not find a single shell galaxy in
their sample of nine early-type galaxies previously verified to exist
in extremely isolated environments, even though, according to the
prognosis, at least four shell galaxies should have been present.
The probability of this (a sample of nine early-type galaxies from
regions of low galaxy density with no shell) is about 1\% if we assume
that 40\% of galaxies in low-density environments have shells.

However, the true abundance of shell galaxies can still be different
from what has been summarized here. It crucially depends on which
galaxies we classify as shell galaxies and on our ability to detect
faint shells in otherwise innocent looking galaxies.

\subsection{Appearance of the shells \label{sub:Vzhled-slupek}}

Shells have been detected in various \textbf{numbers}, appearance
and distributions. Rich systems like NGC\,3923 (Fig.~\ref{fig:3923})
or NGC\,5982 \citep{2007A&A...467.1011S} show about 30 shells, but
it is rather an exception among shell galaxies. A large fraction of
the Malin-Carter catalogue \citeyearpar{1983ApJ...274..534M} consists
of galaxies with less then 4 shells. It is in fact difficult to make
statements about numbers of shells in galaxies, because the detection
of all of them (sometimes even the proof of their existence) is a
delicate matter. Shells actually contain only a \textbf{fraction of
total luminosity} of the host galaxy, mostly from 3 to 6\% (e.g.,
it is 5\% for the famous NGC\,3923; \citealp{1988ApJ...326..596P}).
Shell \textbf{surface brightness contrast} is very low, about 0.1--0.2\,mag
\citep{1986A&A...166...53D}. \citet{1986Sci...231..227S} states
that on the brightness profiles of host galaxies, ripples appear as
minor steps of about 1--10\% in the local light distribution.

\begin{figure}[H]
\begin{centering}
\includegraphics[width=1\textwidth]{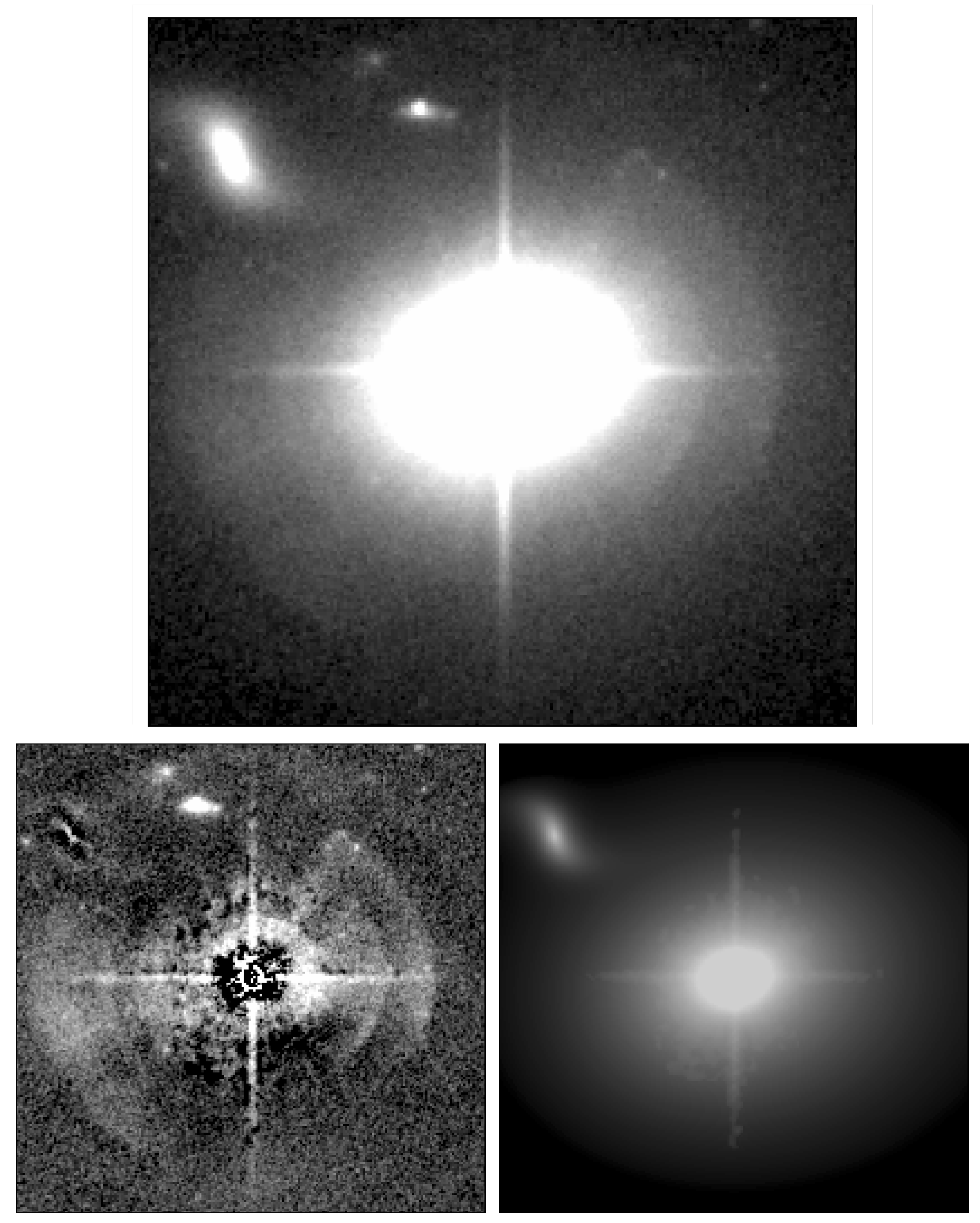}
\par\end{centering}

\caption{\textsf{\small Top: Very deep ACS/WFC image (total integration time
of 11432\,s) of a formerly unknown shell galaxy, the host galaxy
of the quasar MC2\,1635+119 (\citealp{canalizo07,2007AAS...20925104B};
the three images shown here are unpublished and were kindly provided
by G. Canalizo and N. Bennert). The shell structure is already visible
in this final reduced but otherwise unaltered image. The image size
is 10\as$\times$10\as. The residual image is shown in the bottom
left panel and was obtained by subtracting a model -- fitted using
GALFIT \citep{2002AJ....124..266P} -- for the host galaxy light (bottom
right) from the original data (top). Acknowledgment: NASA, STScI.
\label{obr.MC}}}
\end{figure}

To enhance or detect shells and other fine structures in galaxies,
some more or less sophisticated techniques are often used, like \textit{unsharp
masking} for photographic images \citep{1977AASPB..16...10M}, \textit{digital
masking} \citep{1985LNP...232..145S} or s\textit{tructure map} (\citealp{2002ApJ...569..624P};
based on the probabilistic image-restoration method of Richardson
and Lucy; \citealp{1972JOSA...62...55R}). \textit{Host galaxy subtraction}
was used to process the image of a shell galaxy in Fig.~\ref{obr.MC}.

Shells are \textbf{stellar} structures that form arcs in galaxies
(circular or slightly elliptical) that either lie within a specific
double cone on opposite sides of the galaxy, or encircle the galaxy
almost all around. In general, they tend to have sharp outer boundaries,
but many of them are faint and diffuse. \citet{1990dig..book...72P}
and \citet{1987IAUS..127..465W} recognized three different \textbf{morphological
categories} of shell galaxies. 

\begin{figure}[H]
\begin{centering}
\includegraphics[width=12cm]{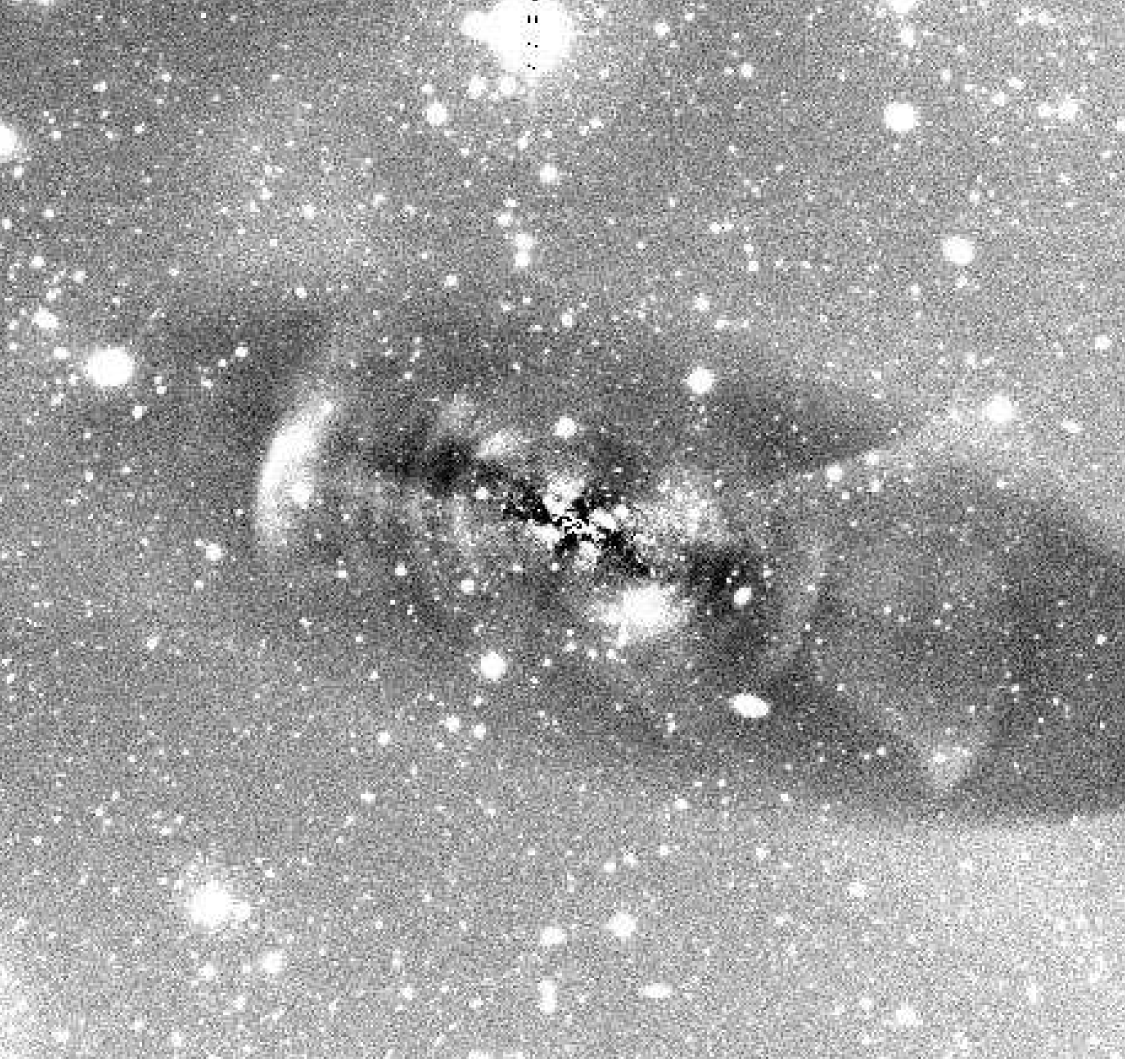}
\par\end{centering}

\caption{\textsf{\small Galaxy-subtracted image of the type}~\textsf{\small I
shell galaxy NGC\,7600 from \citet{1999MNRAS.307..967T}. North is
up and east is to the left. The dark oval shape is an artifact of
the subtraction process. The easternmost shell lies 215\as\ away
from the galaxy center. The field of view is 9\am. Acknowledgment:
The Isaac Newton Group of Telescopes and the Royal Astronomical Society.
\label{obr.7600} }}
\end{figure}

\begin{figure}[H]
\begin{centering}
\includegraphics[width=12cm]{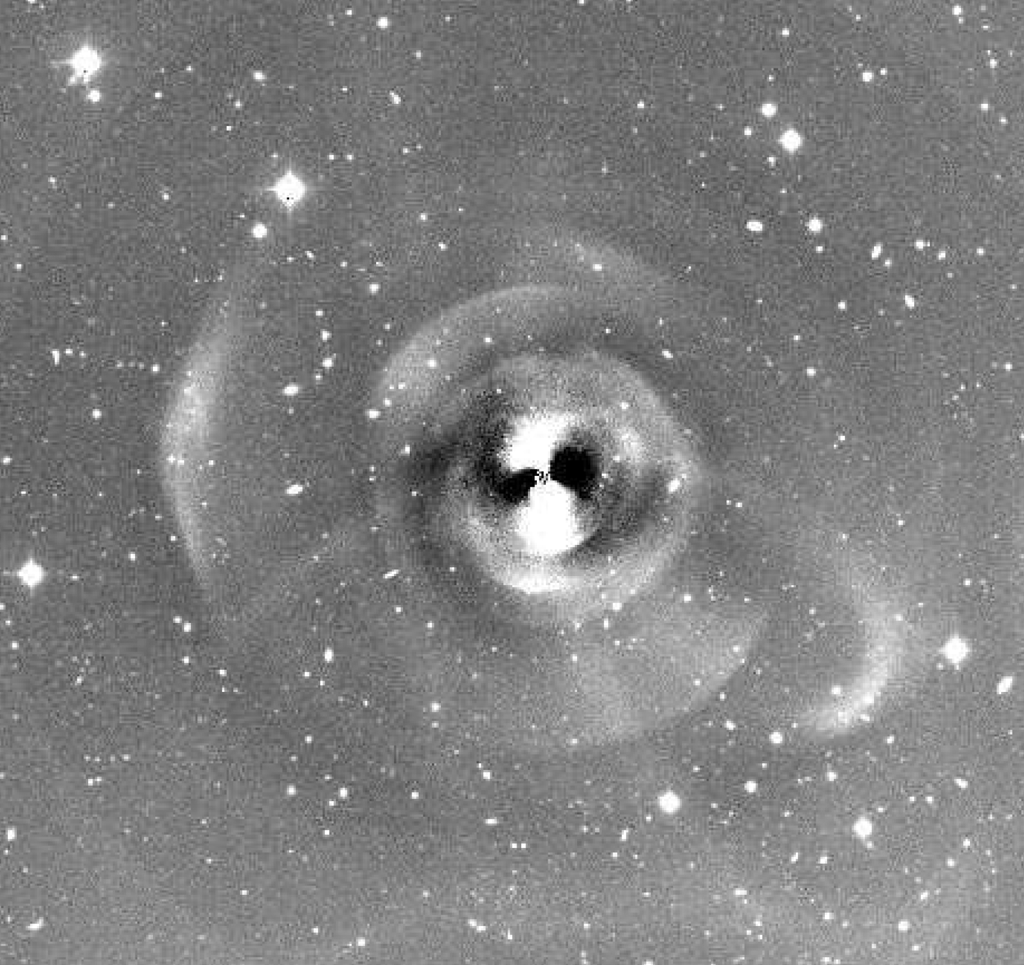}
\par\end{centering}

\caption{\textsf{\small Galaxy-subtracted image of the type}~\textsf{\small II
shell galaxy NGC\,474 from \citet{1999MNRAS.307..967T}. North is
up and east is to the left. The easternmost shell is 202\as\ from
the galaxy center. NGC\,470 is located just off the frame, $\sim$300\as\
west. The field of view is 9\am. Acknowledgment: The Isaac Newton
Group of Telescopes and the Royal Astronomical Society. \label{obr.474} }}
\end{figure}

\begin{itemize}
\item Type~I (Cone) -- shells are interleaved in radius. That is, the next
outermost shell is usually on the opposite side of the nucleus. They
are well-aligned with the major axis of the galaxy. Shell separation
increases with radius. Prominent examples are NGC\,3923 (Fig.~\ref{fig:3923}),
NGC\,5982, NGC\,1344 but also NGC\,7600 in Fig.~\ref{obr.7600}.
\item Type~II (Randomly distributed arcs) -- shell systems that exhibit
arcs which are randomly distributed all around a rather circular galaxy.
A typical example of this kind is NGC\,474 in Fig.~\ref{obr.474}.
\item Type~III (Irregular) -- shell systems that have more complex structure
or have too few shells to be classified.
\end{itemize}
\citet{1990dig..book...72P} has found all three types in approximately
the same fraction.

\citet{1986A&A...166...53D} state that the \textbf{angular distribution}
of the shells is strongly related to the eccentricity of the galaxy.
When the elliptical is nearly E0, the structures are randomly spread
around the galactic center. On the contrary, when the galaxy appears
clearly flattened ($>$E3), the shell system tends to be aligned with
its major axis. In this case, shells are also interleaved on both
sides of the center. Their \textbf{ellipticity} is in general low,
but neatly correlated to the eccentricity of the elliptical. Nearly
E0 galaxies are surrounded by circular shells, while the ellipticity
of the shells is of about 0.15 for E3--E4 galaxies.

When we define the \textbf{radial range} of the shell system as the
ratio between the distance from the galactic center to the outermost
and the innermost shells, then this range of radii, over which shells
are found, is large. The value reaches over 60 for type~I galaxy
NGC\,3923 (the innermost shell is less than 2 kpc from center and
the outermost one $\sim$100\,kpc; \citealp{1988ApJ...326..596P}),
but in most systems, a ratio of 10 or less would be more typical.
The range is lower than 5 for systems where only a few shells are
detected \citep{1986A&A...166...53D}.

In their sample of three shell galaxies, \citet{1986ApJ...306..110F}
found that the characteristic \textbf{thicknesses} of shells are of
the order of 10\% or less of their distance from the center of the
galaxy.

\citet{1987IAUS..127..465W} probed 66 of the 74 galaxies in the range
from 01h 40m to 13h 46m in the \citet{1983ApJ...274..534M} catalogue.
They found that shells commonly occur \textbf{close to the nucleus}.
In roughly 20\% of the systems these innermost shells have \textbf{spiral
morphology}. 

\noindent \vfill{}

\clearpage

\subsection{Colors \label{sub:Barvy}}

At the beginning of the research on shell galaxies, it was widely
believed that shells are rather bluer than the underlying galaxy \citep{1985ARA&A..23..147A}.
But it was rather difficult to obtain relevant data for shells with
only several percent of galaxy's luminosity and the uncertainty was
probably huge.

\citet{1982Natur.295..126C} presented broad-band optical and near-IR
photometry of NGC\,1344. The color indices derived suggest that the
shell comprises a stellar population, perhaps bluer than the main
body of the galaxy. The first CCD photometric observations of shell
galaxies were made in April 1983 at the CFHT (Canada-France-Hawaii
Telescope) by \citet{1986ApJ...306..110F} for their three objects
(NGC\,2865, NGC\,5018, and NGC\,3923). Unlike the shells of NGC\,2865
and NGC\,5018 which were found bluer than the galaxy itself, the
shells of NGC\,923 had similar color indices to those of the galaxy.
The results were obtained from the outer shells of the galaxies.

\citet{1986ApJ...310..597P} got the same result for NGC\,3923 and
in addition for NGC\,3051 as well. On the other hand, \citet{1990AJ....100.1073M}
found both redder and slightly bluer systems of shells among their
three shell galaxies (Arp\,230, NGC\,7010, and Arp\,223 = NGC\,7585).
Multicolor photometry of NGC\,7010 shows a color trend between the
center and the galaxy periphery, red in the center and blue further
out.

Recent observations, using the ever-improving observational capabilities
may turn the old myth of blue shells over. \citet{2007A&A...467.1011S}
wrote: {}``To date, observations give a confusing picture on shell
colors. Examples are found of shells that are redder, similar, or
bluer, than the underlying galaxy. In some cases, different authors
report opposite color differences (shell minus galaxy) for the same
shell. Color even seems to change along some shells; examples are
NGC\,2865 \citep{1986ApJ...306..110F}, NGC\,474 \citep{1990dig..book...72P},
and NGC\,3656 \citep{1997ApJ...486L..87B}. Errors in shell colors
are very sensitive to the correct modeling of the underlying light
distribution. HST images allow for a detailed modeling of the galaxy
light distribution, especially near the centers, and should provide
increased accuracy in the determination of shell colors.'' In their
sample of central parts of six galaxies (NGC\,1344, NGC\,3923, NGC\,5982,
NGC\,474, NGC\,2865, and NGC\,7626) they find only one shell (in
NGC\,474) with blue color. All other shells have similar or redder
colors -- what is just contrary to the results of \citeauthor{1986ApJ...306..110F}
in 80's for NGC\,2865 and \citet{1982Natur.295..126C} for NGC\,1344.
\citeauthor{2007A&A...467.1011S} attribute the red color to dust
which is physically connected to the shell (see Sect.~\ref{sub:Gas-and-dust}).

\citet{1995AJ....109.1576F} measured shell colors of shell galaxy
IC\,1459 and found them to be similar to the underlying galaxy. In
their study of the shell galaxies NGC\,474 and NGC\,7600, \citet{1999MNRAS.307..967T}
found inner shells redder than the outer ones. For the first shells,
colors seem to follow those of the galaxy, for NGC\,7600 three outermost
shells are bluer than the galaxy. In \citet{1999AAS...194.0714L}
it is said that a preliminary reduction of the shell sample shows
that most of the shells have colors that are similar to the elliptical.
The shell colors in the shell galaxy MC\,0422-476%
\footnote{The reference name of object derived from the 1950 coordinates. The
last digit is a decimal fraction of degree, truncated. Notation used
in \citet{1983ApJ...274..534M} catalogue (MC).%
} are scattered around the underlying galaxy value \citep{2000MNRAS.319..977W}.
\citet{2004AN....325..359P} inspected another sample of five galaxies
with shells (NGC\,474, NGC\,6776, NGC\,7010, NGC\,7585, and IC\,1575)
and found the color of the shells being similar to or slightly redder
than that of the host galaxy with the exception of one of the outer
shells in NGC\,474, the only interacting galaxy in the sample.

\subsection{Gas and dust \label{sub:Gas-and-dust}}

\citet{1985ARA&A..23..147A} found that shells are not a good indicator
of the presence of \textbf{dust}. Shell galaxies (64 items) of \citet{1987MNRAS.228..933W}
have rather higher dust contain than normal elliptical. \citet{2007A&A...467.1011S}
detected central dust features out of dynamical equilibrium in all
of their six shell galaxies. Using HST archival data, about half of
all elliptical galaxies exhibit visible dust features (\citealt{2005AJ....129.2138L}:
47\% of 177 in field galaxies). On the other hand, \citet{2001AJ....121..808C}
found evidence for dust features in approximately 75\% of both the
isolated and group galaxies (17 of 22 and 9 of 12, respectively).
But in their sample also all of the galaxies that display shell/tidal
features contain dust. Also \citet{rampazzo07} found all of their
three shell galaxies to show evidence of dust features in their center.

Moreover, \citet{2007A&A...467.1011S} discovered that the shells
contain more dust per unit stellar mass than the main body of the
galaxy. This could explain redder color of shells which is observed
in many cases (Sect.~\ref{sub:Barvy}). Observational evidence for
significant amounts of dust residing in a shell was also found in
NGC\,5128 \citep{2004A&A...415...95S}. 

In general, both the ionized and neutral gas contents of shell galaxies
are thus comparable to those of normal early-type galaxies \citep{1986A&A...166...53D}
or rather higher \citep{1987MNRAS.228..933W}. However, arcs of \textbf{H}\,\textbf{I}
have been discovered \citep{1994ApJ...423L.101S,1995ApJ...444L..77S}
lying parallel to but outside of the outer stellar arcs in a few shell
systems (Cen\,A = NGC\,5128 and NGC\,2865). In Centaurus\,A, gas
has the same arc-like curvature but is displaced 1\am\ ({\small $\sim1$}\,{\small kpc})
to the outside of the stellar shells. A similar discovery has been
made by \citet{2001AJ....122.1758B} in NGC\,3656. The shell, at
9\,kpc from the center, has traces of H\,I with velocities bracketing
the stellar velocities, providing evidence for a dynamical association
of H\,I and stars at the shell. \citet{1997AAS...191.8212P} found
an off-centered H\,I ring in NGC\,1210. A short report about H\,I
in shell galaxies has been done by \citep{1997ASPC..116..362S}.

\citet{2000A&A...356L...1C} reveal the presence of dense \textbf{molecular
gas} in the shells of NGC\,5128 (Cen\,A). Cen\,A, the closest active
galaxy, is a giant elliptical with jets and strong radio lobes on
both sides of a prominent dust lane which is aligned with the minor
axis of the galaxy \citep{1990AJ.....99.1781V,1992ApJ...395..444C,1984ApJ...276..491H}.
A significant amount of gas and dust is situated predominantly in
an equatorial disk where vigorous star formation is occurring \citep{1979AJ.....84..284D}.
\citeauthor{2000A&A...356L...1C} detected CO emission from two of
the fully mapped optical shells with associated H\,I emission, indicating
the presence of $4.3\times10^{7}$\,M\suns of H$_{\text{2}}$, assuming
the standard CO to H$_{\text{2}}$ conversion ratio.

About $5\times10^{8}$\,M\suns of molecular gas is located in the
inner 2\am\ ({\small $\sim13$}\,{\small kpc}) of the NGC\,1316
(Fornax\,A) and is mainly associated with the dust patches along
the minor axis \citep{2001A&A...376..837H}. In addition, the four
H\,I detections in the outer regions are all far outside the main
body of NGC\,1316 and lie at or close to the edge of the faint optical
shells and X-ray emission of NGC\,1316. The location and velocity
structure of the H\,I are reminiscent of other shell galaxies such
as Cen\,A.

Around $8\times10^{7}$\,M\suns of neutral hydrogen, and some $10^{9}$\,M\suns
of molecular hydrogen have been previously found in NGC\,3656 by
\citet{1996AJ....111.1053B}. Roughly 10\% of the total gas content,
one third of the neutral hydrogen, lies in an extension to the south,
what is also similar to Cen\,A. NGC\,3656 also contains a prominent
central dust line \citep{2007iuse.book..383L}.

These galaxies seem to form up an interesting category of shell galaxies
-- aside from the shells, they also contain a prominent central dust
line, good amount of gas (usually both H\,I and CO detected), and
are usually strong radio sources with jets and active nucleus. Galaxies
with these features are suspected of cannibalization of a gas-rich
companion. Some examples of this group are NGC\,5128 (Centaurus\,A),
NGC\,1316 (Fornax\,A), NGC\,3656, NGC\,1275 (Perseus\,A; massive
network of dust, active nucleus; \citealp{1998AJ....115.1778C}),
IC\,1575 (active nucleus in the center drives the jet orthogonally
to the strong central dust lane, producing the two radio lobes; \citealp{2004AN....325..359P}),
and possibly IC\,51 \citep{2013AJ....145...34S}, NGC\,5018 \citep{rampazzo07},
and NGC\,7070A \citep{2003MNRAS.343..819R}.

\citet{1999A&A...343...23P} found that the softer \textbf{X-ray}
component which likely comes from hot gas, is not as large as expected
for a global inflow, in a galaxy of an optical luminosity as high
as that of NGC\,3923. \citet{2000AJ....120.1946S} find that early-type
galaxies with fine structure (e.g. shells) are exclusively X-ray underluminous
and, therefore, deficient in hot gas. 

\citet{2003MNRAS.343..819R} analyzed the \textbf{warm gas} kinematics
in five shell galaxies. They found that stars and gas appear to be
decoupled in most cases. \citet{rampazzo07} , \citet{marino09},
and \citet{trinchieri08} investigated star formation histories and
hot gas content using the NUV and FUV Galaxy Evolution Explorer (GALEX)
observations (and in the latter case also X-ray ones) in a few shell
galaxies.

\subsection{Radio and infrared emission}

\citet{1987MNRAS.224..895W} surveyed a subset of 64 galaxies of the
Malin \& Carter catalogue at 20 and 6\,cm with the VLA. Apart from
Fornax\,A, only two galaxies of their set contained obvious extended
\textbf{radio sources}. 42\% of the galaxies were detected, down to
a 6-cm flux density limit of about 0.6\,mJy. This detection rate
does not differ significantly from normal early-type galaxies. In
a complete sample of 46 southern 2\,Jy radio galaxies at intermediate
redshifts ($0.05<z<0.7$) of \citet{2011MNRAS.410.1550R}, 35\% of
galaxies have shells.

A more interesting discovery was made by \citet{1987MNRAS.228..933W}.
Eight of the previous sample of 64 shell galaxies plus two from \citet{1984AJ.....89...53S}
sample of E and S0 galaxies were \textbf{detected by IRAS}. And here
comes the discovery: All of these galaxies are also radio sources
with 6-cm flux densities $\geq0.6$\,mJy. They noted that according
to the binomial distribution, the probability of finding all 10 galaxies
at both wavelengths by chance would be 0.1\%. From non-shell galaxies
which are detected in the IRAS survey, only 58\% are radio sources.
So, there is a strong \textbf{radio-infrared correlation} for shell
galaxies. In the tree-dimensional radio-infrared-shell space, no significant
correlation is seen in any two dimensions, but a correlation is apparently
found if all three are taken together.

\citet{1989AJ.....97..363T} investigated infrared color-color diagram
of early-type galaxies. On average, shell galaxies appear to have
broadband \textbf{mid- and far-infrared} energy distributions very
similar to those of normal S0 galaxies, although many of them were
classified as ellipticals.

\subsection{Other features of host galaxies\label{sub:Other-features}}

From their sample of 100 shell galaxies, \citet{1988MNRAS.235..813C}
derived that about 15--20\% of shell galaxies have \textbf{nuclear
post-starburst spectra}. \citet{2011MNRAS.410.1550R} found shells
in 15 out of 33 (45\%) of the non-starburst systems, but in only 1
out of 13 (8\%) of the starburst systems. All their objects are powerful
radio galaxies (PRGs) and quasars.

\citet{2000A&A...353..917L} have studied star formation history in
a sample of 21 shell galaxies and 30 early-type galaxies that are
members of pairs, located in very low density environments. The last
star formation event (which involved different percentages of mass)
that happened in the nuclear region of shell galaxies is statistically
old (age of the burst from 0.1 to several Gyr) with respect to the
corresponding one in the sub-sample of the interacting galaxies (age
of the burst $<0.1$\,Gyr or ongoing). This distinction has been
possible only using diagrams involving newly calibrated {}``blue''
indices. Assuming that stellar activity is somehow related to the
shell formation, shells have to be long lasting structures.

There is an obvious strong association between \textbf{kinematically
distinct/decoupled cores} ({}``KDC'' or {}``KDCs'') and shell
galaxies. First example of an elliptical galaxy with a \textbf{KDC}
was NGC\,5813 \citep{1982MNRAS.201..975E}. These galaxies are characterized
by a rotation curve that shows a decoupling in rotation between the
outer and inner parts of the galaxy. In some spectacular cases, the
core can be spinning rapidly in the opposite direction to the outer
part of the galaxy (e.g. IC\,1459). It was found by \citeauthor{1992PhDT.......111F}
(\citeyear{1992PhDT.......111F}; cited in \citealp{1999MNRAS.306..437H})
that all of the nine well-established KDCs and a further four out
of the six {}``possible KDCs'' possess shells.

Some galaxies are known to contain multiple nuclei (e.g. NGC\,4936,
NGC\,7135, MC\,0632-629, MC\,0632-629). \citet{1994AJ....107.1713F}
conducted the first systematic search for \textbf{secondary nuclei}
in a sample of 29 known shell galaxies. They find six (20\%) galaxies
with a possible secondary nucleus, what they concluded to be a probable
upper limit to the true fraction of secondary nuclei. In the sample
of radio galaxies of \citet{2011MNRAS.410.1550R}, five galaxies have
more than one nucleus while also having shells detected. That makes
20\% of their shell galaxies containing the secondary nucleus. Thereof
one double nucleus is uncertain (PKS\,1559+02) and one galaxy has
triple nucleus indicated (PKS\,0117-15). On the other hand, \citet{1999A&A...345..419L}
in their sample of 21 shell galaxies found only one (ESO\,240-100)
to be characterized by the presence of a double nucleus.

According to \citet{1987IAUS..127..465W}, shell galaxies have an
enormous \textbf{diversity} of central surface brightness. In addition,
\citet{1987MNRAS.224..895W} found a wide variety of optical appearances,
suggesting that shell galaxies are not a homogeneous class with uniform
physical characteristics. 

\clearpage

\section{Summary of shell characteristics \label{sec:characterictics}}
\begin{enumerate}
\item Shells are observed in at least 10\% of early-type galaxies (E and
S0) and $\sim$1\% of spirals.\label{enu:cetnost}
\item Shell galaxies occur markedly most often in regions of low galaxy
density.\label{enu:low-density}
\item The number of shells in a galaxy ranges from 1 to $\sim$30.\label{enu:number}
\item The shells contain at most a few per cent of the overall brightness
of the galaxy.\label{enu:brightness}
\item Surface brightness contrast of the shells is very low, about 0.1--0.2\,mag.
\item Shells are of stellar nature.\label{enu:stellar}
\item For type~I shell galaxies (see in Sect.~\ref{sub:Vzhled-slupek}),
shells are interleaved in radius and their separation increases with
radius.\label{enu:interleaved}
\item Shells appear to be aligned with the galaxy's major axis and slightly
elliptical for flattened galaxies, and randomly spread around the
galactic center for nearly E0 galaxies.\label{enu:elipticity}
\item The radial range of shells (the ratio of the radii of the outermost
and the innermost shells) is typically less then 10 but can reach
over 60.\label{enu:Range}
\item Shells commonly occur close to the nucleus.\label{enu:close-to-nucleus}
\item In roughly 20\% of the systems, the innermost shells have spiral morphology.\label{enu:20=000025spiral}
\item Shells can have any color, perhaps they are rather similar to or slightly
redder than the host galaxy.\label{enu:redder}
\item The colors of shells are different even in the same galaxy, tend to
be red in the center and bluer further out.\label{enu:The-colours-grad}
\item It seems that galaxies with shells also contain central dust features.
\label{enu:central-dust}
\item An increased amount of dust has been observed in shells.\label{enu:dust-amount} 
\item Slightly displaced arcs of H\,I, with respect to the stellar shells,
have been discovered in some galaxies. \label{enu:Slightly-displaced-HI}
\item Molecular gas associated with shells was detected in several galaxies.
\label{enu:Molecular-gas-associated}
\item The detection rate of radio emission of shell galaxies is similar
to other early-type galaxies.
\item There is probably a strong radio-infrared correlation for galaxies
which possess shells.
\item 15--20\% of shell galaxies have nuclear post-starburst spectra. \label{enu:post-starburst}
\item There is a strong association between kinematically distinct/decoupled
cores and shells in galaxies.\label{enu:KDC}
\item The shell galaxies have an enormous diversity of central surface brightness
and a wide variety of optical appearances.\label{enu:central-brightness}
\end{enumerate}
\clearpage

\section{Scenarios of shells' origin \label{sec:Scenarios}}

In the eighties and nineties several theories of formation of shell
galaxies were proposed. They can be divided into three categories:
\begin{itemize}
\item \textbf{Gas dynamical theories} (Sect.~\ref{sub:GDM}) -- The first
truly developed theories connect star formation and the formation
of shells. These theories, however, seem to be contradicted by observation
and now they are not usually taken into consideration. 
\item \textbf{Weak Interaction Model} (WIM, Sect.~\ref{sub:WIM}) -- According
to this model, shells are density waves induced in a thick disk population
of dynamically cold stars by a weak interaction with another galaxy.
WIM has nice explanations for many phenomena related to the shells
but suffers from some deficiencies and obscurities.
\item \textbf{Merger model} -- The most widely accepted theory is based
on the idea that the stars in shells come from a cannibalized galaxy.
The entire Sect.~\ref{sec:MM} is devoted to this model.
\end{itemize}
For a more detailed review, see \citet{EbrovaMAT}.

\subsection{Gas dynamical theories \label{sub:GDM}}

The first theory of shell formation has been proposed by \citet{1980Natur.287..613F},
who suggested that shells are regions of recent star formation in
a shocked galactic wind. Gas produced by the evolution of stars in
an elliptical galaxy and driven out of the galaxy in a wind powered
by supernovae would be heated and compressed as it passes through
a shock. As the gas cools, star formation can occur. This scenario
was expanded by \citet{1985ApJS...58...39B} and \citet{1985ApJ...291...80W}.
In the \citet{1985ApJ...291...80W} model, shells are initiated in
a blast wave expelled during an active nucleus phase early in the
history of the galaxy, sweeping the interstellar medium in a gas shell,
in which successive bursts of star formation occur, leading to the
formation of several stellar shells.

This scenario was inspired by the supposedly bluer color of the shells,
but as time and the measurements have shown, shells are composed mostly
of old populations of stars (see Sect.~\ref{sub:Barvy}). As \citeauthor{1985ApJ...291...80W}
mention, star formation is a subject only to local conditions and
is a stochastic process. This is in conflict with the observed interleaving
of shells in many shell galaxies. Further, there is the failure to
detect either ionized or neutral gas associated with the shells except
in a very few cases. \citet{1986A&A...166...53D} argued that the
mechanism of star formation in such a galactic wind is not known;
the galaxy should have possessed a very large amount of interstellar
matter in order to produce stellar mass of a typical shell system;
and the supernovae explosions might rapidly dispel the wind which
would exclude that as much as 20--25 shells form around some shell
galaxies.

\citet{1987MNRAS.229..129L} reconciled previous models with the last
observations at that time. Only a modest outburst is demanded by the
authors to cause a period of star-formation in an outward-moving disturbance
from the galactic core. The newly-formed stars occupy a small volume
in the orbital phase-space of the underlying galaxy. The shells were
produced in the same phase-wrapping mechanism as in the merger model
(Sect.~\ref{sub:phase-wrapping}) producing an interleaved shell
system (point~\ref{enu:interleaved} in Sect.~\ref{sec:characterictics}).
The model does not exclude the merger hypothesis, since a merger can
lead to a burst of star formation in the galactic core that is the
precursor of the initial blast wave. The inner shells are older than
the outer ones in this scenario. This could lead to the color gradient
which seems to be observed in some cases (point~\ref{enu:The-colours-grad}
in Sect.~\ref{sec:characterictics}) and which was not known at the
time.

All these arguments are sound, but other observed aspects of shell
galaxies seem to exclude the model of \citeauthor{1987MNRAS.229..129L}
anyway. Aside from the already mentioned points, \citet{2001AJ....121..808C}
discovered a consistency of the colors of the isolated galaxies with
and without shells and it argues against the picture in which shells
are caused by asymmetric star formation. Again the failure to detect
gas in shells argues against this scenario. Finally, the lack of signs
of recent star formation in the shells is the most fatal reality for
the model discussed here.

A rather different scenario was proposed by \citet{1987ApJ...319..601U},
and was quickly forgotten for its clumsiness and only a little agreement
with observations. They tentatively considered a hot supernova-driven
galactic wind as a process which produces both extended multiple stellar
shells and hot X-ray coronae which have been detected around a number
of early-type galaxies. Few of them also have shells (NGC\,1316,
NGC\,1395, NGC\,3923, and NGC\,5128). This scenario suffers from
much the same diseases as the former ones. Moreover, it gives no explanation
for the increasing separation of shells with radius, since the distribution
of shells is variable with the lapse of time in this scenario. As
previously mentioned, early-type galaxies with fine structure are
X-ray underluminous, thus deficient in hot gas (Sect.~\ref{sub:Gas-and-dust}).
However, this theory seems to be primarily out of game because of
the observed systematic interleaving of shells.

All the models mentioned above more or less fell in condemnation and
oblivion before they even started to try explaining more detailed
characteristics observed in shell galaxies.

\subsection{Weak Interaction Model (WIM) \label{sub:WIM} \foreignlanguage{british}{\label{sub:Observational-WIM}}}

\citet{1990MNRAS.247..122T} came up with an elegant and revolutionary
model of shell formation in elliptical/lenticular galaxies which is
still in the game today. According to them, shells are density waves
induced in a thick disk population of dynamically cold stars by a
weak interaction with another galaxy -- whence the name, the Weak
Interaction Model (WIM). A year later, this hypothesis was further
developed and supported by new simulations of \citet{1991MNRAS.253..256T}.

To support their theory, the authors state that \citet{1989AJ.....97..363T}
pointed out that most of the elliptical galaxies with shells catalogued
by \citet{1983ApJ...274..534M} are classified elsewhere as S0s. As
such, a significant population of dynamically cold stars moving on
nearly circular orbits could be present in these systems. They also
note that faint thick disks could be present in many elliptical galaxies
without detection. The authors noted that a thick-disk population
which makes up only a few per cent of the total mass of a galaxy is
required to explain the faint features seen in most shell galaxies.
But the disk must by heavy enough to produce shells which form a few
per cent of the overall brightness of the galaxy (point~\ref{enu:brightness}
in Sect.~\ref{sec:characterictics}). \citet{2000MNRAS.319..977W}
looked for such a disk in the shell galaxy MC\,0422-476 and found
no sign of an exponential disk, or any thick disk additional to the
short-axis tube orbits already expected within an oblate ellipsoidal
potential.

The WIM has always been simulated with the parabolic encounter of
the secondary galaxy, since more circular orbits would decay rapidly
during a close encounter, resulting in a merger scenario, while more
hyperbolic orbits would result in encounters too quick to be effective.
This fact can also account for the less frequent occurrence of shell
galaxies in clusters than in the field (point~\ref{enu:low-density}
in Sect.~\ref{sec:characterictics}).

Required mass of the secondary is about 0.05--0.2 of the primary mass
and orbital inclination 45\textdegree{} or less with respect to the
thick disk. The total time of the shell structure's visibility is
typically around 10\,Gyr in \citet{1990MNRAS.247..122T}. But in
the simulations of \citet{1991MNRAS.253..256T}, the shells are visible
for only about 3\,Gyr.

Possibly, the age of the shell system can be deduced from its appearance
and thus the presence of a suitable secondary galaxy at an appropriate
distance could be checked. But e.g. around NGC\,3610 no surrounding
galaxies were found \citep{1998AJ....116.2793S}.

In WIM, the host galaxy is an oblate%
\footnote{An \emph{oblate} ellipsoid is rotationally symmetric around its shortest
axis, whereas for a \emph{prolate} ellipsoid the axis of symmetry
is the longest one. A \emph{triaxial} ellipsoid has no rotational
symmetry at all.%
} spheroid, and shells are readily formed as spiral density waves in
the thick disk which is symmetric about the plane of symmetry of the
galaxy. The model also gives the correct relative frequency of two
types of shell galaxies (i.e. 1:1, Sect.~\ref{sub:Vzhled-slupek}),
since the systems appear as type\,II shell galaxies when viewed at
inclination angles less than approximately 60\textdegree{} (0\textdegree{}
is face-on). At inclination angles larger than 60\textdegree{}, the
systems appear as type\,I. As we change the viewing angle, the observed
ellipticity changes from E0 (for 0\textdegree{}) to E4 (90\textdegree{}),
where E4 may be the true ellipticity of the galaxy, since \citet{1990dig..book...72P},
cited in \citet{1991MNRAS.253..256T}, found a strong peak at this
value in the type\,I ellipticity histogram. However, implications
of this would be somewhat strange -- either all elliptical galaxies
are E4 type oblate spheroids seen from different angles, or shells
do occur only in E4 galaxies, what would be probably in contradiction
to their relatively frequent occurrence.

\citet{1988ApJ...326..596P} pointed out that the shells in NGC\,3923
are much rounder than the underlying galaxy and have an ellipticity
which is similar to the inferred equipotential surfaces. This idea
was originally put forward by \citet{1986A&A...166...53D} who found
such a relationship for their merger simulations (Sect.~\ref{sec:MM}).
The same effect can be seen in the simulations presented by \citet{1991MNRAS.253..256T}.

Another advantage of the WIM lies in its ability to explain the occurrence
of the shells over a broad range of radii (point~\ref{enu:Range}
in Sect.~\ref{sec:characterictics}) and close to the nucleus (point~\ref{enu:close-to-nucleus}),
since shells are formed in the thick disk that is required to be already
present in the galaxy.

In his study of the shell galaxy NGC\,3923, \citet{1988ApJ...326..596P}
discussed varying distribution of the shells -- interleaved in outer
region and roughly symmetric in inner parts. According to this model,
in the outer region of the galaxy, the simulations show a predominantly
one-armed trailing spiral density wave which, when viewed edge-on,
gives rise to the interleaving of the outer shells, naturally aligned
with the major axis. Inside the perigalactic radius of the path of
the intruder, the tidal forces produced during the encounter induce
a bi-symmetric kinematic density wave in the thick disk. \citeauthor{1991MNRAS.253..256T}
has achieved an almost breathtaking agreement with the observation
of radial shell distribution, except for the innermost shells that
have not appeared at all in his simulations. But he believes it could
be remedied by shrinking the core radius of primary galaxy. 

The WIM for shells does not predict the existence of a kinematically
distinct nucleus (KDC, point~\ref{enu:KDC} in Sect.~\foreignlanguage{british}{\ref{sec:characterictics}}).
\citet{1994MNRAS.270L..23H} proposed a mechanism whereby a counter-rotating
core could be formed by the retrograde passage of a massive galaxy
past a slowly rotating elliptical with a pre-existing rapidly rotating
central disk. In their study of the shell galaxy NGC 2865, \citet{1999MNRAS.306..437H}
state that the requirement of the WIM for the nuclear disk to be primordial
is in conflict with the observed absorption line indices. It is also
unlikely that a passing galaxy can transfer a large amount of orbital
angular momentum over a period longer than 0.5\,Gyr without being
captured or substantially disrupted, as NGC\,2865 has an extended
massive dark halo \citep{1995ApJ...444L..77S}. Thus a purely interaction
induced origin for the shells and KDC in NGC\,2865 is ruled out.

The observation by \citet{1986ApJ...310..597P} shows that the surface
brightness of shells in NGC\,3923 is a {}``surprisingly constant''
fraction ($\sim$3--5\%) of the surface brightness of the underlying
galaxy. The WIM produces shells with the correct surface brightness,
since they are formed in a thick disk which has the same surface brightness
profile as the underlying galaxy. However, further observations \citep{1988ApJ...326..596P,2007A&A...467.1011S}
revealed more shells in NGC\,3923 that defy this rule. And there
are more disobedient shell galaxies: NGC\,474 and NGC\,7600 \citep{1999MNRAS.307..967T}
and MC\,0422-476 \citep{2000MNRAS.319..977W}. Similarly for NGC\,2865,
the WIM origin is in conflict with the existence of bright outer shells,
their blue colors, and their chaotic distribution \citep{1986ApJ...306..110F}.

Furthermore, \citet{1998MNRAS.294..182C} revealed a minor axis rotation
of the famous NGC\,3923 what suggests a prolate or triaxial potential,
and challenges the requirement of an oblate potential by the WIM.
They noted that it is difficult to induce minor axis rotation in an
oblate potential without inducing any corresponding major axis rotation
that has not been observed.

\citet{1998AJ....116.2793S} note that the spectacular morphological
fine structure of the shell galaxy NGC\,3610 leads to the natural
conclusion that this galaxy has undergone a recent merger event. This
scenario is supported by the existence of a centrally concentrated
intermediate-age stellar population which is a prediction of the dissipative
gas infall models. Furthermore, the central stellar structure could
have been formed by this infalling gas. It seems unlikely that the
structures were formed by a non-merging tidal interaction since there
is no nearby galaxy. 

It is interesting that nobody has ever noticed any general one-armed
spiral in the outer shells of type\,II shell galaxies nor any bi-symmetric
spiral in inner regions. Only \citet{1987IAUS..127..465W} probed
66 shell galaxies and found that in roughly 20\% of the systems these
innermost shells have spiral morphology. But they did not specify
which galaxies they were nor what spiral morphology has been found.
\citet{1991MNRAS.253..256T} explains: {}``The broken appearance
of the shells is actually an interference pattern formed by the leading
and trailing density waves induced during the encounter'', and he
adds that the faint residual one-armed leading spiral feature seen
at the end of some of the simulations is probably an $m$ = 1 kinematic
density wave%
\footnote{Here, a common method of decomposition of a 2D density or potential
to Fourier modes in the azimuthal direction (that is, Fourier transforming
in the angle separately for every radius) is used. The potential is
decomposed as

\[
\phi(R,\theta)=\phi_{0}(R)+\sum_{m=1}^{\infty}\phi_{m}(R)\cos[m(\theta-\theta_{m}(R))],
\]
what means a sum of harmonics with different amplitudes and phase
shifts for every R. The $\phi_{0}$ (\emph{m} = 0) mode is the axisymmetric
part of the potential, the \emph{m} = 1 mode has an azimuthal period
of 360\textdegree{}, the \emph{m} = 2 mode has 180\textdegree{} and
so on. It is most frequently used for spiral galaxies. The \emph{m}
= 1 mode corresponds to one spiral arm ($\theta_{1}$ is dependent
on \emph{R}) or a closed structure (an ellipse when $\theta_{1}$is
a constant) not concentric with the galaxy. The \emph{m} = 2 mode
is the most common, being either a bar (constant $\theta_{2}$) or
two spiral arms. In the WIM case, the \emph{m} = 2 mode (bi-symmetric
spiral density wave) is important for the inner parts of the disk.%
}. The relative importance of this mode for the shell forming process
is not fully understood, but it does play an important role in determining
the shell morphology produced by the more massive encounters.

\citet{2000MNRAS.319..977W} found many arguments for and against
the WIM in their study of the shell galaxy MC\,0422-476.

\citet{1999A&A...345..419L} favor the WIM, since they derived that
in shell galaxies, the age of the last star forming event ranges from
0.1 to several Gyr. If the last burst of stellar activity that affects
the absorption line strength indices, correlates with the dynamical
mechanism forming the shell features, these shells are long lasting
phenomena. The WIM predicts such a long life for the shells, whereas
for the merger model of \citet{1984ApJ...279..596Q}, Sect.~\ref{sec:MM},
guessed a shorter lifetime due to the initial dispersion of velocities
that the stars of the shell inherited. But for example, in the framework
of the merger model, \citet{1986A&A...166...53D} happily simulated
shell systems for 10\,Gyr.

A consequence of the WIM is that the stars which make up the shells
must be in nearly circular orbits. That is almost opposite to the
conclusions of the merger model (Sect.~\ref{sec:MM}). It could be
thus decided from measurements of the shell velocity fields which
model is favored, but this is indeed a formidable task, as the shells
contain at most a few per cent of the overall brightness of the host
galaxy. Some attempts have been already carried out \citep{1996AJ....111.1053B},
but as far as we know, the results are inconclusive.

To conclude, the WIM has nice explanations for many phenomena related
to the shells (inner shells, shell distribution, symmetry of inner
shells, etc.), for which the competing merger model (Sect.~\ref{sec:MM})
seeks explanations with difficulties or has none at all. On the other
side, the WIM suffers from some deficiencies and obscurities (thick
disk, KDC, shells brightness, etc.). Generally, it seems to lack observational
confirmation of phenomena specific to the model. 

\begin{figure}[!b]
\begin{centering}
\includegraphics[width=15cm]{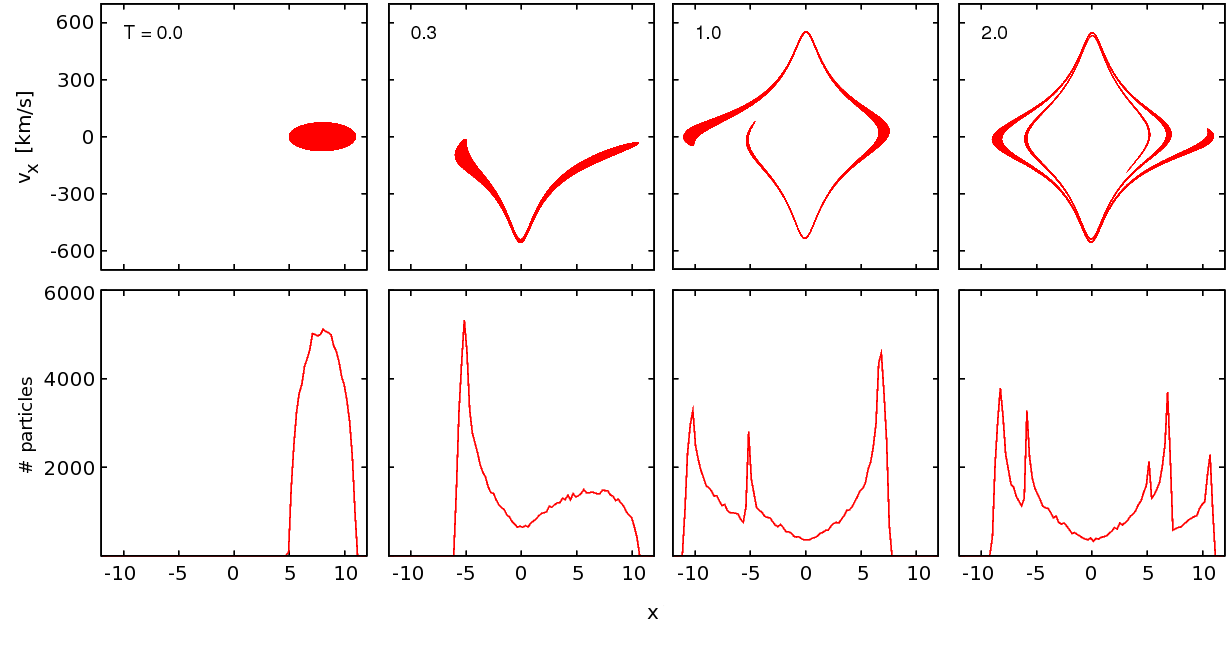}
\par\end{centering}

\caption{\textsf{\small Time evolution of a cloud of test particles falling
into a one dimensional Plummer potential $v-x$ space (upper row),
particle radial density (lower row). The $x$ axis is centered with
the center of the potential and scaled so that 1 on the axis is the
Plummer radius. \label{obr.phase-wrap} }}
\end{figure}

\section{Merger model \label{sec:MM}}

In this section we introduce the merger origin scenario of the shell
galaxies that we consider for the rest of the thesis. For a more detailed
(but slightly outdated) review, see \citet{EbrovaMAT}.

\subsection{Phase wrapping \label{sub:phase-wrapping}}

The idea of a connection between mergers and shells was first published
by \citet{1980ApJ...237..303S} in his study of the shell galaxy NGC\,1316
(Fornax\,A). The presence of shells (or {}``ripples'' as Schweizer
calls them) deep within NGC\,1316 and a surprising number of galaxies
with ripples but no companions fosters his belief that Fornax A, too,
has been shaken by a recent intruder rather than by any of the present
neighbors. \citeauthor{1980ApJ...237..303S} imagined that the ripples
represent a milder version of the strong response that occurs in the
disk of a galaxy when an intruder of comparable mass free-falls through
the center: A circular density wave runs outward, followed sometimes
by minor waves, and give the galaxy the appearance of a ring \citep{1976ApJ...209..382L,1978IAUS...79..109T}.

\citet{1983IAUS..100..347Q,1984ApJ...279..596Q} took up the idea
of a merger origin of shells, but showed it in a slightly different
spirit. When a small galaxy (secondary) enters the scope of influence
of a big elliptical galaxy (primary) on a radial or close to a radial
trajectory, it splits up and its stars begin to oscillate in the potential
of the big galaxy which itself remains unaffected. In their turning
points, the stars have the slowest speed and thus tend to spend most
of the time there, they pile up and produce arc-like structures in
the luminosity profile of the host galaxy. Quinn modeled the formation
of shell galaxies using test-particle and restricted $N$-body codes,
much as many other did later \citep[e.g, ][]{1987ApJ...312....1H,1988ApJ...331..682H,1989ApJ...342....1H,1986A&A...166...53D}
and as we will do in this work as well. It should be also noted that
already \citet{1967MNRAS.136..101L} described something like a pig-trough
dynamics in violent relaxation in stellar systems. 

\begin{figure}[h]
\begin{centering}
\includegraphics[width=15cm]{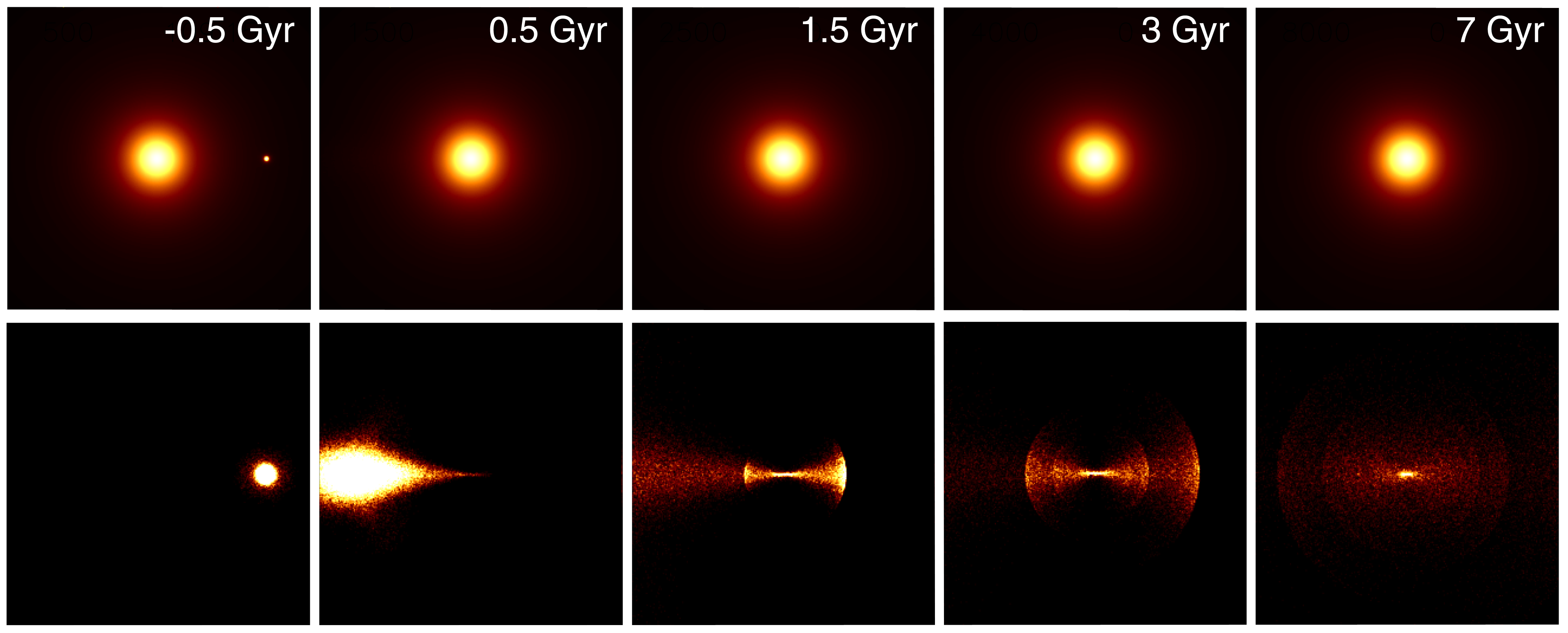}
\par\end{centering}

\caption{\textsf{\small Surface brightness density from the simulation of a
radial minor merger. Top row: both primary and secondary galaxy are
displayed. Bottom row: only the surface density of particles originally
belonging to the secondary is displayed. Panels show an area of $300\times300$\,kpc.
Time-stamps mark the time since the release of the star in the center
of the host galaxy. For parameters of the simulation, see Appendix~\ref{Apx:Videos}
point~\ref{enu:video1}. \label{obr.phase-wrap-3D} }}
\end{figure}

The mechanism is illustrated on the one dimensional example in Fig.~\ref{obr.phase-wrap}.
The density maxima occur near the turnaround points of the particle
orbits. The maximal radial position of the orbit is first reached
by the most tightly bound particles, but as more distant particles
stop and turn around, the density wave propagates slowly in radius
to the outermost turning point set by the least bound particle. The
particles in phase space form a characteristic structure, for which
this mechanism of shell formation is often called {}``phase wrapping''.

In an idealized case, the edges in density are the caustics of the
mapping of the phase density of particles into physical space \citep{1989ApJ...346..690N}.
As a natural consequence, the shells are interleaved in radius and
their separation increases with radius (point~\ref{enu:interleaved}
in Sect.~\ref{sec:characterictics}). Furthermore, the range of the
number of shells present around ellipticals is a simple consequence
of the age of the event. More shells will imply that a longer time
has passed since the merger event. A more detailed explanation and
some equations can be found in Sect.~\ref{sub:TP}. The best insight
on the shell formation is provided by video 1-shells.avi, which is
a part of the electronic attachment. Five snapshots related to the
video can be seen in Fig.~\ref{obr.phase-wrap-3D}. For the description,
see Appendix~\ref{Apx:Videos} point~\ref{enu:video1}.

\subsection{Cannibalized galaxy \foreignlanguage{british}{\label{sub:Companion}}}

The choice of the type of the secondary galaxy initially felt on a
disk galaxy. The authors were probably led to it by two aspects. Firstly,
dynamically cold systems promised to be better in shell formation,
since they occupy a smaller phase volume than velocity dispersion
supported galaxies of comparable masses. In such a process of non-colliding
stars we can assume phase volume conservation according to the Liouville's
theorem. This means that a system with an initially small phase volume
keeps this property and forms sharper shells. So, the visibility of
the shell system is expected to be lower for an elliptical companion
than for a spiral companion of the same mass, since the velocity dispersion
is greater for the elliptical. Secondly, the observations seemed to
suggest that the stars in shells have the color indices of late-type
galaxies (see Sect.~\ref{sub:Barvy}). Later observations have shown
that the shells are not that blue (see also Sect.~\ref{sub:Barvy}),
but even before that the simulations showed that the shell systems
can be formed by a disk as well as an elliptical companion \citep{1986A&A...166...53D,1988ApJ...331..682H}. 

\citet{1988ApJ...331..682H} examined among others the influence of
the phase volume and velocity dispersion of a spherical companion
on shell formation. As was already mentioned above, higher dispersion
means higher blur of resulting shells through the increase of the
phase volume (velocity dispersion is proportional to the square root
of mass of the accreted companion). Another effect brought in by higher
dispersion is that the material can be captured into more tightly
bound orbits, so shells are produced more rapidly, since the shell
production rate is indirectly proportional to the shortest period
of stellar oscillations. This means that for the same potential of
the primary galaxy, we can easily get different shell systems by changing
some parameters of the accreted galaxy, what constituted one of several
serious problems of the idea to explore the potential of the host
galaxy through its shell system.

The disk-like secondary galaxy has some extra options that the spherical
one lacks. By accreting differently inclined disks we can get different
peculiar structures. The resulting configuration of sharp-edged features
is considerably more complex and disordered than for a spherical companion.
For a very flat system, there is also the possibility of forming caustics
through spatial wrapping. That is to say, as the sheet of particles
moves and folds in three-dimensional space, sharp edges can be formed
in its two-dimensional projection onto the plane of the sky. Projection
effects become critical in this context, as evidenced by the different
viewing angles, see \citet{1988ApJ...331..682H}. This effect was
evident already in the simulations by \citet{1984ApJ...279..596Q}.

\subsection{Ellipticity of the host galaxy\label{sub:hostPot}}

\citet{1986A&A...166...53D} tried to explain the observed characteristics
of shell morphology (point~\ref{enu:elipticity} in Sect.~\foreignlanguage{british}{\ref{sec:characterictics}})
with the encounter of a disk galaxy with a prolate or oblate primary
E-galaxy. The secondary galaxy falls into the prolate galaxy around
its symmetry axis and into the oblate galaxy perpendicularly to its
symmetry axis (the symmetry axis is the major axis when the E-galaxy
is prolate, minor axis when oblate). The disk of the secondary galaxy
is always oriented in the direction of the collision. In the prolate
case, the companion stars achieve pendular motion along the major
axis of the E-galaxy. The shells form consequently along this axis,
alternatively on one side and the other (type\,I shell galaxy, see
Sect.~\ref{sub:Vzhled-slupek}). On the contrary, in the oblate case,
the shell system does not possess any symmetry, since there is no
privileged major axis here. The shells appear randomly spread around
the center of the E-galaxy (type\,II shell galaxy).

\citet{1986A&A...166...53D} state that a shell system is found aligned
with the major axis of an elliptical galaxy, only when the E-galaxy
is prolate and the impact angle is likely to be lower than 60\textdegree{}.
A shell system is found aligned with the minor axis of an E-galaxy,
only when the latter is oblate and the impact angle is lower than
$\sim$30\textdegree{}. It is interesting to note that no such system,
with the shell aligned with the minor axis, is known.

However, all this results were negated by \citet{1989ApJ...342....1H},
who also simulated an ellipsoidal potential of the primary galaxy.
Their result is that if the potential well maintains the same shape
at all radii as in the simulations of Dupraz and Combes, then the
shape of the dark matter halo, as well as that of the central galaxy,
is responsible for aligning and confining the shells. If, on the other
hand, the potential is allowed to become spherical at large radii,
the shell alignment and angular extent are less sensitive to the properties
of the potential at small radii. This means that two primaries, one
oblate and the other prolate, can have similar projected shapes and
similar outer shells if the outer isophotentials in each case become
spherical. Hence the shape of the potential at large as well as small
radii needs to be considered when examining the shell extent and alignment.

Even the same authors formerly tried to get some information about
the potential of several chosen shell galaxies \citep{1987ApJ...312....1H},
but for those reasons and the reasons stated in Sects.~\ref{sub:Companion}
and \ref{sub:R-dist.MM}, they were left with nothing to say but:
{}``The shell morphology is sensitive to the shape of the primary
at large and small radii as well as to the detailed structure of the
companion. This would imply that it is difficult, if not impossible,
to infer the form of the primary from the shell geometry alone. In
this conclusion, we disagree with \citet{1986A&A...166...53D}.''

\subsection{Radial distribution of shells \label{sub:R-dist.MM}}

The radial distribution of shells was always probably the most watched
aspect of the merger model. From Sect.~\ref{sub:phase-wrapping}
we already know how easily the merger model reproduces the interleaving
in radii. The shell formation is closely connected to the period of
radial oscillation in the host galaxy potential, what is in any case
an increasing function of radius, see Sect.~\ref{sec:rad_osc}. The
shells as density waves receding from the center, composed in every
moment of different stars, are the older the further from the center
they are. With time, the frequency of the shells increases, thus the
distances between shells decrease towards the center, what is also
in agreement with observations (see Sect.~\ref{sub:Vzhled-slupek}).

The above-mentioned facts suggest a connection of shell distribution
and the potential of the underlying galaxy. But already \citet{1984ApJ...279..596Q}
discovered that the radial distribution of shells derived from the
potential inferred from the observed luminous matter distribution
cannot agree with the observed reality. \citet{1984ApJ...279..596Q}
derived that the potential of the shell galaxy NGC\,3923 must be
less centrally condensed at radii $1<r/r_{\text{e}}<4$ (where $r_{\text{e}}$
is the half-mass radius) than the luminous matter observations predict.
This discovery was reflected by \citet{1986A&A...166...53D,1987ApJ...312....1H}
as they added an extensive dark matter halo in their simulations and
then they were able to better reproduce the observed shape of the
shell distribution. But immediately after that, \citet{1987A&A...185L...1D}
synthesized successfully a similar radial distribution taking into
account the dynamical friction instead of dark matter. Moreover, in
spite of the simplicity of their model, they synthesized a wide variety
of shapes for the shell distribution by varying only the two parameters:
mass ratio of primary and secondary and impact parameter. It all leads
to the conclusion that the shell system is not suitable to study the
potential of a host galaxy. 

Note that in the eighties only photometric data were considered. \citet{mk98}
suggested methods of measurement of the potential using shell kinematics
(Sect.~\foreignlanguage{british}{\ref{sub:Use-of-shells}}). The
method relies on the stars, which form the shell, to be on the close-to-radial
orbits and it is insensitive to the details of the merger such as
the type of cannibalized galaxy and dynamic friction.

The cornerstone of the merger theory is also the huge range of radii
in which the shells occur. A simple merger simulation, as of \citet{1984ApJ...279..596Q}
(see Sect.~\foreignlanguage{british}{\ref{sub:phase-wrapping}}),
is not able to produce shells simultaneously on large and small radii.
The presence of shells deep within the host galaxy (and thus the presence
of deeply bound stars that once were part of the secondary galaxy)
was mysterious from the very beginning. But because at that time the
merger model had no direct competition, it was felt more as a challenge
than a flaw. However, the advent of the WIM (Sect.~\ref{sub:WIM})
that does not have any problems explaining this phenomenon, challenges
the merger model more seriously. 

\citet{1984ApJ...279..596Q} suggested three possible explanations:
First, the infall velocity of the disk may have been small and hence
the disk was initially strongly bound to the elliptical. Second, the
mass ratio may have been closer to unity, and hence energy could have
been transferred from orbital motion to internal velocity dispersion.
But as the most probable explanation he promoted the idea that the
disruption process is a gradual one and that the center-of-mass motion
of the disk is subject to dynamical friction.

Another effect that no one predicted was found by \citet{1990MNRAS.243..199H}.
They self-consistently simulated the secondary galaxy and left the
primary as a rigid potential. During the disruption event there is
a substantial transfer of energy between the various parts of the
satellite. Stars which lead the main body through the encounter are
braked and later form the inner shell system. Stars which lag the
main body are accelerated and turn into an escaping tail. This transfer
is asymmetric and, for the encounters they have studied, the surviving
core suffers a net loss of orbital energy which can shrink the apocenter
of its orbit by a large factor. All these transfer effects increase
with the mass of the satellite. It should be emphasized that this
energy transfer happens only within the original secondary galaxy
and no dynamical friction from the stars of the primary galaxy is
accounted for in this case.

This scenario also allows the shell formation in a larger spread of
radii. If the core of the cannibalized galaxy survives the merger,
new generations of shells are added during each successive passage.
This was predicted by \citet{1987A&A...185L...1D} and successfully
reproduced by \citet{katka11} in self-consistent simulations. Further,
the combination of the loss of orbital energy in this way and the
dynamical friction could bring new results, if properly modeled. This
was also mentioned by \citet{1996A&A...310..757S}, who also simulated
the formation of shell galaxies in a radial merger in a self-consistent
manner, although without any dark matter halo in the primary galaxy.

\subsection{Radiality of the merger \label{sub:Radialnost}}

The assumption of a radial merger is the most awkward and criticized
point of Quinn's model of shell formation. In his work, \citet{1984ApJ...279..596Q}
has shown that if the center-of-mass motion of the infalling disk
is predominantly non-radial, the merger produces confused, often overlapping
shells which appear enclosing. This does not correspond to what we
see in real shell galaxies.

On the other hand, A. Toomre modeled an off-axis release of a non-rotating,
inclined disk into a fixed spherical force field (shown in \citealp{1983IAUS..100..319S})
and his results resemble the observed shapes. The model was similar
to that of Quinn in that the disk was released as a set of test particles
with identical subparabolic velocities. The shells are created via
the mass transfer from the secondary galaxy flying by on a parabolic
trajectory. The captured part forms a complex structure around the
primary galaxy. In this case, a complete merger is not necessary to
produce the shells. \citet{1988ApJ...331..682H} present examples
of objects from the Arp atlas \citep{1966apg..book.....A} that may
well have resulted from such non-merging encounters -- Arp\,92 (NGC\,7603),
103, 104 (NGC\,5216 + NGC\,5218), and 171 (NGC\,5718 + IC\,1042)
all show evidence of interactions as well as diffuse shell-like features
surrounding the more luminous galaxy. \citet{1988ApJ...331..682H}
also note that, as in the strictly planar case, the term \textquotedbl{}shell\textquotedbl{}
can occasionally be a misnomer since the stars near the vicinity of
a sharp edge are not necessarily distributed on a three-dimensional
surface in space. 

However, the requirement of a fairly radial encounter stays valid
to produce type\,I shell galaxies (Sect.~\ref{sub:Vzhled-slupek})
as NGC\,3923 or NGC\,7600 that we have already seen in Fig.~\ref{fig:3923}
and Fig.~\ref{obr.7600}, respectively. A strictly radial merger
of galaxies is improbable, but now cosmological $N$-body simulations
tell us that satellites are preferentially accreted on very eccentric
orbits \citep{2005MNRAS.364..424W,2005MNRAS.358..551B,2006A&A...445..403K}.

\citet{1987A&A...185L...1D} considered that the shell distribution,
from the parabolic encounter with dynamical friction, remains unchanged
for a (small but) significant range of impact parameters. The more
massive the secondary galaxy is (compared with the primary), the larger
range is allowed. \citet{2005MNRAS.361.1030G,2005MNRAS.361.1043G}
carried out $N$-body simulations of encounters between spherical
galaxies with and without a dark halo with $\sim10^{4}$ particles.
Shells are rather a byproduct of their work, but they were able to
get them even for impact parameters enclosing 95\% of the total mass
of the primary. Even earlier, \citet{1989Natur.338..123B} examined
the evolution of a compact group of six disk galaxies in a self-consistent
simulation of 65,536 particles. The result was a giant elliptical
galaxy containing the shells. The shells were created during the final
infall of the last galaxy into the merged body of all other galaxies.
The initial distribution of the disk galaxies and their inclinations
were by no means special, and \citeauthor{1989Natur.338..123B} did
not specifically try to get the shells. This simulation may mean that
during the evolution of a compact group, the shell galaxies are indeed
formed in the final stage of the merger. Similarly, recently \citet{2011ApJ...743L..21C}
found shell galaxies as a product of galaxy formation in Milky Way-mass
dark halo in two from six simulated halos from the Aquarius project
\citep{2008MNRAS.391.1685S}, which builds upon large-scale cosmological
simulations. Furthermore, it is supported by the observed high occurrence
of shells in isolated giant galaxies (Sect.~\ref{sub:V=0000FDskyt-shell-galaxies}).

\subsection{Major mergers\label{sub:MajorM}}

\citet{1992ApJ...399L.117H} published results of their simulation
of a major merger which creates shells. Two identical galaxies with
self-gravitating disks and halos merged following a close collision
from a parabolic orbit. The plane of each disk initially coincides
with the orbital plane. When plotted in phase space, the remnant exhibits
more than 10 clearly defined phase-wraps which can be identified with
shells. Shells also occur near the nucleus and appear to be aligned
with the major axis of the resulting galaxies.

\citet{2005MNRAS.357..753G} examined the creation of elliptical galaxies
from mergers of disks. They used disk-bulge-halo or bulge-less, disk-halo
models with mass ratios of the participants of 1:1, 1:2, and 1:3 and
various impact parameters. As a result of those mergers, shells which
could be identified in phase space occurred sometimes. They found
out that the models without bulges with the mass ratio of 1:2 or 1:3
lead to more prominent shells. But these were always shell systems
of type\,II (all-round) or type\,III (irregular). \citeauthor{2005MNRAS.357..753G}
note the lack of shells in remnants of equal-mass mergers and on all
prograde mergers. This contrasts with the shell system presented by
\citet{1992ApJ...399L.117H}, a prograde merger of two equal-mass,
bulge-less disks. The perfect alignment of the disk spins with the
orbital angular momentum may have favored the formation of shells
in their model.

\citet{2005MNRAS.361.1030G,2005MNRAS.361.1043G} have also carried
out simulations of encounters between spherical galaxies (see Sect.~\ref{sub:Radialnost}):
In their first paper without a dark halo and in the second one with
a dark halo (with mass ratios of 1:1, 1:2, and 1:4). The sharpness
of the occurring shells was higher in models with a halo. A head-on
collision for a run with mass a ratio 4:1 showed the shells even after
5\,Gyr from the first encounter of the galaxy centers. But the shells
showed up also in the merger with 1:2 mass ratio and a nonzero impact
parameter. In any case, the shells are formed from particles of the
less massive galaxy through the same phase wrapping that was established
by \citet{1984ApJ...279..596Q}.

To summarize, shells can be formed via a merger even in the cases
when the mass ratios are not as dramatic as it has been simulated
in the 80s (the big mass of the secondary galaxy could influence the
alignment of shells with the major axis of the host galaxy, but no
one has so far explored it). It is probably not common to have shells
when two disk galaxies of comparable masses merge. \citet{1992ApJ...399L.117H}
got shells in their model maybe only thanks to the very special conditions
of the collision they have chosen. Furthermore, the interleaving structure
and more generally the distribution of shells is not known for such
cases. Some authors have guessed a major-merger origin for the shell
galaxies in their observational studies \citep{1995ApJ...444L..77S,2001AJ....122.1758B,2001MNRAS.322..643G,2006A&A...453..493S}.

\subsection{Simulations with gas \label{sub:Simulations-with-gas}}

Only a few works have been dedicated to modeling the formation of
shell galaxies in the presence of gas, all of them in the framework
of the minor-merger model. \citet{1993ApJ...405..142W} used a variant
of the TREESPH code but self-gravity was strictly ignored. The primary
galaxy was treated as a rigid spherically symmetric potential. They
performed four runs -- two radial and two non-radial; two of them
were prograde with the disk inclined by 45\textdegree{}. Isothermal
processes were assumed (T = $10^{4}$\,K) except for one run where
radiative cooling was allowed, and at the end 94\% particles had temperature
6,000\textendash{}10,000\,K. Main results are that in all cases gaseous
and stellar debris segregated and gas forms dense rings around the
nucleus of the primary galaxy where massive star formation may occur.
Furthermore the diameter of the ring depends on the impact parameter
(the total angular momentum in the ring is 50\% of the initial value
for those particles); radial and inclined encounter forms a \textit{s}-shaped
ring and a counterrotating core; and about a half of all the gas particles
is captured in these rings.

A completely different conclusion was reached by \citet{1997ApJ...481..132K}.
They used the sticky particle method (after collision, the radial
velocity component of the particle is halved and the sign reversed)
and performed four runs of simulation -- radial (twice), prograde,
retrograde (all with zero inclination). Both galaxies were self-gravitating
systems. Star formation was modeled as a probability of a change of
a gas particle to a stellar based on local gas density. They found
defi{}nitely no significant segregation of gas and stars; star formation
was mainly reduced because of scattering on the deep potential well
of the primary (radial and retrograde runs); for slightly prograde
orbit, the inner part of the secondary galaxy survives, a small stellar
bar of the secondary is created which causes bar-driven gas infl{}ow
and a strong starburst. In the radial run with a less concentrated
primary, a larger part of the secondary survives and the oscillating
remnant destroys the shells. They state that the {}``poststarburst''
nature of shell galaxies is due to the cessation of star formation
in the disk galaxies caused by the merger (no massive star formation
is caused by the encounter itself).

The model of \citet{1999ASPC..182..489C,2000ASPC..197..339C,2000ASPC..209..273C}
was based on the belief in two components of galactic gas -- diffuse
H\,I gas ends in center of primary, while the small and dense gas
clouds have an intermediate behavior between stars and H\,I. They
took into account the dynamical friction and a proper treatment of
the dissipation of the gas (using cloud-cloud collision code). The
gaseous component was liberated fi{}rst since it was less bound than
stars. Then stars lose their energy due to the dynamical friction
what causes some displacement of the gaseous and stellar shells. That
was really observed in some shell galaxies, see Sect.~\ref{sub:Gas-and-dust}.

\subsection{Merger model and observations}

Merger models can well explain the interleaving of shells and their
increasing separation with radius (point~\ref{enu:interleaved} in
Sect.~\foreignlanguage{british}{\ref{sec:characterictics}}) and
the number of shells increases with  time. The observed brightness
of shells puts a lower limit to the mass of the original secondary
galaxy that is usually several per cent of the primary (point~\ref{enu:brightness}
in Sect.~\foreignlanguage{british}{\ref{sec:characterictics}}).
The question of an alignment of shells with the major axis of the
host galaxy and the correlation between the type of the shell galaxy
and ellipticity (point~\ref{enu:elipticity} in Sect.~\foreignlanguage{british}{\ref{sec:characterictics}})
remains unsettled for the merger model. The merger model has also
problems explaining the large range of radii where the shells are
found and their occurrence at low radii (points~\ref{enu:Range}
and \ref{enu:close-to-nucleus} in Sect.~\foreignlanguage{british}{\ref{sec:characterictics}}).
Mergers of different secondary galaxies can explain different colors
of shells and their possible difference from the color of the underlying
galaxy (point~\ref{enu:redder} in Sect.~\foreignlanguage{british}{\ref{sec:characterictics}}).

A merger origin of shell systems is supported by many observations,
a list of which would be lengthy. It seems that all the shell galaxies
that have been so far examined in detail contain dust close to the
nucleus (point~\ref{enu:central-dust} in Sect.~\foreignlanguage{british}{\ref{sec:characterictics}}).
These dust features are often found to be out of dynamical equilibrium
(Sect.~\ref{sub:Gas-and-dust}), what clearly points to their external
origin. Shell galaxies contain even more characteristics believed
to be the results of a merger, including tidal tails, multiple nuclei
or nuclear post-starburst spectra.

It seems that about 20\% of shell galaxies could contain a second
nucleus (Sect.~\ref{sub:Other-features}) -- a characteristic that
one would expect in a galaxy after a merger event. \citet{1994AJ....107.1713F}
calculate that this could be an expected frequency due to the short
lifetime of the nucleus of the secondary galaxy as opposed to the
long-living shells. They note that it is also the expected frequency
for the WIM origin of shell galaxies -- the galaxies with the double
nuclei would be those we see at the moment when the secondary galaxy
just passes through the primary. 

A large support for the merger theories comes from the kinematically
distinct cores (KDCs). Even before it was recognized that all known
galaxies with KDCs in 1992 are shell galaxies, (point~\ref{enu:KDC}
in Sect.~\foreignlanguage{british}{\ref{sec:characterictics}}, see
also Sect.~\ref{sub:Other-features}), the origin of KDCs from mergers
of galaxies has been independently anticipated. Already \citet{1984ApJ...287..577K}
proposed this mechanism for the formation of counterrotating cores
in elliptical galaxies and \citet{1990ApJ...361..381B} investigated
this using self-consistent numerical simulations of mergers between
elliptical galaxies of unequal mass, and found that the core kinematics
in the remnant depend mostly upon the orbital angular momentum at
a late stage of the merger, whereas the kinematics of the outer regions
is largely the original kinematics of the primary. Thus, in retrograde
encounters a counter-rotating core can form. \citet{1991Natur.354..210H},
cited in \citet{1999MNRAS.307..967T}, demonstrated the formation
of a counterrotating central gas disk in a merger of two gas-rich
disk galaxies of equal mass. But this model is less widely accepted
than the previous one. \citet{1994MNRAS.270L..23H} suggested a model
that would comply with the WIM, but it is probably even less popular.

Enormous diversity of central surface brightness (point~\ref{enu:central-brightness}
in Sect.~\foreignlanguage{british}{\ref{sec:characterictics}}) and
other characteristic show that shell galaxies are otherwise not a
compact or privileged group of galaxies -- so to say, the secondary
cannot choose on what it falls. Still some selection effect seems
to be there, because shell galaxies are much more often seen in regions
with low galactic density (point~\ref{enu:low-density} in Sect.~\foreignlanguage{british}{\ref{sec:characterictics}}).
That can be explained with velocities in galaxy clusters being too
high for one galaxy to be captured by another, or the influence of
the surrounding galaxies breaks the shells structure or even prevents
it from forming; or both.

Simulations show (Sect.~\foreignlanguage{british}{\ref{sub:Simulations-with-gas}})
that in the framework of the merger model of shells' creation, diffuse
gas is introduced into the center of the host galaxy (point~\ref{enu:post-starburst}
in Sect.~\foreignlanguage{british}{\ref{sec:characterictics}}),
while dense gas clouds form slightly displaced shells with respect
to the stellar shells (points~\ref{enu:Slightly-displaced-HI} and
\ref{enu:Molecular-gas-associated} in Sect.~\foreignlanguage{british}{\ref{sec:characterictics}}).
Both are in agreement with the observations.

As the observations show, shells in galaxies are fairly common (point~\ref{enu:cetnost}
in Sect.~\foreignlanguage{british}{\ref{sec:characterictics}}, see
also Sect.~\ref{sub:V=0000FDskyt-shell-galaxies}). It means that
in fact they occur even more frequently because from the three-dimensional
shape of the shells as introduced by \citet{1984ApJ...279..596Q},
Sect.~\ref{sec:MM}, we can easily understand that we see shells
only when looking from angles close to the plane perpendicular to
the line of the collision. But it is not that improbable as the shells
in mergers are formed in a much larger range of impact parameters
than it was originally believed (see Sect.~\ref{sub:MajorM}) and
interactions between galaxies are quite a common matter.

\section{Measurements of gravitational potential in galaxies}

Before we present our original results, we introduce the reader shortly
to the topic of measuring galactic potentials, particularly in the
case of elliptical and shell galaxies.

\subsection{Insight into methods}

The issue of the determination of the overall potential and distribution
of the dark matter in galaxies is among the most prominent in galactic
astrophysics. In disk galaxies, where stars and gas move on near-circular
orbits, we can derive the potential (at least in the disk plane) directly
up to several tens of kiloparsecs from the center of the galaxy in
question. Early-type galaxies lack such kinematical beacons. 

Several different methods have been used to measure the potentials
and the potential gradients of elliptical galaxies, including strong
gravitational lensing (e.g., \citealp{2006ApJ...649..599K,2009ApJ...703L..51K,auger10}),
weak gravitational lensing (e.g., \citealp{mandelbaum08}), X-ray
observations of hot gas in the massive gas-rich galaxies (e.g., \citealp{fukazawa06,churazov08,das10}),
rotational curves from detected disks and rings of neutral hydrogen
(e.g., \citealp{weijmans08}), stellar-dynamical modeling from integrated
light spectra (e.g., \citealp{thomas11}), as well using tracers such
as planetary nebulae (e.g., \citealp{coccato09}), globular clusters
(e.g., \citealp{norris12}) and satellite galaxies (e.g., \citealp{nierenberg11,deason12}). 

All the methods have various limits, e.g., the redshift of the observed
object, the luminosity profile, gas content, and so forth. In particular,
the use of stellar dynamical modeling is plausible in the wide range
of galactic masses, as far as spectroscopic data are available. However,
it becomes more challenging past few optical half-light radii. Moreover,
the situation is made complex by our insufficient knowledge of the
anisotropy of spatial velocities. Another complementary gravitational
tracers or techniques are required to derive mass profiles in outer
parts of the galaxies. While comparing independent techniques for
the same objects at the similar galactocentric radii, the discrepancies
in the estimated circular velocity%
\footnote{The concept of circular velocity is commonly used even in elliptical
galaxies where none or small amount of the matter is expected to move
on circular orbits. It is a quantity which says what speed would move
the body launched into a circular orbit. Provided spherical symmetry
of the galaxy, it simply denotes the quantity $\sqrt{r\phi'(r)}$
, where $\phi'(r)$ is the first derivative of the galactic potential
with respect to the galactocentric radius $r$.%
} curves were revealed together with several interpretations (e.g.,
\citealp{churazov10,das10}). The compared techniques usually employ
modeling the X-ray emission of the hot gas (assuming hydrostatic equilibrium)
and dynamical modeling of the optical data in the massive early-type
galaxies. Therefore, even for the most massive galaxies with X-ray
observations at disposal, there is a need for other methods to independently
constrain the gravitational potential at various radii.

\subsection{Use of shells \label{sub:Use-of-shells}}

Using the radial distribution of shells to derive the potential of
the host galaxy seems tempting, but it insofar generally failed due
to reasons discussed in Sect.~\ref{sub:R-dist.MM}. The question
remains whether it is better to use the outer shells that are less
affected by the dynamical friction and possible later generations
of shells, or if we could, by careful modeling of all the relevant
physical processes, reproduce the whole observed shell distribution
for a suitable potential.

An alternative hypothetical use of shells to determine the dark matter
content of galaxies is proposed by \citet{sanderson12}. The increased
concentration of matter and its low velocity dispersion in the shells
is favorable for indirect detection of dark matter via gamma-ray emission
from dark matter self-annihilation due to the Sommerfeld effect.

A slightly less exotic, though not less bold method has been proposed
by \citet{mk98}. The method uses shells to constrain the form of
the gravitational potential in the case of validity of the \citet{1984ApJ...279..596Q}
merger model (described in Sect.~\foreignlanguage{british}{\ref{sub:phase-wrapping}}).
They studied theoretically the kinematics of a stationary shell, a
monoenergetic spherically symmetric system of stars oscillating on
radial orbits in a spherically symmetric potential. They predicted
that spectral line profiles of such a system exhibit two clear maxima,
which provide a direct measure of the gradient of the gravitational
potential at the shell radius. 

In practice, the situation is far more complex and the shells themselves
are faint structures in a bright galaxy, so the fulfillment of this
program seems almost impossible. However, the authors state that they
have carried out signal-to-noise ratio calculations for some of the
brighter shell galaxies such as NGC\,3923, and have ascertained that
data of the requisite quality could be obtained with a couple of nights
integration using a 4-m telescope.

Now comes the era when the instrumental equipment begins to allow
us to actually obtain such kind of data and that requires deeper theoretical
understanding of the topic. In Part~\ref{PART II-S.kin}, we extend
the work of \citet{mk98} and we develop methods to better reproduce
parameters of the potential of the host galaxy from measured data.

The first attempt to analyze the kinematical imprint of a shell observationally
was made by \citet{romanowsky12}, who used globular clusters as shell
tracers in the early-type galaxy M87, the central galaxy in the Virgo
cluster. They obtained wide-field (0--200\,kpc from the center) high-precision
(median velocity uncertainties: 14\,km$/$s) spectroscopic data for
488 globular clusters. They found signatures of a cold stream (about
15 globular clusters at 150\,kpc) and a large shell-like pattern
(about 30 globular clusters between 50 and 100\,kpc) and verified
the presence of these features using statistical tests. These features
are the fi{}rst large stellar substructure with a clear kinematical
detection in any type of galaxy beyond the Local Group. The stream
is associated with a known stellar filament but there is no photometric
shell visible in the galaxy. Typical surface brightness in the region
of the shell-like pattern is $\mu_{\mathrm{V}}\sim27$\,mag$/$arcsec\textsuperscript{2}.
Following the calculations of \citet{mk98}, \citet{romanowsky12}
derived circular velocity at the shell radius $v\mathrm{_{c}\sim270}$\,km$/$s
while X-ray data indicate $v\mathrm{_{c}\sim}$650--900\,km$/$s
in the same region. Further analysis done by the authors suggests
that for such a shell to be created, the host galaxy would have to
accrete a large group of dwarf galaxies or a single giant elliptical
or a lenticular galaxy (about 5 times bigger than the entire Milky
Way system). 

\citet{fardal12} obtained radial velocities (median error 3\,km$/$s)
of 363 red giant branch stars in the region of the so-called Western
Shelf in M31, the Andromeda galaxy. The Western Shelf, located about
25\,kpc from the center of the galaxy, is one of several features
in the stellar halo of M31. In the space of line-of-sight velocity
velocity versus projected radius, the data they obtained show a wedge-like
pattern. This is consistent with the previous finding of \citet{fardal07}
who reproduced main photometric structures in the stellar halo using
a simulation of an accretion of a dwarf satellite within the accurate
M31 potential model. They inferred that the Western Shelf is a shell
from the third orbital wrap%
\footnote{If we considered the remains of the accreted satellite to be a shell
system, we would assign number 2 to this shell, see Sect.~\ref{sub:TP}.%
} of a tidal debris stream. Using similar simulation, \citet{fardal12}
derived that the Western Shelf moves with phase velocity of 40\,km$/$s
and that the wedge pattern has a global offset -20\,km$/$s with
respect to the systemic velocity due to the angular momentum.

\clearpage

\newpage{}

\part{Shell kinematics\label{PART II-S.kin}}

A lot of useful information about the shell galaxies can be extracted
from the kinematics of the stars forming the shell system. That it
is by measuring the line-of-sight velocity distribution (LOSVD) near
the edge of the shell. Now comes the era when the instrumental equipment
begins to allow us to actually obtain such kind of data and that requires
deeper theoretical understanding of the topic. First attempts to analyze
such kind of data have been already made, see Sect.~\ref{sub:Use-of-shells}.
The idea to use shell kinematics, has been proposed by \citet{mk98},
hereafter \citetalias{mk98}, and we further developed it in papers
\citet{jilkova10} and \citet{2012A&A...545A..33E}, Appendices~\ref{apx:clanek-lucka}
and~\ref{apx:clanek-huevo}, respectively. 

\begin{figure}[H]
\centering{}\includegraphics[width=12cm]{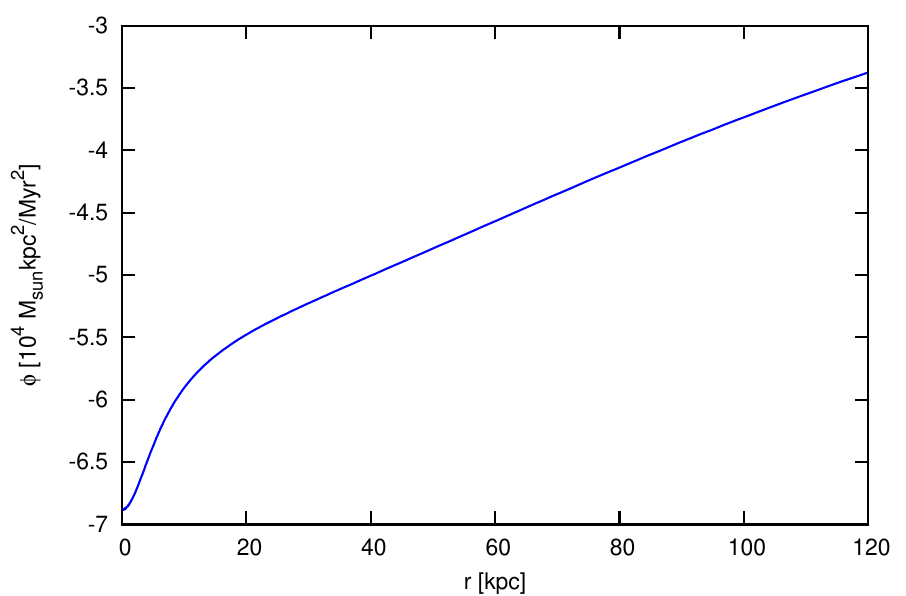}
\caption{\textsf{\small Potential of the host galaxy. The potential is modeled
as a double Plummer sphere with parameters listed in Table~\ref{tab:param}.
\label{fig:pot} }}
\end{figure}

\section{Preliminary provisions}

First we introduce several useful notions to aid the reader.

\subsection{Host galaxy potential model \label{sec:param}}

\begin{figure}[H]
\centering{}\includegraphics[width=7.8cm]{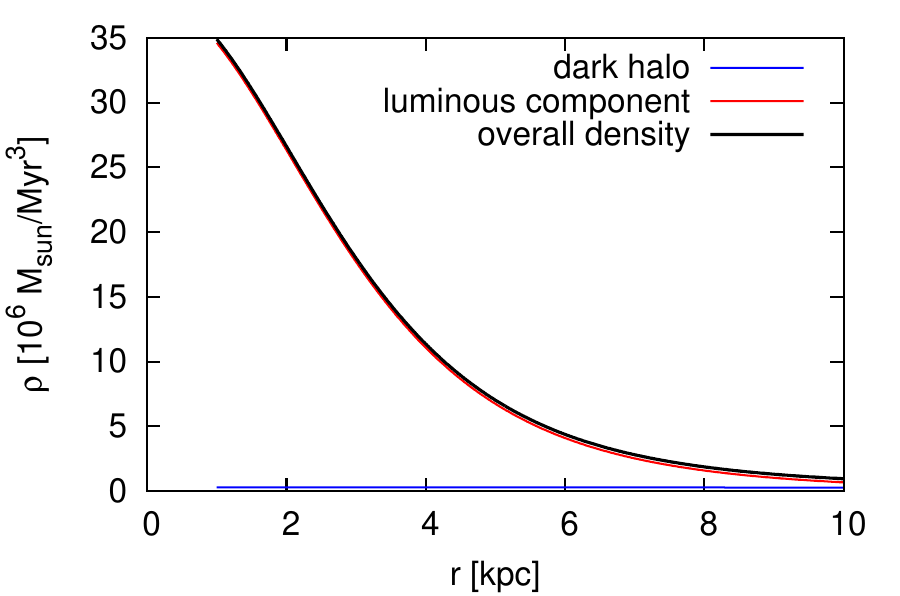}\includegraphics[width=7.5cm]{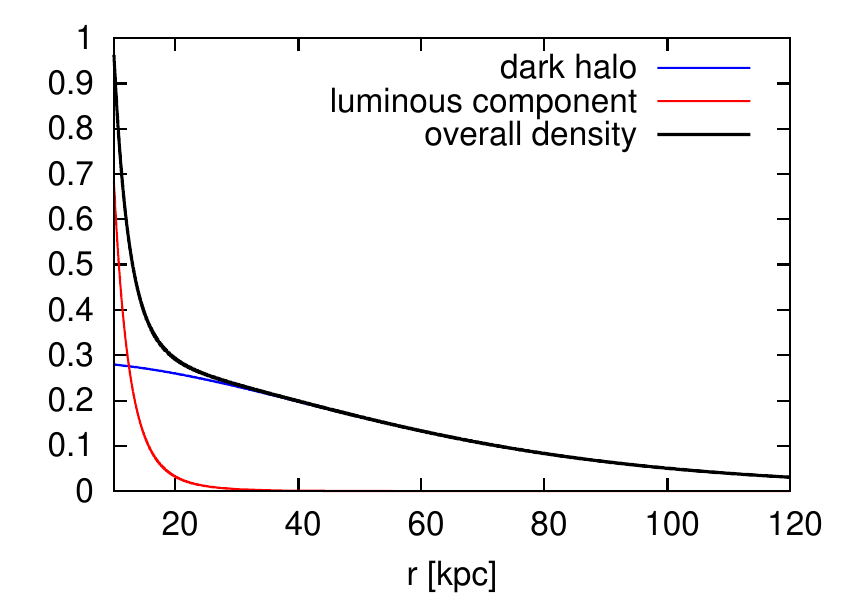}
\caption{\textsf{\small Density of the host galaxy. The potential is modeled
as a double Plummer sphere with parameters listed in Table~\ref{tab:param}.
\label{fig:rho} }}
\end{figure}

In this part of the thesis, we will often need to illustrate the shell
kinematics using specific examples. For this purpose, the potential
of the host galaxy is modeled as a double Plummer sphere with parameters
presented in Table~\ref{tab:host}, unless specified otherwise. This
model has properties consistent with observed massive early-type (and
even shell) galaxies \citep{auger10,nagino09,fukazawa06}. The forms
of the potential and density for the chosen model are shown in Figs.~\ref{fig:pot}
and \ref{fig:rho}, respectively. 

\begin{table}[H]
\centering{}%
\begin{tabular}{ccc}
\hline 
 & Plummer radius  & total mass\tabularnewline
 & kpc & M\suns\tabularnewline
\hline 
\noalign{\vskip0.1cm}
luminous component & 5  & $2\times10^{11}$\tabularnewline
\noalign{\vskip0.1cm}
dark halo & 100 & $1.2\times10^{13}$\tabularnewline
\hline 
\end{tabular}\caption{\textsf{\small Parameters of the potential of the host galaxy used
in Part~\ref{PART II-S.kin}. The potential is modeled as a double
Plummer sphere. \label{tab:host}}}
\end{table}

The potential of a Plummer sphere can be expressed as 
\begin{equation}
\phi(r)=-\frac{\mathrm{G}\, M}{\sqrt{r^{2}+\varepsilon^{2}}},\label{eq:pot-1}
\end{equation}
where G is the gravitational constant, \textit{$M$} is the total
mass of the galaxy, \textit{$r$} is the distance from the center
of the galaxy and $\varepsilon$ is the Plummer radius. The radial
density then reads 
\begin{equation}
\rho(r)=\rho_{0}\frac{1}{(1+r^{2}/\varepsilon^{2})^{5/2}},\label{eq:hust-1}
\end{equation}
where $\rho_{0}=3M/(4\pi\varepsilon^{3})$ is the central density.
The interested reader can find more on the Plummer potential in Sects.~\ref{sec:Plummer}--\ref{sub:Disp-PP}. 

Let us note that such a choice of the potential of the host galaxy
represents a whole class of models. For example, we can express all
distances in the terms of the Plummer radius of the luminous component
and all masses in the terms of the total mass of the luminous component
and then choose these two parameters at will. For clarity, we nevertheless
keep the specific values noted below.

\subsection{Terminology}

In this section, we briefly introduce terms used in next sections.
\begin{itemize}
\item \textbf{Model of radial oscillations} -- through Part~\ref{PART II-S.kin},
the word \textit{model} is assigned to the concept described in Sect.~\ref{sec:rad_osc}
and used for modeling of shell kinematics. The model assumes that
shells are made by stars on strictly radial orbits released at one
moment in the center of the host galaxy. The potential of the host
galaxy is chosen to represent real galaxies reasonably well. In our
work, we restrict ourselves to a double Plummer sphere introduced
in Sect.~\ref{sec:param}.
\item \textbf{Approximation of constant acceleration and shell velocity}
(Sect.~\ref{sec:IOA}) -- it is basically the model of radial oscillations
but the value of acceleration in the host galaxy as well as the value
of the shell phase velocity are always constant. The approximation
is assumed to be valid only in the vicinity of the shell edge. In
the framework of this approximation, the position of line-of-sight
velocity maxima are calculated using either of the following three
methods: the approximative LOSVD (Sect.~\ref{sub:LOSVD-app}); the
approximative maximal LOS velocities (Sect.~\ref{sub:vlos,max});
and and the method using the slope of the LOSVD intensity maxima (Sect.~\ref{sub:Slope}).
Differences between these methods are summarized in Sect.~\ref{sec:Compars}.
\item \textbf{Higher order approximation} (Sect.~\ref{sec:Higher-order-approximation})
-- similarly as previous, but this time we allow the value of acceleration
in the host galaxy to change linearly with galactocentric radius.
\item \textbf{Simulation} -- in this part, we only use this term when we
model shell galaxies in the simulation of a radial minor merger of
galaxies using test particles (Sect.~\ref{sec:N-Simulations}).
\end{itemize}

\subsection{Quantities}
\begin{lyxlist}{00.00.0000}
\item [{$t$}] time; usually indicates the time since the release of stars
at the center of the host galaxy
\item [{$\mathbf{r}=(x,y,z)$}] vector of Cartesian coordinates that are
oriented so that the origin is at the center of the host galaxy; $x-y$
is the projected plane  ({}``the sky'') and the $z$ direction coincides
with the line of sight (LOS); $x$-axis is also the collision axis
although in the model of radial oscillations it is just a virtual
concept, since no collision is actually modeled
\item [{$X,Y$}] coordinates of the projected plane 
\item [{$r$}] galactocentric radius, distance from center of the galaxy;
$r=\sqrt{x^{2}+y^{2}+z^{2}}$
\item [{$R$}] projected radius, the projection of $r$ into the $x-y$
plane; $R=\sqrt{x^{2}+y^{2}}$
\item [{$\phi(r)$}] potential of the host galaxy; in this part, we use
a spherically symmetric potential introduced in Sect.~\ref{sec:param};
parameters of the potential are the total mass $M_{*}$, $M_{\mathrm{DM}}$
and the scale radius $\varepsilon_{*}$, $\varepsilon_{\mathrm{DM}}$
of the luminous and dark component, respectively
\item [{$\rho(r)$}] spatial density (in a spherically symmetric system)
\item [{$v_{\mathrm{c}}$}] circular velocity; provided spherical symmetry
of the galaxy, it simply denotes the quantity $\sqrt{r\phi'(r)}$
, where $\phi'(r)$ is the first derivative of the galactic potential
with respect to the galactocentric radius $r$.
\item [{$a$}] acceleration in the host galaxy; $a_{0}$ is the constant
term and $a_{1}$ is the coefficient of a linear term of the expansion
of the acceleration around the shell edge 
\item [{$T(r)$}] period of radial motion at the galactocentric radius
$r$ in the host galaxy potential; Eq.~(\ref{eq:Tr})
\item [{$n$}] serial number of a shell; shells are traditionally numbered
from the outermost to the innermost ones; Sect.~\ref{sub:TP}
\item [{$r_{\mathrm{TP}}$}] current turning point, i.e. the radius where
the stars are located in their apocenters at a given moment (the moment
of measurement); Eq.~(\ref{eq:Trn})
\item [{$v_{\mathrm{TP}}$}] phase velocity of a current turning point;
Eq.~(\ref{eq:vTP})
\item [{$r_{\mathrm{*}}$}] position of a star at a given time $t$ since
the release of the star in the center of the host galaxy; Eqs.~(\ref{eq:r*-})
and (\ref{eq:r*+}); often plain $r$ also denotes the position of
stars but the meaning is clear from the context
\item [{$r_{\mathrm{ac}}$}] position of the apocenter of a star (uniquely
related to the energy of the star for radial orbits); Eqs.~(\ref{eq:r*-})
and (\ref{eq:r*+})
\item [{$v_{\mathrm{r}}$}] stellar velocity at the galactocentric radius
$r$; in the model of radial oscillations the stellar velocity is
always in the radial direction
\item [{$r_{\mathrm{s}}$}] position of the edge of a shell, a function
of time $r_{\mathrm{s}}(t)$; Sect.~\ref{sub:edge}, Eq.~(\ref{eq:rs-pseudoEq})
\item [{$r_{\mathrm{s0}}$}] position of the shell edge at the moment of
measurement
\item [{$v_{\mathrm{s}}$}] phase velocity of a shell edge; approximately
equal to $v_{\mathrm{TP}}$; Eq.~(\ref{eq:vs})
\item [{$t_{\mathrm{s}}$}] time when a star currently at radius $r$ will
or did reach the corresponding edge of the shell; Sect.~\ref{sub:star-app}
\item [{$v_{\mathrm{los}}$}] line-of-sight velocity; the projection of
the stellar velocity into $z$ direction; $v_{\mathrm{los}}=v_{r}z/r$
\item [{$v_{\mathrm{los,max}}$}] the maximal absolute value of the LOS
velocity 
\item [{$r_{v\mathrm{max}}$}] radius of maximal LOS velocity, radius from
which comes the contribution to the LOSVD at the maximal speed $v_{\mathrm{los,max}}$;
Sect.~\ref{sub:rvmax}
\item [{$z_{v\mathrm{max}}$}] spot at the line of sight from which comes
the contribution to the LOSVD at the maximal speed; $z_{v\mathrm{max}}=\pm\sqrt{r_{v\mathrm{max}}^{2}-R^{2}}$,
Sect.~\ref{sub:Nature-of-4peak}
\item [{$F(v_{\mathrm{los}})$}] line-of-sight velocity distribution (LOSVD);
Eq.~(\ref{eq:Fvlos0})
\item [{$\sigma_{\mathrm{sph}}\left(r_{\mathbf{s}}\right)$}] shell-edge
density distribution; Eq.~(\ref{eq:sigma}), Sects.~\ref{sub:Equations-of-LOSVD},
\ref{sub:sigma}, and \ref{sub:Nature-of-4peak}
\item [{$\Sigma_{\mathrm{sph}}\left(r_{\mathbf{s}}\right)$}] discrete
equivalents of $\sigma_{\mathrm{sph}}\left(r_{\mathbf{s}}\right)$;
Eq.~(\ref{eq:SIGMA})
\item [{$\Sigma_{\mathrm{los}}\left(R\right)$}] projected surface density,
the projection of spacial density into the $x-y$ plane
\end{lyxlist}

\section{Model of radial oscillations \label{sec:rad_osc}}

If we approximate the shell system with a simplified model, we can
describe its evolution completely depending only on the potential
of the host galaxy. The approximation lies in the numerical integration
of radial trajectories of stars in a spherically symmetric potential.
Stars behave as if they were released in the center of the host galaxy
at the same time and their distribution of energies is continuous.
Usually we demand that the distribution is continuous at least in
such a range that stars with apocentra 10--30\,kpc around the edge
of the observed shell are present. Moreover we need that their density
in this region does not go sharply to zero. In some cases, we need
to know the distribution of energies explicitly. We express it in
terms of the shell-edge density distribution (Sects.~\ref{sub:Equations-of-LOSVD}),
which is a quantity more suitable for our situation and which can
be unambiguously converted to the distribution of energies or the
initial velocity distribution (Appendix~\ref{Apx:IVD}). We show
that the particular choice of the function does not affect the results
presented in this work (Sects.~\ref{sub:sigma}).

We call this model the \textit{model of radial oscillations}, and
it corresponds to the notion that the cannibalized galaxy came along
a radial path and disintegrated in the center of the host galaxy.
As a result the stars were released at one moment in the center and
began to oscillate freely on radial orbits. This approach was first
used by \citet{1984ApJ...279..596Q}, followed by \citet{1986A&A...166...53D,1987A&A...185L...1D}
and \citet{hernquist87a,1987ApJ...312....1H}.

This model uses the exact knowledge of the chosen potential of the
host galaxy, but requires it to be spherically symmetric. The potential
can be given analytically or numerically and the stellar trajectories
are usually integrated numerically. It differs from the real shell
galaxies in several aspects but it is still the most exact analytical
model that we can easily construct. We will show that in this model,
the LOSVD of shells exhibits four intensity maxima and how the position
of these maxima are connected with the parameters of the host galaxy
potential. All the following approximations will be compared to the
model of radial oscillations. Later we will show that the model agrees
very well with results of test-particle simulations of the formation
of the shell galaxies (Sect.~\ref{sec:N-Simulations}).

\subsection{Turning point positions and their velocities\label{sub:TP}}

In shell galaxies, the shells are traditionally numbered according
to the serial number of the shell, $n$, from the outermost to the
innermost (which in the model of radial oscillations for a single-generation
shell system corresponds to the oldest and the youngest shell, respectively).
If the cannibalized galaxy comes from the right side of the host galaxy,
stars are released in the center of the host galaxy. After that, they
reach their apocenters for the first time. But a shell does not form
here yet, because the stars are not sufficiently phase wrapped. We
call this the zeroth oscillation (the zeroth turning point) as we
try to match the number of oscillations with the customary numbering
scheme of the shells. We label the first shell that occurs on the
right side (the same side from which the cannibalized galaxy approached)
with $n=1$. Shell no.~2 appears on the left side of the host galaxy,
no.~3 on the right, and so forth.

In the model of radial oscillations, the shells occur close to the
radii where the stars are located in their apocenters at a given moment
(the current turning point, $r_{\mathrm{TP}}$, in our notation).
The shell number $n$ corresponds to the number of oscillations that
the stars near the shell have completed or are about to complete.
The current turning point $r_{\mathrm{TP}}$ must follow the equation
\begin{equation}
t=(n+1/2)T(r_{\mathrm{TP}}),\label{eq:Trn}
\end{equation}
 where $t$ is the time elapsed since stars were released in the center
of the host galaxy. $T(r)$ is the period of radial motion at a galactocentric
radius $r$ in the host galaxy potential $\phi(r)$: 
\begin{equation}
T(r)=\sqrt{2}\int_{0}^{r}\left[\phi(r)-\phi(r')\right]^{-1/2}\mathrm{d}r'.\label{eq:Tr}
\end{equation}
 The radial period is defined as the time required for a star to travel
from apocenter to pericenter and back \citep{1987gady.book.....B}.

The position of the current turning point evolves in time with a velocity
given by the derivative of Eq.~(\ref{eq:Trn}) with respect to radius
\begin{equation}
v_{\mathrm{TP}}(r;n)=\mathrm{d}r/\mathrm{d}t=\frac{1}{\mathrm{d}t/\mathrm{d}r}=\frac{1}{n+1/2}\left(\mathrm{d}T(r)/\mathrm{d}r\right)^{-1}.\label{eq:vTP}
\end{equation}
We can clearly see from this relation, which was first derived by
\citet{1984ApJ...279..596Q}, that any further turning point (turning
point with higher $n$) at the same radius moves more slowly than
the former one. Thus causes a gradual densification of the space distribution
of the shell system with time.

Technically, the reason for this densification is that the time difference
between the moments when two stars with similar energy reach their
turning points is cumulative. Let $\bigtriangleup t$ be the difference
in periods at two different radii $r_{\mathrm{a}}$ and $r_{\mathrm{b}}$
(with $r_{\mathrm{a}}<r_{\mathrm{b}}$, on the right). The radius
where stars complete the first oscillation moves from $r_{\mathrm{a}}$
to $r_{\mathrm{b}}$ in $\bigtriangleup t$. But in the second orbit
on the left, the stars from $r_{\mathrm{b}}$ will already have a
lag of $\bigtriangleup t$ behind those from $r_{\mathrm{a}}$ and
will just be getting a second one, so the third one (the second on
the same side) reaches $r_{\mathrm{b}}$ from $r_{\mathrm{a}}$ in
$3\times\bigtriangleup t$. Every $n$th completed oscillation on
the right side, then moves $n$ times more slowly than the first one.
The situation is similar on the left side, and the shell system is
getting denser. Moreover, the turning point has an additional lag
of $1/2T(r_{\mathrm{TP}})$, because the stars were released in the
center of the host galaxy before their zeroth oscillation. This is
the source of the factor $(n+1/2)$ in Eqs.~(\ref{eq:Trn}) and (\ref{eq:Tr}).

\subsection{Real shell positions and velocities \label{sub:edge} }

Even in the framework of the model of radial oscillations, the position
and velocity of the true edge of the shell cannot be expressed in
a straightforward manner. Photometrically, shells appear as a step
in the luminosity profile of the galaxy with a sharp outer cut-off.
This is because the stars of the cannibalized galaxy occupy a limited
volume in the phase space. With time, the shape of this volume gets
thinner, more elongated, and wrapped around invariant surfaces defined
by the trajectories of the stars in the phase space, increasing its
coincidence with these surfaces. A shell appears close to the points
where the invariant surface is perpendicular to the plane of the sky
\citep{1989ApJ...346..690N}. For the $n$th shell, this is the largest
radius where stars about to complete their $n$th oscillation are
currently located. This radius corresponds to the shell edge (Sect.~\ref{sub:Visage})
and it is always larger than that of the current turning point of
the stars that are completing their $n$th oscillation. Thus, the
shell edge consists of outward-moving stars about to complete their
$n$th oscillation.

\citet{1986A&A...166...53D} state that the stars forming the shell
move with the phase velocity of the shell. While we show that this
holds only roughly, we use this approximation in Sect.~\ref{sec:IOA}
to derive the relation between the shell kinematics and the potential
of the host galaxy.

The position of a star, $r_{\mathrm{*}}$, at a given time $t$ since
the release of the star in the center of the host galaxy is given
by an implicit equation for $r_{\mathrm{*}}$ and is a function of
the star energy, or equivalently the position of its apocenter $r_{\mathrm{ac}}$.%
\footnote{We denote the apocenter of the star corresponding to its energy as
$r_{\mathrm{ac}}$, whereas $r_{\mathrm{TP}}$ (the current turning
point) is the radius at which the stars reach their apocenters at
the time of measurement.%
} For stars with the integer part of $t/[2T(r_{\mathrm{ac}})]$ odd,
the equation reads: 
\begin{equation}
\begin{array}{rcl}
t=(n+1)\sqrt{2} & \int_{0}^{r_{\mathrm{ac}}} & \left[\phi(r_{\mathrm{ac}})-\phi(r')\right]^{-1/2}\mathrm{d}r'-\\
- & \int_{0}^{r_{\mathrm{*}}} & \left[2(\phi(r_{\mathrm{ac}})-\phi(r'))\right]^{-1/2}\mathrm{d}r'.
\end{array}\label{eq:r*-}
\end{equation}
 For stars that have completed an even number of half-periods (only
such stars are found on the shell edge), the equation is 
\begin{equation}
\begin{array}{rcl}
t=n\sqrt{2} & \int_{0}^{r_{\mathrm{ac}}} & \left[\phi(r_{\mathrm{ac}})-\phi(r')\right]^{-1/2}\mathrm{d}r'+\\
+ & \int_{0}^{r_{\mathrm{*}}} & \left[2(\phi(r_{\mathrm{ac}})-\phi(r'))\right]^{-1/2}\mathrm{d}r'.
\end{array}\label{eq:r*+}
\end{equation}
 The first term in Eq.~(\ref{eq:r*+}) corresponds to $n$ radial
periods for the star's energy ($n$ is maximal so that $nT(r_{\mathrm{ac}})<t$),
while the other term corresponds to the time that it takes to reach
radius $r_{\mathrm{*}}$ from the center of the galaxy. Even for the
simplest galactic potentials, these equations are not analytically
solvable and must be solved numerically.

The position of the $n$th shell $r_{\mathrm{s}}$ equals the maximal
radius $r_{\mathrm{*}}$ that solves Eq.~(\ref{eq:r*+}) for the
given $n$.%
\footnote{In the approximation of a constant shell velocity, $v_{\mathrm{s}}$,
and a constant galactocentric acceleration, $a_{0}$ (Sect.~\ref{sec:IOA}),
the distance between the current turning points and the shell radius
is $r_{\mathrm{s}}-r_{\mathrm{TP}}=-v_{\mathrm{s}}^{2}/(2a_{0})$.%
} In symbolic notation 
\begin{equation}
r_{\mathrm{s}}=\mathrm{max}\{r_{\mathrm{*}}(r_{\mathrm{ac}});\left\lfloor t/T(r_{\mathrm{ac}})\right\rfloor =n-1\},\label{eq:rs-pseudoEq}
\end{equation}
 where $r_{\mathrm{*}}(r_{\mathrm{ac}})$ is an implicit function
given by Eq.~(\ref{eq:r*+}). Simultaneously, we require $r_{\mathrm{ac}}$
to satisfy the equation $\left\lfloor t/T(r_{\mathrm{ac}})\right\rfloor =n-1$,
where $\left\lfloor x\right\rfloor $ indicates the integer part of
$x$, so that $\left\lfloor t/T(r_{\mathrm{ac}})\right\rfloor $ is
the number of periods completed by the star since the release of the
star in the center of the host galaxy. Radial period $T(r_{\mathrm{ac}})$
is defined by Eq.~(\ref{eq:Tr}) and $n$ is the serial number of
the shell for which we want to find the edge radius $r_{\mathrm{s}}$.

Such a radius is actually identical to the step in projected surface
density that corresponds to the shell edge (Sect.~\ref{sub:Visage}).
For a shell with nonzero phase velocity the shell edge is always further
from the center than the current turning point, $r_{\mathrm{TP}}<r_{\mathrm{s}}$.
On the other hand, the apocenter $r_{\mathrm{ac}}$ of a star currently
located at the shell edge is obviously further from the center than
the current shell edge position.

The shell velocity $v_{\mathrm{s}}$ is obtained from the numerical
derivative of a set of values of $r_{\mathrm{s}}$ for several close
values of $t$ 
\begin{equation}
v_{\mathrm{s}}=\mathrm{d}r_{\mathrm{s}}/\mathrm{d}t.\label{eq:vs}
\end{equation}

The stellar velocity at the shell edge, $v_{r}(r_{\mathrm{s}})$,
is obtained by inserting $r_{\mathrm{s}}$ with its corresponding%
\footnote{By \textit{corresponding} we mean that the pair of values $r_{\mathrm{s}}=r_{*}$
and $r_{\mathrm{ac}}$ solves Eq.~(\ref{eq:r*+}) for a given time
$t$ since the release of the star in the center of the host galaxy,
a given serial number $n$ of a shell and a given potential of the
host galaxy $\phi(r)$.%
} $r_{\mathrm{ac}}$ into: 
\begin{equation}
v_{r}(r_{\mathrm{*}})=\pm\sqrt{2[\phi(r_{\mathrm{ac}})-\phi(r_{\mathrm{*}})]}.\label{eq:vrr}
\end{equation}
 For the stars following Eq.~(\ref{eq:r*+}), the velocity will be
positive; for the rest, it will be negative. The positive velocity
means that the stars are moving outward. The edge of a shell is exclusively
made up of stars with positive velocities. Recall that the star moves
along radial trajectories.

It is clear that $v_{r}(r_{\mathrm{s}})\leq v_{\mathrm{s}}$. Actually,
$v_{r}(r_{\mathrm{s}})$ is lower than the phase velocity of the shell
(Table~\ref{tab:param}) but the difference between the values of
these velocities is small. At the same time, the position of the shell
for a given time is not far from the current turning point, and their
separation changes slowly in galactic potentials. Thus, the velocity
of the turning points given in Eq.~(\ref{eq:vTP}) is a good approximation
for the shell velocity (Fig.~\ref{fig:vs}). Eq.~(\ref{eq:vTP})
is not generally solvable analytically either, but the numerical calculation
of $v_{\mathrm{TP}}$ is much easier than determining the true shell
velocity $v_{\mathrm{s}}$. The procedure to calculate $v_{\mathrm{s}}$
is described in this section.

\begin{figure}[h]
\centering{}\includegraphics[width=12cm]{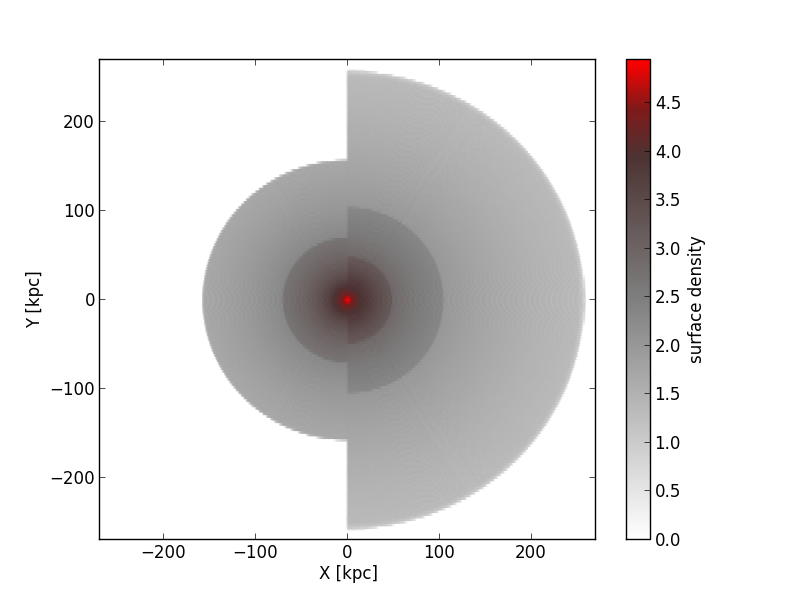}
\caption{\textsf{\small Projected surface density of shells in the host galaxy,
with potential introduced in Sect.~\ref{sec:param}, 2.2\,Gyr after
the release of the stars in the center. The scale bar is logarithmic
in arbitrary units. \label{fig:MRO-SB} }}
\end{figure}

\subsection{Appearance of the shells \label{sub:Visage} }

The model of radial oscillations is primarily used for calculating
the positions of LOSVD maxima. Nevertheless, we can also use it to
derive the spatial and projected surface density of the stars that
form the shell ($\rho(r)$ and $\Sigma_{\mathrm{los}}(R)$, respectively)
and the shape of the LOSVD itself. We do not aim to produce these
quantities with such a precision that would be required for comparison
with observation within this model. But we can still have a look
at them to obtain qualitative insight, although their exact shape
is not important for our work.

To do that, it is not sufficient to know the kinematics as described
in Sect.~\ref{sub:edge} but we need to add an assumption about the
radial dependence of the shell-edge density distribution $\sigma_{\mathrm{sph}}\left(r_{\mathbf{s}}\right)$.
We chose this to correspond to a constant number of stars at the edge
of the shell; for more details, see Sects.~\ref{sub:Equations-of-LOSVD},
\ref{sub:sigma}, and \ref{sub:Nature-of-4peak}. Furthermore we assume
that the density of stars on the shells has uniform angular distribution.
In most cases, we follow the shell kinematics only between $0.9r_{\mathrm{s}}-r_{\mathrm{s}}$
and thus an opening angle of at least $51.7\lyxmathsym{\textdegree}$
is sufficient. 

Fig.~\ref{fig:MRO-SB} shows the projected surface density of the
five outermost shells at 2.2\,Gyr after the release of the stars
in the center of the host galaxy (for parameters of the potential,
see Sect.~\ref{sec:param}). Projected surface density of the host
galaxy itself is not displayed. The opening angle of the shells is
chosen to be the full $180\lyxmathsym{\textdegree}$. Shells with
an odd serial number are to the right, those with an even number to
the left, corresponding to the cannibalized galaxy flying in from
the right hand side of the host galaxy. The whole picture is analogical
to the results of the \textit{N}-particle simulation analyzed in Sect.~\ref{sec:sim-mod}.

\begin{figure}[h]
\centering{}\includegraphics[width=12cm]{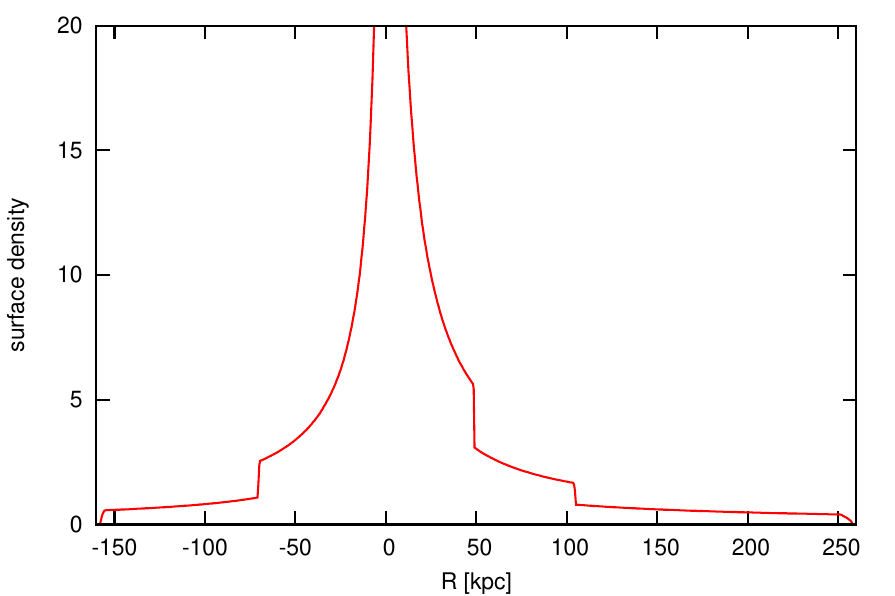}
\caption{\textsf{\small Projected surface density of shells with respect to
the projected radius, the same as in Fig~\ref{fig:MRO-SB}. \label{fig:MRO-SBrad} }}
\end{figure}

In practice, such a projected surface density depends only on the
projected radius $R$ and it is shown also in Fig.~\ref{fig:MRO-SBrad}.
Jumps in the density do indeed correspond to the radius $r_{\mathrm{s}}$
in the sense in which it is introduced in Sect.~\ref{sub:edge},
Eq.~(\ref{eq:rs-pseudoEq}).

\subsection{Kinematics of shell stars \label{sub:LOSVD-rad} \label{sec:4peak}}

In the model of radial oscillations, we can also describe the LOSVD
of a shell at a given time $t$, for a given potential of the host
galaxy $\phi(r)$. Eqs.~(\ref{eq:r*-}) and (\ref{eq:r*+}) determine
the current star position $r_{\mathrm{*}}$ and the shell number $n$
for any apocenter $r_{\mathrm{ac}}$ in a range of energies. The radial
velocity of a star on the particular radius is given by inserting
the corresponding pair of $r_{\mathrm{ac}}$ a $r_{\mathrm{*}}$ in
Eq.~(\ref{eq:vrr}). Naturally, the projections of these velocities
to the selected line of sight (LOS) form the LOSVD, which can be formally
expressed by Eq.~(\ref{eq:Fvlos-final}). To reconstruct the LOSVD,
we have to add an assumption about the radial dependence of the shell-edge
density distribution $\sigma_{\mathrm{sph}}\left(r_{\mathbf{s}}\right)$.
We chose this to correspond to a constant number of stars at the edge
of the shell, $\sigma_{\mathrm{sph}}\left(r_{\mathbf{s}}\right)\propto1/r_{\mathrm{s}}^{2}$.
In Sects.~\ref{sub:Equations-of-LOSVD}, \ref{sub:sigma}, and \ref{sub:Nature-of-4peak},
we deal with this function in detail and show that the particular
choice does not matter much. Here we concisely describe the LOSVD
at the projected radius $R$ which is less than the position of current
turning points, $R<r_{\mathrm{TP}}$. The other case ($r_{\mathrm{TP}}<R<r_{\mathrm{s}}$)
is discussed in Sect.~\ref{sub:Characteristics-of-spectral}.

\begin{figure*}[t]
 \includegraphics[width=7.5cm]{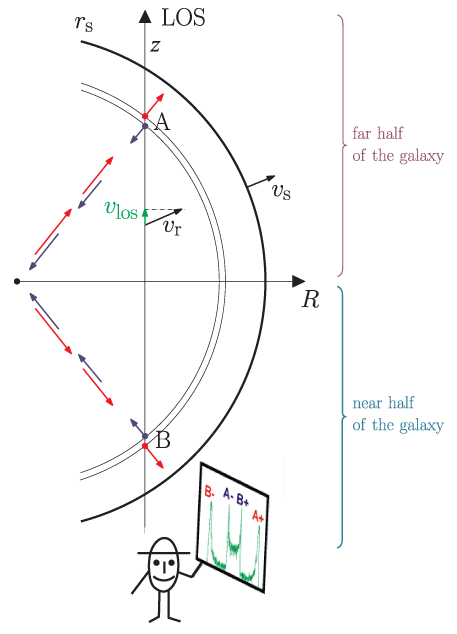}
\includegraphics[width=7.5cm]{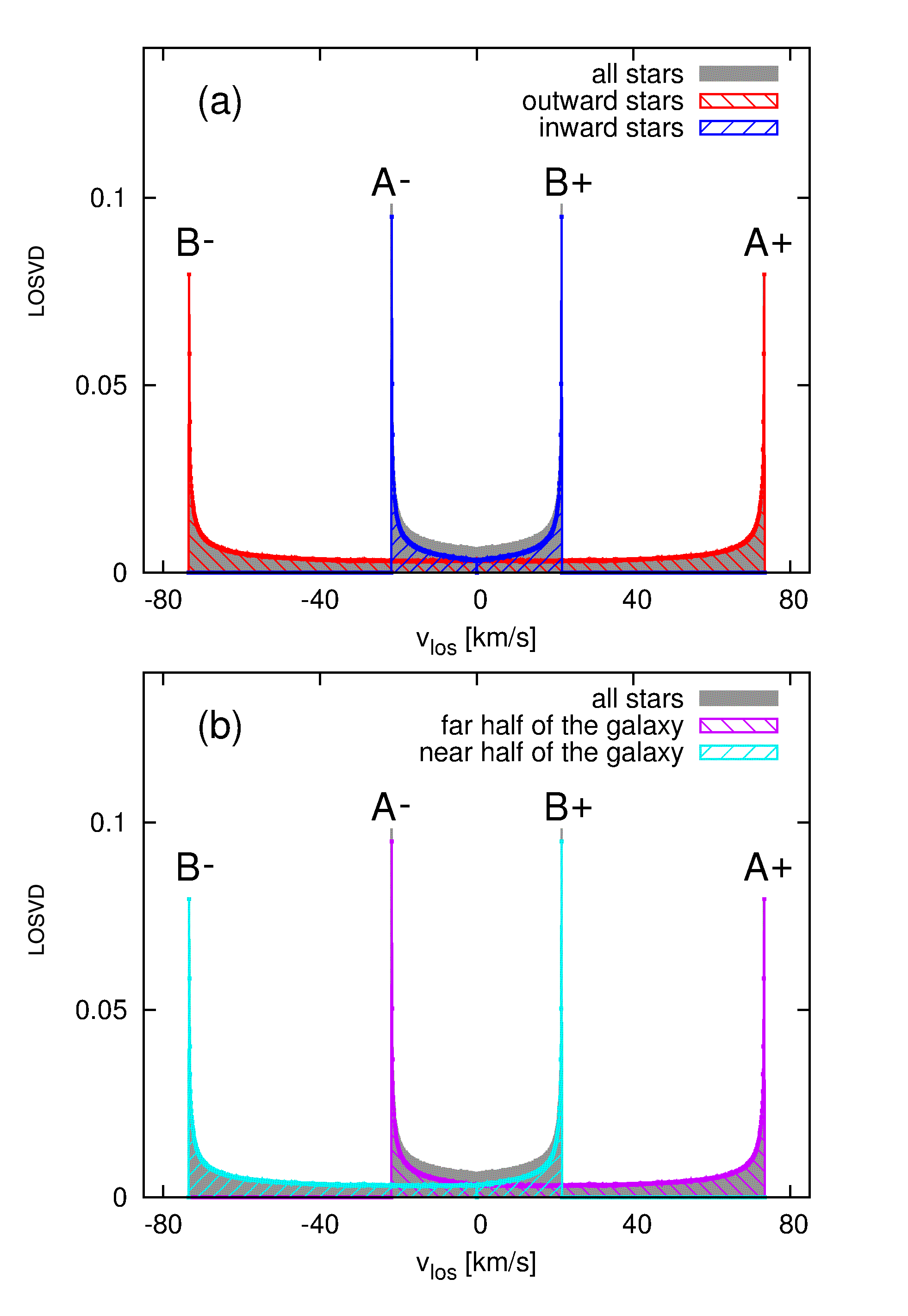}
\caption{\textsf{\small Left: Scheme of the kinematics of a shell with radius
$r_{\mathrm{s}}$ and phase velocity $v_{\mathrm{s}}$. The shell
is composed of stars on radial orbits with radial velocity $v_{\mathrm{r}}$
and LOS velocity $v_{\mathrm{los}}$. Right: The LOSVD at projected
radius $R=0.9r_{\mathrm{s}}$, where $r_{\mathrm{s}}=120$\,kpc (parameters
of the shell are highlighted in bold in Table~\ref{tab:param}),
in the framework of the model of radial oscillations. The profile
does not include stars of the host galaxy, which are not part of the
shell system, and is normalized, so that the total flux equals one.
(a) The LOSVD showing separate contributions from inward and outward
stars; (b) the same profile, separated for contributions from the
near and far half of the host galaxy. \label{fig:Mr.Eggy}\label{fig:anath} }}
\end{figure*}

Mr. Eggy measures the LOSVD of stars in the shell, which is composed
of inward and outward stars on radial trajectories as illustrated
in Fig.~\ref{fig:Mr.Eggy}. The stars near the edge of the shell
move slowly. But it is clear from the geometry that contributions
add up from different galactocentric distances, where the stars are
either still traveling outwards to reach the shell or returning from
their apocenters to form a nontrivial LOSVD. 

For every galactocentric distance $r$ intersected by the line of
sight $z$, there is a different radial stellar velocity $v_{\mathrm{r}}$
and a different projection factor $z/r$. \textcolor{black}{The maximal/minimal
LOS velocity comes from stars at two particular locations along the
line of sight (A and B), both of which are at the same galactocentric
distance for outward or }inward stars (the \textit{radii of maximal
LOS velocity}, Sects.~\ref{sub:Nature-of-4peak} and \ref{sub:Nature-of-4peak};
$r_{\mathrm{A}}^{\mathrm{outward}}=r_{\mathrm{B}}^{\mathrm{outward}}\equiv r_{v\mathrm{max}}^{\mathrm{outward}}$;
$r_{\mathrm{A}}^{\mathrm{inward}}=r_{\mathrm{B}}^{\mathrm{inward}}\equiv r_{v\mathrm{max}}^{\mathrm{inward}}$).
For inward stars, points A and B are closer to the center of the host
galaxy than for outward stars ($r_{v\mathrm{max}}^{\mathrm{inward}}<r_{v\mathrm{max}}^{\mathrm{outward}}$)
as indicated in Fig.~\ref{fig:Mr.Eggy} on the left. This will be
discussed more precisely in Sect.~\ref{sec:Compars} (see also Fig.~\ref{fig:r_vmax}).
\textcolor{black}{The maximal/minimal LOS velocity corresponds to
the intensity maximum of the LOSVD, }as can be seen in the right-hand
panels of Fig.~\ref{fig:anath}. The nature of this correspondence
is explained in Sect.~\ref{sub:Nature-of-4peak}.

The edge of the shell moves outwards with velocity $v_{\mathrm{\mathrm{s}}}$.
At any given instant, the stars that move inwards are returning from
a point where the shell edge was at some earlier time, and so their
apocenter is inside the current shell radius $r_{\mathrm{s}}$. Similarly,
the stars that move outwards will reach the shell edge in the future.
Consequently, the stars that move inwards are always closer to their
apocenter than those moving outwards at the same radius, and their
velocity is thus smaller. The inward stars move toward Mr. Eggy in
the farther of the two points (A) and away from them in the nearer
point (B), while the stars moving outwards behave in the opposite
manner. Together, there are four possible velocities with the maximal
contribution to the LOSVD, resulting in its symmetrical quadruple
shape shown in Fig.~\ref{fig:anath}. In the picture, the intensity
maxima coincide with velocity extremes for separate contributions
to the LOSVD (for more details, see Sect.~\ref{sub:Nature-of-4peak}).

\begin{figure}[h]
\centering{}\includegraphics[width=7.8cm]{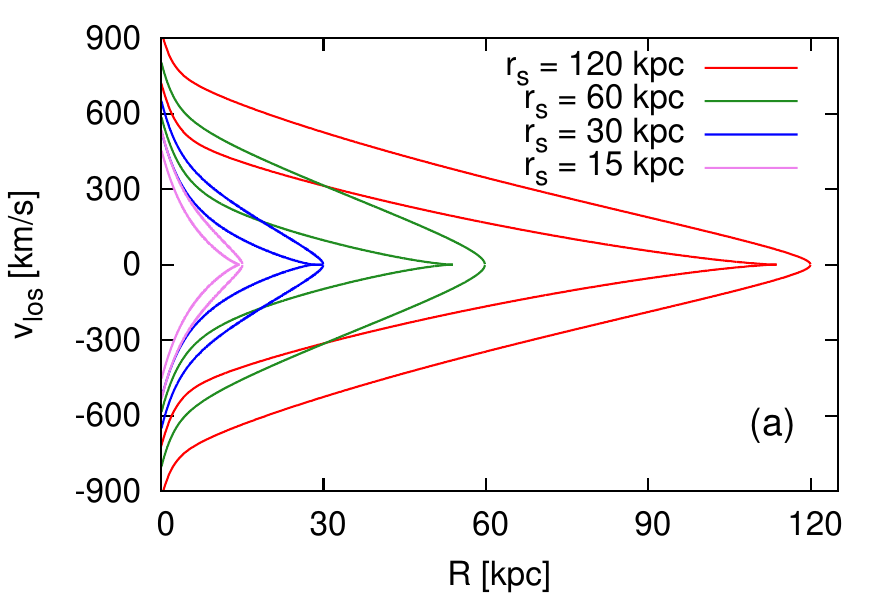}\includegraphics[width=7.5cm]{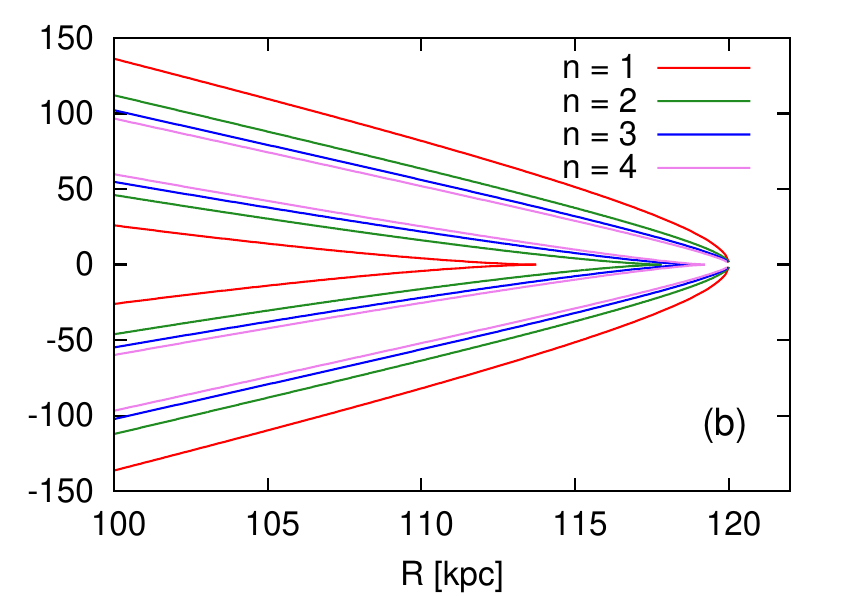}
\caption{\textsf{\small Locations of peaks of the LOSVDs in the framework of
the model of radial oscillations: (a) for the first shell at different
radii, (b) for the first to the fourth shell at the radius of 120\,kpc.
Parameters of all shells are shown in Table~\ref{tab:param}. For
parameters of the host galaxy potential, see Sect.~\ref{sec:param}.
\label{fig:radmax} }}
\end{figure}

\subsection{Characteristics of spectral peaks \label{sub:Characteristics-of-spectral}}

In this section we describe and demonstrate the characteristics of
the LOSVD maxima in the model of radial oscillations using a particular
host galaxy model. We model the potential of the host galaxy as a
double Plummer sphere,\textcolor{black}{{} as described in Sect.~\ref{sec:param}.} 

The separation between peaks of the LOSVD for a given projected radius
$R$ is given by the distance of $R$ from the edge of the shell $r_{\mathrm{s}}$.
The profile shown in Fig.~\ref{fig:anath} corresponds to projected
radius $R=0.9r_{\mathrm{s}}$. The closer to the shell edge, the narrower
the profile is. The separation of the peaks at a given $R$ depends
on the phase velocity of the specific shell, near which we observe
the LOSVD. This velocity is, for a fixed potential, given by the shell
radius and its serial number (Sect.~\ref{sub:TP}). These effects
are illustrated in Fig.~\ref{fig:radmax}, where we show the positions
of the LOSVD peaks for the first shell at different radii $r_{\mathrm{s}}$
and for a shell at 120\,kpc with different serial numbers $n$. Note
that the higher the serial number $n$ at a given radius, the smaller
is the difference in the phase velocity between the two shells with
consecutive serial numbers and thus in the positions of the respective
peaks. Parameters of the corresponding shells can be found in Table~\ref{tab:param}.

\begin{table}[H]
\centering{}%
\begin{tabular}{cccccccc}
\hline 
$t$  & $n$  & $r_{\mathrm{s}}$  & $r_{\mathrm{TP}}$  & $v_{\mathrm{s}}$  & $v_{r}(r_{\mathrm{s}})$  & $v_{\mathrm{TP}}$  & $v_{\mathrm{c}}$ \tabularnewline
Myr  &  & kpc  & kpc  & km$/$s  & km$/$s  & km$/$s  & km$/$s \tabularnewline
\hline 
215  & 1  & 15  & 14.5  & 63.5  & 57.5  & 61.2  & 245\tabularnewline
416  & 1  & 30  & 28.3  & 90.3  & 82.6  & 81.0  & 261\tabularnewline
634  & 1  & 60  & 53.9  & 165.8  & 151.5  & 151.8  & 362\tabularnewline
1006  & 1  & 120  & 113.9  & 142.4  & 133.3  & 141.8  & 450\tabularnewline
\textbf{1722}  & \textbf{2}  & \textbf{120}  & \textbf{117.9}  & \textbf{84.7}  & \textbf{79.4}  & \textbf{84.7}  & \textbf{450}\tabularnewline
2428  & 3  & 120  & 118.9  & 60.3  & 54.6  & 60.3  & 450\tabularnewline
3130  & 4  & 120  & 119.3  & 46.8  & 42.6  & 47.0  & 450\tabularnewline
\hline 
\end{tabular}\caption{\textsf{\small Parameters of shells for which the LOSVD intensity
maxima are shown in Fig.~\ref{fig:radmax}. $t$: time since the
release of stars at the center of the host galaxy, in which the shell
has reached its current radius calculated in the framework of the
model of radial oscillations; $n$: serial number of a shell (Sect.~\ref{sub:TP});
$r_{\mathrm{s}}$: shell radius; $v_{\mathrm{s}}$: shell phase velocity
according to the method described in Sect.~\ref{sub:edge}; $r_{\mathrm{TP}}$:
galactocentric radius of the current turning points of the stars at
this time, given by Eq.~(\ref{eq:Trn}); $v_{r}(r_{\mathrm{s}})$:
radial velocity of the stars at the shell edge; $v_{\mathrm{TP}}$:
phase velocity of the current turning point according Eq.~(\ref{eq:vTP});
$v_{\mathrm{c}}$: circular velocity at the shell-edge radius. For
parameters of the host galaxy, see Sect.~\ref{sec:param}. The shell
that is used in Figs.~\ref{fig:anath}, \ref{fig:zona}--\ref{fig:barevny-apoc},
and \ref{fig:r_vmax}--\ref{fig:120app-Far} is highlighted in bold.
\label{tab:param}}}
\end{table}

\begin{figure}[H]
\centering{}\includegraphics[width=12cm]{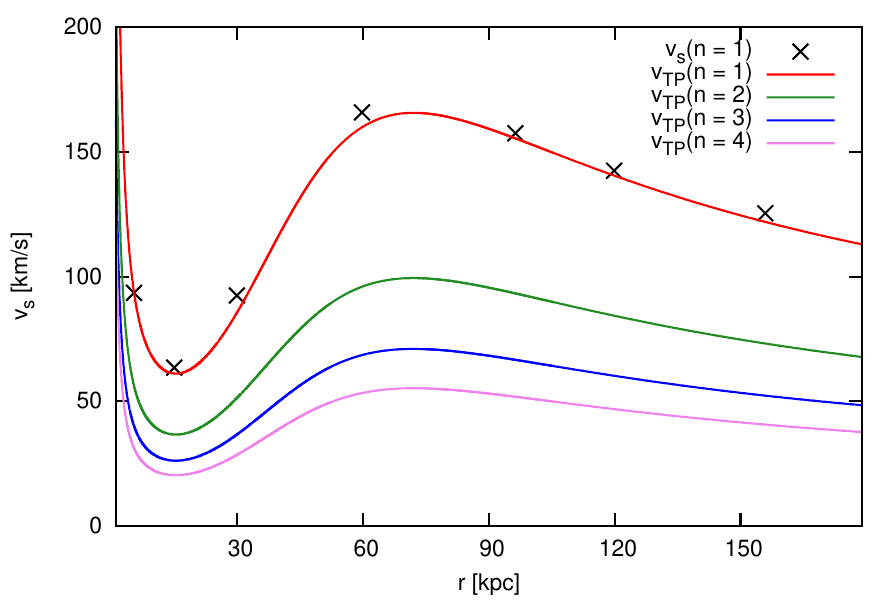}
\caption{\textsf{\small Dependence of the phase velocity of the turning points
on the galactocentric radius for the first four shells according to
Eq.~(\ref{eq:vTP}). For parameters of the host galaxy potential,
see Sect.~\ref{sec:param}. Black crosses show the true velocity
of the first shell calculated for several radii according to the method
described in Sect.~\ref{sub:edge}. \label{fig:vs} }}
\end{figure}

\clearpage

The radial dependence of the phase velocity of the first four shells
in the whole host galaxy is shown in Fig.~\ref{fig:vs}. Using Eq.~(\ref{eq:vTP}),
we see that the velocity of each subsequent shell differs from the
first one only by a factor of $3/(1+2n)$. The large interval of the
galactocentric radii where the shell velocity increases is caused
by the presence of the halo with a large scaling parameter. In fact,
we do not show the shell velocity, but the velocity of the turning
points at the same radius. Nevertheless, these are sufficiently close.
Black crosses show the true velocity of the first shell calculated
for several radii according to the method described in Sect.~\ref{sub:edge}.
For shells of higher $n$, these differences between the phase velocity
of a shell and the corresponding turning point with consecutive serial
numbers are even smaller.

\begin{figure}[H]
\centering{}\includegraphics[width=12cm]{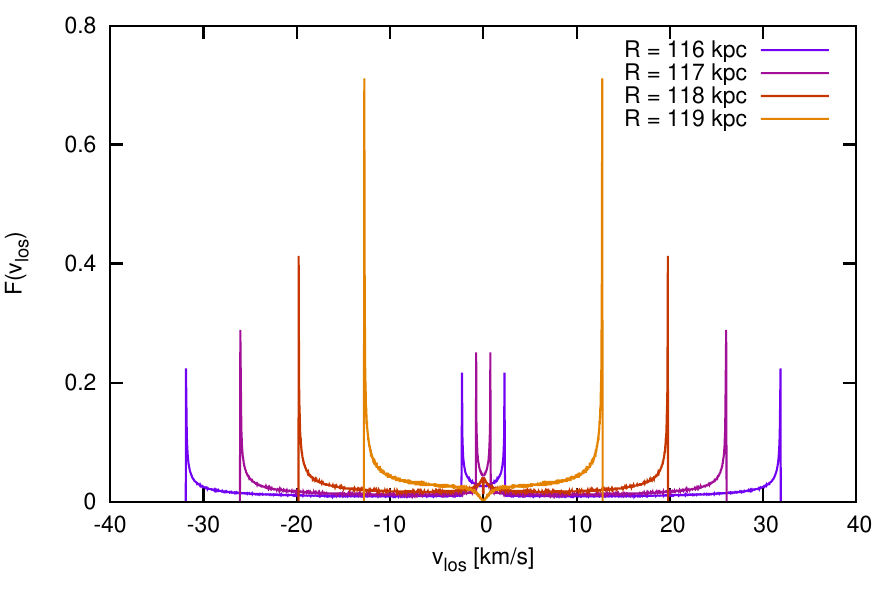}
\caption{\textsf{\small Evolution of the LOSVD near the shell edge for the
second shell at $r_{\mathrm{s}}=120$\,kpc (parameters of the shell
are highlighted in bold in Table~\ref{tab:param}) for the projected
radius 116, 117, 118, and 119\,kpc in the framework of the model
of radial oscillations. In this model, the current turning points
of stars in the shell are at $r_{\mathrm{TP}}=117.9$\,kpc. For $R>r_{\mathrm{TP}}$
the inner maxima disappear. Profiles do not include stars of the host
galaxy, which are not part of the shell system and are normalized
so that the total flux equals one. For parameters of the host galaxy
potential, see Sect.~\ref{sec:param}. \label{fig:zona} }}
\end{figure}

The edge of a moving shell is at the radius, which is always slightly
further from the center than the current turning points. Between these
radii ($r_{\mathrm{TP}}<R<r_{\mathrm{s}}$), there is an intricate
zone, where all the stars of a given shell move outwards. As shown
in Fig.~\ref{fig:zona}, when the LOS radius from lower radii gets
near to the turning points of the stars, the inner maxima of the LOSVD
approach each other until they merge and finally disappear. We actually
see a minimum in the middle of the LOSVD closer to the shell edge
than the current turning points. The intricate zone is much larger
for the first shell. For the shell radius of 120\,kpc in our host
galaxy potential, it occupies 6\,kpc for the first shell, 2\,kpc
for the second one, and less than one kpc for the fourth shell (Table~\ref{tab:param}).

\noindent \vfill{}

\clearpage

\subsection{Equations of LOSVD \label{sub:Equations-of-LOSVD}}

We want to investigate the LOSVD, $F(v_{\mathrm{los}})$, on a given
projected radius $R$ for one particular shell. Assuming cylindrical
symmetry of the shell system, $F^{\mathrm{near}}(v_{\mathrm{los}})=F^{\mathrm{far}}(-v_{\mathrm{los}})$,
where the superscripts indicate the near and far half of the galaxy.
The total LOSVD is obtained adding the two contributions together.

$F^{\mathrm{far}}(v_{\mathrm{los}})$ form the far half of the galaxy
is given by the integral of the distribution of shell stars $f(\mathbf{r},v_{\mathrm{los}})$
along the line of sight

\begin{equation}
F^{\mathrm{far}}(v_{\mathrm{los}})=\int_{0}^{z_{\mathrm{fin}}}f(\mathbf{r},v_{\mathrm{los}})\mathrm{d}z.\label{eq:Fvlos0}
\end{equation}

In the model of radial oscillations, we assume spherical symmetry
of the shell system and thus the distribution function depends only
on galactocentric radius $r$. Moreover, in this model, stars are
located on a three-dimensional hypersurface in the six-dimensional
phase space as they move as if they were released all at once in the
center of the galaxy. In this case $z_{\mathrm{fin}}=\sqrt{r_{\mathrm{s}}^{2}-R^{2}}$.
Furthermore, for a given $r$, in each moment there are only two possible
values for the radial velocity, $v_{r1}$ and $v_{r2}$, therefore
only two possible values for its projection to the line of sight,
thus 
\begin{equation}
f(\mathbf{r},v_{\mathrm{los}})=\rho_{1}(r)\delta[v_{\mathrm{los}}-\frac{z}{r}v_{r1}]+\rho_{2}(r)\delta[v_{\mathrm{los}}-\frac{z}{r}v_{r2}],\label{eq:F-f}
\end{equation}
where $\delta$ is the Dirac delta function; and $\rho_{1}(r)$ and
$\rho_{2}(r)$ are the densities of stars with the velocities $v_{r1}$
and $v_{r2}$, respectively. The values $v_{r1}$ and $v_{r2}$ are
taken from Eq.~(\ref{eq:vrr}), into which we put both pairs $[r;r_{\mathrm{ac}1}]$
and $[r;r_{\mathrm{ac}2}]$, that solve Eqs.~(\ref{eq:r*-}) and
(\ref{eq:r*+}) in Sect~\ref{sub:edge} for given galactic potential
$\phi(r)$, time $t$ since the release of the star, and serial number
$n$ of a shell. In Eqs.~(\ref{eq:r*-}) and (\ref{eq:r*+}) $r$
is substituted for $r_{*}$ and $r_{\mathrm{ac}1}$ or $r_{\mathrm{ac}2}$
for $r_{\mathrm{ac}}$.

To evaluate the density, $\rho(r)$, let us first define $N\left(r_{\mathbf{s}}\right)$
as the probability density for stars to have their shell radius within
an interval $(r_{\mathbf{s}},r_{\mathbf{s}}+\mathrm{d}r_{\mathbf{s}})$.
Then we can define the distribution $\sigma_{\mathrm{sph}}\left(r_{\mathbf{s}}\right)$
as 
\begin{equation}
\sigma_{\mathrm{sph}}\left(r_{\mathbf{s}}\right)=m\frac{N\left(r_{\mathbf{s}}\right)}{r_{\mathbf{s}}^{2}},\label{eq:sigma}
\end{equation}
 where $m$ is the (average) mass of a star. We call $\sigma_{\mathrm{sph}}\left(r_{\mathbf{s}}\right)$
the \textit{shell-edge density distribution}. In this case, $r_{\mathbf{s}}$
is a function of the stellar energy, $r_{\mathbf{s}}(r_{\mathrm{ac}})$,
and stands for the value of the shell edge radius at the moment when
the star with the corresponding energy is at the shell edge.

The radial dependence of $\sigma_{\mathrm{sph}}\left(r_{\mathbf{s}}\right)$
determines the time evolution of the projected surface density of
a shell, Sect.~\ref{sec:Shell-brightness}. The shell-edge density
distribution also determines what the distribution of stellar velocities
was at the time of their release in the center of the host galaxy,
see Appendix~\ref{Apx:IVD}.

The spatial density $\rho(r)$ is given by 
\begin{equation}
\rho(r)=\sum_{i=1}^{2}\frac{r_{\mathrm{s}i}^{2}(r)}{r^{2}}\sigma_{\mathrm{sph}}\left(r_{\mathbf{s}i}(r)\right)\frac{\mathrm{d}r_{\mathrm{s}i}(r)}{\mathrm{d}r},\label{eq:F-rho}
\end{equation}
 where $r_{\mathrm{s}}(r)$ is the location where the stars, currently
situated at the radius $r$, will or did reach their respective shell
edge, and $r_{\mathrm{s}}(r)$ has two solutions, $r_{\mathrm{s1}}(r)$
and $r_{\mathrm{s2}}(r)$, for one $r$, where $0<r<r_{\mathrm{s}}$. 

Eq.~(\ref{eq:F-rho}) is easy to understand: the first fraction,
$r_{\mathrm{s}i}^{2}(r)/r^{2}$, corresponds to the geometrical dilution
of the number of stars during radial movement and the last fraction,
$\mathrm{d}r_{\mathrm{s}i}(r)/\mathrm{d}r$, converts the somewhat
ephemeral distribution function in an artificially chosen parameter
(shell radius) into a coordinate density. The final formal expression
for the LOSVD then reads 

\begin{equation}
F^{\mathrm{far}}(v_{\mathrm{los}})=\int_{0}^{z_{\mathrm{fin}}}\sum_{i=1}^{2}\frac{r_{\mathrm{s}i}^{2}(r)}{r^{2}}\sigma_{\mathrm{sph}}\left(r_{\mathbf{s}i}(r)\right)\frac{\mathrm{d}r_{\mathrm{s}i}(r)}{\mathrm{d}r}\delta[v_{\mathrm{los}}-\frac{z}{r}v_{ri}]\mathrm{d}z.\label{eq:Fvlos-final}
\end{equation}
We call this expression {}``formal'', because -- at least in the
model of radial oscillations -- we are not able to obtain closed analytical
expression for almost any of the terms involved.

\subsection{Shell-edge density distribution and LOSVD \label{sub:sigma}}

For us, the modeling of the shape of the LOSVD is of peripheral importance,
as we will eventually need to know only the positions of the LOSVD
maxima. The peaks occur at the edge of the distribution (Sect~\ref{sub:Nature-of-4peak}).
The determination of the location of the line-of-sight velocity extremes
does not require the knowledge of stellar density profile. We do not
even aim to qualitatively model the shape of the LOSVD, but we can
still show it to obtain a qualitative insight.

\begin{figure}[!b]
\centering{}\includegraphics[width=12cm]{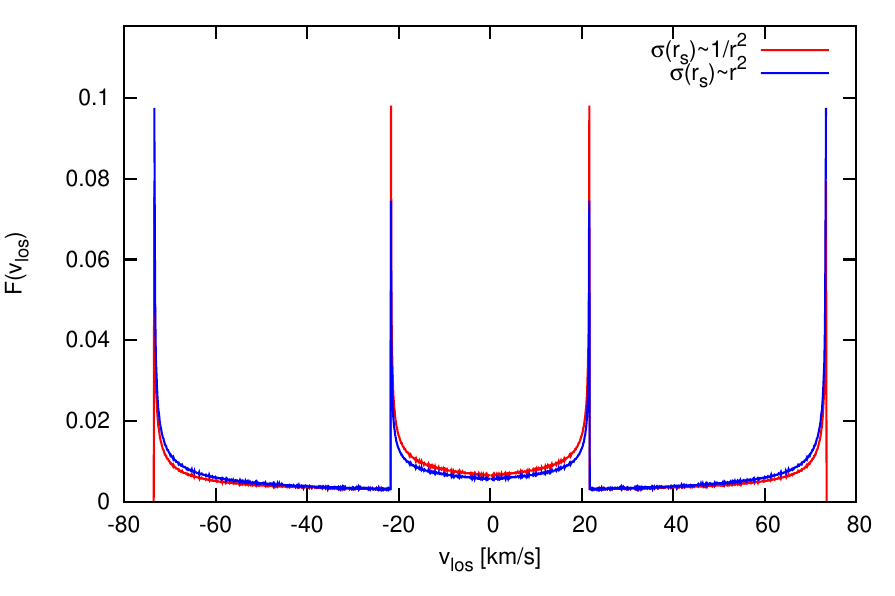}
\caption{\textsf{\small LOSVD of the second shell at $r_{\mathrm{s}}=120$\,kpc
(parameters of the shell are highlighted in bold in Table~\ref{tab:param})
for the projected radius 108\,kpc in the framework of the model of
radial oscillations, where the shell-edge density distribution is
$\sigma_{\mathrm{sph}}\left(r_{\mathbf{s}}\right)\propto r_{\mathrm{s}}^{2}$
for the blue curve and $\sigma_{\mathrm{sph}}\left(r_{\mathbf{s}}\right)\propto1/r_{\mathrm{s}}^{2}$
for the red one. The profiles do not include stars of the host galaxy,
which are not part of the shell system, and are normalized, so that
the total flux equals one. \label{fig:sigma} }}
\end{figure}

If we want to obtain the full LOSVD, we have to choose the radial
dependence of the shell-edge density distribution $\sigma_{\mathrm{sph}}\left(r_{\mathbf{s}}\right)$.
In the framework of the radial-minor-merger origin of shell galaxies,
$\sigma_{\mathrm{sph}}\left(r_{\mathbf{s}}\right)$ depends on the
parameters of the merger that has produced the shells. It is determined
by the energy distribution of stars of the cannibalized galaxy in
the instant of its decay in the center of the host galaxy. But the
energy distribution is principle unknown for real shell galaxies and
it can be very different for various collisions even if we consider
only radial mergers. Thus you need to choose $\sigma_{\mathrm{sph}}\left(r_{\mathbf{s}}\right)$
somehow arbitrary.

For simplicity, we choose the shell-edge density distribution to be
\begin{equation}
\sigma_{\mathrm{sph}}\left(r_{\mathbf{s}}\right)\propto1/r_{\mathrm{s}}^{2},
\end{equation}
corresponding to a shell containing the same number of stars at each
moment. It turns out that no reasonable choice of this function has
an effect on the general characteristics of the LOSVD and the principles
of its formation that we describe in Sect.~\ref{sub:Nature-of-4peak}. 

For illustration, we demonstrate the LOSVD of $\sigma_{\mathrm{sph}}$
increasing as $r^{2}$ and $\sigma_{\mathrm{sph}}$ decreasing as
$1/r^{2}$ in Fig.~\ref{fig:sigma}. For the profiles shown, the
ratio of the inner and outer peaks changes with the change of the
$\sigma_{\mathrm{sph}}$, but the peak positions are unaffected and
the overall shape of the profile does not change significantly. For
shells that were created in a radial minor merger, we can expect the
shell-edge density distribution to rise in the inner part of the host
galaxy, followed by an extensive area of its decrease. The fact that
the main features of the LOSVD do not depend on the choice of $\sigma_{\mathrm{sph}}$
means that our method of measuring the potential of shell galaxies
is not sensitive to the details of the decay of the cannibalized galaxy
It also means that, for the purposes of the modeling the LOSVD of
shells, we can safely pick $\sigma_{\mathrm{sph}}$ of our choice.

\subsection{Nature of the quadruple-peaked profi{}le \label{sub:Nature-of-4peak}}

\begin{figure}[h]
\begin{centering}
\includegraphics[width=12cm]{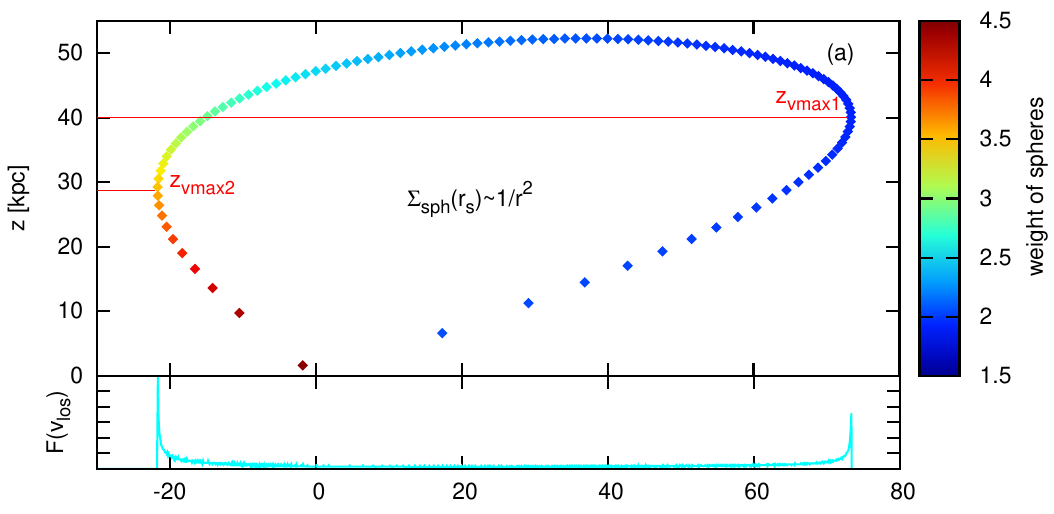}
\par\end{centering}

\centering{}\includegraphics[width=12cm]{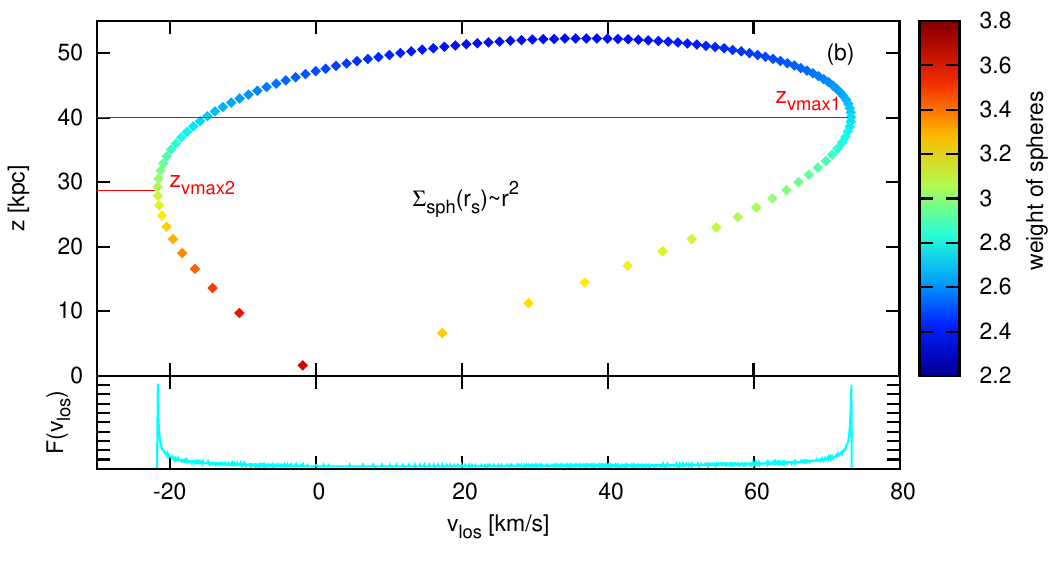}
\caption{\textsf{\small The LOSVD and its different contributions along the
line of sight $z$ for the far half of the host galaxy for the second
shell at $r_{\mathrm{s}}=120$\,kpc (parameters of the shell are
highlighted in bold in Table~\ref{tab:param}) for the projected
radius 108\,kpc in the framework of the model of radial oscillations.
Graphs (a) and (b) differ in the choice of }$\Sigma_{\mathrm{sph}}\left(r_{\mathbf{s}}\right)$:\textsf{\small{}
(a) $\Sigma_{\mathrm{sph}}(r_{\mathrm{s}})\propto1/r_{\mathrm{s}}^{2}$,
(b) $\Sigma_{\mathrm{sph}}(r_{\mathrm{s}})\propto r_{\mathrm{s}}^{2}$.
\label{fig:barevny} }}
\end{figure}

Now we will show, why the LOSVD is so insensitive to the choice of
the radial dependence of the shell-edge density distribution $\sigma_{\mathrm{sph}}\left(r_{\mathbf{s}}\right)$.
Fig.~\ref{fig:barevny} shows the formation of the quadruple-peaked
profile for the far half of the galaxy (that is, for positive values
of $z$) at particular projected radius $R$. The inner peak is located
to the left, the outer one to the right (Fig.~\ref{fig:barevny}
-- lower panels). For the near half of the galaxy, the graph is simply
reflected along the axis $v_{\mathrm{los}}=0$. To help visualize
the problem, we show the individual contributions to the LOSVD from
stars with different shell radii that correspond to different points
along the line of sight. To allow that, we discretize their continuous
distribution $\sigma_{\mathrm{sph}}\left(r_{\mathbf{s}}\right)$ to
a set of equidistant spheres. Each of the spheres carries a density
of stars obtained by integration of the distribution $\sigma_{\mathrm{sph}}\left(r_{\mathbf{s}}\right)$
over a small range in shell radii as follows: 
\begin{equation}
\Sigma_{\mathrm{sph}}(r_{\mathrm{s}})=\int_{r_{\mathrm{s}}-\Delta r_{\mathrm{s}}/2}^{r_{\mathrm{s}}+\Delta r_{\mathrm{s}}/2}\sigma_{\mathrm{sph}}(r)\mathrm{d}r\label{eq:SIGMA}
\end{equation}
to represent the given part of the distribution. To each of the spheres,
we associate the weight 
\begin{equation}
I=(r_{\mathrm{s}}/r)^{2}\Sigma_{\mathrm{sph}}(r_{\mathrm{s}})\frac{r}{\sqrt{r^{2}-R^{2}}},
\end{equation}
which shows its contribution to the LOSVD. Similarly to Eq.~(\ref{eq:F-rho}),
the term $(r_{\mathrm{s}}/r)^{2}$ simply takes into account the geometric
dilution of the sphere with radius. The factor $r/\sqrt{r^{2}-R^{2}}$
reflects the fact that spheres with different radii are intersected
by the line of sight under different angles. The color of the point
encodes the weight $I$ for each contributing sphere -- the upper
panels of both figures (a) and (b) in Fig.~\ref{fig:barevny}. Note
that to each value of $z$ we can assign the corresponding $r=\sqrt{z^{2}+R^{2}}$.

To evaluate which spheres contribute to the observed shell profile,
we let them evolve (either backwards or forwards) from the point in
time when they will reach or have reached their shell radii to the
time of the observation and we place them on the exact locations they
reach after this evolution. This operation is a discrete analog of
the term $\mathrm{d}r_{\mathrm{s}}(r)/\mathrm{d}r$ in Eq.~(\ref{eq:F-rho})
which transfers the distribution in $r_{\mathrm{s}}$ into the distribution
in actual positions at the time of observation. In the figures, we
can see its effects as dilution and thickening of the distribution
of the colored points in different parts of the plane. The points
are located at a curve in the $v_{\mathrm{los}}-z$ plane. The shape
of the curve is determined by the $\delta$ functions in Eq.~(\ref{eq:F-f}).
Finally, we count the spheres in bins of $v_{\mathrm{los}}$ irrespective
of their $z$ coordinate to obtain the LOSVD (the lower panels of
both figures (a) and (b) in Fig.~\ref{fig:barevny}).

$\sigma_{\mathrm{sph}}\left(r_{\mathbf{s}}\right)$ and $\Sigma_{\mathrm{sph}}\left(r_{\mathbf{s}}\right)$
are different quantities, but from Eq.~(\ref{eq:SIGMA}) it is clear
that once we choose the radial dependence of one of them, the other
has to have the same dependence. In Fig.~\ref{fig:barevny}~(a),
this function is chosen to be $\Sigma_{\mathrm{sph}}(r_{\mathrm{s}})\propto1/r_{\mathrm{s}}^{2}$,
which is the formula we generally use for $\sigma_{\mathrm{sph}}\left(r_{\mathbf{s}}\right)$
or $\Sigma_{\mathrm{sph}}\left(r_{\mathbf{s}}\right)$ unless specifically
noted otherwise. In Fig.~\ref{fig:barevny}~(b) we show that the
quadruple-peaked shape appears even for a completely reversed density
function $\Sigma_{\mathrm{sph}}(r_{\mathrm{s}})\propto r_{\mathrm{s}}^{2}$.
The densities are calculated relative to the density at the radius
of current turning points, $\Sigma_{\mathrm{sph}}(r_{\mathrm{TP}})=1$. 

The bottom panels of both figures in Fig.~\ref{fig:barevny} show
the LOSVD itself. Although the weights of every point are different
for the different choices of $\Sigma_{\mathrm{sph}}(r_{\mathrm{s}})$,
the dominant effect is the bending of the curve in the $v_{\mathrm{los}}-z$
plane around $z_{v\mathrm{max1/2}}$ at the LOS velocity extremes
and thus the points around these extremes are much denser for a unit
of the $v_{\mathrm{los}}$ than in the inner part of the distribution.
This effect is completely the same for both (a) and (b). The change
of the weight causes relative differences in the heights of the LOSVD
peaks, but in no way casts any doubts over their existence at the
extremes of the projected velocity.

\begin{figure}[!b]
\centering{}\includegraphics[width=12cm]{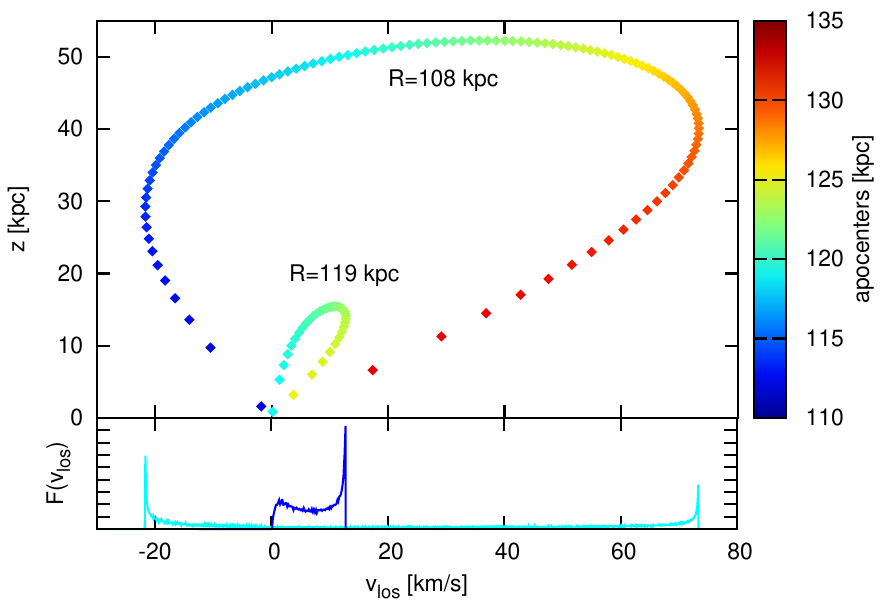}
\caption{\textsf{\small LOSVD and its individual contributions along the line
of sight for the far half of the host galaxy for the second shell
at $r_{\mathrm{s}}=120$\,kpc (parameters of the shell are highlighted
in bold in Table~\ref{tab:param}) for the projected radius 108\,kpc
(light blue curve in the lower panel) and 119\,kpc (dark blue curve
in the lower panel) in the framework of the model of radial oscillations.
\label{fig:barevny-apoc} }}
\end{figure}

The points $z_{v\mathrm{max1/2}}$ correspond to the radii of maximal
LOS velocity $r_{v\mathrm{max1/2}}$ (points A and B from Sect.~\ref{sec:4peak})
through the equation $r_{v\mathrm{max}}=\sqrt{z_{v\mathrm{max}}^{2}+R^{2}}$.
If the density in the vicinity of these points quickly dropped towards
zero, the peaks could disappear. This should certainly not happen
at all projected radii around the shell edge, because then there would
be no shell at all. Moreover, such a gap has no physical foundation
for shells of radial-minor-merger origin. On the other hand, if the
shell has rather stream-like nature, the stars may be present in only
one half of the galaxy. Then just one inner and one outer peak would
be observable (i.e. the inner peak at negative velocities and the
outer at positive or vice versa). This is probable the case of the
so-called Western Shelf in the Andromeda galaxy \citep{fardal12}.

The only case of disappearance of peaks, which is natural to the model
of radial oscillations, occurs for the inner peaks in the zone between
the current turning points $r_{\mathrm{TP}}$ and the shell edge.
The reason is evident from Fig.~\ref{fig:barevny-apoc} where we
show the contributions along the line of sight at projected radii
$R=108$\,kpc and $R=119$\,kpc, while the edge of the shell is
at $r_{\mathrm{s}}=120$\,kpc and the current turning points at $r_{\mathrm{TP}}=118$\,kpc.
The color code in this case encodes the positions of the apocenters
of the stars contributing to the respective LOSVD. The location of
the apocenters $r_{\mathrm{ac}}$ roughly corresponds to the radii
$r_{\mathrm{s}}(r)$ where the stars will or have been located during
their passage through the edge of the shell. The radius $r_{\mathrm{s}}(r)$
is obviously always slightly closer to the center of the host galaxy
than the apocenters of the respective stars. For the shell that we
show (the second shell at $r_{\mathrm{s}}=120$\,kpc) the difference
of these radii is (for the chosen potential of the host galaxy) approximately
$r_{\mathrm{ac}}-r_{\mathrm{s}}(r)=2$\,kpc.%
\footnote{In the approximation of a constant shell velocity, $v_{\mathrm{s}}$,
and a constant galactocentric acceleration, $a_{0}$ (Sect.~\ref{sec:IOA}),
the following holds: $r_{\mathrm{ac}}-r_{\mathrm{s}}(r)=-v_{\mathrm{s}}^{2}/(2a_{0})$.
This is an expression for the difference of the radius of apocenter
of a star and the radius of the passage of the very same star through
the edge of the shell. Incidentally (and only in this approximation),
the same expression holds for the difference of the current turning
point and the shell radius $r_{\mathrm{TP}}-r_{\mathrm{s}}$ even
though the current turning point represents the apocenter for stars
that have already been on the shell edge.%
}

\section{Stationary shell}

\citetalias{mk98} studied the kinematics of a stationary shell
\textendash{} a monoenergetic spherically symmetric system of stars
oscillating on radial orbits in a spherically symmetric potential.
They derived an analytic approximation for the LOSVD in the vicinity
of the shell edge, predicting a double-peaked spectral-line profile,
where the locations of these peaks are connected via a simple relation
to the gradient of the potential of the host galaxy at the shell edge.

As our work expands the analysis of \citetalias{mk98}, we also
show the derivation of their results. In Sect.~\ref{sec:N-Simulations},
we apply also their method to the simulated data and compare the results
with the results of our methods. Furthermore, the approximation of
a stationary shell allows some calculations that prove impossible
for a moving shell, such as the calculation of an explicit analytical
shape of the LOSVD.

The stationary shell differs qualitatively from the model of radial
oscillations, because it requires stars to appear at all radii between
$R$ and $r_{\mathrm{s}}$, where $R$ is the projected radius at
which we observe the LOSVD. But because all the stars in this system
have the same energy, it is impossible to create such a situation
by releasing all of the stars at one time from one point.

\subsection{Motion of stars in a shell system}

Let the shell edge be again $r_{\mathrm{s}}$. Stars at this radius
are in their apocenters and thus stationary. We assume following:
\begin{itemize}
\item stars are on strictly radial orbits
\item all stars have the same energy
\item stars are near the shell edge, so $1-r/r_{\mathrm{s}}\ll1$ 
\end{itemize}
The radial velocity of stars at a given galactocentric radius $r$
is then given by the difference of the host galaxy potential $\phi$
at this radius and at the edge of the shell

\begin{equation}
v_{r\pm}=\pm\sqrt{2\left[\phi(r_{\mathrm{s}})-\phi(r)\right]}.
\end{equation}
The velocity projected to the line of sight is 
\begin{equation}
v_{\mathrm{los}}^{2}=\left(1-R^{2}/r^{2}\right)2\sqrt{\phi(r_{\mathrm{s}})-\phi(r)}.\label{eq:vlos^2}
\end{equation}
 Expanding this function around $r=r_{\mathrm{s}}$ we obtain 
\begin{equation}
\begin{array}{rcl}
v_{\mathrm{los}}^{2}= &  & -2\left(r-r_{\mathrm{s}}\right)\phi'(r_{\mathrm{s}})\left(1-R^{2}/r_{\mathrm{s}}^{2}\right)-\\
 &  & -\left(r-r_{\mathrm{s}}\right)^{2}\frac{1}{r_{\mathrm{s}}^{3}}\left[4R^{2}\phi'(r_{\mathrm{s}})+r_{\mathrm{s}}\left(r_{\mathrm{s}}^{2}-R^{2}\right)\phi''(r_{\mathrm{s}})\right]+\\
 &  & +o\left[\left(r-r_{\mathrm{s}}\right)^{3}\right],
\end{array}\label{eq:o3}
\end{equation}
 where $\phi'(r_{\mathrm{s}})$ and $\phi''(r_{\mathrm{s}})$ are
the first and the second derivative of the potential of the host galaxy
with respect to the radius at $r_{\mathrm{s}}$. Near the edge of
the shell ($\left|R-r_{\mathrm{s}}\right|\ll r_{\mathrm{s}}$), the
following holds: 
\begin{equation}
\left(1-R^{2}/r_{\mathrm{s}}^{2}\right)\simeq2\frac{r_{\mathrm{s}}-R}{r_{\mathrm{s}}}.\label{eq:1-R2/rs2}
\end{equation}
Using Eq.~(\ref{eq:1-R2/rs2}) and neglecting all terms of the order
$o\left[\left(R-r_{\mathrm{s}}\right)^{3}\right]$, Eq.~(\ref{eq:o3})
takes the form 
\begin{equation}
v_{\mathrm{los}}^{2}\simeq4\left(r_{\mathrm{s}}-r\right)\left(r-R\right)\frac{\phi'(r_{\mathrm{s}})}{r_{\mathrm{s}}}.\label{eq:vlos-Mk98}
\end{equation}
The derivative of this expression is zero when 
\begin{equation}
r=\frac{1}{2}(R+r_{\mathrm{s}}).\label{eq:rvmaxMK}
\end{equation}
 thus the extremes of the projected velocity, $v_{\mathrm{los,max}\pm}$,
must follow 
\begin{equation}
v_{\mathrm{los,max}\pm}=\pm v_{\mathrm{c}}(1-R/r_{\mathrm{s}}),\label{eq:vlos,maxMK}
\end{equation}
where $v_{\mathrm{c}}=\sqrt{r_{\mathrm{s}}\phi'(r_{\mathrm{s}})}$
is the circular velocity in the potential of the host galaxy at the
radius of the shell. If we call $\bigtriangleup v_{\mathrm{los}}=2\left|v_{\mathrm{los,max}\pm}\right|$
the difference between the minimal and maximal LOS velocity at the
given galactocentric radius, the derivative of this variable directly
gives the derivative of the gravitational potential of the galaxy
at the radius of the shell edge (equation (7) in \citetalias{mk98}):
\begin{equation}
\frac{\mathrm{d}(\bigtriangleup v_{\mathrm{los}})}{\mathrm{d}R}=-2\frac{v_{\mathrm{c}}}{r_{\mathrm{s}}}.\label{eq:sklonMK}
\end{equation}

\subsection{Constant acceleration \label{sub:z/r}}

Alternatively, we may assume that the stars move in a gravitational
field of a constant acceleration $a_{0}=-\phi'(r_{\mathrm{s}})$.
In such a case, the radial velocity $v_{r}$ of a star at radius $r$
will by given by 
\begin{equation}
v_{r\pm}=\pm\sqrt{2a_{0}(r-r_{\mathrm{s}})}\label{eq:vrMK}
\end{equation}
and its projection to the line of sight 
\begin{equation}
v_{\mathrm{los}}^{2}=\left(v_{r\pm}z/r\right)^{2}=-2a_{0}(r_{\mathrm{s}}-r)\left(1-R^{2}/r^{2}\right),\label{eq:vlos-a0}
\end{equation}
 where $R$ and $z$ denote the projected radius and the distance
along the line of sight, respectively. The center of the host galaxy
is located at $R=0$ and $z=0$. Comparing Eq.~(\ref{eq:vlos-Mk98})
and Eq.~(\ref{eq:vlos-a0}) , we obtain an approximative relation
for the projection factor $z/r$ near the edge of the shell

\begin{equation}
z/r=\sqrt{1-R^{2}/r^{2}}\simeq\sqrt{2(r/r_{\mathrm{s}}-R/r_{\mathrm{s}})}.\label{eq:z/r}
\end{equation}
We use this relation in Sect.~\ref{sub:vlosFardal} in order to calculate
the extremes of the LOS velocity in the approximation of a shell with
a constant phase velocity.

\subsection{LOSVD}

Eq.~(\ref{eq:sklonMK}) shows, that by measuring the width of the
projected velocity distribution at different radii near the shell
edge we can easily obtain the gradient of the potential of the host
galaxy at the shell edge. Measuring the extremes of the LOS velocity
may prove very difficult in practice, particularly because of the
contamination of the signal from the shell by the light of the host
galaxy. For the stationary shell, we can however calculate the shape
of the LOSVD explicitly and it turns out that the extremes of the
LOS velocity correspond to the maxima of the intensity in the LOSVD,
as shown below in this section.

\begin{figure}[h]
\centering{}\includegraphics[width=12cm]{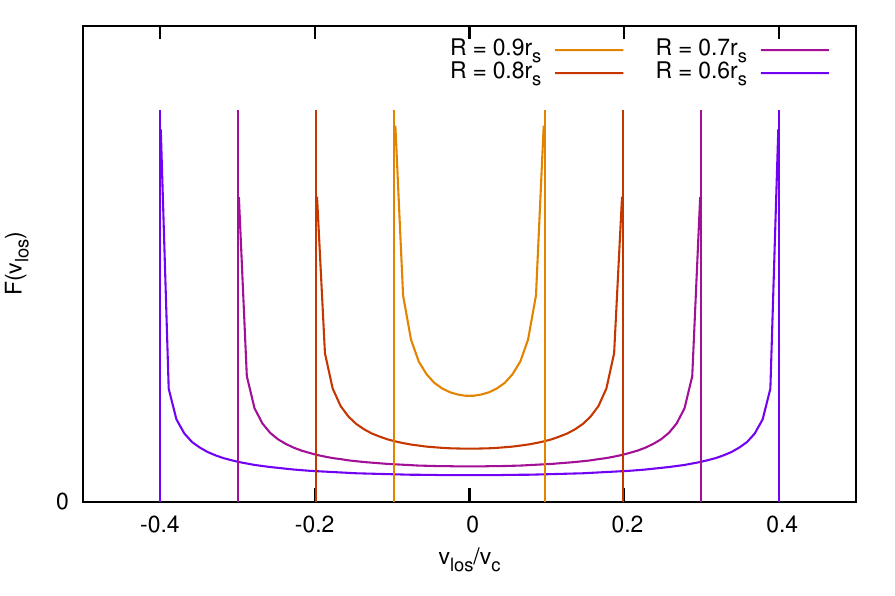}
\caption{\textsf{\small LOSVD of the stationary shell at four projected radii
according to Eq.~(\ref{eq:Fvlos}). }\label{fig:Fvlos} }
\end{figure}

\citetalias{mk98} derived the analytical form of LOSVD, $F(v_{\mathrm{los}})$,
in the approximation for the projected radius close to the edge of
a stationary shell $r_{\mathrm{s}}$. For the construction of the
LOSVD, we start with Eq.~(\ref{eq:Fvlos0}) -- the integration of
the stellar distribution function in the shell along the line of sight
at the chosen projected radius $R$. The problem is again spherically
symmetric, thus the distribution depends only on the radius $r$.
Moreover, for a stationary shell, the spatial density near the shell
edge is proportional to $\rho(r)\varpropto(v_{r}r^{2})^{-1}$, thus
it is useful to express the distribution function in radial velocity

\begin{equation}
f(\mathbf{r},v_{\mathrm{los}})=f(r,v_{r})\frac{\mathrm{d}v_{r}}{\mathrm{d}v_{\mathrm{los}}}.
\end{equation}
It follows from Eq.~(\ref{eq:vlos-Mk98}) that a particular value
of the projected velocity can be found only at two specific galactocentric
radii $r_{\pm}$ along the line of sight 

\begin{equation}
r_{\pm}=r_{\mathrm{s}}/2\sqrt{R/r_{\mathrm{s}}+1\pm\left[(1-R/r_{\mathrm{s}})^{2}-\left(v_{\mathrm{los}}/v_{\mathrm{c}}\right)^{2}\right]}.
\end{equation}
 Note that at a particular galactocentric radius, the value of the
radial velocity is fully determined in the case of a stationary shell,
see Eq.~(\ref{eq:vrMK}). Thus 
\begin{equation}
f(r,v_{r})=\frac{k}{v_{r}r^{2}}\delta(v_{r}-v_{r\pm}),
\end{equation}
 where $\delta$ is the Dirac delta function and $k$ is a constant
of proportionality of the density at the given shell radius. The LOSVD
then take the form

\begin{equation}
F(v_{\mathrm{los}})=\int\frac{k}{v_{r}r^{2}}\delta(v_{r}-v_{r\pm})\frac{\mathrm{d}z}{\mathrm{d}v_{\mathrm{los}}}\mathrm{d}v_{r}
\end{equation}
yielding after the integration
\begin{equation}
F(v_{\mathrm{los}})=\frac{kr_{\mathrm{s}}^{2}\left|v_{\mathrm{los}}\right|}{2v_{\mathrm{c}}}\left[\frac{1}{r_{+}z_{+}v_{r+}\left|R+r_{\mathrm{s}}-2r_{+}\right|}+\frac{1}{r_{-}z_{-}v_{r-}\left|R+r_{\mathrm{s}}-2r_{-}\right|}\right],\label{eq:Int(Fvlos)}
\end{equation}
 where $z_{\pm}=(r_{\pm}^{2}-R^{2})$. Eq.~(\ref{eq:Int(Fvlos)})
can be further simplified for $r_{\pm}$ near $r_{\mathrm{s}}$ and
assuming $1-R/r_{\mathrm{s}}\ll1$ to obtain a final relation (equation
(15) in \citetalias{mk98}) 
\begin{equation}
F(v_{\mathrm{los}})\propto1/\left[r_{\mathrm{s}}\sqrt{(1-R/r_{\mathrm{s}})^{2}-\left(v_{\mathrm{los}}/v_{\mathrm{c}}\right)^{2}}\right].\label{eq:Fvlos}
\end{equation}
 The function $F(v_{\mathrm{los}})$ has a clear double-peaked profile,
symmetric around zero (or rather the overall velocity of the system).
Examples of such a profile are shown in Fig.~\ref{fig:Fvlos}.

\subsection{Comparison with the model of radial oscillations}

\begin{figure}[H]
\begin{centering}
\includegraphics[width=12cm]{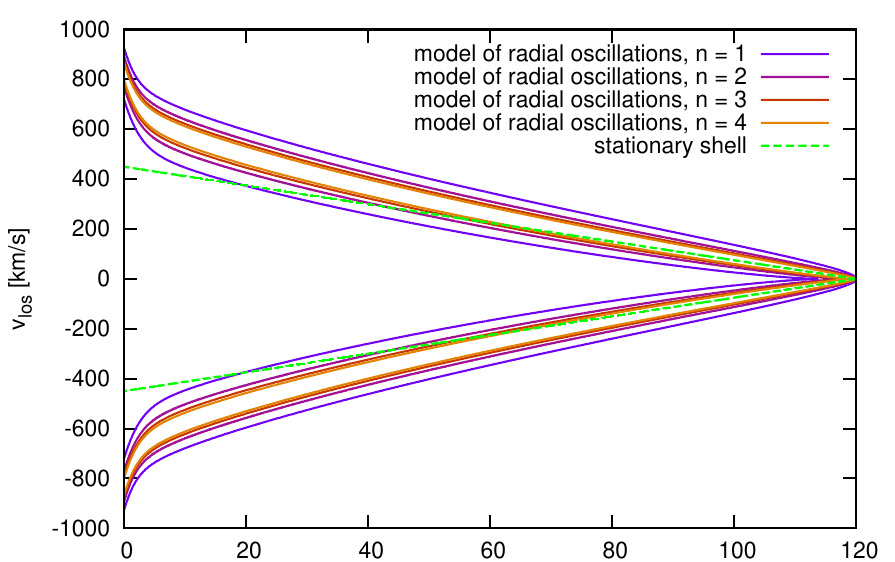}
\par\end{centering}

\centering{}\includegraphics[width=12cm]{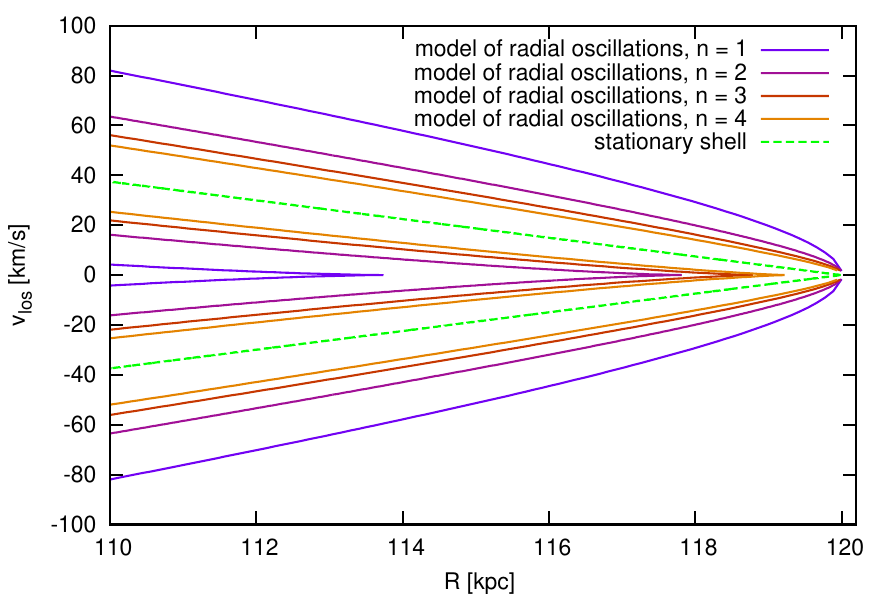}
\caption{\textsf{\small LOSVD peak locations for the stationary shell at the
radius of 120\,kpc according to Eq.~(\ref{eq:vlos,maxMK}) (green
dashed lines); and for the first four shells at the radius of 120\,kpc
(parameters of the shells are listed in Table~\ref{tab:param}) according
to the model of radial oscillations (Sect.~\ref{sub:LOSVD-rad}).
The upper panel shows the whole range of radii, the lower zooms in
on the edge of the shell. \label{fig:MK98} }}
\end{figure}

The approximation of the stationary model differs qualitatively from
the model of radial oscillations in that there is only a double-peaked
profile (instead of a quadruple-peaked one). If the real shell galaxies
are of radial-minor-merger origin, they would rather exhibit a profile
with four peaks (Sect.~\ref{sec:4peak}). Nevertheless, we can compare
the locations of the two peaks of the stationary shell with the model
(Sect.~\ref{sub:LOSVD-rad}) in Fig.~\ref{fig:MK98}. We have inserted
the values of the shell radius $r_{\mathrm{s}}=120$\,kpc and the
circular velocity at the edge of the shell in the chosen potential
$v_{\mathrm{c}}=450$\,km$/$s (for parameters of the host galaxy
potential, see Sect.~\ref{sec:param}) into Eq. (\ref{eq:vlos,maxMK}).

On the other hand the model of radial oscillations uses the complete
knowledge of the potential and the velocity of the shell at different
times derived from it. The higher is the number of the shell, the
lower is its velocity and the closer are the peaks of the quadruple-peaked
profile to each other and to the green line of the stationary shell.
However this holds only near the edge of the shell. For lower radii,
the approximation of a stationary shell causes the positions of the
peaks to diverge from the model of the radial oscillations.

\section{Constant acceleration and shell velocity \label{sec:IOA}}

Now we will leave the stationary case and look at the kinematics of
a moving shell. The nonzero velocity of the shell complicates the
kinematics of shells in two aspects. Due to the energy difference
between inward and outward stars at the same radius, the LOSVD peak
is split into two, see Fig.~\ref{fig:Mr.Eggy}, and the shell edge
is not at the radius of the current turning point, but slightly further
from the center of the host galaxy. In this section, we describe
the LOSVD of such a shell using the assumption of a locally constant
galactic acceleration together with the assumption of a locally constant
shell phase velocity. We call it the \textit{approximation of constant
acceleration and shell velocity}. In addition, we assume that the
velocity of stars at the edge of the shell is equal to the phase velocity
of the shell.

This approximation is nothing but a modification of the model of
radial oscillations for a constant acceleration and shell velocity
and thus the concept that the stars behave as if they were released
in the center of the host galaxy at the same time and their distribution
of energies is continuous is still valid in this approximation.

\subsection{Motion of a star in a shell system \label{sub:star-app}}

The galactocentric radius of the shell edge is a function of time,
$r_{\mathrm{s}}(t)$, where $t=0$ is the moment of measurement and
$r_{\mathrm{s}}(0)=r_{\mathrm{s0}}$ is the position of the shell
edge at this time. We assume following:
\begin{itemize}
\item stars are on strictly radial orbits
\item locally constant value of the radial acceleration $a_{0}$ in the
host galaxy potential%
\footnote{By \textit{locally constant} we mean that we apply one constant value
of radial acceleration or shell velocity to the calculation of the
stellar kinematics for one shell in the whole range of radii of interest.
Nevertheless, we use a different value for different shells, even
when considering stars at the same radii. Moreover note, that for
stars that give the highest contribution to the LOSVD peaks, the range
$0-r_{\mathrm{s0}}$ in projected radii corresponds approximately
to $1/2r_{\mathrm{s0}}-r_{\mathrm{s0}}$ in galactocentric radii.
\label{fn:loc.const}%
}
\item a locally constant velocity of the shell edge $v_{\mathrm{s}}$\textsuperscript{\ref{fn:loc.const}}
\item stars at the shell edge have the same velocity as the shell%
\footnote{In Sect.~\ref{sub:edge} we have discussed that the stars at the
shell edge in fact do not have the same velocity as the shell, but
in Table~\ref{tab:param} we show using examples that these velocities
are very similar.%
}
\end{itemize}
The galactocentric radius of each star is at any time $r(t)$, while
$t_{\mathrm{s}}$ is the time when the star could be found at the
shell edge $r_{\mathrm{s}}(t_{\mathrm{s}})$. Then the equation of
motion and the initial conditions for the star near a given shell
radius are 
\begin{equation}
\frac{\mathrm{d}^{2}r(t)}{\mathrm{d}t^{2}}=a_{0},
\end{equation}
 
\begin{equation}
\left.\frac{\mathrm{d}r(t)}{\mathrm{d}t}\right|_{t=t_{\mathrm{s}}}=v_{\mathrm{s}},
\end{equation}
 
\begin{equation}
r(t_{\mathrm{s}})=r_{\mathrm{s}}(t_{\mathrm{s}})=v_{\mathrm{s}}t_{\mathrm{s}}+r_{\mathrm{s0}}.
\end{equation}
 The solution of these equations is 
\begin{equation}
r(t)=a_{0}(t-t_{\mathrm{s}})^{2}/2+v_{\mathrm{s}}(t-t_{\mathrm{s}})+r_{\mathrm{s}}(t_{\mathrm{s}}),
\end{equation}
 
\begin{equation}
v_{r}(t)=v_{\mathrm{s}}+a_{0}(t-t_{\mathrm{s}}),
\end{equation}
 and the actual position of the star $r(0)$ and its radial velocity
$v_{r}(0)$ at time of measurement ($t=0$) are 
\begin{equation}
r(0)=t_{\mathrm{s}}^{2}a_{0}/2+r_{\mathrm{s0}},\label{eq:r(0)IOA}
\end{equation}
 
\begin{equation}
v_{r}(0)=v_{\mathrm{s}}-a_{0}t_{\mathrm{s}}.
\end{equation}
 Eliminating $t_{\mathrm{s}}$ from the two previous equations, we
get 
\begin{equation}
v_{r}(0)_{\pm}=v_{\mathrm{s}}\pm v_{\mathrm{c}}\sqrt{2\left(1-r(0)/r_{\mathrm{s0}}\right)},\label{eq:v0}
\end{equation}
 where $v_{\mathrm{c}}=\sqrt{-a_{0}r_{\mathrm{s0}}}$ is the circular
velocity at the shell-edge radius.

\subsection{Approximative LOSVD \label{sub:LOSVD-app}}

The projection of the velocity given by Eq.~(\ref{eq:v0}) to the
LOS at a projected radius $R$ will be 
\begin{equation}
\begin{array}{rcl}
v_{\mathrm{los}\pm} & = & \sqrt{1-R^{2}/\left(r\left(0\right)\right)^{2}}v_{r}(0)_{\pm}=\\
 & = & \sqrt{1-R^{2}/\left(r\left(0\right)\right)^{2}}\left[v_{\mathrm{s}}\pm v_{\mathrm{c}}\sqrt{2\left(1-r(0)/r_{\mathrm{s0}}\right)}\right].
\end{array}\label{eq:vlos}
\end{equation}
 Using this expression, we can model the LOSVD at a given projected
radius for a given shell. For the proper choice of a pair of values
$v_{\mathrm{c}}$ and $v_{\mathrm{s}}$, we can find a match with
observed and modeled peaks of the LOSVD. When we use this approach,
we call it the \textit{approximative LOSVD}.

To model the approximative LOSVD by Eq.~(\ref{eq:vlos}), we have
to add an assumption about the radial dependence of the shell-edge
density distribution, Eq.~(\ref{eq:sigma}). We chose this function
in the same manner as in the model of radial oscillations that is
in a way that corresponds to constant number of stars at the edge
of the shell. In Sect.~\ref{sub:sigma} we have shown in the model
of radial oscillations that a different choice of the radial dependence
of the shell brightness changes neither the quadruple-peaked shape
of the LOSVD of the shells, nor the positions of the maximal/minimal
velocity which corresponds to the peaks of the LOSVD. This holds also
for the approximative LOSVD, because the approximative LOSVD is very
close to the LOSVD from the model of the radial oscillations, see
Fig.~\ref{fig:app-rez}. For the approximative LOSVD also holds that
the inner peaks of the LOSVD disappear in the zone between the current
turning points and the edge of the shell.

\subsection{Radius of maximal LOS velocity \label{sub:rvmax}}

\citetalias{mk98} proved that near the edge of a stationary shell,
$r_{s}$, the maximum intensity of the LOSVD is at the edge of the
distribution. They also proved that the maximal absolute value of
the LOS velocity $v_{\mathrm{los,max}}$ comes from stars at the galactocentric
radius 
\begin{equation}
r_{v\mathrm{max}}=\frac{1}{2}(R+r_{\mathrm{s0}}),\label{eq:1/2(R+rs)}
\end{equation}
 at each projected radius $R$.

For a moving shell, analogous equations are significantly more complex
and a similar relation cannot be easily proven. Nevertheless, when
we apply both results of \citetalias{mk98} we can show in examples
(Figs.~\ref{fig:120app}, \ref{fig:app-rez}, and others) that their
use is valid, even for nonstationary shells. In the framework of the
radial oscillations model (Sect.~\ref{sub:LOSVD-rad}), we have shown
that the peaks of the LOSVD occur at the edges of distributions of
the near or the far half of the galaxy (Sect.~\ref{sub:Nature-of-4peak}).
The inner peak corresponds to inward-moving stars and the outer one
to outward-moving ones. This approach is used in the equations in
Sect.~\ref{sub:vlos,max}. The maximal LOS velocity corresponds to
the outer peak and the minimal to the inner one. Reasons and justification
for use of Eq.~(\ref{eq:1/2(R+rs)}) for $r_{v\mathrm{max}}$ are
discussed in Sect.~\ref{sec:Compars}, point~\ref{item:app-vmax}
(see also Fig.~\ref{fig:r_vmax}).

\subsection{Approximative maximal LOS velocity \label{sub:vlos,max}}

Using the results of \citetalias{mk98}, we derive an expression
for the maxima/minima of the LOS velocity corresponding to locations
of the LOSVD peaks in observable quantities (i.e., the maxima/minima
of the LOS velocity, the projected radius, and the shell radius) by
substituting $r_{v\mathrm{max}}$ given by Eq.~(\ref{eq:1/2(R+rs)})
for $r(0)$ in Eq.~(\ref{eq:vlos}) 
\begin{equation}
\begin{array}{rcl}
v_{\mathrm{los,max}\pm}\! & = & \!\left(v_{\mathrm{s}}\pm v_{\mathrm{c}}\sqrt{1-R/r_{\mathrm{s0}}}\right)\,\times\\
 &  & \times\sqrt{1-4\left(R/r_{\mathrm{s0}}\right)^{2}\left(1+R/r_{\mathrm{s0}}\right)^{-2}}.
\end{array}\label{eq:vlos,max}
\end{equation}
 For the measured locations of the LOSVD peaks $v_{\mathrm{los,max}+}$,
$v_{\mathrm{los,max}-}$, projected radius $R$, and shell-edge radius
$r_{\mathrm{s0}}$, we can express the circular velocity $v_{\mathrm{c}}$
at the shell-edge radius and the current shell velocity $v_{\mathrm{s}}$
by using inverse equations: 
\begin{equation}
v_{\mathrm{c}}=\frac{\left|v_{\mathrm{los,max}+}-v_{\mathrm{los,max}-}\right|}{2\sqrt{\left(1-R/r_{\mathrm{s0}}\right)\left[1-4\left(R/r_{\mathrm{s0}}\right)^{2}\left(1+R/r_{\mathrm{s0}}\right)^{-2}\right]}},\label{eq:vc,obs}
\end{equation}
 
\begin{equation}
v_{\mathrm{s}}=\frac{v_{\mathrm{los,max}+}+v_{\mathrm{los,max}-}}{2\sqrt{1-4\left(R/r_{\mathrm{s0}}\right)^{2}\left(1+R/r_{\mathrm{s0}}\right)^{-2}}}.\label{eq:vs,obs}
\end{equation}
 We call this approach the \textit{approximative maximal LOS velocity}.

\subsection{Slope of the LOSVD intensity maxima \label{sub:Slope}}

Alternatively, the value of the circular velocity $v_{\mathrm{c}}$
at the shell-edge radius could be inferred from measurements of positions
of peaks at two or more different projected radii for the same shell:
let $\bigtriangleup v_{\mathrm{los}}=v_{\mathrm{los,max}+}-v_{\mathrm{los,max}-}$,
where $v_{\mathrm{los,max}\pm}$ satisfy Eq.~(\ref{eq:vlos,max}).
Then, in the vicinity of the shell edge, 
\begin{equation}
\begin{array}{rcl}
\bigtriangleup v_{\mathrm{los}} & = & 2v_{\mathrm{c}}\sqrt{\left(R/r_{\mathrm{s0}}-1\right)\left[1-4\left(R/r_{\mathrm{s0}}\right)^{2}\left(1+R/r_{\mathrm{s0}}\right)^{-2}\right]}\simeq\\
 & \simeq & 2(1-R/r_{\mathrm{s0}})v_{\mathrm{c}},
\end{array}
\end{equation}
 and taking the derivative with respect to the projected radius 
\begin{equation}
\frac{\mathrm{d}(\bigtriangleup v_{\mathrm{los}})}{\mathrm{d}R}=-2\frac{v_{\mathrm{c}}}{r_{\mathrm{s0}}},\label{eq:sklon}
\end{equation}
 which happens to be the same expression as Eq.~\ref{eq:sklonMK}
(equation (7) in \citetalias{mk98}). Nevertheless, for a stationary
shell, $\bigtriangleup v_{\mathrm{los}}$ is the distance between
the two LOSVD intensity maxima of a stationary shell, whereas in this
framework, it is the distance between the outer peak for positive
velocities and the inner peak for negative velocities or vice versa.
This equation allows us to measure the circular velocity in shell
galaxies using the slope of the LOSVD intensity maxima in the $R\times v_{\mathrm{los}}$
diagram.

When we use this approach, we call it the use of the \textit{slope
of the LOSVD intensity maxima}. It requires us to measure the LOSVD
for at least two different projected radii. In exchange, as we show
in Sect.~\ref{sub:Recover}, that it promises a more accurate derivation
of $v_{\mathrm{c}}$. However it does not allow the derivation of
the shell velocity $v_{s}$. For this purpose, we can use Eq.~(\ref{eq:vlos,max})
to derive a hybrid relation between the positions of the LOSVD peaks,
the circular velocity at the shell-edge radius $v_{\mathrm{c}}$,
and the shell velocity: 
\begin{equation}
v_{\mathrm{s}}^{2}=v_{\mathrm{c}}^{2}(1-R/r_{\mathrm{s0}})+\frac{v_{\mathrm{los,max}+}v_{\mathrm{los,max}-}}{4\left(R/r_{\mathrm{s0}}\right)^{2}\left(1+R/r_{\mathrm{s0}}\right)^{-2}-1}.\label{eq:vs-vc}
\end{equation}
If we insert the value of $v_{\mathrm{c}}$ derived from the measurement
of the LOSVD intensity maxima into this equation, we can expect a
better estimate of the phase velocity of the shell.

\subsection{Comparison of approaches \label{sec:Compars}}

\begin{figure}[h]
\centering{}\includegraphics[width=12cm]{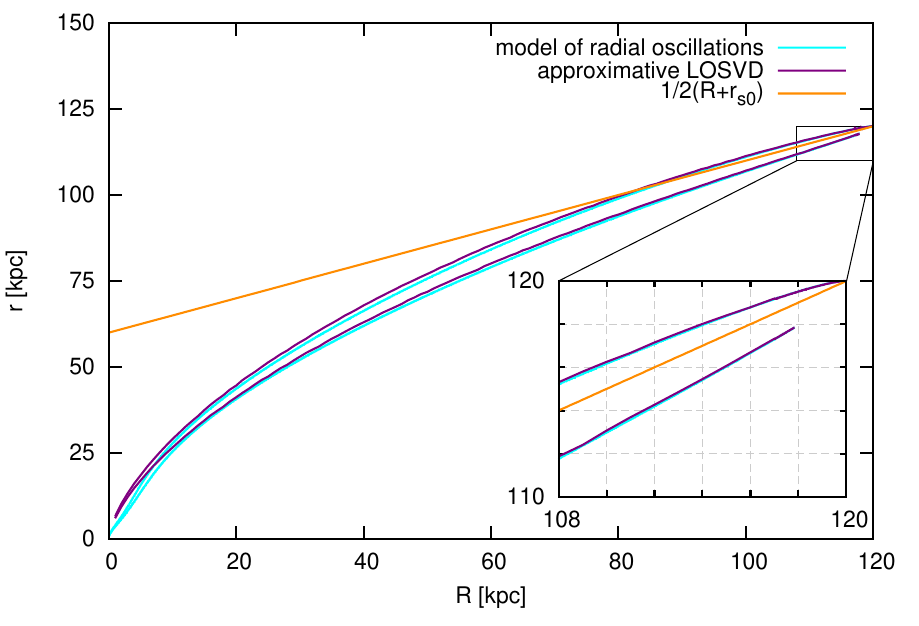}
\caption{\textsf{\small Galactocentric radii $r_{v\mathrm{max}}$ that contribute
to the LOSVD maxima according to Eq.~(\ref{eq:1/2(R+rs)}), which
was used in the derivation of the approximative maximal/minimal LOS
velocities (Sect.~\ref{sec:Compars}, point~\ref{item:app-vmax})\,--\,orange
curve, according to the approximative LOSVD (Sect.~\ref{sec:Compars},
point~\ref{item:app-LOSVD})\,--\,purple curves, and according
to the model of radial oscillations (Sect.~\ref{sub:LOSVD-rad})\,--\,light
blue curves for the second shell at\,120 kpc (parameters of the shell
are highlighted in bold in Table~\ref{tab:param}). For parameters
of the host galaxy potential, see Sect.~\ref{sec:param}. \label{fig:r_vmax} }}
\end{figure}

\begin{figure}[!t]
\begin{centering}
\includegraphics[width=12cm]{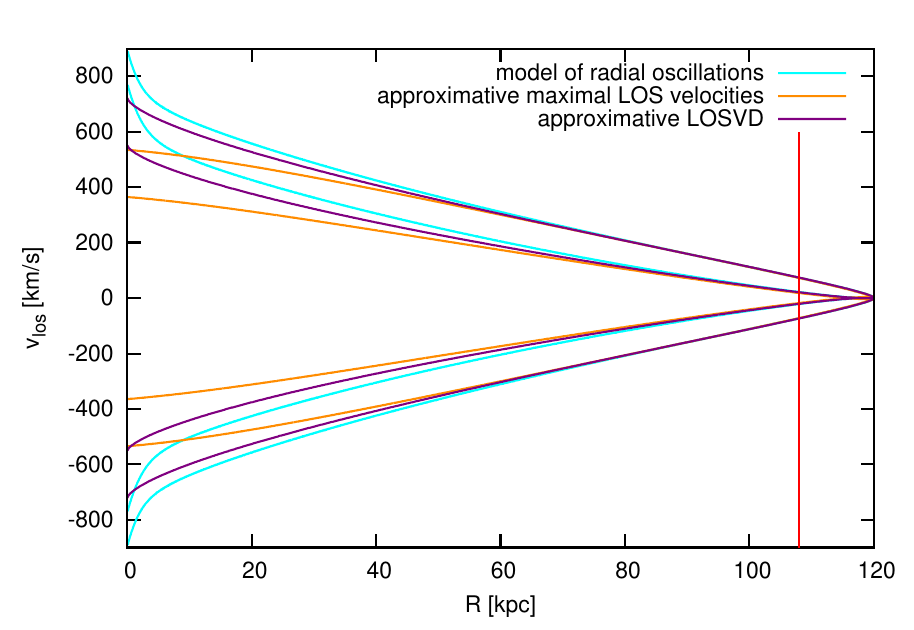}
\par\end{centering}

\centering{}\includegraphics[width=12cm]{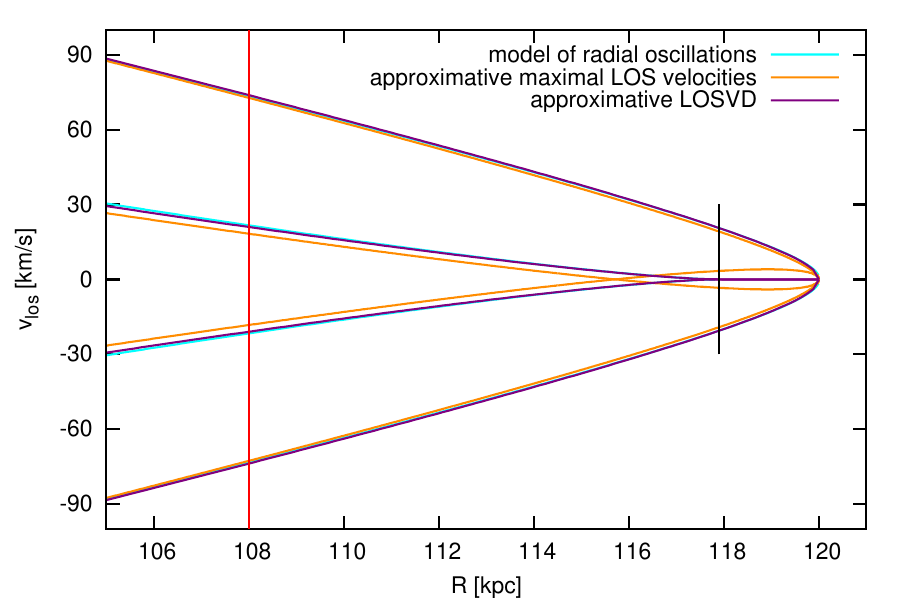}
\caption{\textsf{\small LOSVD peak locations for the second shell at the radius
of 120\,kpc (parameters of the shell are highlighted in bold in Table~\ref{tab:param})
according to the approximative maximal LOS velocities (Sect.~\ref{sec:Compars},
point~\ref{item:dis-los}) given by Eq.~(\ref{eq:vlos,max}) (orange
curves); the approximative LOSVD (Sect.~\ref{sec:Compars}, point~\ref{item:dis-LOSVD})
given by Eq.~(\ref{eq:vlos}) (purple curves); and the model of radial
oscillations (Sect.~\ref{sub:LOSVD-rad}) (light blue curves which
almost merged with the purple ones near the shell edge). The red line
shows the position of the LOSVD from Fig.~\ref{fig:app-rez}, the
black one shows the position of the current turning points. The upper
panel shows the whole range of radii, the lower zooms in on the edge
of the shell. For parameters of the host galaxy potential, see Sect.~\ref{sec:param}.
\label{fig:120app} }}
\end{figure}

\begin{figure}[h]
\centering{}\includegraphics[width=12cm]{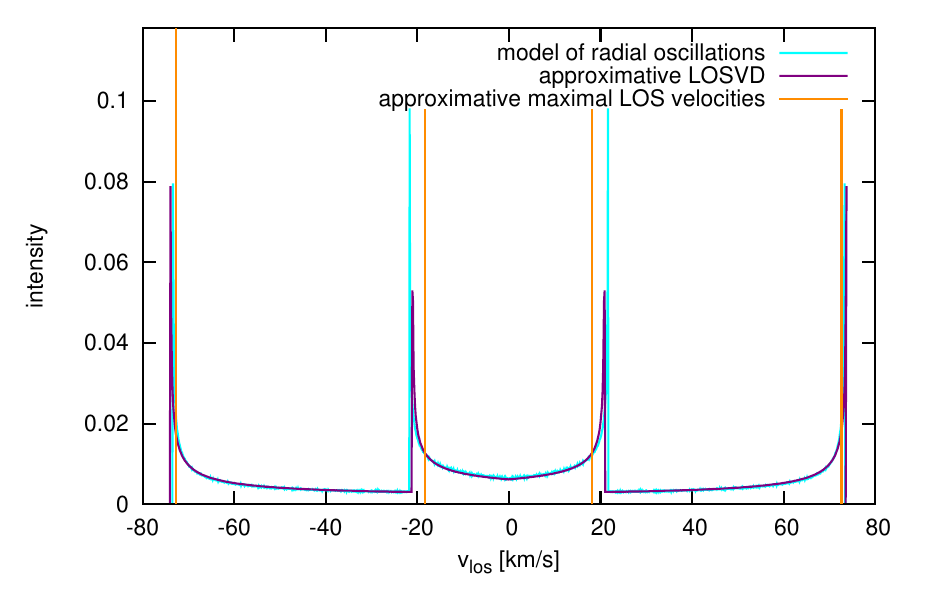}
\caption{\textsf{\small LOSVD of the second shell at $r_{\mathrm{s}}=120$\,kpc
(parameters of the shell are highlighted in bold in Table~\ref{tab:param})
for the projected radius $R=0.9r_{\mathrm{s}}=108$\,kpc according
to the approximative LOSVD (Sect.~\ref{sec:Compars}, point~\ref{item:app-LOSVD})
given by Eq.~(\ref{eq:vlos}) (purple curve) and the model of radial
oscillations (Sect.~\ref{sub:LOSVD-rad}) (light blue curve almost
merged with the purple one). Locations of peaks as given by the approximative
maximal LOS velocities (Sect.~\ref{sec:Compars}, point~\ref{item:app-vmax})
given by Eq.~(\ref{eq:vlos,max}) are plotted with orange lines.
Profiles do not include stars of the host galaxy that are not part
of the shell system and are normalized, so that the total flux equals
to one. For parameters of the host galaxy potential see Sect.~\ref{sec:param}.
\label{fig:app-rez} }}
\end{figure}

The approximation of a constant radial acceleration in the host galaxy
potential and shell phase velocity (Sect.~\ref{sec:IOA}) splits
into three different analytical and semi-analytical approaches for
obtaining values of the circular velocity $v_{\mathrm{c}}$ at the
shell-edge radius and the shell phase velocity $v_{\mathrm{s}}$.
Different approaches/models have a different color assigned. This
color is used in Figs.~\ref{fig:r_vmax}--\ref{fig:vlosmax-a1-10}
and \ref{fig:vel.map}--\ref{fig:minfit} to represent the output
of the corresponding approach. Here we summarize differences, advantages
and disadvantages in these three approaches:
\begin{enumerate}
\item \textbf{\textit{The approximative LOSVD}} (purple curves): For the
given shell at the chosen projected radius, Eq.~(\ref{eq:vlos})
is a function of only two parameters -- $v_{\mathrm{c}}$ and $v_{\mathrm{s}}$.
Assuming a radial dependence of the shell-edge density distribution,
Eq.~(\ref{eq:vlos}) allows us to plot the whole LOSVD (Sect.~\ref{sub:LOSVD-app}).
However, computing the LOSVD and the positions of peaks requires a
numerical approach in this framework. When deriving $v_{\mathrm{c}}$
and $v_{\mathrm{s}}$ from the observed LOSVD, we need to find a numerical
solution to Eq.~(\ref{eq:vlos}) and to search for a pair of $v_{\mathrm{c}}$
and $v_{\mathrm{s}}$, which matches the (simulated) data best. \label{item:dis-LOSVD}
\label{item:app-LOSVD}
\item \textbf{\textit{The approximative maximal LOS velocities}} (orange
curves): Eq.~(\ref{eq:vlos,max}) supplies the positions of the peaks
directly. It differs from the previous approximation in the assumption
about the galactocentric radius $r_{v\mathrm{max}}$, from which comes
the contribution to the LOSVD at the maximal speed. The assumption
is that $r_{v\mathrm{max}}$ is given by Eq.~(\ref{eq:1/2(R+rs)}),
which was derived by \citetalias{mk98} for a stationary shell.
This equation is actually only very approximate (see Fig.~\ref{fig:r_vmax}),
but allows us to analytically invert Eq.~(\ref{eq:vlos,max}) to
obtain formulae for the direct calculation of $v_{\mathrm{c}}$ and
$v_{\mathrm{s}}$ from the measured peak positions in the spectrum
of the shell galaxy near the shell edge -- Eqs.~(\ref{eq:vc,obs})
and (\ref{eq:vs,obs}). Nevertheless, when measuring in the zone between
the radius of the current turning points and the shell radius, we
can expect very bad estimates of $v_{\mathrm{c}}$ and $v_{\mathrm{s}}$.
\label{item:dis-los} \label{item:app-vmax} 
\item Using the \textbf{\textit{slope of the LOSVD intensity maxima}} in
the \mbox{$R\times v_{\mathrm{los}}$} diagram: Eq.~(\ref{eq:sklon})
cannot be used to draw theoretical LOSVD maxima for the given potential
of the host galaxy, because it connects only the circular velocity
in the host galaxy and the difference of the slopes of the LOSVD maxima.
Moreover, the difference of the slopes alone does not allow us to
determine the shell velocity, but we can use Eq.~(\ref{eq:vs-vc})
as it is described in Sect.~\ref{sub:Slope}. Nevertheless it is
this approach that gives the most accurate estimate of $v_{\mathrm{c}}$
when applied to simulated data, Sect.~\ref{sub:Recover}. \label{item:dis-slope}
\end{enumerate}
These methods can be compared with the model of radial oscillations
described in Sect.~\ref{sub:LOSVD-rad} (plotted with light blue
curves in the relevant figures). The model of radial oscillations
uses thorough knowledge of the potential of the host galaxy. From
it we extract the circular velocity at the shell-edge radius and the
current shell velocity and we put them in the approximative relations
derived in Sect.~\ref{sec:IOA}. We apply all the three approximations
to the simulated data in  Sect.~\ref{sub:Recover}.

Fig.~\ref{fig:r_vmax} shows a comparison of the radii that contribute
to the LOSVD maxima according to the model of radial oscillations,
the approximative LOSVD, and the approximative maximal LOS velocities.
For the first two methods, the radius corresponding to the inner maxima
of the LOSVD (which are the maxima created by the inward stars) is
lower than that for the outer maxima, whereas Eq.~(\ref{eq:1/2(R+rs)})
assumes the same $r_{v\mathrm{max}}$ for both inward and outward
stars.

Fig.~\ref{fig:120app} shows locations of the LOSVD peaks for the
second shell at the radius of 120\,kpc near the shell-edge radius.
The purple curve is calculated using the approximative LOSVD (Sect.~\ref{sec:Compars},
point~\ref{item:app-LOSVD}) given by Eq.~(\ref{eq:vlos}), into
which we inserted the velocity of the second shell according to the
model of radial oscillations and the circular velocity in the potential
of the host galaxy (see Sect.~\ref{sec:param} for parameters of
the potential). The purple curve does not differ significantly from
the light blue curve calculated in the model of radial oscillations
(Sect.~\ref{sub:LOSVD-rad}). The more important deviations in the
orange curve of the approximative maximal LOS velocities (Sect.~\ref{sec:Compars},
point~\ref{item:app-vmax}) given by Eq.~(\ref{eq:vlos,max}), are
caused by Eq.~(\ref{eq:1/2(R+rs)}) for $r_{v\mathrm{max}}$. With
this assumption, approximative maximal LOS velocities (the orange
curve) predict that around the zone between the current turning point
and the shell edge, the inner peaks change signs. This means that
for the part of the galaxy closer to the observer, both inner and
outer peaks will fall into negative values of the LOS velocity and
vice versa. However, from the model of the radial oscillations, we
know that the signal from the inner peak in a given (near or far)
part of the galaxy is always zero or has the opposite sign to that
of the outer peak.

The model of the radial oscillations and the approximative LOSVD given
by Eq.~(\ref{eq:vlos}) were also used to construct the LOSVD for
the second shell located at 120\,kpc, at the projected radius of
108\,kpc in Fig.~\ref{fig:app-rez}. The graph also shows the locations
of the peaks using the approximative maximal LOS velocities given
by Eq.~(\ref{eq:vlos,max}).

To wrap up, all three approaches give a good agreement with the model
of radial oscillations. The first approach is practically identical
to this model in the vicinity of the shell edge, but it requires numerical
solution of equations. The second approach is more approximative and
gives worse results particularly in the zone between the current turning
point and the shell edge, but allows direct expression of $v_{\mathrm{c}}$
and $v_{\mathrm{s}}$. The third approach gives only the relation
between the slopes of the LOSVD maxima and $v_{\mathrm{c}}$, but
we have already announced that it gives the best results when calculating
$v_{\mathrm{c}}$ from the simulated data.

\subsection{Projection factor approximation\label{sub:vlosFardal}}

In Sect.~\ref{sub:z/r} we have derived an approximative relation
for the factor $z/r$ that projects the galactocentric velocity of
the stars at radial trajectories to the line of sight, Eq.~(\ref{eq:z/r}),
which has been already used by \citet{fardal12} to derive the relation
for $v_{\mathrm{los,max}\pm}$. Inserting this equation to the expression
for the projected velocity of the stars of the shell, Eq.~(\ref{eq:vlos})
in Sect.~\ref{sub:LOSVD-app}, we get 
\begin{equation}
v_{\mathrm{los}\pm}(r)\simeq\sqrt{2(r/r_{\mathrm{s0}}-R/r_{\mathrm{s0}})}\left[v_{\mathrm{s}}\pm v_{\mathrm{c}}\sqrt{2\left(1-r/r_{\mathrm{s0}}\right)}\right].\label{eq:vlos-far}
\end{equation}
\begin{figure}[h]
\centering{}\includegraphics[width=12cm]{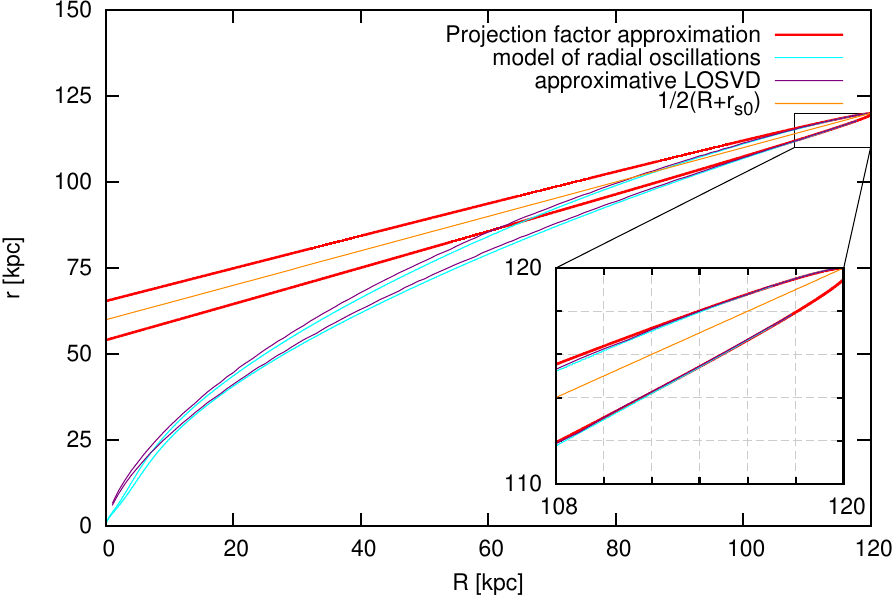}
\caption{\textsf{\small Galactocentric radii $r_{v\mathrm{max\pm}}$ that contribute
to the LOSVD maxima for the second shell at\,120 kpc (parameters
of the shell are highlighted in bold in Table~\ref{tab:param}) according
to Eq.~(\ref{eq:rvmax-Far})\,--\,red curves. For comparison, we
show the radii $r_{v\mathrm{max}}$ according to the model of radial
oscillations (Sect.~\ref{sub:LOSVD-rad})\,--\,light blue curves\,--\,and
according to the approximation of Sect.~\ref{sec:Compars} (orange
and purple curves). See also Fig.~\ref{fig:r_vmax}. \label{fig:rmax-Far} }}
\end{figure}
 The derivative of this expression is zero for $r=r_{v\mathrm{max}\pm}$,
where 
\begin{equation}
r_{v\mathrm{max\pm}}=r_{\mathrm{s0}}\left(\frac{v_{\mathrm{s}}}{4v_{\mathrm{c}}}\right)^{2}\left[\frac{1}{2}\left(\frac{4v_{\mathrm{c}}}{v_{\mathrm{s}}}\right)^{2}\left(1+\frac{R}{r_{\mathrm{s0}}}\right)-1\pm\sqrt{\left(\frac{4v_{\mathrm{c}}}{v_{\mathrm{s}}}\right)^{2}\left(1-\frac{R}{r_{\mathrm{s0}}}\right)+1}\right].\label{eq:rvmax-Far}
\end{equation}
Near the edge of the shell, the values $r_{v\mathrm{max\pm}}$ are
in good coincidence with the galactocentric radii that contribute
to the LOSVD maxima according to the model of radial oscillations
(Sect.~\ref{sub:LOSVD-rad}), whereas at lower radii they differ
substantially, Fig.~\ref{fig:rmax-Far}.

\begin{figure}[h]
\centering{}\includegraphics[width=7.8cm]{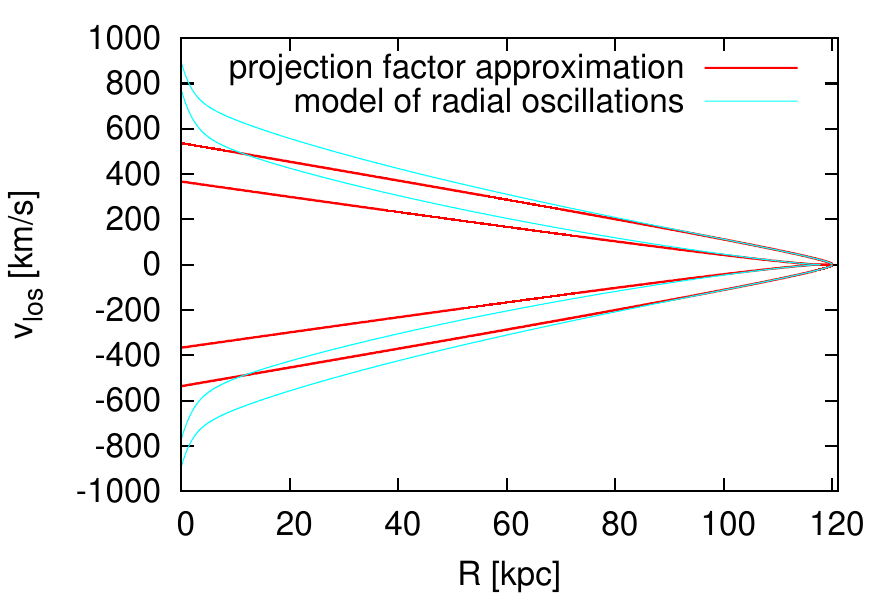}\includegraphics[width=7.5cm]{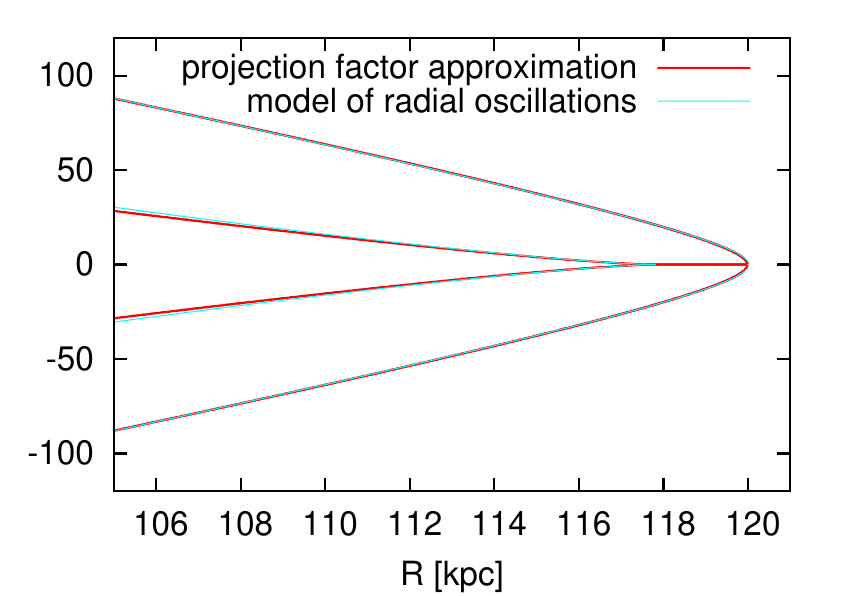}
\caption{\textsf{\small LOSVD peak locations for the second shell at the radius
of 120\,kpc (parameters of the shell are highlighted in bold in Table~\ref{tab:param}).
The red curves show the values of the functions $v_{\mathrm{los+}}(r_{v\mathrm{max+}})$
and $v_{\mathrm{los-}}(r_{v\mathrm{max-}})$, where $v_{\mathrm{los\pm}}(r)$
is given by Eq.~(\ref{eq:vlos-far}) and $r_{v\mathrm{max\pm}}$
follows Eq.~(\ref{eq:rvmax-Far}). The light blue curves are LOSVD
peak locations according to the model of radial oscillations (Sect.~\ref{sub:LOSVD-rad}).
The left panel shows the whole range of radii, the right zooms in
on the edge of the shell. For parameters of the host galaxy potential,
see Sect.~\ref{sec:param}. \label{fig:120app-Far} }}
\end{figure}

The position of the outer LOSVD peaks is expressed as the function
$v_{\mathrm{los+}}(r_{v\mathrm{max+}})$, the position of the inner
peaks as $v_{\mathrm{los-}}(r_{v\mathrm{max-}})$, Fig.~\ref{fig:120app-Far}.
The equations have a solution only for $r_{v\mathrm{max}}<R$. The
radius, where $r_{v\mathrm{max-}}=R$, is the radius of the current
turning point $r_{\mathrm{TP}}$ in this approximation and for $R>r_{\mathrm{TP}}$
the inner peaks disappear. Eq.~(\ref{eq:rvmax-Far}) implies 
\begin{equation}
r_{\mathrm{TP}}=r_{\mathrm{s0}}\left[1-\frac{1}{2}\left(\frac{v_{\mathrm{s}}}{v_{\mathrm{c}}}\right)^{2}\right].
\end{equation}

The functions $v_{\mathrm{los+}}(r_{v\mathrm{max+}})$ and $v_{\mathrm{los-}}(r_{v\mathrm{max-}})$
are a good approximation to the LOSVD peak locations near the edge
of the shell, as can be seen in Fig.~\ref{fig:120app-Far}. Using
these functions are a better way to calculate these than the approximative
LOSVD (Sect.~\ref{sec:Compars}, point~\ref{item:dis-LOSVD}), because
their values are given analytically. Nevertheless they are such a
complicated function of the circular velocity $v_{\mathrm{c}}$ at
the shell-edge radius and the current shell velocity $v_{\mathrm{s}}$
that they do not allow the expression of these variables as a simple
function of observable quantities, unlike the approximative maximal/minimal
LOS velocities (Sect.~\ref{sec:Compars}, point~\ref{item:app-vmax}).
Thus we will not use these function in the following and show them
only for the sake of completeness and comparison with \citet{fardal12}.

\section{Higher order approximation \label{sec:Higher-order-approximation}}

The approximation of a locally constant galactic acceleration $a_{0}$
and shell phase velocity $v_{\mathrm{s}}$, Sect.~\ref{sec:IOA},
describes the positions of the LOSVD peaks fairly well and allows
a good determination of the parameters of the potential of the host
galaxy. Nevertheless we try to have a look outside the realm of constant
$a_{0}$ and $v_{\mathrm{s}}$ using the same concept that stars
behave as if they were released in the center of the host galaxy at
the same time and their distribution of energies is continuous.

\subsection{Motion of a star in a shell system}

The galactocentric radius of the shell edge is a function of time,
$r_{\mathrm{s}}(t)$, where $t=0$ is the moment of measurement and
$r_{\mathrm{s}}(0)=r_{\mathrm{s0}}$ is the position of the shell
edge at this time. Let us define a new coordinate system $s$, where
the radial coordinate is the distance from the edge of the shell,
in the same direction as the galactocentric radius 
\begin{equation}
s(t)=r(t)-r_{\mathrm{s0}}.
\end{equation}
The position of the stars of the given shell in this system is always
negative. 

We assume the following:
\begin{itemize}
\item stars are on strictly radial orbits
\item radial acceleration in the potential of the host galaxy is given as
$a(s)=a_{0}+a_{1}s$, where $a_{0}$ and $a_{1}$ are constant for
a given shell
\item position of the shell edge is (insofar) a general function of time
$s_{\mathrm{s}}(t)$
\item stars at the shell edge have the same velocity as the shell
\end{itemize}
The position of each star is at any time $s(t)$, while $t_{\mathrm{s}}$
is the time when the star could be found at the shell edge $s_{\mathrm{s}}(t_{\mathrm{s}})$.
Then the equation of motion and the initial conditions for the star
near a given shell radius are 
\begin{equation}
\frac{\mathrm{d}^{2}s(t)}{\mathrm{d}t^{2}}=a_{0}+a_{1}s,\label{eq:acc-a1}
\end{equation}
 
\begin{equation}
\left.\frac{\mathrm{d}s(t)}{\mathrm{d}t}\right|_{t=t_{\mathrm{s}}}=v_{\mathrm{s}},
\end{equation}
 
\begin{equation}
s(t_{\mathrm{s}})=s_{\mathrm{s}}(t_{\mathrm{s}}).
\end{equation}

The solution to these equation differs for negative and positive values
of $a_{1}$. The position of a star in a general time $t$ is given
by 
\begin{equation}
\ensuremath{s(t,a_{1}>0)=\frac{\left[a_{1}s_{\mathrm{s}}(t_{\mathrm{s}})+a_{0}\right]\cosh\left[\sqrt{a_{1}}\left(t-t_{\mathrm{s}}\right)\right]+\sqrt{a_{1}}v_{\mathrm{s}}\sinh\left[\sqrt{a_{1}}\left(t-t_{\mathrm{s}}\right)\right]-a_{0}}{a_{1}},}
\end{equation}
\begin{equation}
\ensuremath{s(t,a_{1}<0)=\frac{\left[\left|a_{1}\right|s_{\mathrm{s}}(t_{\mathrm{s}})-a_{0}\right]\cos\left[\sqrt{\left|a_{1}\right|}\left(t-t_{\mathrm{s}}\right)\right]+\sqrt{\left|a_{1}\right|}v_{\mathrm{s}}\sin\left[\sqrt{\left|a_{1}\right|}\left(t-t_{\mathrm{s}}\right)\right]+a_{0}}{\left|a_{1}\right|}},
\end{equation}
 where $\sinh(x)=1/2\left[\exp(x)-\exp(-x)\right]$ and $\cosh(x)=1/2\left[\exp(x)+\exp(-x)\right]$.
For $a_{1}=0$, the solution of Sect.~\ref{sub:star-app} holds.
At the time of the measurement $t=0$ we obtain two pairs of equations
for the position of the star $s(0)$ and its radial velocity $v_{r}(0)=\left.\mathrm{d}s(t)/\mathrm{d}t\right|_{t=0}$,
depending on the sign of $a_{1}$ 
\begin{equation}
\begin{array}{rcl}
s(0,a_{1}>0) & = & 1/a_{1}\left\{ \left[a_{1}s_{\mathrm{s}}(t_{\mathrm{s}})+a_{0}\right]\cosh\left(t_{\mathrm{s}}\sqrt{a_{1}}\right)-\sqrt{a_{1}}v_{\mathrm{s}}\sinh\left(t_{\mathrm{s}}\sqrt{a_{1}}\right)-a_{0}\right\} ,\\
v_{r}(0,a_{1}>0) & = & 1/\sqrt{a_{1}}\left\{ \sqrt{a_{1}}v_{\mathrm{s}}\cosh\left(t_{\mathrm{s}}\sqrt{a_{1}}\right)-\left[a_{1}s_{\mathrm{s}}(t_{\mathrm{s}})+a_{0}\right]\sinh\left(t_{\mathrm{s}}\sqrt{a_{1}}\right)\right\} ,
\end{array}\label{eq:+a1}
\end{equation}
\begin{equation}
\begin{array}{rcl}
s(0,a_{1}<0) & = & 1/\left|a_{1}\right|\left\{ \left[\left|a_{1}\right|s_{\mathrm{s}}(t_{\mathrm{s}})-a_{0}\right]\cos\left(t_{\mathrm{s}}\sqrt{\left|a_{1}\right|}\right)-\sqrt{\left|a_{1}\right|}v_{\mathrm{s}}\sin\left(t_{\mathrm{s}}\sqrt{\left|a_{1}\right|}\right)+a_{0}\right\} ,\\
v_{r}(0,a_{1}<0) & = & 1/\sqrt{\left|a_{1}\right|}\left\{ \sqrt{a_{1}}v_{\mathrm{s}}\cos\left(t_{\mathrm{s}}\sqrt{\left|a_{1}\right|}\right)+\left[\left|a_{1}\right|s_{\mathrm{s}}(t_{\mathrm{s}})-a_{0}\right]\sin\left(t_{\mathrm{s}}\sqrt{\left|a_{1}\right|}\right)\right\} .
\end{array}\label{eq:-a1}
\end{equation}
 For galactic potentials, one value of $s(0)$ will yield solutions
for two different values of $t_{\mathrm{s}}$ and correspondingly
two values of $v_{r}(0)$ and its projection to the line of sight.
The minimal and maximal LOS velocities show the positions of LOSVD
peaks.

\subsection{Comparison of approximations}

Now we compare this higher order approximation with the approximation
of a constant acceleration (Sect.~\ref{sec:IOA}) and the model of
radial oscillations (Sect.~\ref{sub:LOSVD-rad}). For higher accuracy,
we can obviously introduce the acceleration of the shell $a{}_{\mathrm{s}}$
and express the shell position as $s_{\mathrm{s}}(t_{\mathrm{s}})=v_{\mathrm{s}}t_{\mathrm{s}}+a{}_{\mathrm{s}}t_{\mathrm{s}}^{2}/2$.
However, for observation data it would mean to fit 4 parameters ($a_{0}$,
$a_{1}$, $v_{\mathrm{s}}$, and $a{}_{\mathrm{s}}$), what could
prove difficult in practice. To compare the approximations, we thus
restrict ourselves to a shell of constant velocity, that is $s_{\mathrm{s}}(t_{\mathrm{s}})=v_{\mathrm{s}}t_{\mathrm{s}}$,
like in Sect.~\ref{sec:IOA}.

\begin{figure}[H]
\begin{centering}
\includegraphics[width=12cm]{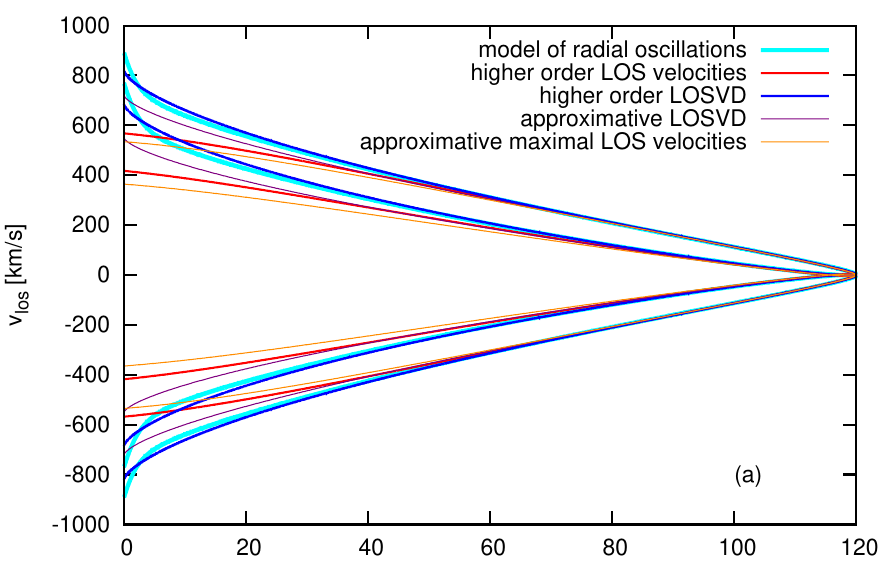}
\par\end{centering}

\begin{centering}
\includegraphics[width=12cm]{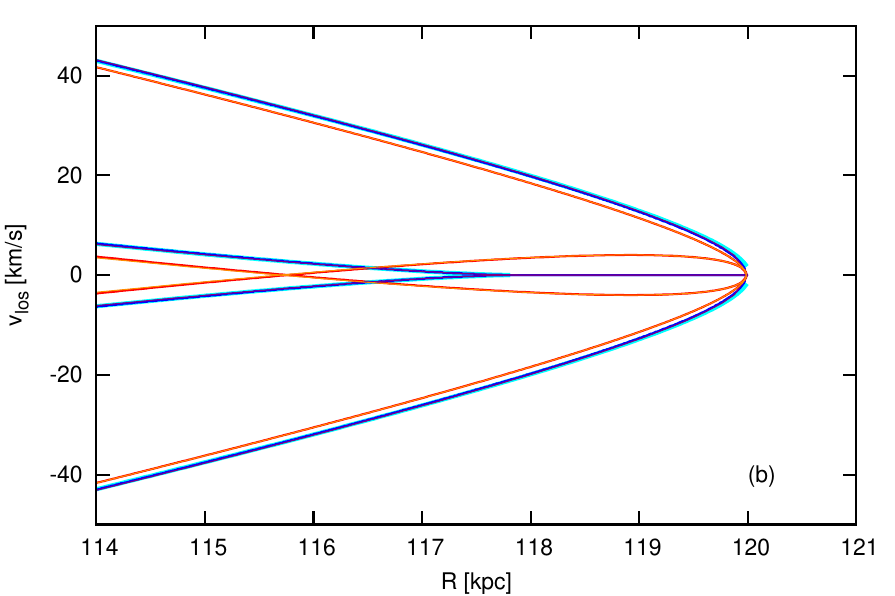}
\par\end{centering}

\centering{}\caption{\textsf{\small Comparison of LOSVD peak locations in different approximations
for the second shell at the radius of 120\,kpc, $a_{1}=1.2\times10^{-5}$\,Myr\textsuperscript{-2}.
The upper panel shows the whole range of radii, the lower zooms in
on the edge of the shell. For parameters of the host galaxy potential,
see Sect.~\ref{sec:param}. \label{fig:vlosmax-a1-120} }}
\end{figure}

\begin{figure}[H]
\begin{centering}
\includegraphics[width=12cm]{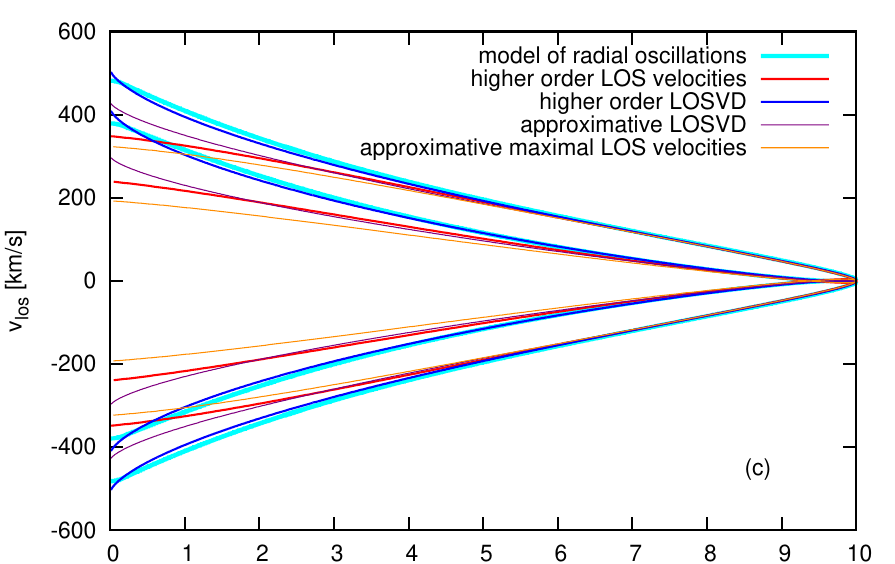}
\par\end{centering}

\centering{}\includegraphics[width=12cm]{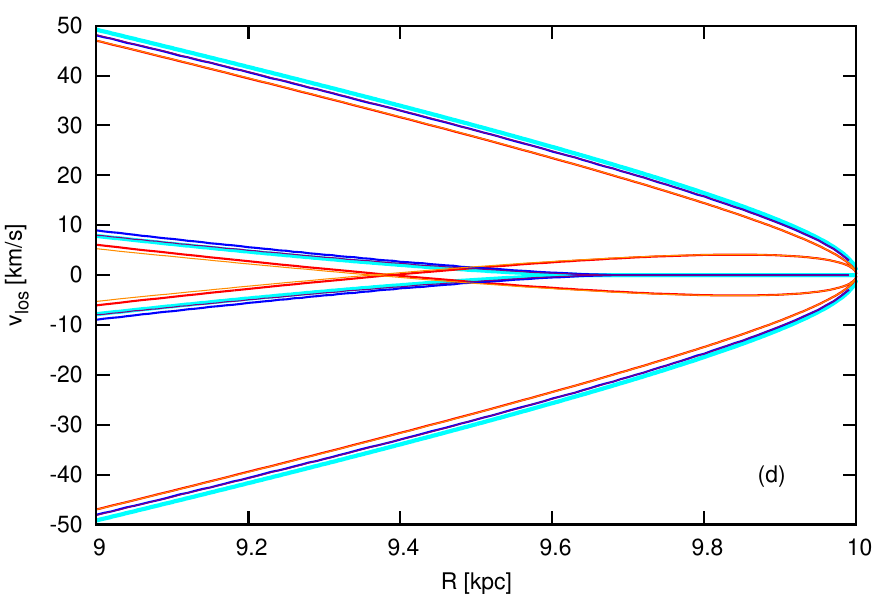}\caption{\textsf{\small Comparison of LOSVD peak locations in different approximations
for the first shell at the radius of 10\,kpc, $a_{1}=8.5\times10^{-4}$\,Myr\textsuperscript{-2}.
The upper panel shows the whole range of radii, the lower zooms in
on the edge of the shell. For parameters of the host galaxy potential,
see Sect.~\ref{sec:param}. \label{fig:vlosmax-a1-10} }}
\end{figure}
Besides the usual second shell at 120\,kpc showed in Fig.~\ref{fig:vlosmax-a1-120},
we show also the first shell at 10\,kpc in Fig.~\ref{fig:vlosmax-a1-10}.
In our case, the value of $a_{1}$ at the galactocentric distance
of 10 kpc is almost two orders of magnitude larger than the corresponding
value at 120\,kpc (see Fig.~\ref{fig:a1}). For the approximations,
we have used values of parameters calculated from the potential of
the host galaxy (for parameters of the host galaxy potential, see
Sect.~\ref{sec:param}). The model of radial oscillations (thick
light-blue curves) requires the knowledge of the potential at all
radii. The maxima/minima of the LOS velocities (that correspond to
the locations of the peaks of the LOSVD) are shown in purple for the
approximation of a constant acceleration (or, as we call it, using
the \textquotedbl{}approximative LOSVD\textquotedbl{} by Eq.~(\ref{eq:vlos})),
and in dark blue for a LOS projection of the solution of Eq.~(\ref{eq:+a1})
with a nonzero $a_{1}$, which is positive for both shells. At the
edge of the shell, both approximations are almost identical to the
model of radial oscillations. On the other hand, at lower galactocentric
radii, only the approximation with a nonzero $a_{1}$ follows the
model of radial oscillations reasonably well. In general, the shell
will be difficult to observe in real galaxies at lower projected radii,
but for the case of observations of individual stars, star clusters
and planetary nebulae, the kinematical imprint of the shell could
be observed considerably far from its edge.

The purple and blue curves are calculated by finding maxima/minima
of the LOS velocities at each projected radius. It is possible to
obtain these in a much easier, but less accurate manner using the
approximation for the radius of maximal LOS velocity $r_{v\mathrm{max}}=\frac{1}{2}(R+r_{\mathrm{s0}})$,
as described in Sect.~\ref{sub:rvmax}. The orange and red curves
in Fig.~\ref{fig:vlosmax-a1-120} and Fig.~\ref{fig:vlosmax-a1-10}
show the result of this procedure in the approximation of a constant
acceleration (the \textquotedbl{}approximative maximal LOS velocity\textquotedbl{},
Eq.~(\ref{eq:vlos,max})) and in the approximation with a nonzero
value of $a_{1}$, respectively. Again, both approximations merge
near the edge of the shell. For lower projected radii, the two curves
separate again, but taking into account their overall difference from
the model of radial oscillations, we cannot in this case consider
the approximation of a nonzero $a_{1}$ to be a significant improvement.
The approximative maximal LOS velocity with constant acceleration
has the advantage that it allows a direct expression of basic variables
(the circular velocity $v_{\mathrm{c}}$ at the shell-edge radius
and shell phase velocity $v_{\mathrm{s}}$) in terms of observable
quantities, facilitating and easy application to measured data. The
same cannot be done in the approximation with a nonzero value of $a_{1}$.

\subsection{$\boldsymbol{a{}_{1}}$ }

\begin{figure}[h]
\centering{}\includegraphics[width=12cm]{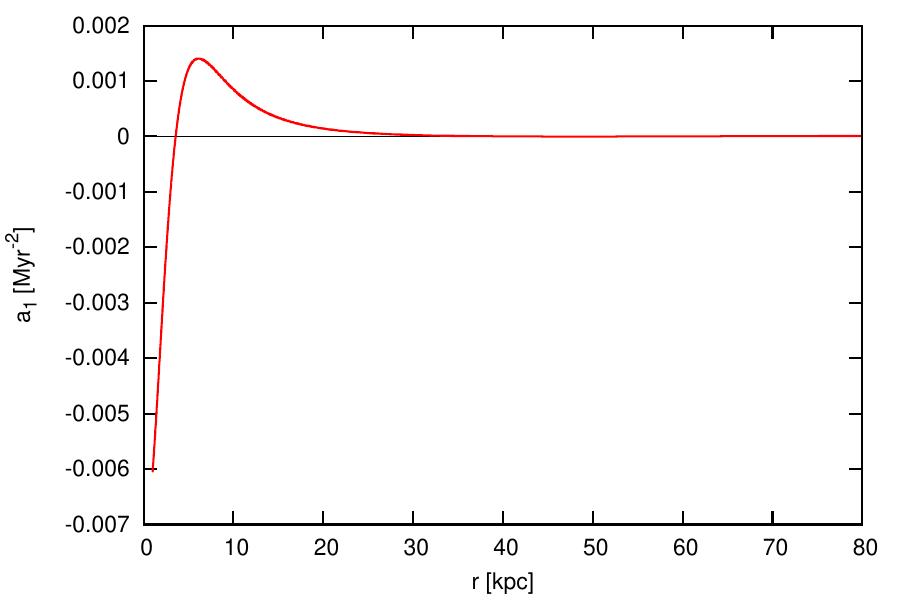}
\caption{\textsf{\small The radial dependence of $a_{1}$ in the host galaxy.
For parameters of the host galaxy potential, see Sect.~\ref{sec:param}.
\label{fig:a1} }}
\end{figure}

The assumption about the function $a(r)$ in the host galaxy is in
fact an assumption on the radial dependence of the density of the
host galaxy, by 
\begin{equation}
a(r)=\frac{4\pi\mathrm{G}}{r^{2}}\int_{0}^{r}\rho(r')r'^{2}\mathrm{d}r',\label{eq:a-rho}
\end{equation}
 where $\rho(r)$ is the density in the host galaxy and G is the gravitational
constant. For the case of constant acceleration $a=a_{0}$ the derivative
of Eq.~(\ref{eq:a-rho}) with respect to $r$ shows that the density
goes to zero for large $r$ as 
\begin{equation}
\rho(r)=\frac{a_{0}}{2\pi\mathrm{G}}r^{-1},
\end{equation}
 whereas for $a=a_{0}+a_{1}(r-r_{\mathrm{s0}})$ the density goes
to $\frac{3a_{1}}{4\pi\mathrm{G}}$ for large $r$ as 
\begin{equation}
\rho(r)=\frac{3a_{1}}{4\pi\mathrm{G}}+\frac{a_{0}+a_{1}r_{\mathrm{s0}}}{2\pi\mathrm{G}}r^{-1}.
\end{equation}
It is important to note that this approximation of the acceleration
is applied only locally, although this word may sometimes mean a fairly
large span of radii. The parameter $a_{1}$ may, in real galaxies,
assume both positive and negative values. In Fig.~\ref{fig:a1} we
show the radial dependence of \textbf{$a_{1}$ }in the host galaxy
modeled as a double Plummer sphere (for parameters of the host galaxy
potential, see Sect.~\ref{sec:param}). 

\section{Test-particle simulation \label{sec:N-Simulations}}

We performed a simplified simulation of formation of shells in a radial
minor merger of galaxies. Both merging galaxies are represented by
smooth potential. Millions of test particles were generated so that
they follow the distribution function of the cannibalized galaxy at
the beginning of the simulation. The particles then move according
to the sum of the gravitational potentials of both galaxies. When
the centers of the galaxies pass through each other, the potential
of the cannibalized galaxy is suddenly switched off and the particles
continue to move only in the fixed potential of the host galaxy. We
use the simulation to demonstrate the validity of our methods of recovering
the parameters of the host galaxy potential by \textit{measuring}%
\footnote{By \textit{measuring,} we mean that the data measured are the output
of our simulation.%
} the positions of the peaks in the LOSVD of simulated data. 
\begin{figure}[h]
\centering{}\includegraphics[width=12cm]{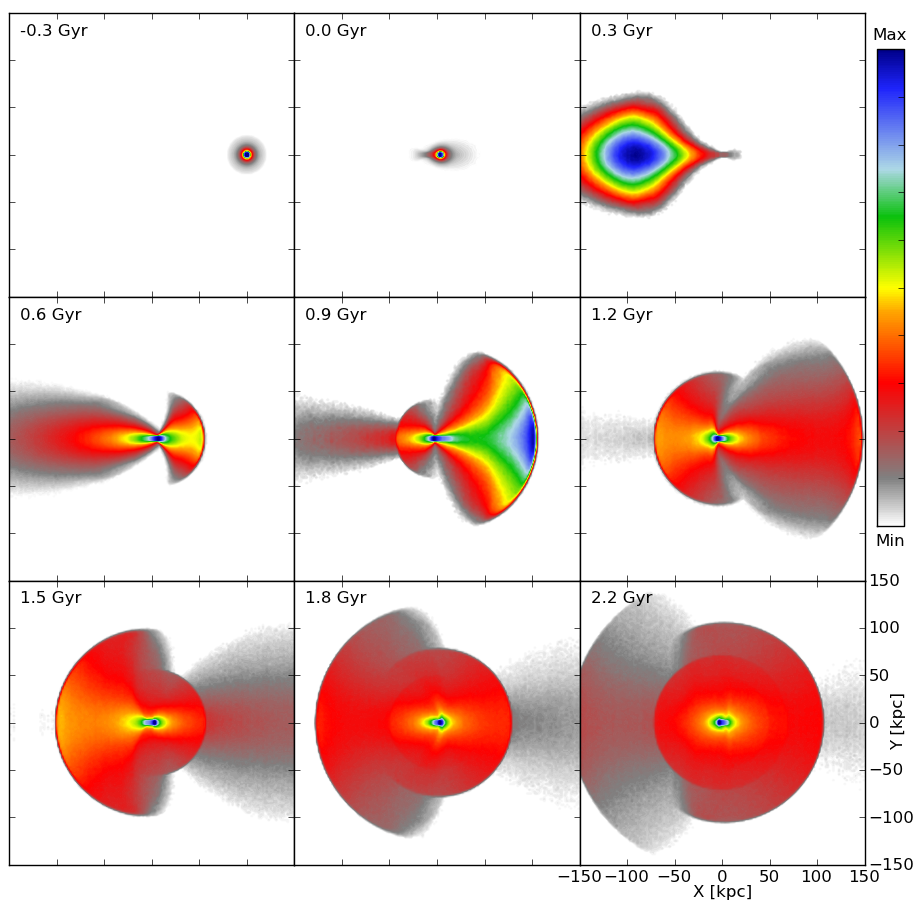}
\caption{\textsf{\small Snapshots from our test-particle simulation of the
radial minor merger, leading to the formation of shells. Each panel
covers 300$\times$300\,kpc and is centered on the host galaxy. Only
the surface density of particles originally belonging to the satellite
galaxy is displayed. The density scale varies between frames, so that
the respective range of densities is optimally covered. Time-stamps
mark the time since the release of the star in the center of the host
galaxy. \label{fig:movie} }}
\end{figure}

In all cases, we look at the galaxy from the view perpendicular to
the axis of collision, so that the cannibalized galaxy originally
flew in from the right.%
\footnote{We use the term \textit{cannibalized galaxy }even before and during
the merger process.%
} Information on details of the simulation process can be found in
Sect.~\ref{sub:Configuration}.

\subsection{Parameters of the simulation \label{sec:param-sim}}

The potential of the host galaxy is the same as the one described
in Sect.~\ref{sec:param}. Let us only recall that it is a double
Plummer sphere with respective masses $M_{*}=2\times10^{11}$\,M\suns
and $M_{\mathrm{DM}}=1.2\times10^{13}$\,M\suns, and Plummer radii
$\varepsilon_{*}=5$\,kpc and $\varepsilon_{\mathrm{DM}}=100$\,kpc
for the luminous component and the dark halo, respectively. The potential
of the cannibalized galaxy is chosen to be a single Plummer sphere
with the total mass $M=2\times10^{10}$\,M\suns and Plummer radius
$\varepsilon_{*}=2$\,kpc.

The details of the simulations are described in Sect.~\ref{sub:Configuration}.
In the simulations that we present in this part, neither the gradual
decay of the cannibalized galaxy nor the dynamical friction is included.
The cannibalized galaxy is released from rest at a distance of 100\,kpc
from the center of the host galaxy. When it reaches the center of
the host galaxy in 306.4\,Myr, its potential is switched off and
its particles begin to oscillate freely in the host galaxy. The shells
start appearing visibly from about 50\,kpc of galactocentric distance
and disappear at around 200\,kpc, as there are very few particles
with apocenters outside these radii, Fig.~\ref{fig:movie}. Video
from the simulation is part of the electronic attachment. For the
description of the video, see Appendix \ref{Apx:Videos} point \ref{enu:video2-kinem}
and \ref{enu:video3-proj}.

\subsection{Comparison of the simulation with models \label{sec:sim-mod}}

In the simulations, some of the assumptions that we used earlier (the
model of radial oscillations, Sect.~\ref{sec:rad_osc}) are not fulfilled.
First, the particles do not move radially, but on more general trajectories,
which are, in the case of a radial merger, nevertheless very eccentric.
Second, not all the particles are released from the cannibalized galaxy
right in the center of the host galaxy; when the potential is switched
off, the particles are located in the broad surroundings of the center
and some are even released before the decay of the galaxy. These effects
cause a smearing of the kinematical imprint of shells, as the turning
points are not at a sharply defined radius, but rather in some interval
of radii for a given time.

\begin{figure}[H]
\centering{}\includegraphics[width=12cm]{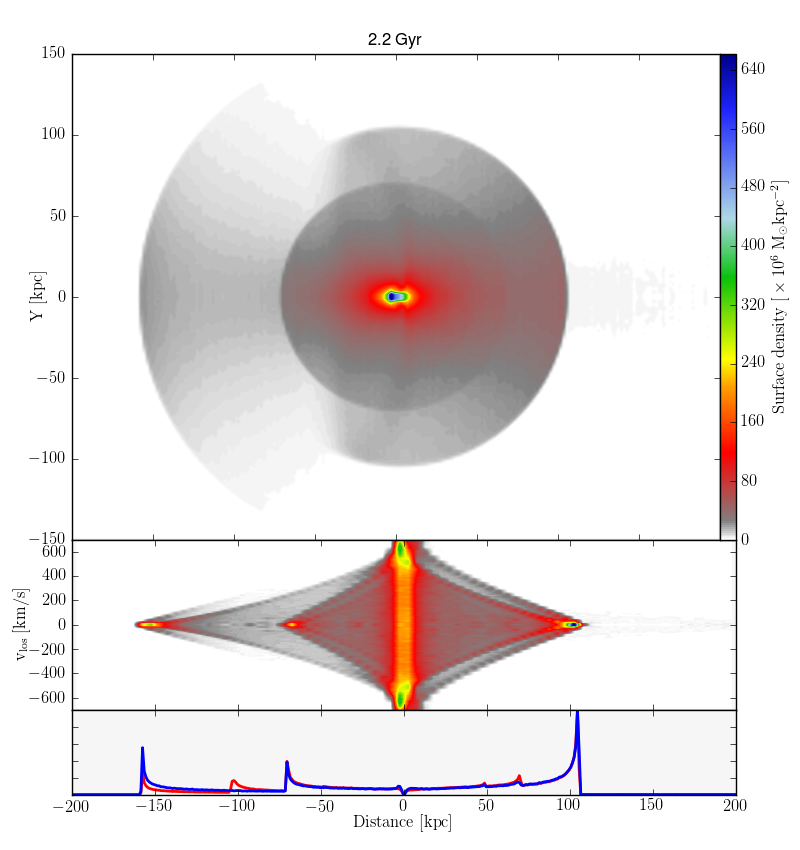}
\caption{\textsf{\small Simulated shell structure 2.2\,Gyr after the decay
of the cannibalized galaxy. Only the particles originally belonging
to the cannibalized galaxy are taken into account. Top: surface density
map; middle: the LOSVD density map of particles in the $\pm1$\,kpc
band around the collision axis; bottom: histogram of galactocentric
distances of particles. The angle between the radial position vector
of the particle and the $x$-axis (the collision axis) is less than
90$\degr$ for the blue curve and less than 45$\degr$ for the red
curve. The horizontal axis corresponds to the projected distance $X$
in the upper panel, to the projected radius $R$ in the middle panel,
and to the galactocentric distance $r$ in the lower panel. \label{fig:2200Myr} }}
\end{figure}

\clearpage

\begin{table}[H]
\centering{}%
\begin{tabular}{cccccc}
\hline 
{\small $r_{\mathrm{s}}$ } & {\small $n$ } & {\small $r_{\mathrm{TP,model}}$ } & {\small $v_{\mathrm{s,sim}}$ } & {\small $v_{\mathrm{s,model}}$ } & {\small $v\mathrm{_{c,model}}$}\tabularnewline
{\small kpc } &  & {\small kpc } & {\small km$/$s } & {\small km$/$s } & {\small km$/$s}\tabularnewline
\hline 
{\small 48.8 } & {\small 5 } & {\small 48.5 } & {\small 38.7$\pm$2.1 } & {\small 38.7 } & {\small 326}\tabularnewline
{\small $-$70.6 } & {\small 4 } & {\small $-$69.9 } & {\small 59.8$\pm$1.6 } & {\small 54.3 } & {\small 390}\tabularnewline
{\small 105.0 } & {\small 3 } & {\small 103.9 } & {\small 68.1$\pm$1.9 } & {\small 63.5 } & {\small 441}\tabularnewline
{\small $-$157.8 } & {\small 2 } & {\small $-$155.7 } & {\small 74.3$\pm$1.2 } & {\small 72.4 } & {\small 450}\tabularnewline
{\small 257.4 } & {\small 1 } & {\small 251.0 } & {\small 97.5$\pm$1.4 } & {\small 95.7 } & {\small 406}\tabularnewline
\hline 
\end{tabular}\caption{\textsf{\small Parameters of the shells in a simulation 2.2\,Gyr
after the decay of the cannibalized galaxy. The shell positions $r_{\mathrm{s}}$
are taken from the simulation. The values of $r_{\mathrm{TP,model}}$
and $v_{\mathrm{s,model}}$ are calculated for the shell position
$r_{\mathrm{s}}$ and its corresponding serial number $n$ according
to the model of radial oscillations (Sect.~\ref{sec:rad_osc}). The
shell velocity $v_{\mathrm{s,sim}}$ is derived from 20 positions
between the times 2.49--2.51\,Gyr for each shell. The value $v\mathrm{_{c,model}}$
corresponds to the circular velocity at the shell-edge radius $r_{\mathrm{s}}$
for the chosen potential of the host galaxy (Sect.~\ref{sec:param}).
\label{tab:param-sim} }}
\end{table}

\begin{figure}[H]
\centering{}\includegraphics[width=12cm]{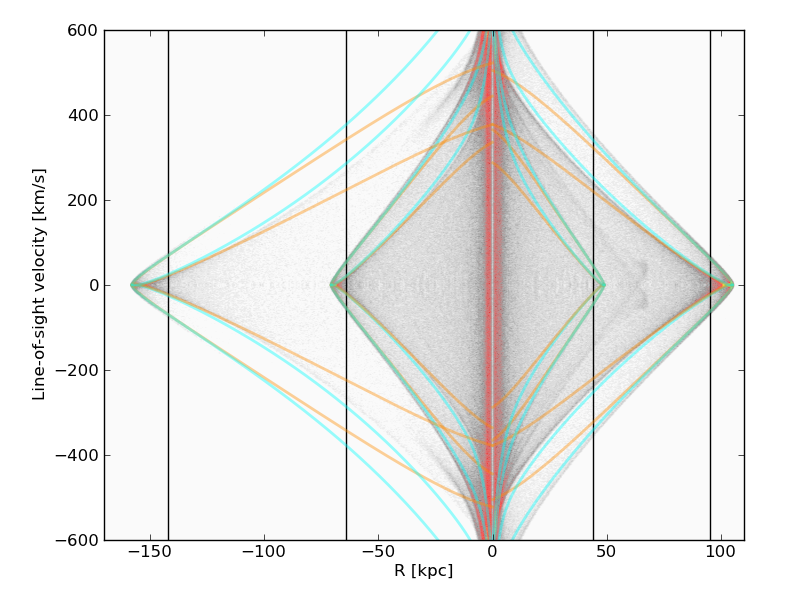}
\caption{\textsf{\small LOSVD map of the simulated shell structure 2.2\,Gyr
after the decay of the cannibalized galaxy (middle panel in Fig.~\ref{fig:2200Myr}).
Light blue curves show locations of the maxima according to the model
of radial oscillations (Sect.~\ref{sub:LOSVD-rad}) for shell radius
$r_{\mathrm{s}}$, corresponding serial number $n$, and the known
potential of the host galaxy (Sect.~\ref{sec:param}). Orange curves
are derived from the approximative maximal LOS velocities (Sect.~\ref{sec:Compars},
point~\ref{item:app-vmax}) given by Eq.~(\ref{eq:vlos,max}) for
$r_{\mathrm{s}}$, $v_{\mathrm{s,model}}$, and $v\mathrm{_{c,model}}$.
Parameters of the shells are shown in Table~\ref{tab:param-sim}.
Black lines mark the location at $0.9r_{\mathrm{s}}$ for each shell.
The LOSVD for these locations are shown in Fig.~\ref{fig:rezy-sim}.
The map includes only stars originally belonging to the cannibalized
galaxy. \label{fig:vel.map} }}
\end{figure}

The model of radial oscillations presented in Sect.~\ref{sec:rad_osc}
predicts that 2.2\,Gyr after the decay of the cannibalized galaxy,
five outermost shells should lie at the radii of 257.3, $-$157.8,
105.1, $-$70.5, and 48.8\,kpc. The negative radii refer to the shell
being on the opposite side of the host galaxy with respect to the
direction from which the cannibalized galaxy flew in. These radii
agree surprisingly well with the radii of the shells \textit{measured}%
\footnote{Recall that by \textit{measuring,} we mean that the data measured
are the output of our simulation.%
} in the simulation 2.2\,Gyr after the decay of the cannibalized galaxy,
see Fig.~\ref{fig:2200Myr} and Table~\ref{tab:param-sim}. The
position of the shell edge $r_{\mathrm{s}}$ in the simulation was
determined as the position of a sudden decrease of the projected surface
density (see Figs.~\ref{fig:FR} and ~\ref{fig:FR158}). These values
are shown in Table~\ref{tab:param-sim}.

In the simulation, the first shell at 257.4\,kpc is composed of only
a few particles, and therefore we will not consider it (its parameters
are listed in Table~\ref{tab:param-sim} for completeness). Thus,
the outermost relevant shell in the system lies at $-$157.8\,kpc
and has a serial number $n=2$. Also, the shell at 48.8\,kpc suffers
from lack of particles, but we will include it nevertheless.

\begin{figure}[!t]
\centering{}\includegraphics[width=15cm]{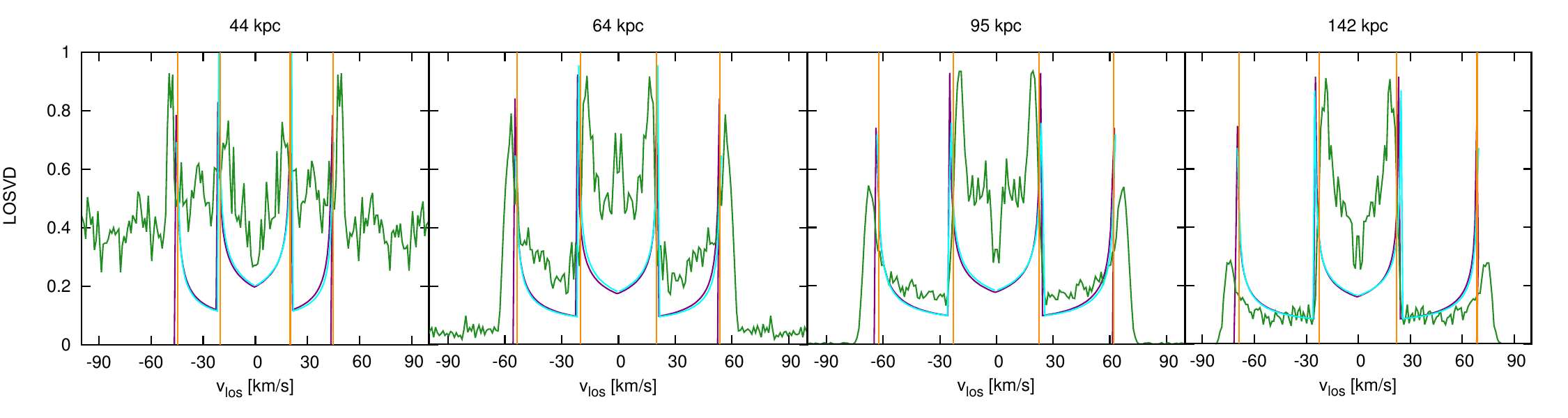}
\caption{\textsf{\small LOSVDs of four shells at projected radii $0.9r_{\mathrm{s}}$
(indicated as the title of each plot) 2.2\,Gyr after the decay of
the cannibalized galaxy (parameters of the shells are shown in Table~\ref{tab:param-sim}).
The simulated data are shown in green, the LOSVDs according to the
approximative LOSVD (Sect.~\ref{sec:Compars}, point~\ref{item:app-LOSVD})
given by Eq.~(\ref{eq:vlos}) in purple, and LOSVDs according to
the model of radial oscillations (Sect.~\ref{sub:LOSVD-rad}) in
light blue. The graph also shows the locations of the peaks using
the approximative maximal LOS velocities (Sect.~\ref{sec:Compars},
point~\ref{item:app-vmax}) given by Eq.~(\ref{eq:vlos,max}) by
orange lines. Profiles do not include stars of the host galaxy, which
are not part of the shell system. The theoretical profiles are scaled
so that the intensity of their highest peak approximately agrees with
the highest peak of the simulated data. LOSVD is given in relative
units, so maxima of the profiles have values of about 0.9. \label{fig:rezy-sim} }}
\end{figure}

Fig.~\ref{fig:vel.map} shows the comparison between the LOSVD in
the simulation, the peaks of the LOSVD computed in the model of radial
oscillations (light blue curves), and the approximative maximal LOS
velocities\,--\,Eq.~(\ref{eq:vlos,max}) (orange curves). To evaluate
the approximative maximal LOS velocities, we obtained the shell velocity
$v_{\mathrm{\mathrm{s,model}}}$ from the model of radial oscillations
(Sect.~\ref{sec:rad_osc}) for the respective serial number $n$
of the shell and circular velocity $v\mathrm{_{c,model}}$ at the
shell-edge radius, using our knowledge of the potential of the host
galaxy. The values of all the respective shell quantities are listed
in Table~\ref{tab:param-sim}. Within the resolution of Fig.~\ref{fig:vel.map},
the theoretical positions of the LOSVD maxima agree very well with
the simulated data, even further from the shell edge than the usual
limit of $0.9r_{\mathrm{s}}$.

Fig.~\ref{fig:vel.map} also shows the locations that correspond
to the radii of $0.9r_{\mathrm{s}}$ for each individual shell (black
lines). The LOSVD for these locations is shown in Fig.~\ref{fig:rezy-sim}.
The data are taken from an area spanning $0.5\times2$\,kpc centered
at $(R,0)$ in the projected $X-Y$ plane, where $R$ is the number
indicated above the corresponding panel in Fig.~\ref{fig:rezy-sim}.
The positions of simulated LOSVD peaks largely agree with the approaches
of the approximation of constant acceleration and shell velocity described
in Sect.~\ref{sec:Compars} and with the model of radial oscillations
(Sect.~\ref{sec:rad_osc}).

\begin{figure}[!t]
\centering{}\includegraphics[width=7.5cm]{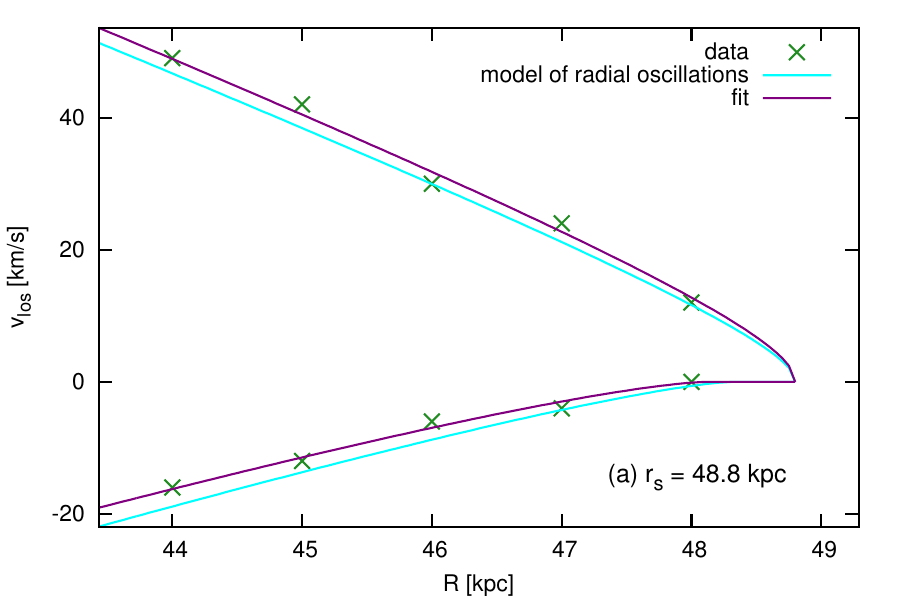}\includegraphics[width=7.5cm]{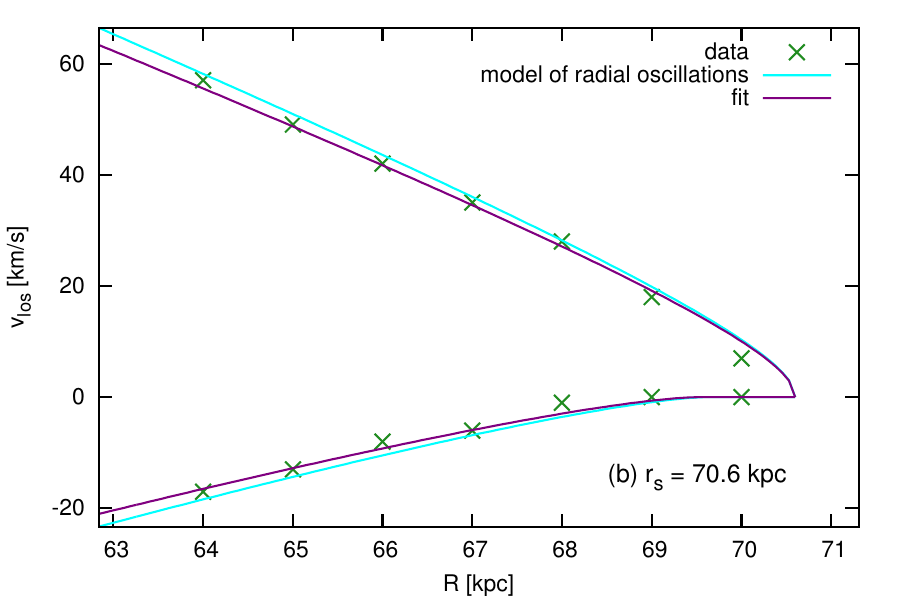}\\
 \includegraphics[width=7.5cm]{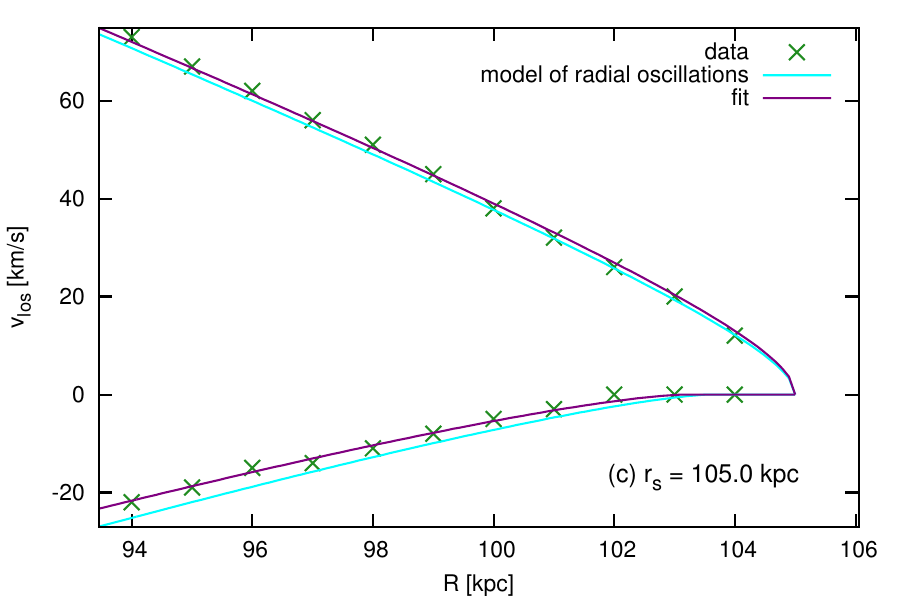}\includegraphics[width=7.5cm]{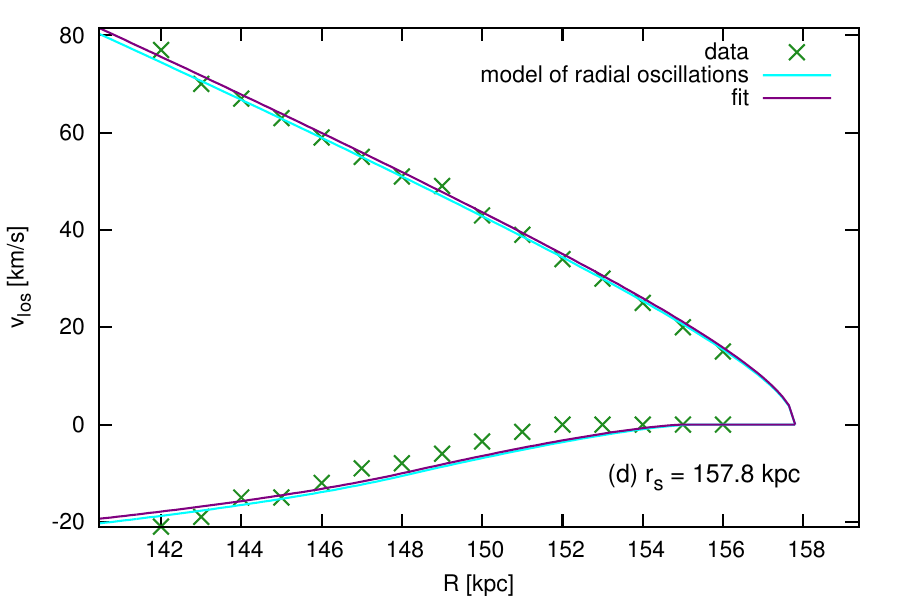}
\caption{\textsf{\small Fits for circular velocity $v_{\mathrm{c}}$ and shell
velocity $v_{\mathrm{s}}$ using the approximative LOSVD (Sect.~\ref{sec:Compars},
point~\ref{item:app-LOSVD}) given by Eq.~(\ref{eq:vlos}) for four
shells ($r_{\mathrm{s}}$ indicated in bottom right corner of each
plot) in the simulation 2.2\,Gyr after the decay of the cannibalized
galaxy. The best fit is the purple curve, and its parameters are shown
in Tables~\ref{tab:param-vc} and \ref{tab:param-vs} in the columns
labeled $v_{\mathrm{c,fit}}$ and $v_{\mathrm{s,fit}}$. The green
crosses mark the }\textsf{\textit{\small measured}}\textsf{\small{}
maxima in the LOSVD, and the light blue curves show the locations
of the theoretical maxima derived from the host galaxy potential according
to the model of radial oscillations (Sect.~\ref{sub:LOSVD-rad}).
Note that the values of $v_{\mathrm{c}}$ and $v_{\mathrm{s}}$ used
in the approximative LOSVD for the purple line were obtained by fitting
the parameters to the simulated data, whereas in Figs.~\ref{fig:120app},
\ref{fig:app-rez}, and \ref{fig:rezy-sim}, the values are known
from the model of the host galaxy potential. \label{fig:minfit} }}
\end{figure}

\subsection{Recovering the potential from the simulated data \label{sub:Recover}}

We used a snapshot from our simulation, which 2.2\,Gyr after the
decay of the cannibalized galaxy, as a source of the simulated data
and tried to reconstruct the parameters of the potential of the host
galaxy from the locations of the LOSVD peaks \textit{measured} from
the simulated data by using the the approximation of constant acceleration
and shell velocity (Sect.~\ref{sec:IOA}).

For a given host galaxy, the signal-to-noise (S$/$N) ratio in the
simulated data is a function of the number of simulated particles,
the age of the shell system, the distribution function of the cannibalized
galaxy, and the impact velocity. For a given radius in the simulated
data, we can obtain arbitrarily good or bad S$/$N ratios by tuning
these parameters. Thus, we adopted the universal criteria: 1) the
LOSVD of each shell is observed down to 0.9 times its radius; 2) we
\textit{measured} the positions of the LOSVD peaks in different locations
within the shell, sampled by 1\,kpc steps. These criteria give us
between 7 and 15 \textit{measurements} for a shell. Each \textit{measurement}
contains two values: the positions of the outer and inner peaks, $v_{\mathrm{los,max}+}$
and $v_{\mathrm{los,max}-}$, respectively, for each projected radius
$R$ (see green crosses in Fig.~\ref{fig:minfit}). 

\begin{figure}[h]
\centering{}\includegraphics[width=12cm]{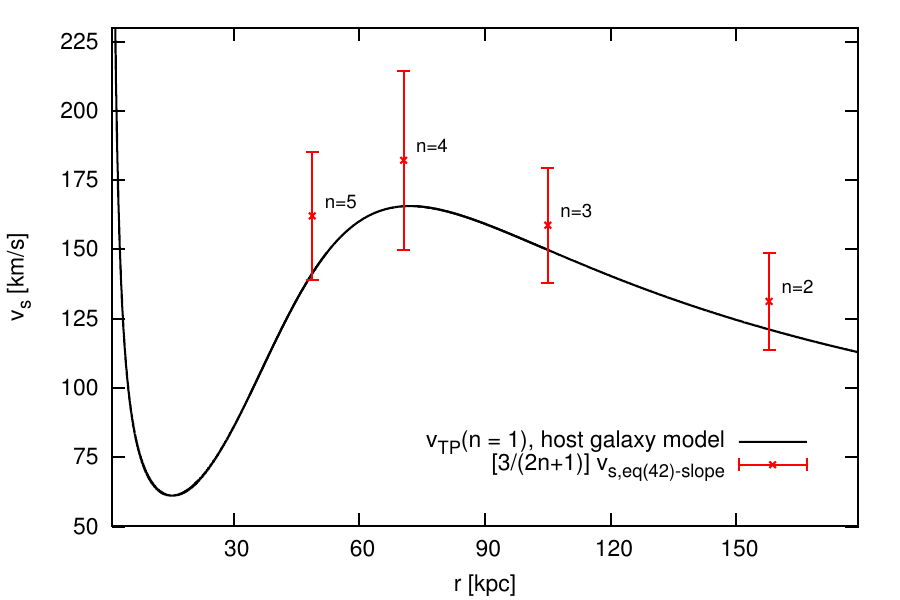}
\caption{\textsf{\small Comparison of velocity of the shell as a function of
radius from the model and the simulated data. Velocity for the first
shell ($n=1$) in the host galaxy model is shown by the black line.
Red crosses show $v_{\mathrm{s,eq(\ref{eq:vs-vc})-slope}}$ (Table\,\ref{tab:param-vs})
as they result from the analysis of the simulated LOSVD. Values are
corrected for shell number $n$ by the factor $3/(2n+1)$, so they
correspond to velocity of the first shell, e.g., Eq.\,(\ref{eq:vTP}).
\label{fig:vscomp} }}
\end{figure}

We do not estimate the errors, since the real data will be dominated
by other sources, such as the contamination of the signal from the
light of the host galaxy and the accuracy of the subtraction of this
background light, night-sky background in the case of ground-based
telescopes, detector noise, instrumental dispersion, accuracy in the
determination of the systemic velocity and so forth. So we decided
to quote only the mean square deviation and the standard error of
the linear regression.

First we used the approximative maximal LOS velocities given by Eqs.~(\ref{eq:vc,obs})
and (\ref{eq:vs,obs}) for a direct calculation of the circular velocity
$v_{\mathrm{c,eq(\ref{eq:vc,obs})}}$ at the shell-edge radius $r_{\mathrm{s}}$
and the current shell velocity $v_{\mathrm{s,eq(\ref{eq:vs,obs})}}$.
These equations are the inverse of Eq.~(\ref{eq:vlos,max}), which
corresponds to the model shown in orange lines in pictures throughout
the text (Sect.~\ref{sec:Compars}, point~\ref{item:app-vmax}).
Mean values from all the \textit{measurements} for each shell are
shown in Tables~\ref{tab:param-vc} and~\ref{tab:param-vs} in the
end of the section.

\begin{figure}[h]
\centering{}\includegraphics[width=12cm]{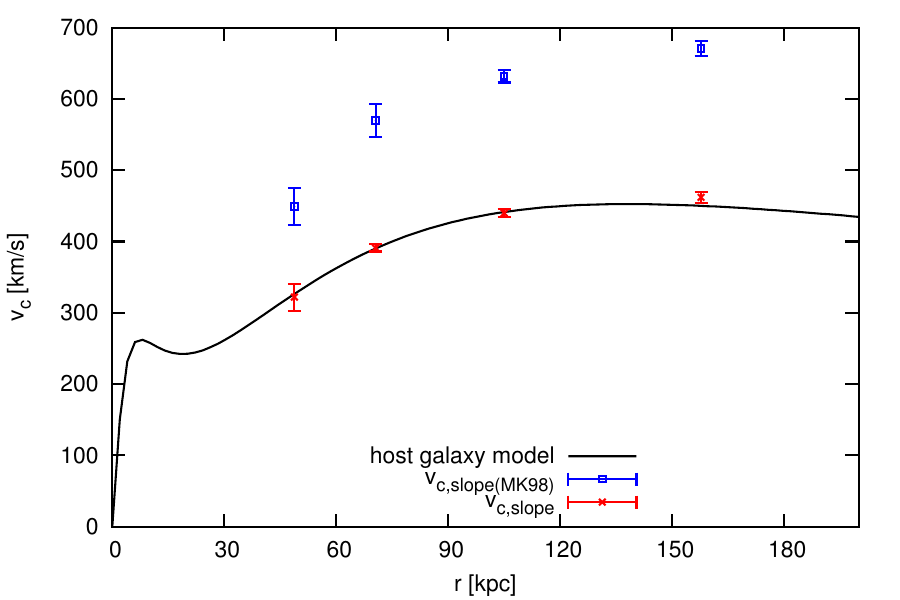}
\caption{\textsf{\small Circular velocity of the model and values derived from
the simulated data: $v\mathrm{_{c,model}}$ of the host galaxy model
is shown by the black line; blue and red points show values of circular
velocity as they result from the analysis of the simulated LOSVD (see
Sect.\,\ref{sec:sim-mod} and Table\,\ref{tab:param-vc} for the
numbers). \label{fig:vccomp} }}
\end{figure}

Compared with the approximative maximal LOS velocities, we obtain
a better agreement with the circular velocity of our host galaxy potential
when using the slope of the LOSVD intensity maxima (Sect.~\ref{sec:Compars},
point~\ref{item:dis-slope}) given by Eq.~(\ref{eq:sklon}), where
we fit the linear function of the \textit{measured} distance between
the outer and the inner peak on the projected radius ($v_{\mathrm{c,slope}}$
in Table~\ref{tab:param-vc} and in Fig.~\ref{fig:vccomp}). To
estimate the shell velocity, we use a hybrid relation Eq.~(\ref{eq:vs-vc})
between the positions of the LOSVD peaks, the circular velocity at
the shell-edge radius $v_{\mathrm{c}}$, and the shell velocity. We
substitute the values of $v_{\mathrm{c,slope}}$ derived from the
\textit{measurements} (that we know better describe the real circular
velocity of the host galaxy) into this relation, thus obtaining the
improved \textit{measured} shell velocity $v_{\mathrm{s,eq(\ref{eq:vs-vc})-slope}}$
(Table~\ref{tab:param-vs} and Fig.~\ref{fig:vscomp}).

In the zone between the current turning points and the shell edge,
the inner peaks coalesce and gradually disappear (Fig.~\ref{fig:zona}).
The simulated data do not show a disappearance of the inner peaks
as abrupt and clear as the theoretical LOSVD profiles predict, so
that in this zone, we can usually \textit{measure} one inner peak
at 0\,km$/$s. The information from these \textit{measurements} is
degenerate, and thus we defined a subsample of simulated \textit{measurements}
with all four clear peaks in the LOSVD (in the columns labeled SS
in Tables~\ref{tab:param-vc} and \ref{tab:param-vs}). The spread
of the values derived using the approximative maximal LOS velocities
given by Eqs.~(\ref{eq:vc,obs}) and (\ref{eq:vs,obs}) is significantly
lower for the subsample ($v_{\mathrm{c,eq(\ref{eq:vs,obs})}}^{\mathrm{SS}}$
and $v_{\mathrm{s,eq(\ref{eq:vc,obs})}}^{\mathrm{SS}}$) due to the
exclusion of areas where these equations do not hold well. On the
contrary, the slope of the linear regression in Eq.~(\ref{eq:sklon})
using the slope of the LOSVD intensity maxima gives a worse result
(with a larger error) for the subsample $v_{\mathrm{c,slope}}^{\mathrm{SS}}$
than the approximative maximal LOS velocities.

The third option to derive the circular velocity $v_{\mathrm{c}}$
at the shell-edge radius $r_{\mathrm{s}}$ and shell velocity $v_{\mathrm{s}}$
from the simulated data is to use the approximative LOSVD given by
Eq.~(\ref{eq:vlos}), which corresponds to the model shown in purple
lines in pictures throughout the text (Sect.~\ref{sec:Compars},
point~\ref{item:app-LOSVD}). However, this requires a numerical
solution of the equation for a given pair of $v_{\mathrm{c}}$ and
$v_{\mathrm{s}}$. We have calculated two sums of squared differences
between $v_{\mathrm{los,max}}(v_{\mathrm{c}},v_{\mathrm{s}})$ as
given by the approximative LOSVD and the simulated data. One for $v_{\mathrm{los,max-}}(v_{\mathrm{c}},v_{\mathrm{s}})$
and a second one for $v_{\mathrm{los,max+}}(v_{\mathrm{c}},v_{\mathrm{s}})$
. Then we have searched for the minimum of the sum of these two values
to obtain best fitted values $v_{\mathrm{c,fit}}$ and $v_{\mathrm{s,fit}}$
(see Tables~\ref{tab:param-vc} and \ref{tab:param-vs} for the results).
Errors were estimated using the ordinary least squared minimization
as if the functions $v_{\mathrm{los,max}+}(v_{\mathrm{c,fit}},v_{\mathrm{s,fit}})$
and $v_{\mathrm{los,max}-}(v_{\mathrm{c,fit}},v_{\mathrm{s,fit}})$
were fitted separately; quoted is the larger of the two errors. 

\begin{table}[h]
\centering{}%
\begin{tabular}{cccccccccc}
\hline 
\noalign{\vskip-0.1cm}
 &  &  &  &  &  &  &  &  & \tabularnewline[-0.1cm]
{\footnotesize $r_{\mathrm{s}}$ } & {\footnotesize $v\mathrm{_{c,model}}$ } & {\footnotesize $N$ } & {\footnotesize $N^{\mathrm{SS}}$ } & {\footnotesize $v_{\mathrm{c,eq(\ref{eq:vc,obs})}}$ } & {\footnotesize $v_{\mathrm{c,eq(\ref{eq:vc,obs})}}^{\mathrm{SS}}$ } & {\footnotesize $v_{\mathrm{c,slope}}$ } & {\footnotesize $v_{\mathrm{c,slope}}^{\mathrm{SS}}$ } & {\footnotesize $v_{\mathrm{c,fit}}$ } & {\footnotesize $v_{\mathrm{c,slope(MK98)}}$}\tabularnewline
{\footnotesize kpc } & {\footnotesize km$/$s } &  &  & {\footnotesize km$/$s } & {\footnotesize km$/$s } & {\footnotesize km$/$s } & {\footnotesize km$/$s } & {\footnotesize km$/$s } & {\footnotesize km$/$s}\tabularnewline
\hline 
{\footnotesize 48.8 } & {\footnotesize 326 } & {\footnotesize 5 } & {\footnotesize 4 } & {\footnotesize 346$\pm$130 } & {\footnotesize 340$\pm$94 } & {\footnotesize 322$\pm$19 } & {\footnotesize 314$\pm$32 } & {\footnotesize 318$\pm$51 } & {\footnotesize 449$\pm$26}\tabularnewline
{\footnotesize $-$70.6 } & {\footnotesize 390 } & {\footnotesize 7 } & {\footnotesize 5 } & {\footnotesize 394$\pm$85 } & {\footnotesize 390$\pm$53 } & {\footnotesize 391$\pm$5 } & {\footnotesize 392$\pm$11 } & {\footnotesize 368$\pm$60 } & {\footnotesize 570$\pm$23}\tabularnewline
{\footnotesize 105.0 } & {\footnotesize 441 } & {\footnotesize 11 } & {\footnotesize 8 } & {\footnotesize 478$\pm$144 } & {\footnotesize 452$\pm$64 } & {\footnotesize 440$\pm$5 } & {\footnotesize 447$\pm$7 } & {\footnotesize 427$\pm$28 } & {\footnotesize 632$\pm$9}\tabularnewline
{\footnotesize $-$157.8 } & {\footnotesize 450 } & {\footnotesize 15 } & {\footnotesize 10 } & {\footnotesize 497$\pm$236 } & {\footnotesize 472$\pm$79 } & {\footnotesize 462$\pm$8 } & {\footnotesize 484$\pm$14 } & {\footnotesize 460$\pm$32 } & {\footnotesize 671$\pm$11}\tabularnewline
\hline 
\end{tabular}\caption{\textsf{\small Circular velocity at the shell-edge radius $r_{\mathrm{s}}$
derived from the }\textsf{\textit{\small measurement}}\textsf{\small{}
of the simulated data 2.2\,Gyr after the decay of the cannibalized
galaxy. $r_{\mathrm{s}}$ and $v\mathrm{_{c,model}}$ have the same
meaning as in Table~\ref{tab:param-sim}. $N$: number of }\textsf{\textit{\small measurements}}\textsf{\small{}
for each shell; $v_{\mathrm{c,eq(\ref{eq:vc,obs})}}$: the mean of
values derived from the approximative maximal LOS velocities given
by Eq.~(\ref{eq:vc,obs}) with its mean square deviation; $v_{\mathrm{c,slope}}$:
a value derived from the linear regression using the slope of the
LOSVD intensity maxima given by Eq.~(\ref{eq:sklon}) and its standard
error (see also Fig.~\ref{fig:vccomp}); $v_{\mathrm{c,fit}}$: a
value derived by fitting a pair of $v_{\mathrm{c}}$ and $v_{\mathrm{s}}$
in the approximative LOSVD given by Eq.~(\ref{eq:vlos}) (Sect.~\ref{sec:Compars},
point~\ref{item:app-LOSVD} and Fig.~\ref{fig:minfit}); $v_{\mathrm{c,slope(MK98)}}$:
the mean of values derived from the slope of the LOSVD intensity maxima
given by Eq.~(\ref{eq:sklon}) with its standard error (see also
Fig.~\ref{fig:vccomp}). In the equation, however, $\bigtriangleup v_{\mathrm{los}}$
is substituted with the distance between the two outer peaks of the
LOSVD intensity maxima in order to mimic the }\textsf{\textit{\small measurement}}\textsf{\small{}
as originally proposed by \citetalias{mk98} for double-peaked profile.
The quantities with the superscript SS correspond to the subsample,
where only }\textsf{\textit{\small measurements}}\textsf{\small{} with
two discernible inner peaks in the LOSVD are used. \label{tab:param-vc} }}
\end{table}

\begin{table}[H]
\centering{}%
\begin{tabular}{cccccccc}
\hline 
\noalign{\vskip-0.1cm}
 &  &  &  &  &  &  & \tabularnewline[-0.1cm]
{\footnotesize{} $r_{\mathrm{s}}$ } & {\footnotesize $v_{\mathrm{s,model}}$ } & {\footnotesize $v_{\mathrm{s,sim}}$ } & {\footnotesize $v_{\mathrm{s,eq(\ref{eq:vs,obs})}}$ } & {\footnotesize $v_{\mathrm{s,eq(\ref{eq:vs,obs})}}^{\mathrm{SS}}$ } & {\footnotesize $v_{\mathrm{s,eq(\ref{eq:vs-vc})-slope}}$ } & {\footnotesize $v_{\mathrm{s,eq(\ref{eq:vs-vc})-slope}}^{\mathrm{SS}}$ } & {\footnotesize $v_{\mathrm{s,fit}}$}\tabularnewline
{\footnotesize kpc } & {\footnotesize km$/$s } & {\footnotesize km$/$s } & {\footnotesize km$/$s } & {\footnotesize km$/$s } & {\footnotesize km$/$s } & {\footnotesize km$/$s } & {\footnotesize km$/$s}\tabularnewline
\hline 
{\footnotesize 48.8 } & {\footnotesize 38.7 } & {\footnotesize 38.7$\pm$2.1 } & {\footnotesize 50.7$\pm$2.3 } & {\footnotesize 51.7$\pm$1.1 } & {\footnotesize 44.2$\pm$6.5 } & {\footnotesize 44.9$\pm$6.3 } & {\footnotesize 53$\pm$16}\tabularnewline
{\footnotesize $-$70.6 } & {\footnotesize 54.3 } & {\footnotesize 59.8$\pm$1.6 } & {\footnotesize 60.8$\pm$9.8 } & {\footnotesize 65.6$\pm$2.0 } & {\footnotesize 60.7$\pm$10.8 } & {\footnotesize 66.0$\pm$2.9 } & {\footnotesize 66$\pm$19}\tabularnewline
{\footnotesize 105.0 } & {\footnotesize 63.5 } & {\footnotesize 68.1$\pm$1.9 } & {\footnotesize 74.8$\pm$4.6 } & {\footnotesize 76.5$\pm$1.4 } & {\footnotesize 68.0$\pm$8.9 } & {\footnotesize 71.3$\pm$2.5 } & {\footnotesize 79$\pm$9}\tabularnewline
{\footnotesize $-$157.8 } & {\footnotesize 72.4 } & {\footnotesize 74.3$\pm$1.2 } & {\footnotesize 84.4$\pm$5.4 } & {\footnotesize 86.7$\pm$2.0 } & {\footnotesize 78.7$\pm$10.5 } & {\footnotesize 82.$\pm$3.5 } & {\footnotesize 85$\pm$14}\tabularnewline
\hline 
\end{tabular}\caption{\textsf{\small Velocity of the shell at the radius $r_{\mathrm{s}}$
derived from the }\textsf{\textit{\small measurement}}\textsf{\small{}
of the simulated data 2.2\,Gyr after the decay of the cannibalized
galaxy. $r_{\mathrm{s}}$, $v_{\mathrm{s,model}}$, and $v_{\mathrm{s,sim}}$
have the same meaning as in Table~\ref{tab:param-sim}. $v_{\mathrm{s,eq(\ref{eq:vs,obs})}}$:
the mean of values derived from the approximative maximal LOS velocities
given by Eq.~(\ref{eq:vs,obs}) with its mean square deviation; $v_{\mathrm{s,eq(\ref{eq:vs-vc})-slope}}$:
the mean of values derived from the hybrid relation given by Eq.~(\ref{eq:vs-vc})
with its mean square deviation (see also Fig.~\ref{fig:vscomp});
$v_{\mathrm{s,fit}}$: a value derived by fitting a pair of $v_{\mathrm{c}}$
and $v_{\mathrm{s}}$ in the approximative LOSVD given by Eq.~(\ref{eq:vlos})
(Sect.~\ref{sec:Compars}, point~\ref{item:app-LOSVD} and Fig.~\ref{fig:minfit}).
The quantities with the superscript SS correspond to the subsample,
where only }\textsf{\textit{\small measurements}}\textsf{\small{} with
two discernible inner peaks in the LOSVD are used. Number of }\textsf{\textit{\small measurements}}\textsf{\small{}
is the same as in Table~\ref{tab:param-vc} for each shell. \label{tab:param-vs} }}
\end{table}

The LOSVD intensity maxima resulting from this procedure are plotted
in Fig.~\ref{fig:minfit}, together with the fitted data and the
maxima given by the model of radial oscillations (Sect.~\ref{sub:LOSVD-rad}).
All three agree fairly well. The remaining two methods (the approximative
maximal LOS velocities and using the slope of the LOSVD intensity
maxima) use only equations to derive $v_{\mathrm{c}}$ and $v_{\mathrm{s}}$
and thus we do not show them in the plot. On the other hand, in Figs.~\ref{fig:vscomp}
and \ref{fig:vccomp}, we show the comparison of values extracted
from the simulated data with model values only for the most successful
approach -- using the slope of the LOSVD intensity maxima.

For the sake of comparison with the method of \citetalias{mk98},
we calculated the circular velocity $v_{\mathrm{c,slope(MK98)}}$
at the shell-edge radius $r_{\mathrm{s}}$ using the slope of the
LOSVD intensity maxima given by Eq.~(\ref{eq:sklon}). To mimic the
\textit{measurement} of the circular velocity according to the Eq.~(\ref{eq:sklonMK}),
which was derived for the double-peaked profile, we assume $\bigtriangleup v_{\mathrm{los}}$
is the distance between the two outer peaks of the LOSVD intensity
maxima. In Table~\ref{tab:param-vc} and Fig.~\ref{fig:vccomp},
we can easily see that the values $v_{\mathrm{c,slope(MK98)}}$ differ
from the actual circular velocity of the host galaxy $v\mathrm{_{c,model}}$
by a factor of 1.3--1.5. 

The main message of this section is that in order to obtain the value
of the circular velocity $v_{\mathrm{c}}$ at the shell-edge radius
and shell phase velocity $v_{\mathrm{s}}$ from kinematical data near
the shell edge, the best approach to use is the method based on the
slope of the LOSVD intensity maxima given by Eq.~(\ref{eq:sklon})
without limiting the data to a subsample.

\subsection{Notes about observation \label{sec:fwhm}}

This work is a theoretical one, dealing with simulations and models.
Obtaining and analyzing real data requires preparation, knowledge
and experience that are beyond the goals we have set in this research.
Nevertheless, we will make some remarks regarding potential observation
of shell kinematics.

\begin{figure}[H]
\centering{}\includegraphics[width=15cm]{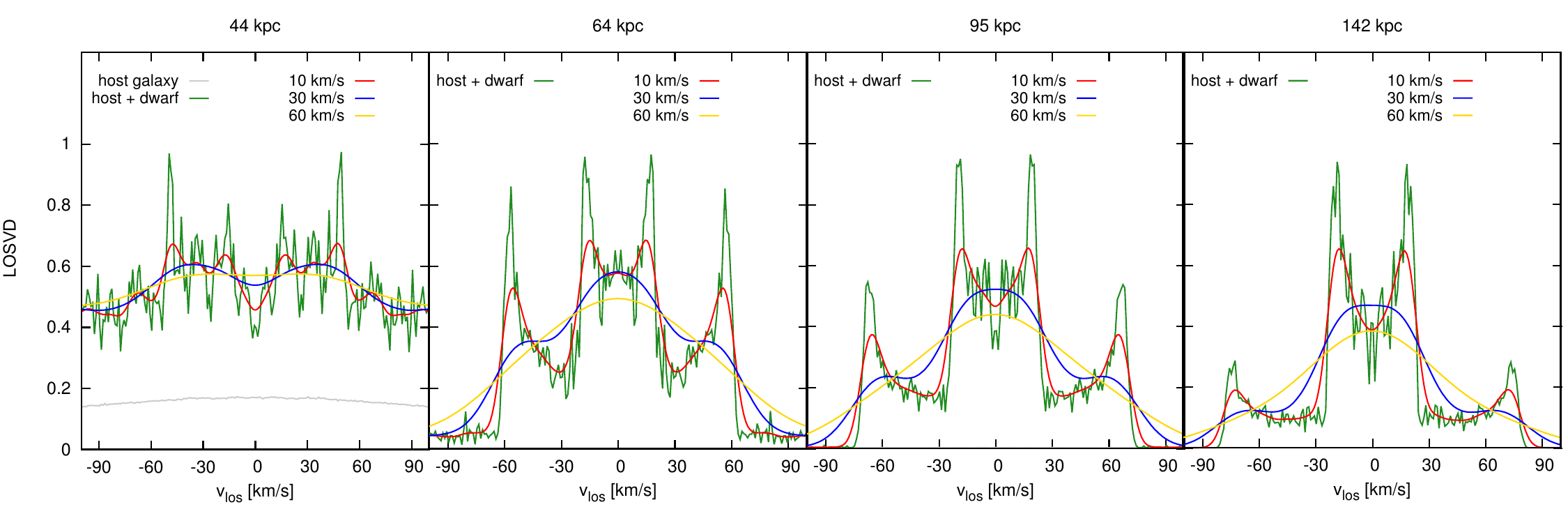}
\caption{\textsf{\small Line profiles of four shells at projected radii $0.9r_{\mathrm{s}}$
(indicated as the title of each plot, same as in Fig.~\ref{fig:rezy-sim})
2.2\,Gyr after the decay of the cannibalized galaxy: gray lines show
the LOSVDs for the host galaxy at a given radius (except for the radius
of 44\,kpc the signal of the host galaxy is negligible comparing
to the signal from the cannibalized galaxy); green lines show the
total LOSVDs from the host and the cannibalized galaxy together; red,
blue, and yellow lines show convolutions of the total simulated data
with different Gaussians representing the instrumental profiles having
the FWHM 10, 30, and 60\,km$/$s, respectively. Scaling is relative,
similar as in Fig.~\ref{fig:rezy-sim}. \label{fig:konv} }}
\end{figure}

When it comes to real observational data, there will be additional
issues to deal with, night-sky background, detector noise, instrumental
dispersion and so forth. \citetalias{mk98} estimated the data of
the requisite quality could be obtained with a couple of nights integration
using a 4-m telescope. 

The situation gets more complex when the LOSVD assumes the quadruple-peaked
profile instead of a double-peaked one. Not only becomes the intensity
of a single peak smaller, but a higher spectral resolution is also
needed to distinguish all four peaks. The instrumental dispersion
naturally smooths features of the spectral profile. In Fig.\,\ref{fig:konv},
we show the LOSVDs from the simulated data smoothed with different
Gaussians representing the instrumental profiles having the full width
at half maximum (FWHM) of 10, 30, and 60\,km$/$s. It is obvious
that relatively high spectral resolution is necessary for observing
an imprint of shell peaks in line profiles. 

We have done our own simplified estimations of the observability
of the LOSVD of shells. First, we used archival data of long-slit
spectroscopy of the outermost shell in NGC\,3923. The data were taken
in July 2001 (about 10 hours of exposure time) and in March 2005 (about
20 hours) with FORS2 instrument at the Very Large Telescope (VLT,
8.2 meter diameter) of the European Southern Observatory. We processed
a part of the data from 2005 using the FORS pipeline.%
\footnote{The procedure was done mostly by Lucie J{\'{\i}}lkov{\'a}, Ivana Orlitov{\'a}, and
Tereza Skalick{\'a}%
} The spectra are generally of a very low signal-to-noise ratio (S$/$N).
We were particularly looking for the magnesium triplet around 5200\,\AA{}
(taken into account the redshift of NGC\,3923, about 30\,\AA{})
and we found no sign of it, so the analysis of kinematics was not
possible. We conclude that the estimate of \citetalias{mk98} was
probably a bit of an understatement.

Furthermore, we used exposure time calculators to determine expected
S$/$N at available instruments (VLT/FORS2, VLT/FLAMES, Calar Alto/PPAK)
assuming the exposure time 20 hours and the surface brightness of
shells between 25 and 28\,mag$/$arcsec\textsuperscript{2} in V
filter. The resulting S$/$N ranges from $\sim0.3$ to $\sim4.4$.
This is not very satisfactory but using the integral field spectroscopy
or the multi object spectroscopy, S$/$N could be increased by a factor
of up to $\sim10$ by summing the signal from all fibers. Moreover,
one can use some kind of a cross-correlation technique (e.g., \citealp{1974A&A....31..129S,1979AJ.....84.1511T})
which allows to extract more accurate kinematic measurements than
the actual resolution of the data is or extract more information from
data with low S$/$N. Eventually, the situation should be much better
with the next generation of telescopes, like the European Extremely
Large Telescope or the James Webb Space Telescope.

Another important issue is the background light of the host galaxy.
It is possible to model the LOSVD of the host galaxy, subtract it
from the overall LOSVD and obtain the clear quadruple-peaked profile,
but it may not be even necessary, because the velocity dispersion
of the stars in the host galaxy would be likely significantly broader
than the distance between the peaks and thus the peaks should be clearly
visible already in the overall LOSVD. 

Moreover, for shells at large radii, the contribution from the stars
of the host galaxy becomes negligible -- and it is exactly the shells
at large radii that are the most interesting because our knowledge
of the potential of the host galaxy is the worst in the outer parts
of the galaxy, where the potential is expected to be dominated by
the dark matter. In our simulated data, the host galaxy light is negligible
already for the shell at 70\,kpc, see Fig.\,\ref{fig:konv}. The
surface brightness of observed shells goes from 24.5\,mag$/$arcsec\textsuperscript{2}
(in V filter) up to the current detection limit of the deepest photometric
observation $\sim29$\,mag$/$arcsec\textsuperscript{2} \citep{1990AJ....100.1073M,1999MNRAS.307..967T,2004AN....325..359P}.
The surface brightness of giant elliptical galaxies at $\sim100$\,kpc
(the position of the outermost shell in NGC\,3923) is 28--30\,mag$/$arcsec\textsuperscript{2}
(in g and r filters; \citealp{2011ApJ...731...89T}).

A category on its own is the measurement of LOS velocities of individual
objects, such as globular clusters, planetary nebulae and individual
giant stars \citep{fardal12,romanowsky12}, where the result is dependent
only on the accuracy of the measurement and the number of measured
objects.

The positions of LOSVD maxima should be symmetric around the systemic
velocity which we can measure or assume to be in the middle between
the peaks. We also need photometric data to find the center of the
host galaxy and to measure the distance of the point of the spectroscopic
observation and the shell edge from the center. As soon as we measure
the locations of the LOSVD peaks $v_{\mathrm{los,max}+}$, $v_{\mathrm{los,max}-}$,
the projected radius $R$ of the measurement, and the shell-edge radius
$r_{\mathrm{s0}}$, we can calculate the value of the circular velocity
$v_{\mathrm{c}}$ at the shell-edge radius and shell phase velocity
$v_{\mathrm{s}}$ using one of the three approaches described in Sect.~\ref{sec:Compars}.
Using the simulated data (Sect.~\ref{sub:Recover}), we found the
derived $v_{\mathrm{c}}$ to be the most accurate when using the slope
of the LOSVD intensity maxima given by Eq.~(\ref{eq:sklon}), which
requires the peak locations to be measured at several different radii.
When a measurement from only one projected radius is available, Eqs.~(\ref{eq:vc,obs})
and (\ref{eq:vs,obs}) can be used to derive $v_{\mathrm{c}}$ and
$v_{\mathrm{s}}$ , respectively.

\section{Shell density \label{sec:Shell-brightness}}

In this section we take an apparent detour from the shell kinematics
to explore the projected and volume densities of a shell. In Sect.~\ref{sub:sigma_los}
we express the projected surface density of the shell edge $\Sigma_{\mathrm{los}}(r_{\mathrm{s}})$
(that is, the projected surface density at the projected radius $R=r_{\mathrm{s}}$)
as a function of $\Sigma_{\mathrm{sph}}$ (Sects.~\ref{sub:Equations-of-LOSVD},
\ref{sub:sigma}, and \ref{sub:Nature-of-4peak}) and the shell-edge
radius $r_{\mathrm{s}}$. In Sect.~\ref{sub:Time-evolution} we investigate
the evolution of $\Sigma_{\mathrm{los}}(r_{\mathrm{s}})$ as a function
of time, as the position of the shell edge is a function of time.
In Sect.~\ref{sub:Volume-density} we show the volume density of
a shell at a frozen moment and finally in Sect.~\ref{sub:Projected-surface-brightness},
we explore the projected surface density of shells near the shell
edge at a given time as a function of the projected radius $R$.

\subsection{Projected surface density of the shell edge \label{sub:sigma_los}}

Each time we needed to model an LOSVD, we have used the assumption
that the shell-edge density distribution $\sigma_{\mathrm{sph}}\left(r_{\mathbf{s}}\right)$
or rather  $\Sigma_{\mathrm{sph}}\left(r_{\mathbf{s}}\right)$ decreases
as $1/r_{\mathrm{s}}^{2}\left(t\right)$, see Sects.~\ref{sub:Equations-of-LOSVD},
\ref{sub:sigma}, and \ref{sub:Nature-of-4peak}. Now we show how
is this value related to an observable quantity, the projected surface
density of the shell edge $\Sigma_{\mathrm{los}}(r_{\mathrm{s}})$.
If we knew or assumed the mass-to-light ratio, $\Sigma_{\mathrm{los}}$
could be easily converted to the projected surface brightness.

\begin{figure}[!t]
\centering{}\includegraphics[width=7.5cm]{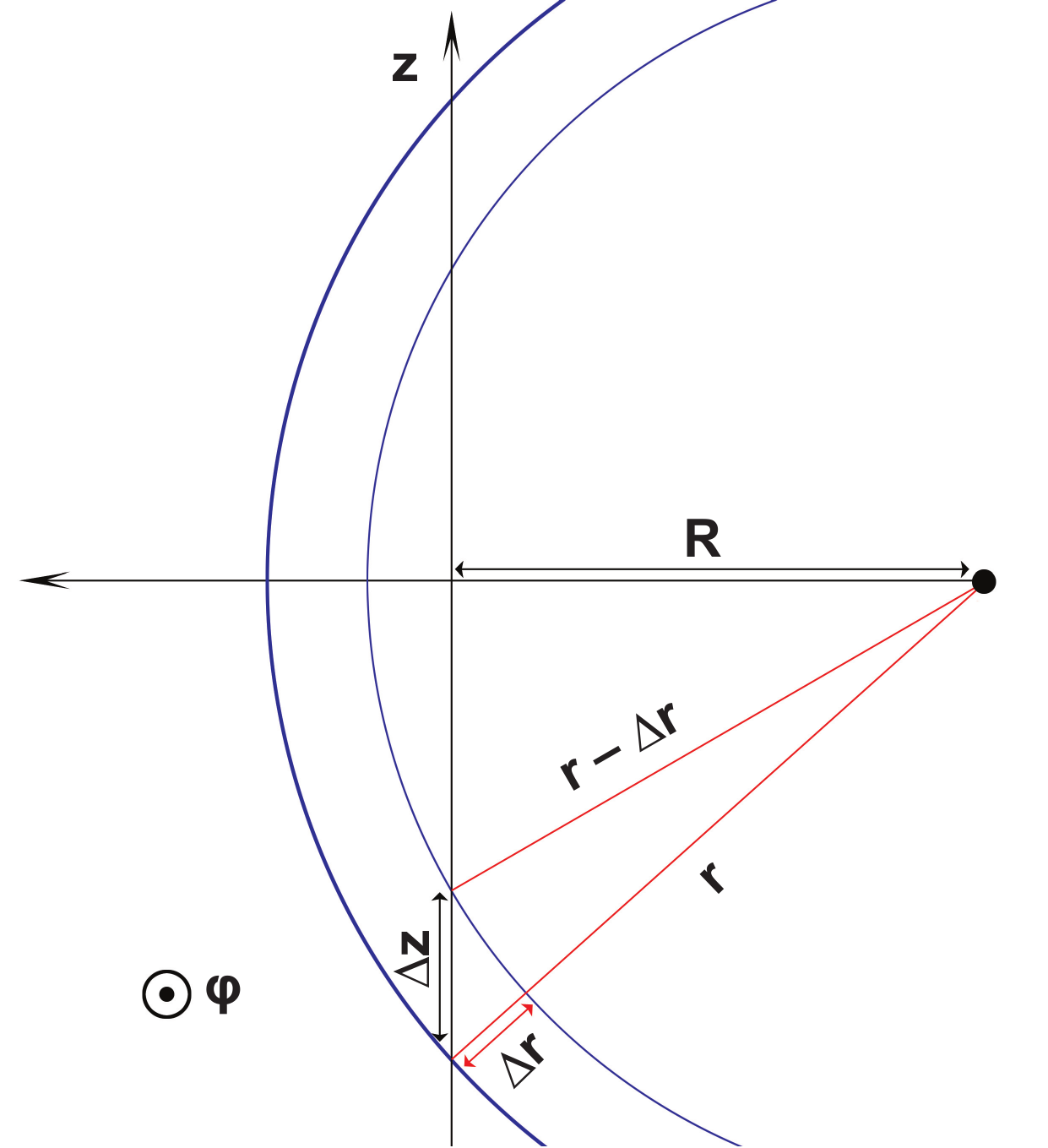}\includegraphics[bb=0bp 0bp 601bp 490bp,width=7.5cm]{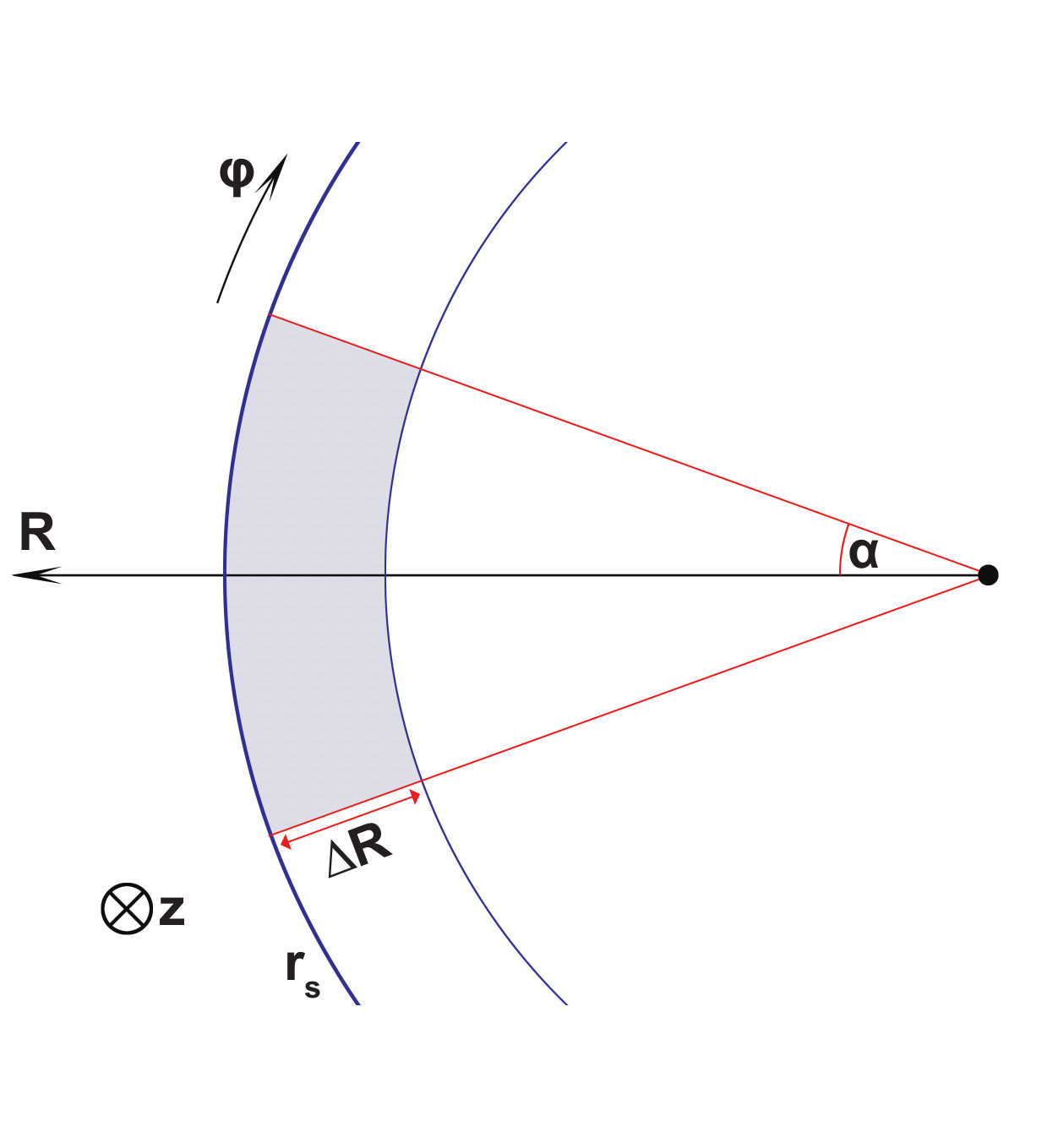}
\caption{\textsf{\small Schema for the calculation of the projected surface
density. \label{fig:bright} }}
\end{figure}

Consider a thin sphere of mass with a uniform spatial density $\text{\ensuremath{\rho}}$
and radius $r_{s}$, Fig.~\ref{fig:bright}. When observed along
the line of sight $z$, the amount of light registered from a point
with a projected radius $R$ in the sphere's image is proportional
to the expression

\begin{equation}
\rho\Delta z=\rho\left(\sqrt{r_{\mathrm{s}}^{2}-R^{2}}-\sqrt{\left(r_{\mathrm{s}}-\Delta r\right)^{2}-R^{2}}\right),
\end{equation}
which for an infinitesimally thin sphere ($\Delta r\rightarrow0$)
reduces to

\begin{equation}
\rho\Delta z\rightarrow\frac{r_{\mathrm{s}}\Sigma_{\mathrm{sph}}}{\sqrt{r_{\mathrm{s}}^{2}-R^{2}}}.
\end{equation}
 This expression diverges when the sphere is observed tangentially
to its surface, that is on the shell edge -- thus to talk about the
projected surface density of the shell edge, we have to integrate
the flux over a small observation area. As the shape of the area is
irrelevant for infinitesimal sizes, we choose an area that is the
easiest to integrate over in spherical coordinates that are convenient
for a radially-symmetric density. Note that the angular size of the
area is approximately $2\Delta R/r_{s}$ and thus the integrated flux
is 
\begin{equation}
\Sigma_{\mathrm{los}}=\frac{2}{S}\Sigma_{\mathrm{sph}}r_{\mathrm{s}}\intop_{0}^{\frac{\Delta R}{r_{\mathrm{s}}}}\intop_{r_{\mathrm{s}}-\Delta R}^{r_{\mathrm{s}}}\frac{R}{\sqrt{r_{\mathrm{s}}^{2}-R^{2}}}\mathrm{d}R\mathrm{d}\phi,
\end{equation}
 where $S=2\Delta R^{2}+o(\Delta R^{3})$ is the size of the integration
area. Since $\intop_{a}^{b}\frac{x}{\sqrt{r^{2}-x^{2}}}\mathrm{d}x=\sqrt{r^{2}-b^{2}}-\sqrt{r^{2}-a^{2}}$,
the integral reads 
\begin{equation}
\Sigma_{\mathrm{los}}\simeq\Sigma_{\mathrm{sph}}\sqrt{\left(2r_{\mathrm{s}}-\Delta R\right)/\Delta R}\propto r_{\mathrm{s}}^{1/2}\Sigma_{\mathrm{sph}}.\label{eq:sigma_los}
\end{equation}

\begin{figure}[!t]
\centering{}\includegraphics[width=12.5cm]{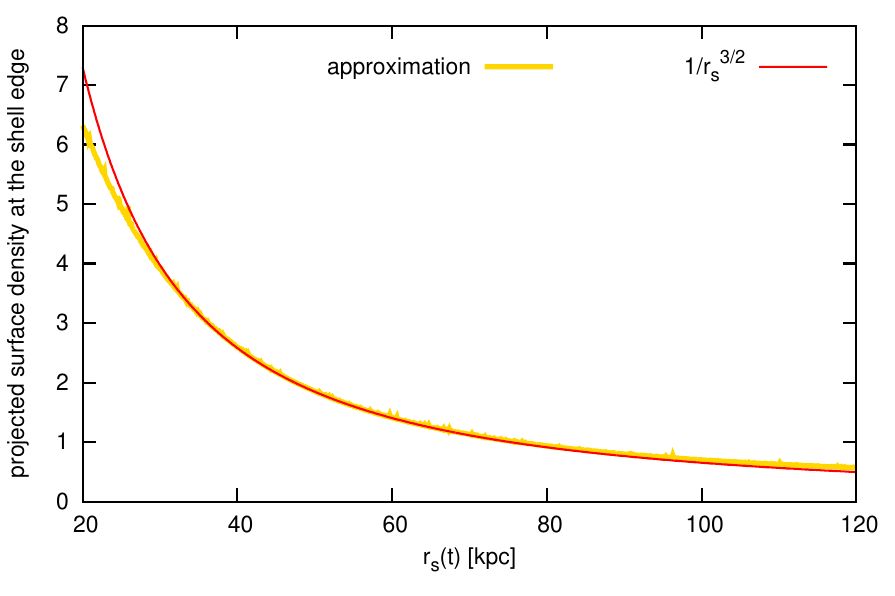}
\caption{\textsf{\small Time evolution of the projected surface density of
the shell edge (0.01\,kpc) in the approximation of a constant radial
acceleration in the host galaxy potential and shell phase velocity
(Sect.~\ref{sec:IOA}) -- yellow curve, in arbitrary units. The red
curve represents a function $r^{-3/2}$ normalized so that it has
the same value at $R=60$\,kpc as the yellow curve. For the parameters
of the host galaxy potential, see Sect.~\ref{sec:param}. \label{fig:bright-IOA} }}
\end{figure}

\begin{figure}[!b]
\centering{}\includegraphics[width=12.5cm]{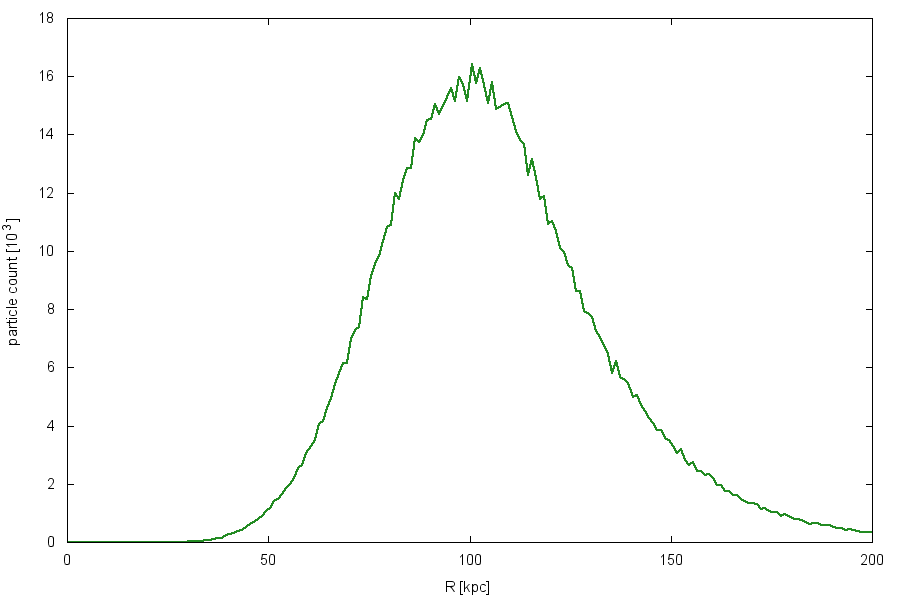}
\caption{\textsf{\small Histogram of apocenters of particles in the simulation
used in Sect.~\ref{sec:N-Simulations}. \label{fig:apoc} }}
\end{figure}

\subsection{Time evolution \label{sub:Time-evolution}}

The radial dependence of $\Sigma_{\mathrm{sph}}$ is chosen, as usual,
as $\Sigma_{\mathrm{sph}}(r_{\mathrm{s}}(t))\propto1/r_{\mathrm{s}}^{2}(t)$.
Then from Eq.~(\ref{eq:sigma_los}) it follows that
\begin{equation}
\Sigma_{\mathrm{los}}(r_{\mathrm{s}}(t))\propto r_{\mathrm{s}}^{-3/2}(t).\label{eq:sigma_-3/2}
\end{equation}
However, the calculation leading to Eq.~(\ref{eq:sigma_los}) assumes
that all the stars are located at the sphere with the radius of the
shell. We have thus examined the time evolution of the projected surface
density of the shell edge in the framework of the approximation of
a constant radial acceleration in the host galaxy potential and shell
phase velocity (Sect.~\ref{sec:IOA}, in this section, Sect.~\ref{sec:Shell-brightness},
hereafter \textit{the approximation}) -- Fig.~\ref{fig:bright-IOA}.
For each shell radius we calculate the motion of stars under a constant
acceleration, but we update this acceleration for different shell
radii according to the chosen potential of the host galaxy (for the
parameters of the potential, see Sect.~\ref{sec:param}). The time
evolution of the projected surface density of the shell edge in this
approximation does not depend on its velocity and thus on its serial
number, see Sect.~\ref{sub:Projected-surface-brightness}. In this
approximation, stars are present at all radii, 0--$r_{\mathrm{s}}$,
in contrast to the calculation that lead us to Eq.~(\ref{eq:sigma_-3/2}),
where we assumed the stars to be located only at the shell radius
(in a given time). Nevertheless, the time evolution of $\Sigma_{\mathrm{sph}}(r_{\mathrm{s}}(t))$,
Fig.~\ref{fig:bright-IOA}, turns out to be essentially identical
when calculated by either of these approaches.

Both the calculation of Eq.~(\ref{eq:sigma_-3/2}), and the approximation
assume $\Sigma_{\mathrm{sph}}$ to decrease as $1/r_{\mathrm{s}}^{2}\left(t\right)$,
corresponding to constant number of stars at the edge of the shell,
$N\left(r_{\mathbf{s}}\right)$. Fig.~\ref{fig:apoc} shows the distribution
of apocenters of particles in the simulation from Sect.~\ref{sec:N-Simulations},
which is a good approximation to real $N\left(r_{\mathbf{s}}\right)$.
We have to honestly admit that this function is anything but constant,
but it is difficult to devise any approximation as the shape of the
distribution significantly varies with parameters of the collision.
Moreover, we do apply this function usually only in a small range
of radii and as we have already shown, the character of the LOSVD
does not depend much on its choice (Sects.~\ref{sub:sigma} and \ref{sub:Nature-of-4peak}).
Converting the histogram of apocenters of the particles to the shell
brightness is not straightforward as, both in the simulation and real
shell galaxies, the distribution of particles is not uniform in azimuth,
contrary to what he assumed in modeling the LOSVD both in the approximation
and in the model of radial oscillations (Sect.~\ref{sub:LOSVD-rad}).

\subsection{Volume density \label{sub:Volume-density}}

\begin{figure}[!t]
\centering{}\includegraphics[width=12cm]{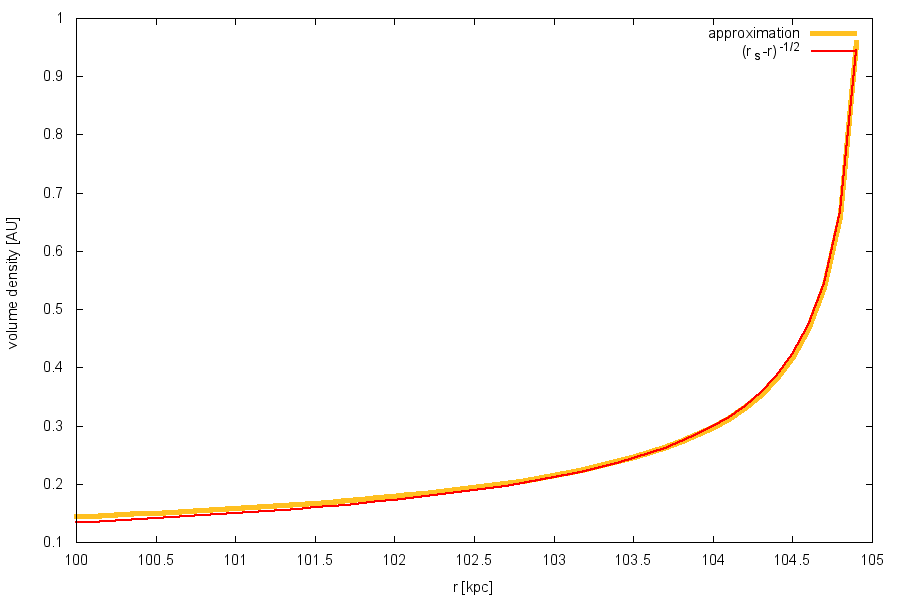}
\caption{\textsf{\small Volume density for the third shell at $105$\,kpc
in the approximation of a constant radial acceleration in the host
galaxy potential and shell phase velocity (Sect.~\ref{sec:IOA})
-- yellow curve, in arbitrary units. The red curve represents a function
$(r_{\mathrm{s}}-r)^{-1/2}$ normalized so that at $r_{\mathrm{s}}-r=1.1$\,kpc
it has the same value as the yellow curve. For the parameters of the
host galaxy potential, see Sect.~\ref{sec:param}. \label{fig:volume-dens} }}
\end{figure}

The calculation in Sect.~\ref{sub:sigma_los} assumes that stars
are at each moment located only on a sphere with the radius of the
shell. Nevertheless it gives good results when compared to the approximation
(Fig.~\ref{fig:bright-IOA}), where this assumption does not hold.
The reason is that the volume density decreases quickly inward from
the shell edge (it obviously decreases outward in a jump, but that
is not of concern at the moment). In their work, \citet{1988ApJ...331..682H}
recall that \citet{1984cath.book.....A} states that for phase wrapped
shells, that are just caustics in the mapping of the particle density
from phase space into three-dimensional space, it holds that the density
behind a caustic should scale as $(r_{\mathrm{s}}-r)^{-1/2}$. This
behavior should be independent of the used potential of the host galaxy.
In Fig.~\ref{fig:volume-dens} we have compared the volume density
near the shell edge in the approximation with this function and they
indeed show a pretty good agreement.

For a stationary shell, the volume density near the shell edge holds
\begin{equation}
\rho(r)=\frac{k}{v_{r}r^{2}},
\end{equation}
where $k$ is a constant for the given shell and $v_{r}$ is the radial
velocity of the shell. In a field of constant acceleration $a_{0}$
Eq.~(\ref{eq:vrMK}) holds -- $v_{r}=\sqrt{2a_{0}(r-r_{\mathrm{s}})},$
thus the volume density is 
\begin{equation}
\rho(r)\propto\frac{1}{r^{2}\sqrt{r-r_{\mathrm{s}}}}.\label{eq:x^-1/2}
\end{equation}
In the vicinity of the shell, the term $(r_{\mathrm{s}}-r)^{-1/2}$
dominates. For a moving shell it is difficult to make such analysis,
but we have seen on an example, in Fig.~\ref{fig:volume-dens}, that
this holds even in such case.

\subsection{Projected surface density \label{sub:Projected-surface-brightness}}

Finally we reach a really observable quantity that is the projected
surface density on the sky for a shell in a given time. For volume
density following Eq.~(\ref{eq:x^-1/2}) the projected surface density
turns out to be constant after integration. Thus we can assume constant
projected surface density/brightness immediately behind the shell.
The sharp-edged appearance of shells is caused by the abrupt decrease
of their brightness outside the shell radius, as we already demonstrated
in Sect.~\ref{sub:Visage}.

\begin{figure}[!t]
\centering{}\includegraphics[width=7.9cm]{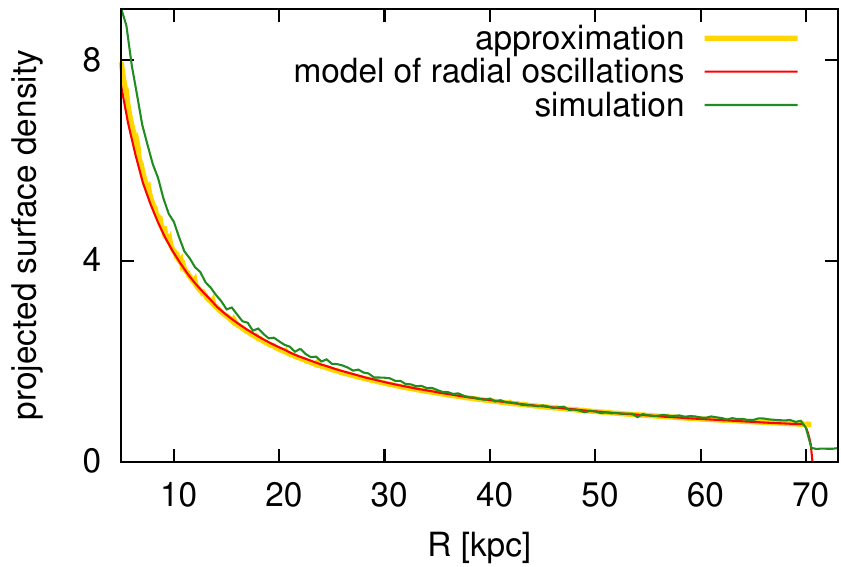}\includegraphics[width=7.2cm]{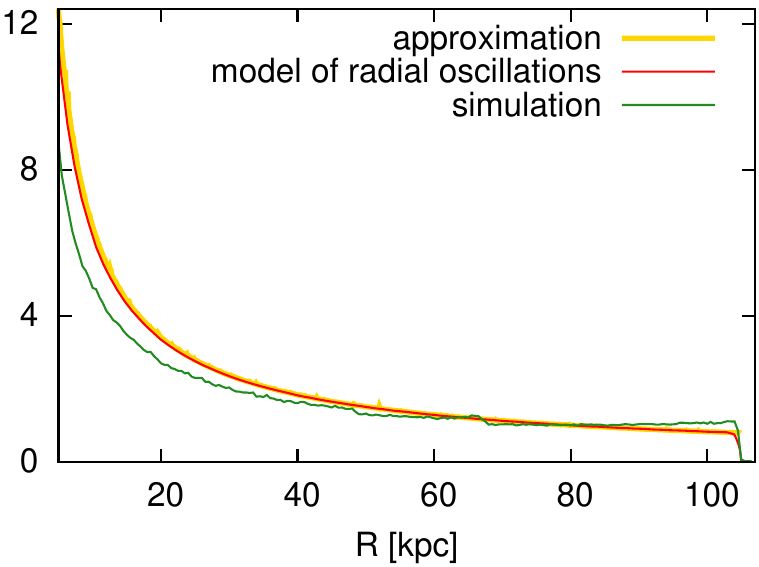}
\caption{\textsf{\small Surface brightness profile for two shells from simulation
used in Sect.~\ref{sec:N-Simulations} -- green curve; for equivalent
shells using the approximation (Sect.~\ref{sec:IOA}) -- yellow curve,
and the model of radial oscillations (Sect.~\ref{sec:rad_osc}) --
red curve. The curves are normalized so that they coincide and assume
unit value at 50 and 80\,kpc for shells with radii 70 and 105\,kpc,
respectively. \label{fig:FR}}}
\end{figure}

Fig.~\ref{fig:FR} shows the projected surface density profile for
two shells from the simulation (Sect.~\ref{sec:N-Simulations}) and
for shells on same radii (70 and 105\,kpc) using the approximation
and the model of radial oscillations (Sect.~\ref{sec:rad_osc}).
The approximation departs from the model of radial oscillations slightly
only in the vicinity of the center of the host galaxy. In the approximation,
the current location of a star for different $t_{\mathrm{s}}$ does
not depend on the shell velocity, see Eq.~(\ref{eq:r(0)IOA}), where
$t_{\mathrm{s}}$ is the time where the star was or will be at the
shell edge. Thus even the projected surface density calculated in
the approximation does not depend on the serial number of the shell.
The character of the profile immediate behind the shell is however
slowly rising toward the center of the host galaxy, rather than constant.
The shapes of the profile from the simulation and the approximation
or the model of the radial oscillations coincide fairly well, even
though the approximation and the model of radial oscillations assume
uniform azimuthal distribution of particles which is obviously not
valid in the simulation (see e.g. Figs.~\ref{fig:2200Myr} or \ref{fig:movie}).

\begin{figure}[!t]
\centering{}\includegraphics[width=12cm]{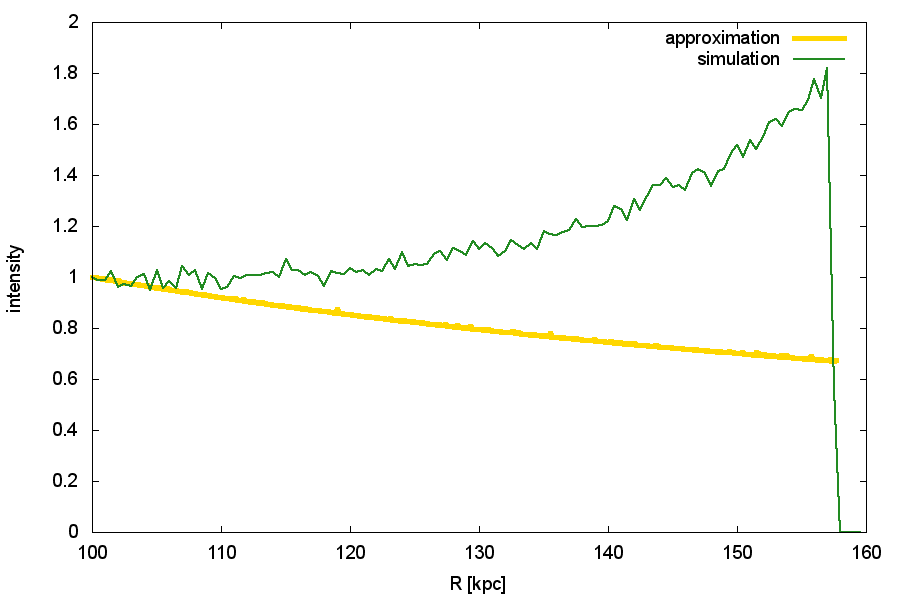}
\caption{\textsf{\small Surface brightness profile near the shell edge for
the outer shell from simulation used in Sect.~\ref{sec:N-Simulations}
-- green curve; and for equivalent shells using the approximation
-- yellow curve, and the model of radial oscillations -- red curve.
The curves are normalized so that they coincide and assume unit value
at 100\,kpc. \label{fig:FR158}}}
\end{figure}

On the other hand, no agreement at all is found for the outermost
shell from the simulation at 158\,kpc near its edge with the approximation
or the model of radial oscillations, Fig.~\ref{fig:FR158}. The simulated
shell even significantly decreases in brightness just at its edge.
The reason for this is that the shell is nearing its demise and stars
to arrive at higher radii are missing (see Figs.~\ref{fig:2200Myr}).
Another factor is the azimuthal development of brightness, as the
shell is the brightest near the axis of the merger and at higher angles
(measured from the axis of the merger) the number of stars decreases.
That, together with a large shell radius causes a decrease in the
projected surface density  at radii lower than the shell radius. A
universal profile of the projected  surface density/brightness for
phase wrapped shells thus does not exist, but in general a rather
constant or rising behavior can be expected for the inner shells,
whereas the outer shell can show decrease toward the center of the
host galaxy.

All the profiles of the projected surface density have been drawn
for a band $\pm1$\,kpc around the merger axis in the projected plane
perpendicular to the merger axis.

\section{Discussion \label{sec:Discussion-Part-II}}

In this part of the thesis, we developed a method to measure the potential
of shell galaxies from kinematical data, extending the work of \citetalias{mk98},
assuming a constant shell phase velocity and a constant radial acceleration
in the host galaxy potential for each shell. The method splits into
three different analytical and semi-analytical approaches (Sect.~\foreignlanguage{british}{\ref{sec:Compars}})
for obtaining the circular velocity in the host galaxy, $v_{\mathrm{c}}$,
and the current shell phase velocity, $v_{\mathrm{s}}$ -- the approximative
LOSVD, the approximative maximal LOS velocities, and the slope of
the LOSVD intensity maxima. In Sect.~\ref{sec:Compars}, the first
two approaches are compared to the model of radial oscillations (numerical
integration of radial trajectories of stars in the host galaxy potential,
Sect.~\ref{sec:rad_osc}). All three approaches are then applied
to data for the four shells obtained from a test-particle simulation
and compared to the theoretical values (Sect.~\ref{sec:sim-mod}).

The approximative LOSVD requires a numerical solution to Eq.~(\ref{eq:vlos})
and the search for a pair of $v_{\mathrm{c}}$ and $v_{\mathrm{s}}$,
which matches the (simulated) data best. Although this approach is
not limited by any assumptions about the radius of the maximal LOS
velocity (Sect.~\ref{sub:rvmax}), it does not give a better estimate
of $v_{\mathrm{c}}$ and $v_{\mathrm{s}}$ for our simulated shell
galaxy than the other two methods. The deviation from the real value
of $v_{\mathrm{c}}$ is between 2\,\% and 6\,\%.

Using the approximative maximal LOS velocities results in simple analytical
relations and is the only one that can in principle be used for a
LOSVD measured at only one projected radius. Nevertheless, when measuring
in the zone between the radius of the current turning points and the
shell radius, we can expect very bad estimates of $v_{\mathrm{c}}$
and $v_{\mathrm{s}}$. The mean value from more measurements of the
LOSVD peaks for each shell of our simulated shell galaxy has similar
accuracy to those of the approximative LOSVD, provided that we include
only the measurements outside the zone between the radius of the current
turning points and the shell radius.

The best method for deriving the circular velocity in the potential
of the host galaxy seems to be to use the slope of the LOSVD intensity
maxima, with a typical deviation in the order of units of km$/$s
when fitting a linear function over all the measured positions of
the LOSVD peaks for each shell. This circular velocity is then used
in the hybrid relation, Eq.~(\ref{eq:vs-vc}), to obtain the best
estimate of the shell phase velocity.

All the approaches, however, derive the shell phase velocity systematically
larger than the prediction of the model of radial oscillations $v_{\mathrm{s,model}}$
and the value derived from positions between the times 2.49--2.51\,Gyr
in the simulation $v_{\mathrm{s,sim}}$ (Table~\ref{tab:param-vs}).
This is because the simulated LOSVD peaks lie too far out (for the
outer peaks) or too far in (for the inner peaks) when compared to
the model of radial oscillations. That can be caused by nonradial
trajectories of the stars of the cannibalized galaxy or by poor definition
of the shell radius in the simulation.

Nevertheless, the shell phase velocity depends, even in the simplified
model of an instant decay of the cannibalized galaxy in a spherically
symmetric host galaxy (Sec.~\ref{sec:rad_osc}), on the serial number
of the shell $n$ and on the whole potential from the center of the
galaxy up to the shell radius, Eq.~(\ref{eq:vTP}). A comparison
of its measured velocity to theoretical predictions is possible only
for a given model of the potential of the host galaxy and the presumed
serial number of the observed shells. In such a case, however, it
can be used to exclude some parameters or models of the potential
that would otherwise fit the observed circular velocity.

The first shell has a serial number equal to one. A higher serial
number means a younger shell. On the same radius, the velocity of
each shell is always smaller than that of the previous one. In practice,
it is difficult to establish whether the outermost observed shell
is the first one created, or whether the first shell (or even the
first couple of shells) is already unobservable. Here, we can use
the potential derived from our method or a completely different one
in a reverse way: to determine the velocity of the first shell on
the given radius and to compare it to the velocity derived from the
positions of the LOSVD peaks. Knowing the serial number of the outermost
shell and its position allows us then to determine the time from the
merger and the impact direction of the cannibalized galaxy. Moreover,
the measurement of shell velocities can theoretically distinguish
the shells from different generations, which can be present in a shell
galaxy \citep{katka11}.

Our method for measuring the potential of shell galaxies has several
limitations. Theoretical analyses were conducted over spherically
symmetric shells, while the test-particle simulation was run for a
strictly radial merger and analyzed in a projection plane parallel
to the axis of the merger. In addition, both analytical analyses and
simulations assume spherical symmetry of the potential of the host
galaxy. In reality, the regular shell systems with higher number of
shells in a single host galaxy are more often connected to galaxies
with significant ellipticity \citep{1986A&A...166...53D}. Moreover,
in cosmological simulations with cold dark matter, halos of galaxies
are described as triaxial ellipsoids \citep[e.g.,][]{jing02,bailin05,allgood06}.
However, the effect of the ellipticity of the isophotes of the host
galaxy on the shell kinematics need not be dramatic, as the shells
have the tendency to follow equipotentials that are in general less
elliptical than the isophotes. \citet{1986A&A...166...53D} concluded
that while the ellipticity of observed shells is generally low, it
is neatly correlated to the eccentricity of the host galaxy. \citet{1988ApJ...326..596P}
pointed out that the shells in NGC\,3923 are much rounder than the
underlying galaxy and have an ellipticity that is similar to the inferred
equipotential surfaces. This idea was originally put forward by \citet{1986A&A...166...53D},
who found such a relationship for their merger simulations. Our method
is in principle applicable even to shells spread around the galactic
center, which are usually connected to rounder elliptical galaxies
if they were created in a close-to-radial merger. Nevertheless, the
combination of the effects of the projection plane, merger axis, and
ellipticity of the host galaxy can modify our results and require
further analyses.

Because the kinematics of the stars that left the cannibalized galaxy
is in the first approximation a test-particle problem, they should
not be much affected by self-gravity of the cannibalized galaxy and
the dynamical friction that this galaxy undergoes during the merger,
both of which have been neglected in Part~\ref{PART II-S.kin}.

Another complication is that the spectral resolution required to distinguish
all four peaks is probably quite high (Sect.~\ref{sec:fwhm} and
Fig.~\ref{fig:konv}) and the shell contrast is usually small. The
higher order approximation, Sect.~\ref{sec:Higher-order-approximation},
is sensible only when kinematical data are available to larger distances
from the shell edge. In the application to simulated data, we considered
a shell that is observable down to 0.9 shell radii. Nevertheless,
there is the possibility to measure shell kinematics using the LOS
velocities of individual globular clusters, planetary nebulae, and,
in the Local group of galaxies, even of individual stars. It is even
possible that the shell kinematics will be detectable in H\,I and
CO emission, see Sects.~\ref{sub:Gas-and-dust} and \foreignlanguage{british}{\ref{sub:Simulations-with-gas}}.

We have also explored the projected surface density of shells, Sect.~\ref{sub:Projected-surface-brightness}.
In the model of radial oscillations, the shells show constant projected
surface density near the shell edge, whereas outside the shell radius,
there is a step-like decrease of the density, creating the sharp-edged
feature of the shells. This behavior can be expected from shells with
a large development in azimuth and a sufficient supply of stars at
different energies. Already in our simple simulation of a radial minor
merger with test particles and instantaneous decay of the cannibalized
galaxy, we can observe a shell with a projected surface density that
defies this description. We can assume that the self-gravity and gradual
decay of the cannibalized galaxy can disrupt the observed profile
even further. Moreover, we worked only in strictly spherical potentials
and any non-zero ellipticity of the host galaxy can play a significant
role. For the moment, all we can say is how the projected surface
density of shells looks within the model of radial oscillations --
any stronger statement would require more detailed simulations.

\clearpage

\newpage{}

\part{Dynamical friction and gradual disruption\label{PART III-DF}}

In the same spirit as in Part~\ref{PART II-S.kin}, we will consider
the formation of the shell structure during a radial minor merger.
This time, we will try to get closer to real shell galaxies by introducing
into the test-particle simulations the gradual decay of the secondary
galaxy as well as its braking by dynamical friction against the primary.%
\footnote{In this section, we use the terms secondary or satellite, rather than
cannibalized galaxy. The host galaxy will be usually referred to as
primary. In related papers, one may also find the notation dwarf or
small galaxy for the secondary and giant elliptical or big galaxy
for the primary.%
}

\section{Motivation }

In Sect.~\ref{sub:TP} we have shown how are the positions of the
shells related to the potential of the host galaxy at different times
from the merger that created the shells. In practice, nevertheless,
it has proven difficult to reproduce the space distribution of the
shells in the observed shell galaxies using sensible potentials (Sect.~\ref{sec:MM}).
The main suspects of making the relation more complex are the dynamical
friction and the possibility to have shells from multiple generations.
In the case when the measurement of shell kinematics is not available,
we can pose a goal less ambitions than the derivation of the potential
of the host galaxy, that is to determine the age of the shell system
(the time of the merger). To this end, the measurement of the position
of the outermost shell could be sufficient, as this shell is the one
which is the least effected by those additional effects. 

As we have mentioned in Sect.~\ref{sec:Objectives-and-motivation},
this is the approach chosen by \citet{canalizo07}. They presented
observations of shells in a quasar host galaxy and, by simulating
the position of the outermost shell by means of restricted $N$-body
simulations, attempted to put constraints on the age of the merger.
They concluded that it occurred a few hundred Myr to $\sim2$\,Gyr
ago, supporting a potential causal connection between the merger,
the post-starburst ages in nuclear stellar populations, and the quasar.
A typical delay of 1--2.5\,Gyr between a merger and the onset of
quasar activity is suggested by both $N$-body simulations by \citet{2005MNRAS.361..776S}
and observations by \citet{2008ApJ...678..751R}. It might therefore
appear reassuring to find a similar time lag between the merger event
and the quasar ignition in a study of an individual spectacular object. 

The issue here is that noone has studied in detail the effects assumed
to complicate the shell distribution (the dynamical friction and the
gradual decay of the secondary galaxy) and thus it is not clear how
exactly they change the shell structure and how they influence the
position of the outermost shell. We try to include the dynamical friction
and the gradual decay of the cannibalized galaxy in test-particle
simulations. The manifestation of these processes in self-consistent
simulations is difficult to separate and sometimes they may even be
confused with non-physical outcomes of used methods. Test-particle
simulations helped us to separate and better understand the roles
of the dynamical friction and gradual tidal decay in the shell formation.
Moreover, self-consistent simulations become demanding on computation
time when we want to explore a significant part of the parameter space.

We look at what these enhanced test-particle simulations tell us about
the potential and merger history of shell galaxies with the focus
on the plausibility of the use of the outermost shell for dating the
merger.

\section{Description of simulation \label{sec:Description-of-simulation}}

In this and the previous part we show results of test-particle simulations
and in this section we describe the procedure of their calculation
in detail.

\subsection{Configuration \label{sub:Configuration}}

The test (i.e. mass-less) particles of the secondary galaxy are generated
(usually in counts from $10^{4}$ to $10^{7}$) so that they follow
the density profile of the secondary galaxy. The particles then move
according to a smooth gravitational potential of both galaxies, which
move with respect to each other based on their masses, shape of potentials,
positions and velocities; Eq.~(\ref{eq:Fgal}). Figures and videos
are generally oriented so that the secondary galaxy approaches originally
from the right hand side.

In the simplest case, when the centers of the galaxies pass through
each other, the potential of the secondary galaxy is suddenly switched
off and the particles continue to move only in the fixed potential
of the primary galaxy. This approach is applied in simulation in Part~\ref{PART II-S.kin}
and in some simulations in Part~\ref{PART III-DF}. In the simulations
with dynamical friction and gradual disruption, the smooth potential
of the secondary galaxy is kept for whole time and its mass is progressively
lowered during each successive passage. The dynamical friction is
added in the form of an (semi-)analytical prescription into the equations
of motion of galaxies.

All the simulations in the thesis are, for the sake of simplicity,
carried out for spherical galaxies, i.e. elliptical galaxies with
zero ellipticity. The secondary (cannibalized) galaxy is always modeled
as a single Plummer sphere. The primary (host) galaxy is modeled as
a single or double Plummer sphere in Part~\ref{PART III-DF}, while
in Part~\ref{PART II-S.kin} its potential has always two components,
both Plummer spheres.

For the numerical integration of the motion of the test particles
and the galaxies, the \textit{Leapfrog} method was chosen. In this
method, velocities derived for a time half step earlier (or later)
than the current position are used to update the position. Conversely,
to update the half-step velocity one step forward, the positions for
the round position in between are used. In so doing the velocities
can be seen to {}``leapfrog'' over the current time step. This simple
enterprise improves the accuracy of the numerical computation by an
order compared to when the position $x$ and velocities $v$ are taken
simultaneously. The error is of the order of $(\bigtriangleup t)^{3}$,
where $\Delta t$ is the time step. For the longest time step used
in our simulations (1\,Myr), the error for the trial circular motion
was only 11 revolutions after 10,000 (compared to the simple analytical
solution that is available in this case), what is only 1 per mille.

\subsection{Plummer sphere\label{sec:Plummer} \label{sub:Force} }

The gravitational potential of each of the galaxies in this part,
Part~\ref{PART III-DF}, is modeled with the Plummer profile with
varying parameters in different simulations:

\begin{equation}
\phi(r)=-\frac{\mathrm{G}\, M}{\sqrt{r^{2}+\varepsilon^{2}}},\label{eq:pot}
\end{equation}
 where G is the gravitational constant, \textit{$M$} is the total
mass of the galaxy, \textit{$r$} is the distance from the center
of the galaxy and $\varepsilon$ is the Plummer radius -- a scale
parameter that determines the compactness of the galaxy. For $\varepsilon$
= 0 the Eq.~(\ref{eq:pot}) represents a simple potential of a point
mass. The Plummer radius corresponds to the effective radius%
\footnote{The effective radius is the radius at which one half of the total
light of the galaxy is emitted interior to this radius.%
} of the galaxy.

While the Plummer model follows the profile of the real spherical
galaxies only approximately, we use it here -- as was the case of
numerous other studies of galaxies -- because of its simple expressions
of dynamical quantities. It was first used by \citet{1911MNRAS..71..460P}
to fit the observations of globular clusters and now is often used
as a stellar distribution model in simulations.

From the Poisson equation $\bigtriangleup\phi=4\pi\mathrm{G}\rho$,
we can easily infer the radial density distribution $\rho$ that acts
as the source for the Plummer potential:

\begin{equation}
\rho(r)=\rho_{0}\frac{1}{(1+r^{2}/\varepsilon^{2})^{5/2}},\label{eq:hust}
\end{equation}
where $\rho_{0}=3M/(4\pi\varepsilon^{3})$ is the central density.
About $\sqrt{2}/4$ (approx. 35\%) of the total mass of the galaxy
is enclosed inside the $r=\varepsilon$ radius. 

The force $F(r)$ acting on a test particle (of a mass \textit{m})
is calculated from the potential by the equation $F(r)=-\bigtriangledown\phi(r)$,
what reads in Plummer potential as:

\begin{equation}
F(r)=-\mathrm{G}\, M\, m\frac{r}{(r^{2}+\varepsilon^{2})^{3/2}}.\label{eq:Ftest}
\end{equation}

The particles in our model then move according to an acceleration
$\mathbf{a}(\mathbf{r})$ given by the potentials of both galaxies

\begin{equation}
\mathbf{a}(\mathbf{r})=-\mathrm{G}\sum_{i}\frac{M_{i}\mathbf{r}_{i}}{(r_{i}^{2}+\varepsilon_{i}^{2})^{3/2}},
\end{equation}
 where the summation goes over pres quantities corresponding to the
secondary galaxy, and one or two components of the primary galaxy.
In simulations where the potential of secondary galaxy is switched
off, the particles continue to move only in the fixed potential of
the primary galaxy. $\mathbf{r}_{i}$ is the vector of distance between
the center of the primary or secondary galaxy and the particle: $\mathbf{r}_{i}=\mathbf{r_{\mathrm{particle}}-r_{\mathrm{galaxy}}}$,
where $\mathbf{r_{\mathrm{particle}}}$ is a position vector of the
particle and $\mathbf{r_{\mathrm{galaxy}}}$ is the position vector
of the center of the primary or the secondary galaxy. 

The action of two Plummer spheres on each other is a little more intricate.
The non-zero radius reduces their attraction compared to two point
masses. This interaction cannot be appropriately described by simple
means, but we approximate the attraction by keeping the form of the
Plummer potential and by defining a common softening parameter in
order to fulfill the law of the action and reaction. The definition
of the common softening parameter is derived from both Plummer radii
and then we use it in the equation of motion as:

\begin{equation}
F(r)=-\mathrm{G}\, M_{\mathrm{p}}M_{\mathrm{s}}\frac{r}{(r^{2}+\varepsilon_{\mathrm{p}}^{2}+\varepsilon_{\mathrm{s}}^{2})^{3/2}},\label{eq:Fgal}
\end{equation}
where \textit{$r$} is the relative distance of centers of masses
of galaxies. The indexes \textit{$\mathrm{p}$} and \textit{$\mathrm{s}$}
mark the quantities corresponding to the primary and the secondary
galaxy. The common softening parameter is then $\varepsilon_{\mathrm{common}}=\sqrt{\varepsilon_{\mathrm{p}}^{2}+\varepsilon_{\mathrm{s}}^{2}}$.
In the case of a two-component primary galaxy, we use in Eq.~(\ref{eq:Fgal})
with $M_{\mathrm{p}}=M_{*}+M_{\mathrm{DM}}$ and $\varepsilon_{\mathrm{p}}^{2}=\varepsilon_{*}^{2}+\varepsilon_{\mathrm{DM}}^{2}$,
where $*$ stands for luminous component and $\mathrm{DM}$ for the
dark halo.

\subsection{Velocity dispersion in Plummer potential \label{sub:Disperze} }

For computation of dynamic friction we will need to know the velocity
dispersion in the Plummer potential, so let's derive it briefly now.
Applying the Jeans equations (see \citealp{1987gady.book.....B},
Ch.~4.2) to our spherically symmetric galaxy without any systematical
movement, we get

\begin{equation}
\frac{\partial\left(\rho(r)\sigma^{2}(r)\right)}{\partial r}=-\rho(r)\frac{\partial\phi(r)}{\partial r},
\end{equation}
where $\sigma$ stands for the velocity dispersion, which is assumed
isotropic at any given \textit{$r$}. Applying the assumption $\sigma(\infty)=0$
we get the solution:

\begin{equation}
\sigma^{2}(r)=\frac{1}{\rho(r)}\intop_{r}^{\infty}\rho(r^{\prime})\frac{\mathrm{d}\phi(r^{\prime})}{\mathrm{d}r^{\prime}}\mathrm{d}r^{\prime}.\label{eq:sigma1P}
\end{equation}
The density $\rho$ and potential $\phi$ of the Plummer sphere are
given by the Eq.~(\ref{eq:hust}) and Eq.~(\ref{eq:pot}), respectively.
The final formula for the velocity dispersion of the galaxy with mass
\emph{$M$} and Plummer radius $\varepsilon$ is thus

\begin{equation}
\sigma^{2}(r)=\frac{\mathrm{G}\, M}{6\,\sqrt{\varepsilon^{2}+r^{2}}}.\label{eq:VD}
\end{equation}

\begin{figure}[!h]
\begin{centering}
\includegraphics[width=0.8\textwidth,keepaspectratio]{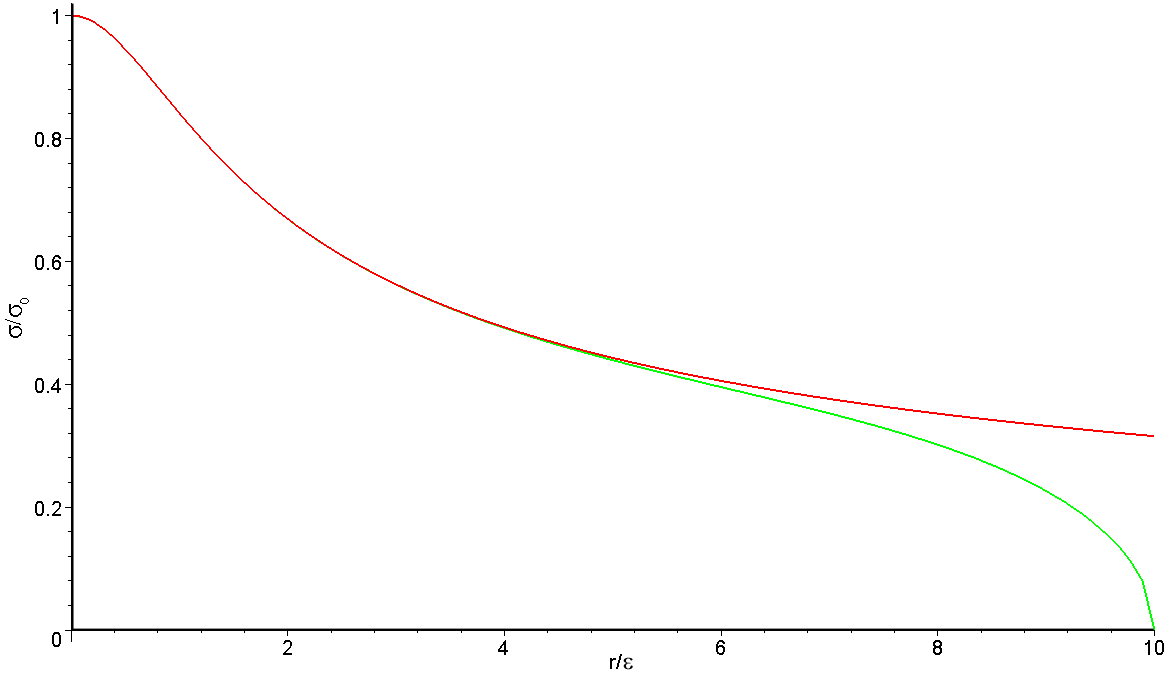}
\par\end{centering}

\caption{\textsf{\small The radial dependence of the velocity dispersion in
a Plummer sphere galaxy extending to infinity (red line) and a galaxy
having the same Plummer profile truncated in 10 times its scale radius
(green line). The distance is in multiples of the scale and the velocity
dispersion in the units of the dispersion in the center $\sigma_{0}$
($\sigma_{0}$ differs negligibly between the two cases). \label{obr.disp}}}
\end{figure}

For the galaxies in our model, we use the Eq.~(\ref{eq:VD}) in a
slightly modified from, because in the previous derivation we considered
an isolated Plummer sphere extending to the infinity. In reality,
the size of a single galaxy is limited (by tidal forces) and so we
assume that at some distance \textit{$R_{\mathrm{tc}}$} it ends and
here, $\sigma(R_{\mathrm{tc}})=0$. With this assumption we get:

\begin{equation}
\sigma^{2}(r)=\frac{\mathrm{G}\, M}{6\,\varepsilon}(1+r^{2}/\varepsilon^{2})^{5/2}\left[\frac{1}{(1+r^{2}/\varepsilon^{2})^{3}}-\frac{1}{(1+R_{\mathrm{tc}}^{2}/\varepsilon^{2})^{3}}\right].\label{eq:VDR}
\end{equation}

The radial dependence of the velocity dispersion for the truncated
and the infinite galaxy are compared in Fig.~\ref{obr.disp}.

\subsection{Velocity dispersion in a double Plummer sphere \label{sub:Disp-PP}}

For a galaxy modeled as two Plummer spheres -- one for the luminous
component and another one for the dark halo -- the situation with
the velocity dispersion is more complex. The presence of one component
influences the dispersion in the other one and vice versa. Eq.~(\ref{eq:sigma1P})
changes to

\begin{equation}
\sigma_{1}^{2}(r)=\frac{1}{\rho_{1}(r)}\intop_{r}^{\infty}\rho_{1}(r^{\prime})\frac{\mathrm{d}\left[\phi_{1}(r^{\prime})+\phi_{2}(r^{\prime})\right]}{\mathrm{d}r^{\prime}}\mathrm{d}r^{\prime}.
\end{equation}
Using Eq.~(\ref{eq:hust}) and Eq.~(\ref{eq:pot}) and after a partial
integration, we obtain 
\begin{equation}
\sigma_{1}^{2}(r)=\frac{\mathrm{G}\, M_{1}}{6\,\sqrt{\varepsilon_{1}^{2}+r^{2}}}+\frac{\mathrm{G}\, M_{2}}{\varepsilon_{2}^{3}}\left(1+r^{2}/\varepsilon_{1}^{2}\right)^{5/2}I(r,\varepsilon_{1},\varepsilon_{2}),\label{eq:dispPP}
\end{equation}
where the first term is identical to the dispersion of the first component
without in the absence of the second one. The integral $I(r,\varepsilon_{1},\varepsilon_{2})$
is solved as follows 
\begin{equation}
I(r,\varepsilon_{1},\varepsilon_{2})=\intop_{r}^{\infty}\frac{r^{\prime}}{\left(1+r'^{2}/\varepsilon_{1}^{2}\right)^{5/2}\left(1+r'^{2}/\varepsilon_{2}^{2}\right)^{3/2}}\mathrm{d}r^{\prime}=\label{eq:dispPP-I}
\end{equation}
 
\[
=\frac{1}{3\left(\varepsilon_{2}^{2}-\varepsilon_{1}^{2}\right)}\left[\frac{1}{\left(r^{2}+\varepsilon_{1}^{2}\right)^{3/2}\left(r^{2}+\varepsilon_{2}^{2}\right)^{1/2}}+\frac{4}{\left(\varepsilon_{2}^{2}-\varepsilon_{1}^{2}\right)^{2}}\left(2-\sqrt{\frac{r^{2}+\varepsilon_{2}^{2}}{r^{2}+\varepsilon_{1}^{2}}}-\sqrt{\frac{r^{2}+\varepsilon_{1}^{2}}{r^{2}+\varepsilon_{2}^{2}}}\right)\right].
\]

\begin{figure}[H]
\begin{centering}
\includegraphics[width=0.8\textwidth]{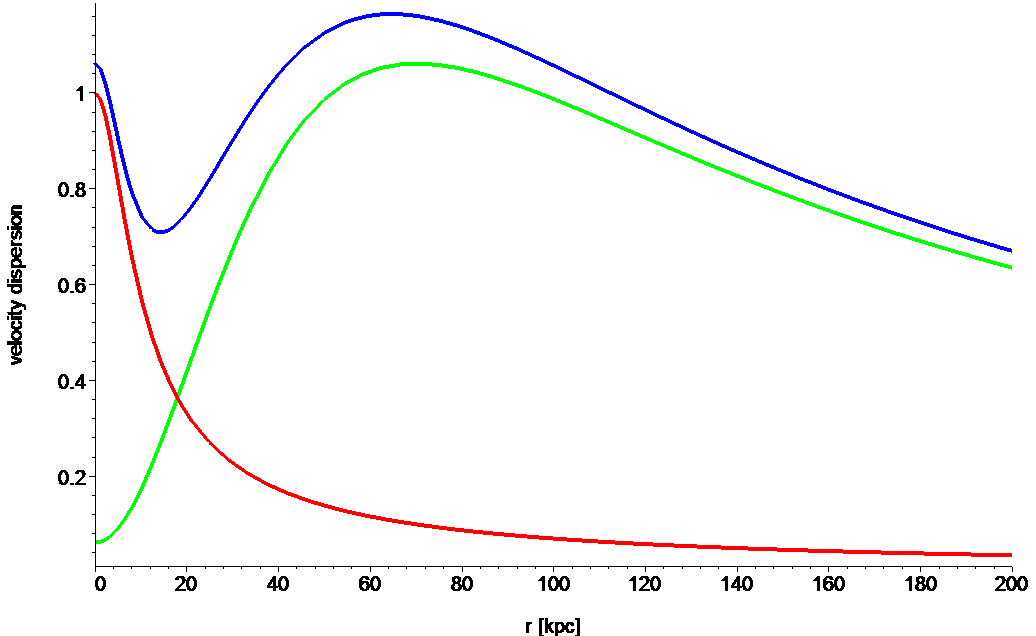}
\par\end{centering}

\caption{\textsf{\small An illustration of the effect of a second component
on the dispersion of a Plummer sphere ($M_{*}=3.2\times10^{11}$\,M\suns,
$\varepsilon_{*}=7$\,kpc). Red: the dispersion of the isolated sphere,
green: additional dispersion caused by the presence of a second component
of large mass ($M_{\mathrm{DM}}=6.4\times10^{12}$\,M\suns) and
large Plummer radius ($\varepsilon_{\mathrm{DM}}=60$\,kpc), blue:
the sum of the two. The dispersion is normalized so that the dispersion
in the center of the first component in the absence of the second
one (181\,km$/$s) equals 1. \label{obr.dispPP}}}
\end{figure}

Fig.~\ref{obr.dispPP} illustrates the effect of the presence of
the other component on the velocity dispersion of a Plummer sphere. 

\subsection{Standard set of parameters\label{sub:SSoP}}

For the future reference, let us define the standard set of parameters
for simulations (used in this Part) as the following set of values:

\noindent \begin{center}
The mass of the primary galaxy: $M_{\mathrm{p}}=3.2\times10^{11}$\,M\suns\\
The secondary to primary mass ratio: 0.02\\
Plummer radius of the primary galaxy: $\varepsilon_{\mathrm{p}}=20$\,kpc\\
The cut-off diameter for the primary galaxy: $R_{\mathrm{tc}}=200$\,kpc\\
Plummer radius of the secondary galaxy: $\varepsilon_{\mathrm{s}}=2$\,kpc\\
The initial radial distance of the secondary galaxy: 180\,kpc\\
The initial velocity of the secondary galaxy: $125$\,km$/$s, the
escape velocity for the initial distance
\par\end{center}

\noindent These values are used as the usual setup of the presented
simulations and we will refer to them often, so we do not have to
repeat them.

Let us only remark that the escape velocity, $v_{\mathrm{esc}}$,
is computed only approximately, on the same grounds as the force between
the galaxy (see Eq.~\ref{eq:Fgal}), i.e. we put

\begin{equation}
v_{\mathrm{esc}}=\sqrt{\frac{2\, G\,(M_{\mathrm{p}}+M_{\mathrm{s}})}{\sqrt{r^{2}+\varepsilon_{\mathrm{common}}^{2}}}.}
\end{equation}
The results of our simulation show that, in the relevant range of
radii, its difference from reality is negligible.

\section{Dynamical friction }

Dynamical friction is a braking force of gravitational origin acting
on a body that moves through the field of stars or any other matter.
We will be interested in the dynamical friction incurred on the secondary
galaxy by the stars and dark matter of the primary galaxy. We encourage
the reader to consult Appendix~\ref{Apx:Intro-DF} -- Introduction
to dynamical friction -- which is a modified chapter from \citet{EbrovaMAT}.
It explains in detail the nature of this phenomenon and it is likely
to be of interest even to a reader already familiar with the topic.

Appendix~\ref{Apx:Intro-DF} also contains a derivation of the Chandrasekhar
formula (Sect.~\ref{sApx:ChF}). Chandrasekhar formula is an analytical
expression derived by \citet{1943ApJ....97..255C} that is still often
used to calculate the dynamic friction. The formula is a good approximation
for the dynamical friction and is easy to use in test-particle simulations.

There are several different simplifi{}cations done during its derivation
(see Sect.~\ref{sApx:Maple-sekce}). One is the assumption of homogeneity
of the stellar field around the braked body (both density and velocity
dispersion are taken as constants). This leads to a relatively simple
expression that contains the so-called Coulomb logarithm. The exact
value of this logarithm is unknown and is usually roughly estimated
and taken as a constant.

In \citet{EbrovaMAT}, we have devised an alternative way to calculate
the dynamical friction in radial mergers (that are the most likely
to produce shell structures). We call it \textit{our modification
of the Chandrasekhar formula} and a detailed description and derivation
can be found in Appendix~\ref{Apx:OurDF}. Here we summarize only
the main ideas.

The homogeneity of density and velocity dispersion is not assumed
during the derivation of the Chandrasekhar formula. Instead, a more
realistic stellar distribution function is used, varying both the
density and velocity dispersion based on the chosen model of the host
galaxy. Using the radial symmetry, the originally 5-dimensional problem
is reduced to a 2-dimensional one, Eq.~(\ref{eq:TENint}), which
is analytically insolvable and so numerical integration is used to
calculate the final result for the dynamical friction. In this approach,
no estimated values are needed as an input, only the distribution
function chosen for the galaxy determines the friction. 

In Sect.~\ref{sApx:compare-to-Chf}, we compare the result of Eq.~(\ref{eq:TENint})
to the Chandrasekhar formula. It is shown that using a constant as
the Coulomb logarithm is completely inadequate for the problem at
hand.

\section{Multiple Three-Body Algorithm (MTBA) \label{sec:MTBA}}

We now investigate another alternative method to calculate the dynamical
friction in radial minor merger. The method is described in the paper
\citet{1994A&A...290..709S} and it is also suitable for test-particle
simulations.

\subsection{Principle and characteristics \label{sub:MTBAprinc}}

\citet{1994A&A...290..709S} used restricted tree-body simulations
to examine dynamical friction in head-on encounter. They adopted the
Multiple Three-Body Algorithm which was originally proposed by \citet{1984ApJ...287..503B}.
The basis of the method is to calculate the motion of the satellite
galaxy from the gravitational influence of the particles in the primary
galaxy. However, it is not a self-consistent simulation, as the particles
are otherwise treated as test particles -- their motion is calculated
as the motion of massless particles in the sum of the gravitational
potentials of both galaxies, in the same manner as in our simulations
of the creation of the shell structure (Sect.~\ref{sec:N-Simulations}
and Sect.~\ref{sec:Simulations-of-shell}). In the case of the MTBA,
the particles are generated so that they follow the distribution function
of the primary galaxy. Only when the motion of the secondary galaxy
is calculated, these particles are used as if each of them had a mass
of $m=M_{\mathrm{p}}/N$, where $M_{\mathrm{p}}$ is the total mass
of the primary galaxy and $N$ is the total number of particles used.
The force/acceleration acting upon the secondary galaxy in each step
is fully determined by the action of all particles in the primary
galaxy upon a chosen smooth potential of the secondary galaxy. Having
also the potential of the satellite act on these particles naturally
perturbs their trajectories and from their force exerted back on the
satellite galaxy the dynamical friction naturally arises.

To summarize, this method to calculate the dynamical friction requires
a model for the potentials of the primary and the secondary galaxy
and the use of particles in the primary galaxy. The particles are
treated in two different ways: as massless when their motion is calculated
and as massive when the motion of the secondary galaxy is calculated.

\citet{1994A&A...290..709S} have directly compared the results of
a MTBA simulation with the coupled solution of the linearized Poison
and collisionless Boltzmann equations for the first passage of the
satellite. They found MTBA to be equivalent to the analytical method.
Compared to their analytical method, the MTBA has the advantage of
easier and faster calculation. Moreover the MTBA is more flexible
so it can follow the whole process until a complete merger. Both these
methods show that the dynamical friction in radial merger is not strictly
proportional to the local density -- contrary to what is assumed in
the Chandrasekhar formula. Moreover, it is a time-dependent process
which depends on the full past history of the merger, contrary to
a satellite on a circular orbit in the co-rotating frame. This observation
cannot be reproduced in any modification of the Chandrasekhar formula
(including ours) which is fundamentally local.

In \citet{1996A&A...310..757S} the MTBA has been compared with a
self-consistent \textit{Particle-Mesh} simulation. The MTBA gives
an accurate estimate of the decay rate of orbital energy of the satellite,
within 10\% of the $N$-body simulation during the first orbit. But
it fails to reproduce the ultimate phase of the merger.

\subsection{Merger parameters \label{sub:Param.MTBA}}

To compare different methods for the calculation of the dynamical
friction, we have modeled the secondary as a point mass (eventually
with a very small softening -- 0.01\,kpc) and have chosen the following
parameters of the collision:

\noindent \begin{center}
The mass of the primary galaxy: $M_{\mathrm{p}}=10^{12}$\,M\suns\\
The secondary to primary mass ratio: 0.01\\
Plummer radius of the primary galaxy: $\varepsilon_{\mathrm{p}}=10$\,kpc\\
The cut-off diameter for the primary galaxy: $R=200$\,kpc\\
The initial radial distance of the secondary galaxy: 100\,kpc\\
The initial velocity of the secondary galaxy: 0\,km$/$s
\par\end{center}

\subsection{Results of simulations}

\begin{figure}[H]
\begin{centering}
\includegraphics[width=12cm]{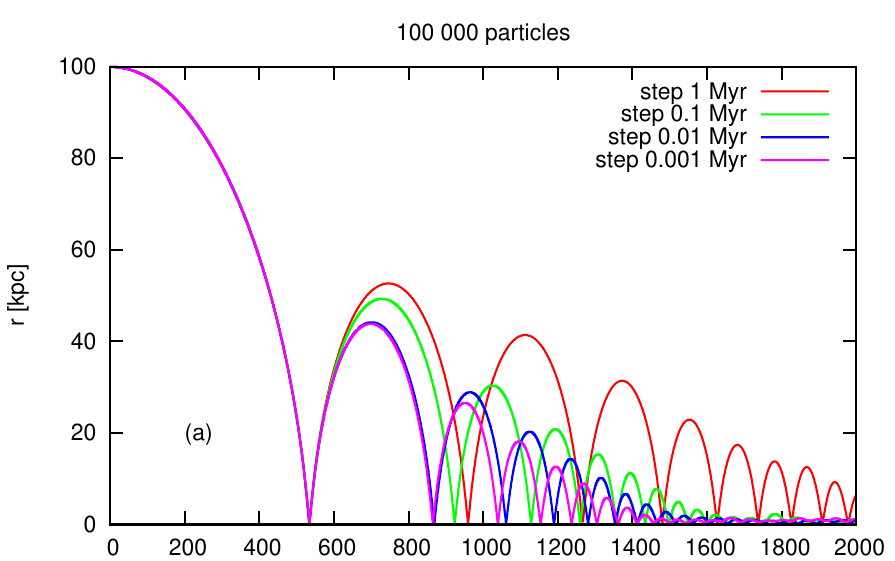}
\par\end{centering}

\centering{}\includegraphics[width=12cm]{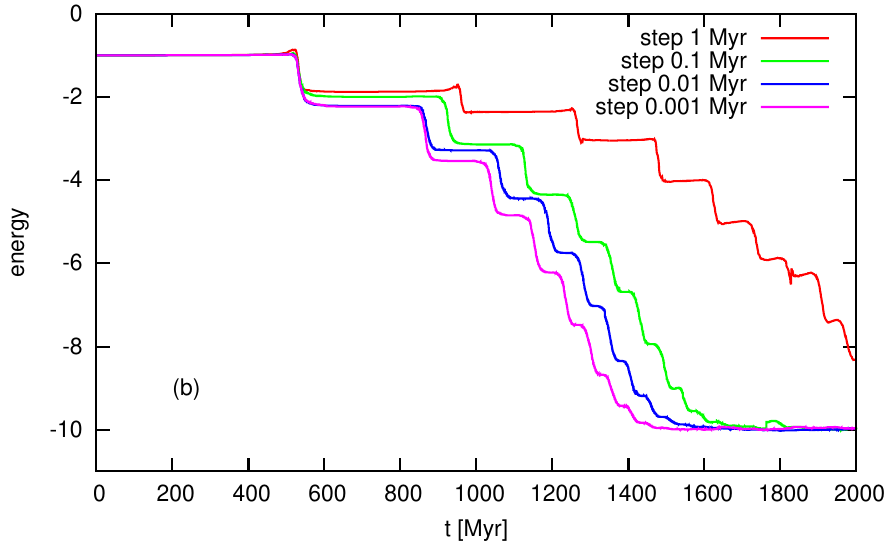}\caption{\textsf{\small (a) Distance of the secondary from the center of the
primary galaxy; (b) energy of the secondary. The motion was calculated
using the MTBA with 100,000 particles for time steps of 0.001--1\,Myr.
Parameters of the collision are given in Sect.~\ref{sub:Param.MTBA}.
\label{fig:MTBA100kp} }}
\end{figure}

\begin{figure}[H]
\begin{centering}
\includegraphics[width=12cm]{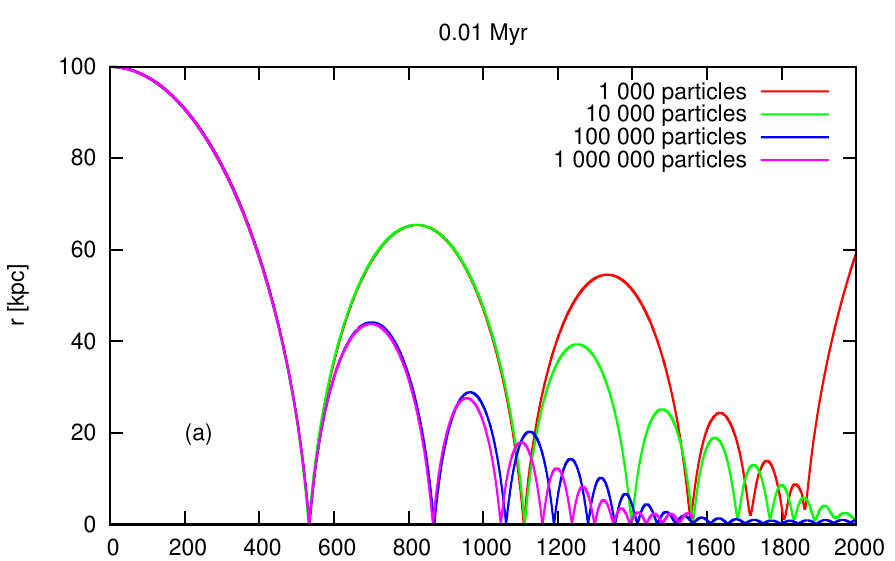}
\par\end{centering}

\begin{centering}
\includegraphics[width=12cm]{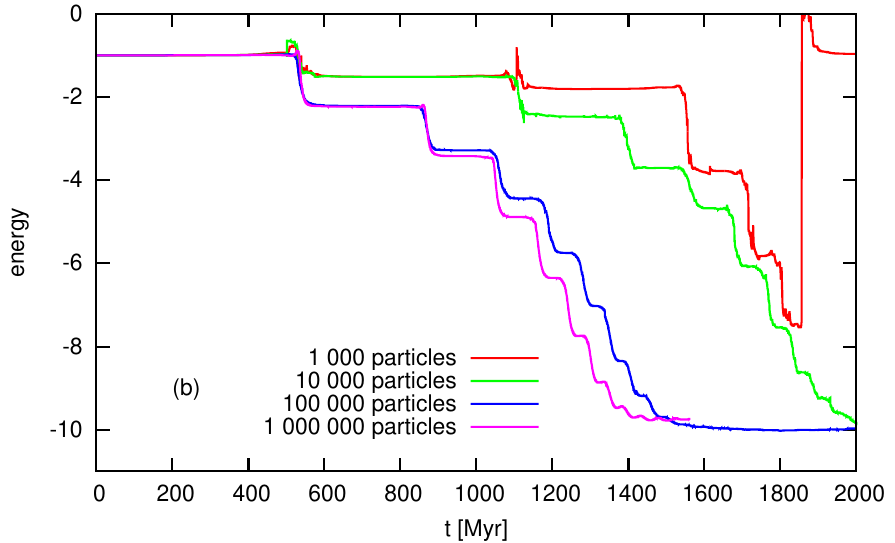}
\par\end{centering}

\centering{}\caption{\textsf{\small (a) Distance of the secondary from the center of the
primary galaxy; (b) energy of the secondary.The motion was calculated
using the MTBA with time step 0.01\,Myr for 1,000-1,000,000 particles.
Parameters of the collision are given in Sect.~\ref{sub:Param.MTBA}.
\label{fig:MTBA0.01} }}
\end{figure}

It turns out that for a successful application of the MTBA it is necessary
to use a high enough number of particles in the primary galaxy and
a small enough time step of integration. The simulation for the chosen
set of parameters (Sect.~\ref{sub:Param.MTBA}) stabilizes for about
100,000 particles with time step of 0.01\,Myr, but even then there
are noticeable differences mainly in the later part of the merger
as we further increase the number of particles and decrease the time
step, see Fig.~\ref{fig:MTBA100kp} and Fig.~\ref{fig:MTBA0.01}.
On the other hand, the introduction of the slight softening in the
interaction of the secondary does not influence the results provided
that enough particles and a small enough time step are used.

\section{Comparison with self-consistent simulations \label{sec:GADGET}}

To compare the calculation of the dynamical friction using the methods
mentioned earlier (Appendix~\ref{Apx:OurDF} and Sect.~\ref{sec:MTBA})
with the self-consistent simulations, we use the simulations performed
by Kate\v{r}ina Barto{\v s}kov{\'a} using GADGET-2. GADGET-2 is free software,
distributed under the GNU General Public License. The code can be
used for studies of isolated systems, or for simulations that include
the cosmological expansion of space. It computes gravitational forces
with a hierarchical tree algorithm (optionally in combination with
a particle-mesh scheme for long-range gravitational forces) and represents
fluids by means of smoothed particle hydrodynamics (SPH). Both the
force computation and the time stepping are fully adaptive. The code
is written in highly portable C and uses a spatial domain decomposition
to map different parts of the computational domain to individual processors.
GADGET-2 was publicly released in 2005 \citep{2005MNRAS.364.1105S}
and presently is the most widely employed code for the cosmic structure
formation.

\subsection{Altering GADGET-2 computational setting \label{sub:AlteringGADGET}}

The parameters of the collision have been set the same as in the previous
case, Sect.~\ref{sub:Param.MTBA}, but with no cut-off diameter.
$10^{5}$ particles have been used to represent the primary galaxy.
The results differ for different settings of computational parameters
in GADGET-2. Here we present results of five simulations that differ
in settings for three chosen parameters and in the accuracy of variables
during the calculation.

During the calculation of the gravitation force, spline softening
is used. \textit{$SoftPar$} is the magnitude of the softening used
for mutual interactions of the particles of the primary galaxy. \textit{$SoftSec$}
is the softening for the secondary and in an interaction between the
secondary and a particle of the primary galaxy, the larger value from
$SoftPar$ and $SoftSec$ is used. $ETIA$ (ErrorTolIntAccuracy) influences
the accuracy of the integration method. It is used in the estimation
of the adaptive integration step $\Delta t$ 
\begin{equation}
\Delta t=\sqrt{\frac{2\, ETIA\, SoftPar}{a}},
\end{equation}
where $a$ is the amount of acceleration the particle has been subjected
to in the previous step. Thus the smaller $ETIA$ we choose, the shorter
will be the time step.\textit{ $Precision$} refers to the type of
the floating-point precision used during numerical calculations.

\begin{figure}[H]
\begin{centering}
\includegraphics[width=12cm]{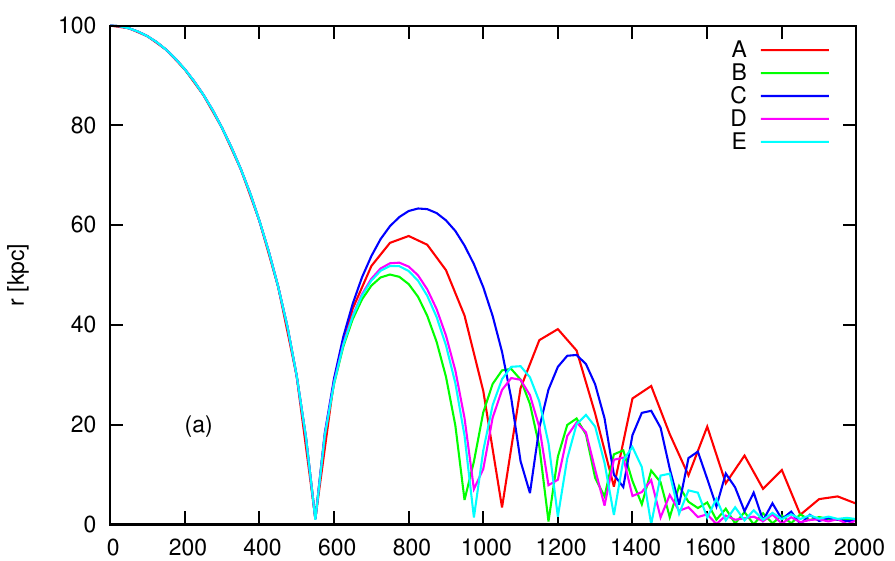}
\par\end{centering}

\centering{}\includegraphics[width=12cm]{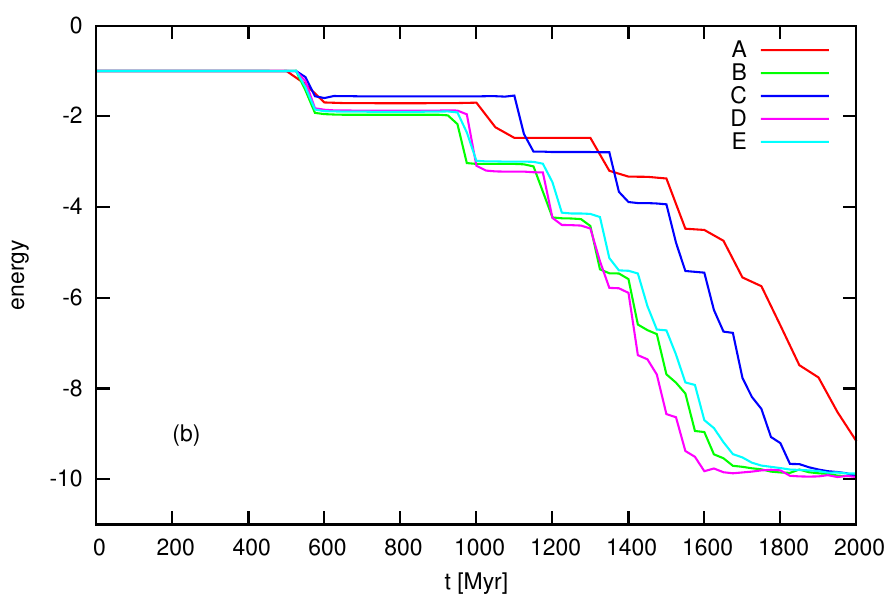}\caption{\textsf{\small (a) Distance of the secondary from the center of the
primary galaxy; (b) energy of the secondary.The motion has been calculated
using GADGET-2. The parameters of the collision are given in Sect.~\ref{sub:Param.MTBA},
the settings for each simulation in Table~\ref{tab:GADGET}. \label{fig:GADGET} }}
\end{figure}

The values we have used in the five different simulations and the
labels of the simulations are shown in Table~\ref{tab:GADGET}. The
orbital decay of the satellite for all the runs is shown in Fig.~\ref{fig:GADGET}.
Run D has been calculated with the highest precision and we thus use
it as a reference in the following section.

\begin{table}[h]
\centering{}%
\begin{tabular}{ccccc}
\hline 
{\small run} & {\small $ETIA$} & {\small $SoftPar$} & {\small $SoftSec$} & {\small $Precision$}\tabularnewline
 &  & kpc & kpc & \tabularnewline
\hline 
{\small A} & {\small 0.002} & 0.21 & 0.05 & Single\tabularnewline
{\small B } & 0.008 & 0.05 & 0.05 & Single\tabularnewline
{\small C} & 0.04 & 0.01 & 0.01 & Single\tabularnewline
{\small D } & 0.04 & 0.01 & 0.01 & Double\tabularnewline
{\small E } & 0.002 & 0.05 & 0.05 & Single\tabularnewline
\hline 
\end{tabular}\caption{\textsf{\small The settings for the GADGET-2 simulations. The meaning
of the parameters is explained in Sect.~\ref{sub:AlteringGADGET}.
\label{tab:GADGET}}}
\end{table}

\subsection{Comparison of methods \label{sub:Comparison-of-methods}}

\begin{figure}[H]
\begin{centering}
\includegraphics[width=12cm]{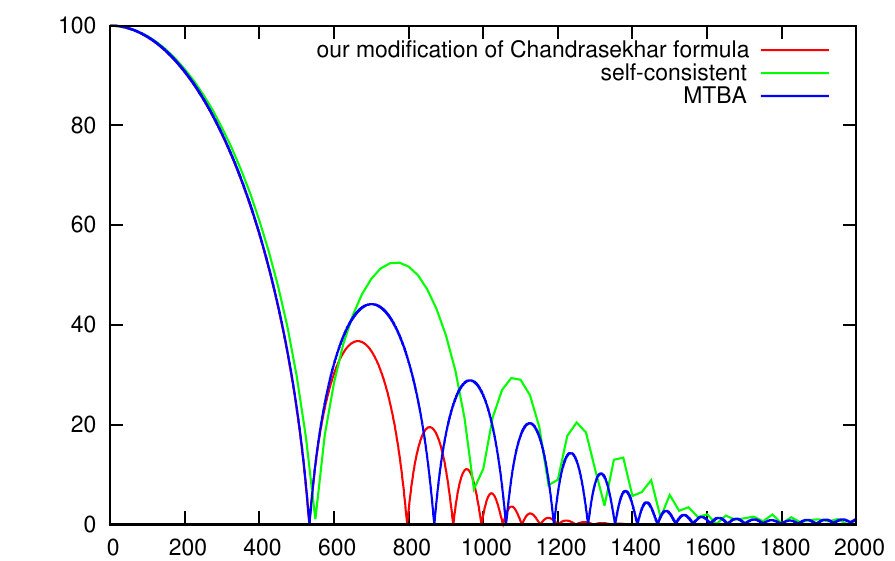}
\par\end{centering}

\centering{}\includegraphics[width=12cm]{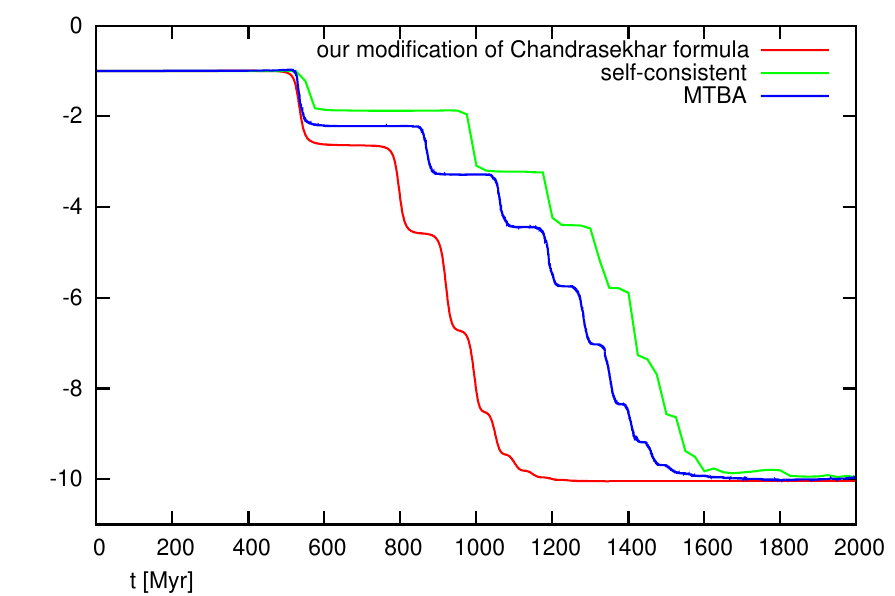}\caption{\textsf{\small (a) Distance of the secondary from the center of the
primary galaxy; (b) energy of the secondary in three different methods:
our modification of Chandrasekhar formula (Sect.~\ref{sApx:Maple-sekce},
red curve); inconsistent simulation with GADGET-2 (Sect.~\ref{sub:AlteringGADGET},
green curve); and MTBA (Sect.~\ref{sub:MTBAprinc}, blue curve).
Parameters of the collision are given in Sect.~\ref{sub:Param.MTBA}.
\label{fig:DFcomp} }}
\end{figure}

Fig.~\ref{fig:DFcomp} shows the orbital decay of the secondary in
the merger with parameters given in Sect.~\ref{sub:Param.MTBA} for
three different methods of calculation of dynamical friction. Our
modification of Chandrasekhar formula adds to the equations of motion
of the secondary the dynamical friction calculated using a numerically
integrated analytical formula as described in Appendix~\ref{Apx:OurDF}.
The MTBA method (Sect.~\ref{sec:MTBA}) is represented by a simulation
with 100,000 particles and time step of 0.01\,Myr. From the self-consistent
simulation with GADGET-2 we show run D (see Sect.~\ref{sub:AlteringGADGET}).

Our modification of Chandrasekhar formula gives by far the fastest
loss of the orbital energy of the satellite, but even the MTBA gives
a significantly larger value of the dynamical friction than the self-consistent
simulation.

In Sect.~\ref{sec:Simulations-of-shell} we will however use our
modification of Chandrasekhar formula for the calculation of the dynamical
friction, as we have conducted a sizable number of simulation using
this method before we became familiar with the MTBA. The MTBA is also
more computationally demanding. It requires a small enough time step
and the inclusion of test particles in the primary galaxy, which are
otherwise of no interest for us. Our modification of Chandrasekhar
formula, on the contrary, gives the same results for the motion of
the secondary galaxy for the time step of 1\,Myr as for any shorter
step.

Doing self-consistent simulations is not an option because of the
number of different simulations required for this study (most of which
we do not show explicitly in this thesis). Because it seems that our
modification of Chandrasekhar formula significantly overestimates
the real value of the friction, the results have to be considered
an upper bound for the influence of the dynamical friction on the
shell structure. At the end, it turns out that the differences in
the shell structure related to the choice of a method to calculate
the dynamical friction is smaller than the uncertainty in the models
of the tidal decay of the secondary galaxy (Sect.~\ref{sec:Tidal-disruption}).

\section{Tidal disruption \label{sec:Tidal-disruption}}

Together with the dynamical friction, the tidal disruption is another
effect that is important for the galactic merger. The tidal disruption
gradually lowers the mass of the cannibalized galaxy and thus mitigates
the effect of the dynamical friction. During shell formation, it is
of particular importance, because the gradual release of stars from
the secondary galaxy has an important effect on the growing shell
structure. The introduction of the tidal disruption into test-particle
simulation is nevertheless a difficult task.

\subsection{Massloss of the secondary\label{sub:TD-Implement.}}

In the context of the tidal disruption of an object in the gravitation
field of another body, the notion of the \emph{tidal radius }is frequently
introduced. This is an approximative approach to the tidal forces,
assuming that under the tidal radius the matter is still bound to
the disrupted body, but it is not the case anymore outside the tidal
radius. The reader may find more details on the concept in Appendix~\ref{Apx:Tidal-radius}.
Here we will only show how we used it in our test-particle simulations.

\begin{figure}[!b]
\centering{}\includegraphics[width=7.5cm]{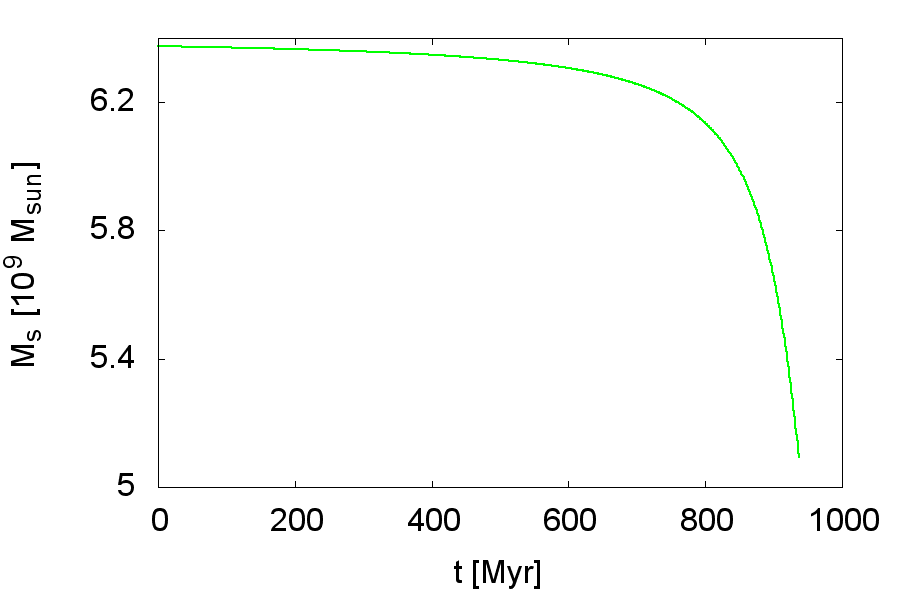}\includegraphics[width=7.5cm]{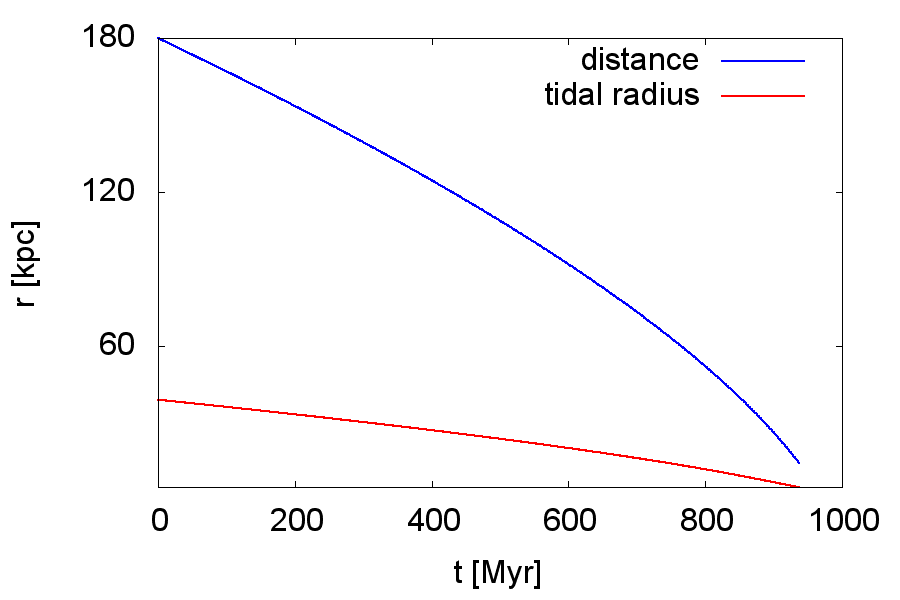}
\caption{\textsf{\small The purely analytical approach to the decay of the
secondary galaxy during the first passage for the standard set of
parameters (Sect.~\ref{sub:SSoP}). Left: the evolution of the mass
of the secondary galaxy. Rights: Distance of the secondary from the
center of the primary galaxy (blue curve) and tidal radius of the
secondary (red curve). \label{fig:tidal-anal} }}
\end{figure}

First we have implemented a purely \textit{analytical approach}, where
we calculate the current tidal radius in every step using Eq.~(\ref{eq:t-aprox})
and update the mass of the secondary galaxy accordingly to the mass
of a Plummer sphere with the original parameters of the secondary
galaxy but restricted to the tidal radius. But this leads to us only
lowering the satellite mass during the first passage through the center
of the primary galaxy, see Fig.~\ref{fig:tidal-anal}. Particles
are released in limited amount also during further passages, but this
mechanism obviously does not reflect the real situation for multiple
passages. 

\begin{figure}[!t]
\begin{centering}
\includegraphics[width=15cm]{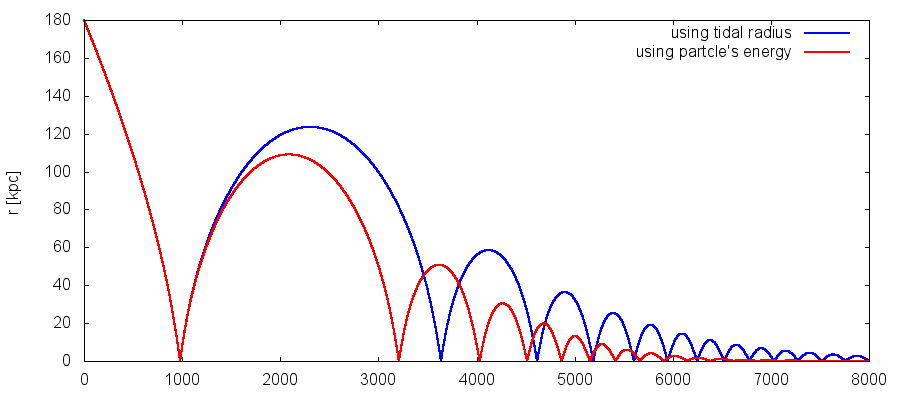}
\par\end{centering}

\begin{centering}
\includegraphics[width=15cm]{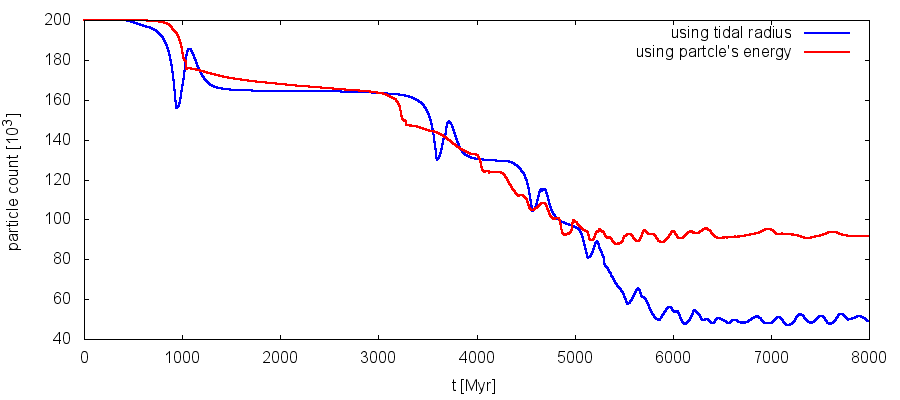}
\par\end{centering}

\caption{\textsf{\small Gradual decay of the secondary galaxy calculated using
test particles. Top: distance between the centers of the primary and
the secondary galaxy. Bottom: the number of particles bound to the
secondary galaxy. Blue curves show the development for the simulation
where we consider as bound particles those inside the sphere of the
tidal radius, the red curves correspond to keeping particles with
lower than escape velocity. Both simulations are carried out for the
standard set of parameters (Sect.~\ref{sub:SSoP}), the dynamical
friction is calculated using our modification of the Chandrasekhar
formula (}Appendix~\ref{Apx:OurDF}\textsf{\small ). \label{fig:N-Ne} }}
\end{figure}

To describe the decay of the satellite during further passages, we
have included in its calculation the test particles of the secondary
galaxy. We count particles that we still consider bound with the satellite
galaxy. The ratio between their number and the number of particles
that we have put in the secondary galaxy at the beginning of the simulation
determines its current mass. As a criterion for bound particles we
consider that 1) the \textit{distance of the particle} from the center
of the secondary galaxy is lower than the current tidal radius; 2)
the \textit{velocity of the particle} with respect to the secondary
galaxy does not exceed the escape velocity for its given distance
from the center of the secondary galaxy. Fig.~\ref{fig:N-Ne} shows
how these two approaches differ for otherwise identical initial conditions.

The use of the tidal radius causes large fluctuations of the number
of bound particles near the passage of the secondary galaxy through
the center of the primary galaxy, when many particles suddenly find
themselves outside the tidal radius. When later the secondary galaxy
retreats from the center of the primary, the tidal radius quickly
increases and more particles are included. Some of them eventually
escape before the secondary reaches its apocenter, but still more
particles stay bound to the secondary than there were during its passage
through the center of the primary. In the other simulation the loss
of particles is more monotonous, the orbital decay slightly faster,
and more particles are caught in the center of the host galaxy.

\begin{figure}[!t]
\begin{centering}
\includegraphics[width=15cm]{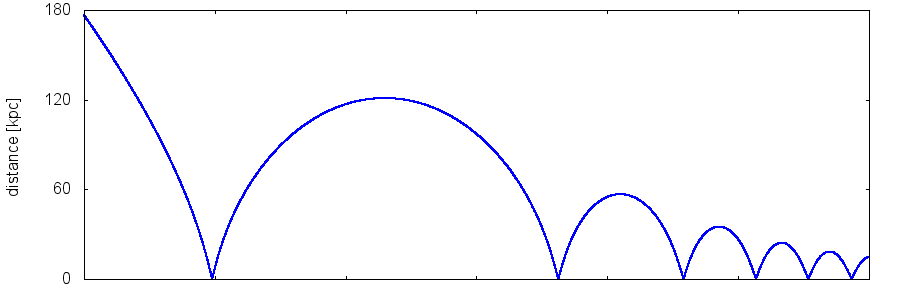}
\par\end{centering}

\begin{centering}
\includegraphics[width=15cm]{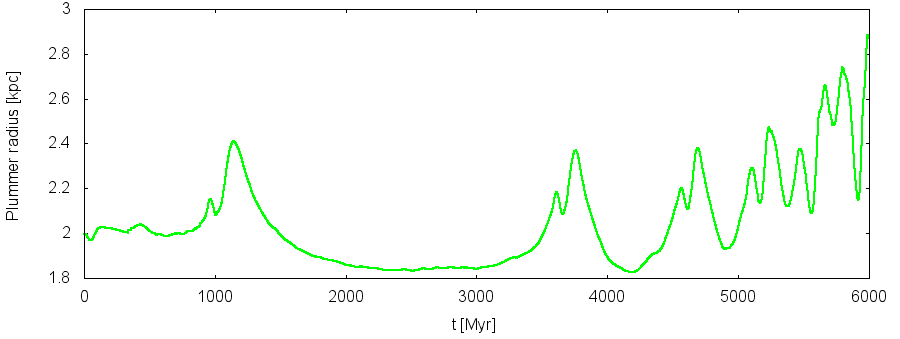}
\par\end{centering}

\caption{\textsf{\small Development of the distance between the primary and
the secondary galaxy (top) and the Plummer radius of the secondary
galaxy (bottom). The simulations carried out for the standard set
of parameters (Sect.~\ref{sub:SSoP}), the dynamical friction is
calculated using our modification of the Chandrasekhar formula (}Appendix~\ref{Apx:OurDF}\textsf{\small ).
The radial density of the secondary galaxy at the beginning of the
simulation and in 5\,Gyr is shown Fig.~\ref{fig:b-sec}. \label{fig:b} }}
\end{figure}

The use of the two different methods to model the tidal disruption
of the secondary does not have a dramatic impact on the merer. Nevertheless,
the times of the passages of the secondary through the center of the
host galaxy and the volume of particles released in each passage differ
between the two models, mainly in the later phases of the merger.
This may have a noticeable impact on the appearance of the shell system
in different time, that is the positions of the shells, their number,
brightness, opening angle and so forth.

The problem is that we have no hint as to which of the methods is
a better approximation for the true decay of the secondary galaxy.
If we were to compare the results with self-consistent simulations,
we would likely get different results depending mainly on the configuration
of the merger. Thus we compare the test-particle simulations done
with different methods for the tidal disruption of the secondary galaxy
and focus on features of the shell system that are independent of
the method used (Sect.~\ref{sec:Sim-DF&TD}).

\subsection{Deformation of the secondary galaxy \label{sub:b}}

Another thing going on during the merger that is difficult to reproduce
in test-particle simulations is the deformation if the cannibalized
galaxy. We model components of galaxies with spherically symmetric
Plummer spheres. Thus we have tried at least to change the profile
of the sphere of the secondary galaxy during the simulation.

\begin{figure}[H]
\centering{}\includegraphics[width=7.5cm]{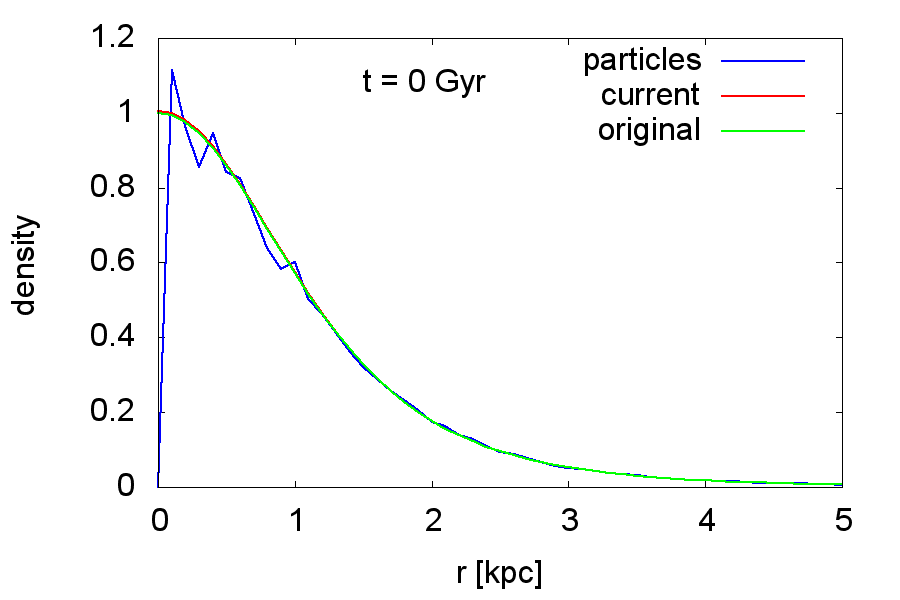}\includegraphics[width=7.5cm]{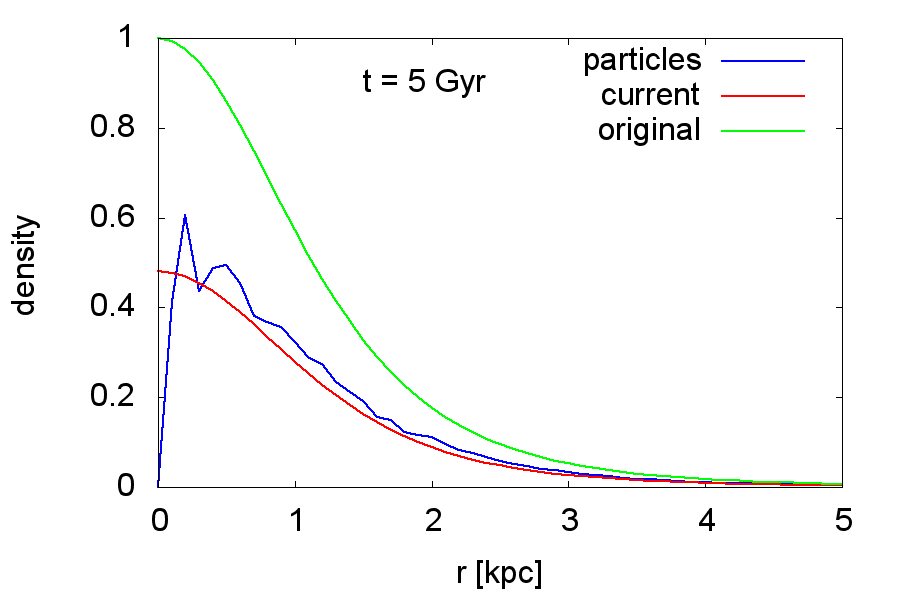}
\caption{\textsf{\small The radial density of the secondary galaxy at the beginning
of the simulation and in 5\,Gyr for the standard set of parameters
(Sect.~\ref{sub:SSoP}). In blue is the density calculated from the
test particles of the secondary galaxy, in green the model of the
secondary chosen at the start of the simulation and in red the density
of the Plummer sphere that corresponds to the changing Plummer radius
which is calculated from the distribution of the test particles. The
density is normalized so that the central density of the initially
chosen Plummer sphere of the secondary galaxy is one. \label{fig:b-sec} }}
\end{figure}

The mean value of the radial distance of a particle $\left\langle r\right\rangle $
in a Plummer sphere is given as 
\begin{equation}
\left\langle r\right\rangle =\frac{\intop_{0}^{R_{\mathrm{tc}}}r'^{3}\rho(r')dr'}{\intop_{0}^{R_{\mathrm{tc}}}r''^{2}\rho(r'')dr''},
\end{equation}
where $\rho(r')$ is the density of the Plummer sphere Eq.~(\ref{eq:hust})
and we express the cut-off in multiplies of the Plummer radius $R_{\mathrm{tc}}=p\varepsilon$.
The mean value of the radial distance is then 
\begin{equation}
\left\langle r\right\rangle =\varepsilon\frac{2\left(1+p^{2}\right)^{3/2}-2-3p^{2}}{p^{3}}.\label{eq:<r>}
\end{equation}
Thus if we calculate the mean value radial distance from the center
of the secondary galaxy for the particles that we consider bound to
it in the simulation 
\begin{equation}
\left\langle r\right\rangle =\sum_{i=1}^{N}r_{i}/N,\label{eq:<r>N}
\end{equation}
we can easily convert it to a new Plummer radius for the secondary
galaxy $\varepsilon_{\mathrm{s}}$. Fig.~\ref{fig:b} shows the development
of the Plummer radius of the secondary galaxy in a simulation with
the standard set of parameters (Sect.~\ref{sub:SSoP}). The Plummer
radius is calculated using Eq.~(\ref{eq:<r>}), where $\left\langle r\right\rangle $
is the mean radial distance of particles under the current tidal radius.
The radial density of the secondary galaxy at the beginning of the
simulation and in 5\,Gyr is shown in Fig.~\ref{fig:b-sec}. It is
important to keep in mind that the density is calculated only from
radial distances from the center of the satellite even though the
spherical symmetry was surely broken during the simulation.

%\clearpage

\begin{figure}[h]
\centering{}\includegraphics[width=12cm]{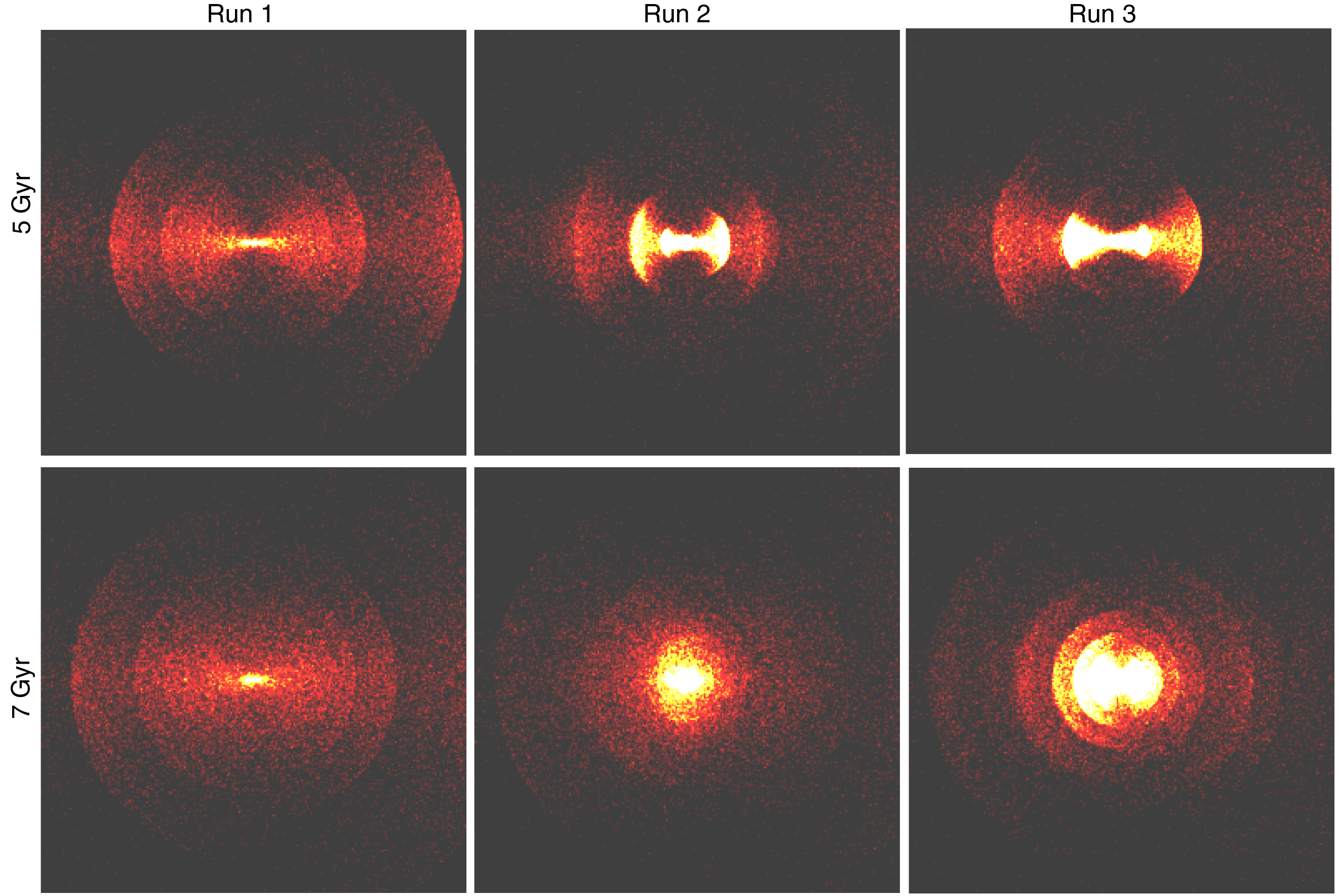}
\caption{\textsf{\small Snapshots of simulations. For description of all runs
see text in Sect.~\ref{sec:Sim-DF&TD}. Time 0 is defined as the
moment then the secondary galaxy reaches the center of the primary
galaxy for the first time, which is (for all three runs) almost exactly
1\,Gyr after it has been released from the distance of 180\,kpc
with escape velocity. Only the surface density of particles originally
belonging to the satellite galaxy is displayed corresponding to the
subtraction of the profile of the primary galaxy. Each box, centered
on the host galaxy, shows 300$\times$300\,kpc. Radial histogram
of particles in 5\,Gyr is shown in Fig.~\ref{fig:DF-hist}. \label{fig:DF-snap} }}
\end{figure}

\section{Simulations of shell structure \label{sec:Simulations-of-shell} }

Now we finally show the combined effect that the inclusion of both
the dynamical friction and gradual decay of the secondary galaxy in
the simulations has on the shell formation. The simulations are carried
out using the method described in Sect.~\ref{sec:Description-of-simulation},
i.e. millions of test particles were generated so that they follow
the distribution function of the secondary galaxy at the beginning
of the simulation. The particles then move according to the sum of
the gravitational potentials of both galaxies that are both represented
by a smooth potential. The galaxies move with respect to each other
as dictated by their masses, shape of potentials, positions and velocities.

The dynamical friction, when included, is calculated using our modification
of the Chandrasekhar formula, see Appendix~\ref{Apx:OurDF}, and
the gradual decay of the secondary galaxy, when included, is calculated
using some of the methods from Sect.~\ref{sub:TD-Implement.}. In
Sect.~\ref{sec:Sim-DF&TD&DM}, we have added the dark halo to the
primary galaxy and Sect.~\ref{sub:Self-cons.} shows the shell formation
in a self-consistent simulations. All the outputs are oriented so
that the secondary originally approached the primary galaxy from the
right hand side.

\subsection{Dynamical friction and tidal disruption \label{sec:Sim-DF&TD}}

\begin{figure}[!t]
\centering{}\includegraphics[width=15cm]{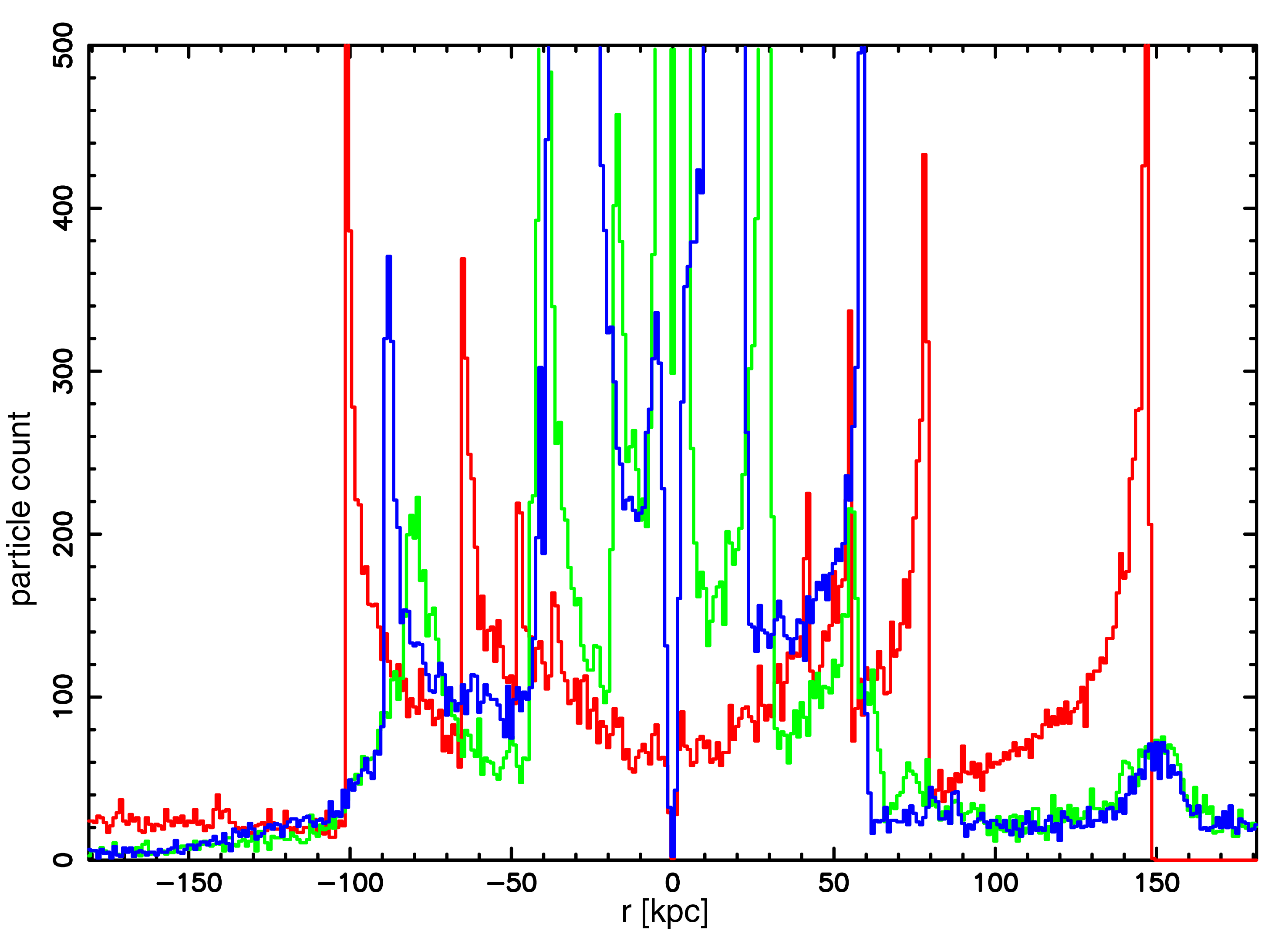}
\caption{\textsf{\small Radial histogram of stars of the secondary galaxy,
centered on the primary 5\,Gyr after the first passage of the secondary
galaxy through the center of the primary galaxy for the three different
simulations -- run}\,\textsf{\small 1 (red), run}\,\textsf{\small 2
(green) and run}\,\textsf{\small 3 (blue). For description of all
runs, see text in Sect.~\ref{sec:Sim-DF&TD}. \label{fig:DF-hist} }}
\end{figure}

We have compared three simulations, all of them for the standard set
of parameters (Sect.~\ref{sub:SSoP}). 
\begin{itemize}
\item Run\,1 -- without dynamical friction and with instant disruption
of the secondary. 
\item Run\,2 -- dynamical friction is calculated using our modification
of the Chandrasekhar formula and the tidal disruption using the analytical
approach based on the tidal radius as described at the beginning Sect.~\ref{sub:TD-Implement.}.
\item Run\,3 -- dynamical friction is again calculated using our modification
of the Chandrasekhar formula, the tidal disruption is based on the
counting of particles inside/outside the current tidal radius. Additionally,
the Plummer radius of the secondary galaxy is constantly recalculated
as described in Sect.~\ref{sub:b}.
\end{itemize}
Snapshot from all the runs for two different times are shown in Fig.~\ref{fig:DF-snap},
radial histograms of particles in Fig.~\ref{fig:DF-hist}. Video
from run\,1 and run\,2 is part of the electronic attachment. For
the description of the video, see Appendix \ref{Apx:Videos} point
\ref{enu:video4-fri}.

We compare a simple simulation (Run\,1) with a pair of simulations
(Runs\,2\,\&\,3), where the tidal decay of the secondary galaxy
is modeled using two different methods. However, we can see a qualitative
shift in the same direction between both Runs\,2\,\&\,3 and the
simple simulation. The result of both Runs\,2\,\&\,3 is a multi-generation
shell system, whereas Run\,1 can in principle give rise only to one
generation of shells.

For both Runs\,2\,\&\,3 there were more particles trapped in the
gravitational field of the host galaxy and a large part of them has
been transported to the vicinity of the center of the host galaxy.
The outer shells (of the first generation) are more diffuse and significantly
less luminous when compared to Run\,1, whereas their positions remain
essentially the same. On the other hand, in the later generations,
there are brighter shells, some of which can overlap with the first-generation
shells. The shells of the later generations appear on smaller radii
and are often bright, whereas in Run\,1 shells at small radii are
completely missing. The evolution of the shell brightness in Run\,1
is somehow calmer, whereas the shells of the later generations in
Runs\,2\,\&\,3 have a tendency to reach very high brightness in
certain small range of radii.

Runs\,2\,\&\,3 are more consistent with observations (Sect.~\ref{sec:Observational-knowledge})
in the sense that their contain shells on both small and large radii.
An important thing to notice is that within our model, any subsequent
passage of the secondary galaxy through the center of the primary
galaxy does not lead to a complete destruction of the shells from
the previous passages. Towards the center of the host galaxy, we find
shells with larger surface brightness, also a feature found in real
shell galaxies. At the same time, in Runs\,2\,\&\,3 we can find
faint shells surrounded by brighter ones from both sides, another
effect observed in real galaxies and impossible to reproduce in a
simple simulation.

The main difference between Run\,2 and Run\,3 lies in the positions
of the shells from the later generations -- those shells that dominate
the system in later times thanks to their brightness. The timing of
the second passage of the secondary galaxy through the center of the
host galaxy is very similar for Run\,2 and Run\,3 but the difference
in energy, mass and decay of the secondary galaxy is sufficient to
produce shells at different radii. Run\,3 also differs significantly
from Run\,2 (and also Run\,1) in that a bright shells system persist
even a long time after the first approach of the secondary galaxy
(7\,Gyr). However, we cannot say whether it is Run\,2 or Run\,3
that better describes the real merger of two galaxies under given
initial conditions. This indicates that quantitative modeling of a
shell system using test-particle simulation is very difficult or even
impossible.

In spite of the difficulties, we dare to state qualitative conclusions
independently on the method chosen for the tidal decay of the secondary
galaxy: the introduction of the dynamical friction and the gradual
decay to our simulations dramatically changes the appearance of shell
structures. Only the outermost shell of the first generation is not
overlayed by later, brighter generations of shells added during next
passages of the satellite through the center of the primary. While
the position of the outermost shell is not much affected by the dynamical
friction, its brightness is rapidly lowered due to the many particles
staying trapped in the weakened but remaining potential of the small
galaxy.

\subsection{Dark halo \label{sec:Sim-DF&TD&DM}}

To be even more realistic, we present a two-component model of the
galaxy \textendash{} a luminous component with a dark halo. The velocity
dispersion of each component is under the influence of the other (Sect.~\ref{sub:Disp-PP}).
The velocity dispersion is an important parameter of the dynamical
friction a thus values of the friction induced by each component slightly
differ (the amount depends on parameters) from the values we get when
the component is isolated (Sect.~\ref{sApx:Maple-sekce}). 

We performed three simulations with parameters listed in Table~\ref{tab:param-halo}.
In all the cases, the mass of the secondary galaxy is 0.02 of the
total mass of the primary; and the secondary approaches with escape
velocity. Dynamical friction is calculated using our modification
of the Chandrasekhar formula (Appendix~\ref{Apx:OurDF}). The mass
of the secondary galaxy was gradually lowered during the simulation
according to the number of test particles under the current tidal
radius (Sect.~\ref{sub:TD-Implement.}) and its Plummer radius was
being adjusted according to the method described in Sect.~\ref{sub:b}.

\noindent \vfill{}

\begin{table}[h]
\centering{}%
\begin{tabular}{ccccccccc}
\hline 
run & $\varepsilon_{*}$ & $M_{*}$ & $\varepsilon_{\mathrm{DM}}$ & $M_{\mathrm{DM}}$ & $\varepsilon_{\mathrm{s}}$ & $M_{\mathrm{s}}$ & $D_{\mathrm{ini}}$ & $v_{\mathrm{ini}}$\tabularnewline
 & kpc & M\suns & kpc & M\suns & kpc & M\suns & kpc & km$/$s\tabularnewline
\hline 
\noalign{\vskip0.1cm}
M0B0 & 7  & $3.2\times10^{11}$ & - & - & 2 & $6.4\times10^{9}$ & 180 & 125\tabularnewline
\noalign{\vskip0.1cm}
M2B6 & 7 & $3.2\times10^{11}$ & 60 & $6.4\times10^{12}$ & 2 & $1.344\times10^{11}$ & 300 & 443\tabularnewline
\noalign{\vskip0.1cm}
M6B10 & 7 & $3.2\times10^{11}$ & 100 & $1.92\times10^{13}$ & 2 & $3.904\times10^{11}$ & 300 & 756\tabularnewline
\hline 
\end{tabular}\caption{\textsf{\small Parameters of simulations. The potentials of the galaxies
are modeled as a single Plummer sphere for the secondary galaxy in
all runs and the primary galaxy in the run M0B0; and as a double Plummer
sphere for the primary in runs M2B6 and M6B10. Indices {*}, DM and
S refer to the luminous and dark components of the primary galaxy
and the secondary galaxy, respectively. $\varepsilon$ is Plummer
radius, $M$ total mass of the Plummer sphere, $D_{\mathrm{ini}}$
initial distance between centers of the secondary and primary galaxies
and $v_{\mathrm{ini}}$ their mutual velocity. \label{tab:param-halo}}}
\end{table}

\noindent \vfill{}

\clearpage

\begin{figure}[H]
\begin{centering}
\includegraphics[width=15cm]{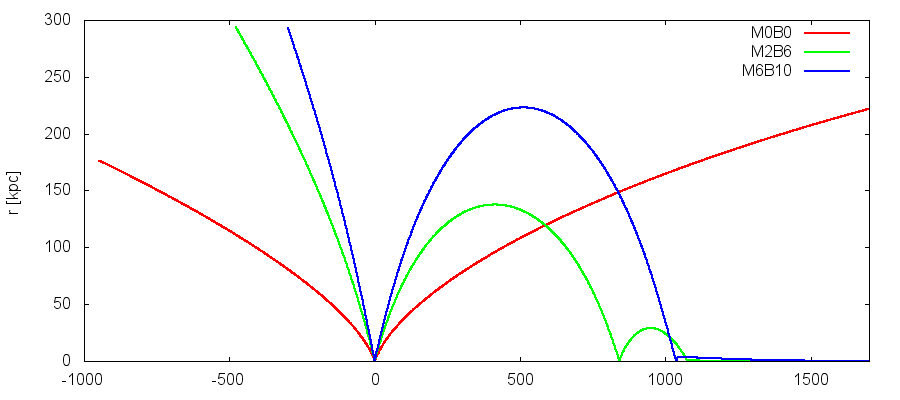}
\par\end{centering}

\begin{centering}
\includegraphics[width=15cm]{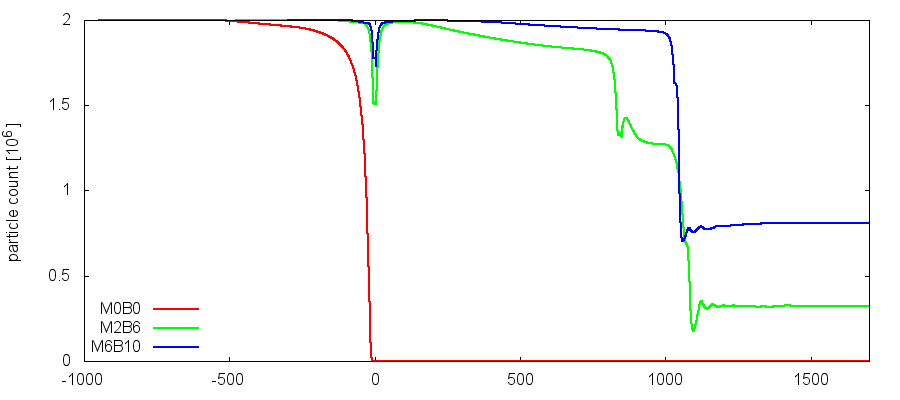}
\par\end{centering}

\begin{centering}
\includegraphics[width=15cm]{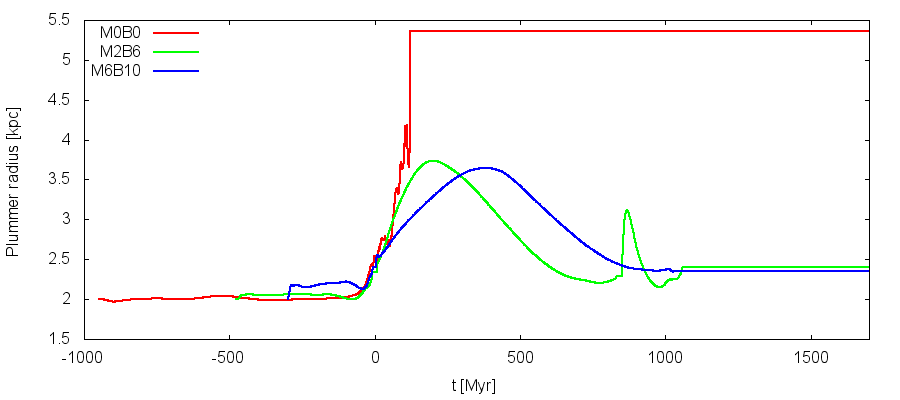}
\par\end{centering}

\caption{\textsf{\small Evolution of the merger for three different configurations
of the dark halo of the primary galaxy -- distance between galaxies,
number of particles bound to the secondary galaxy and its Plummer
radius. For the parameters of the mergers, see Table~\ref{tab:param-halo}.
\label{fig:halo-r-n-b} }}
\end{figure}

\begin{figure}[!h]
\begin{centering}
\includegraphics[width=14.5cm]{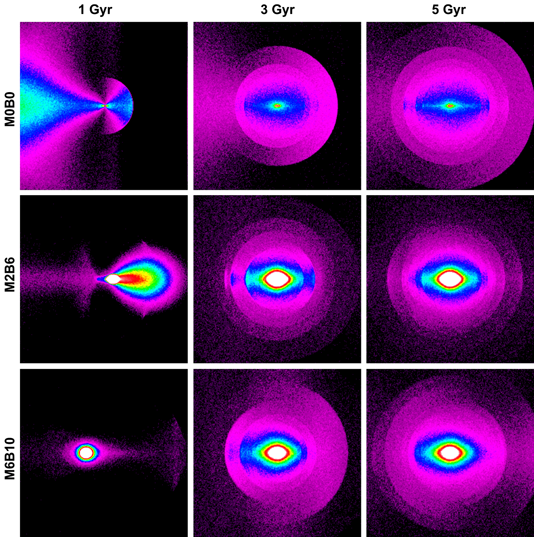}
\par\end{centering}

\caption{\textsf{\small Snapshots from three simulations, for the parameters
of the mergers, see Table~\ref{tab:param-halo}. Time stamps refer
to the time elapsed since the first passage of the secondary galaxy
through the center of the primary galaxy. Each panel covers 300$\times$300\,kpc
and is centered on the host galaxy. Only the surface density of particles
originally belonging to the satellite galaxy is displayed. The density
scale varies between frames, so that the respective range of densities
is optimally covered. \label{fig:halo-film} }}
\end{figure}

Fig.~\ref{fig:halo-r-n-b} illustrates the evolution of the distance
between the galaxies and the gradual decay of the secondary galaxy.
Time stamps of each run have been shifted so that in each case the
secondary galaxy reaches the center of the primary galaxy at time
0. In the first case (run M0B0 without any halo), the secondary galaxy
lost all particles during the first passage and this simulation is
rather equivalent to simulations with instant disruption. In the configurations
that include the halo (runs M2B6 and M6B10), the velocity is such
on the other hand that the primary galaxy catches only very few particles
in the first passage and a significant growth of the shell structure
is observed only in later phases of the merger. 

\begin{figure}[!h]
\begin{centering}
\includegraphics[width=15cm]{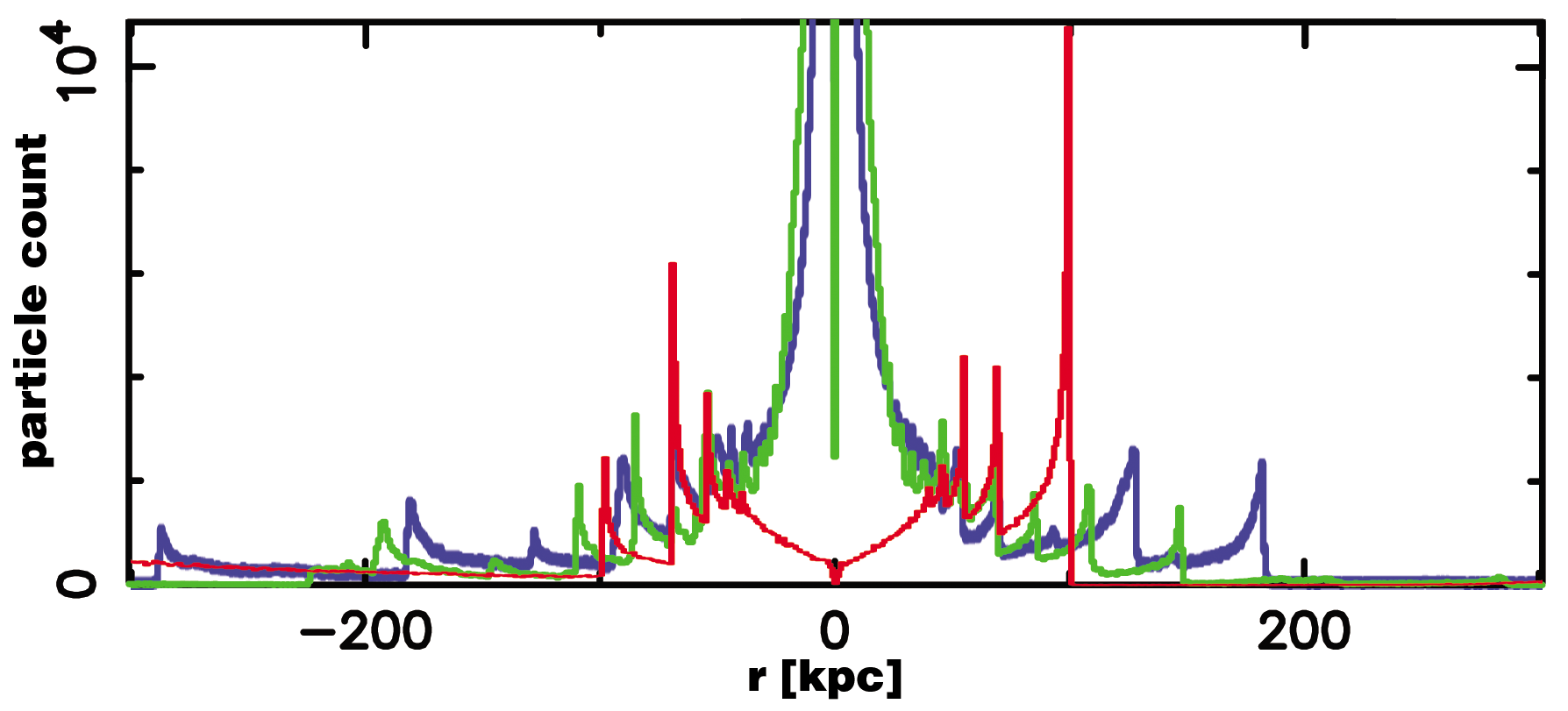}
\par\end{centering}

\caption{\textsf{\small Radial histogram of stars of the secondary galaxy,
centered on the primary 3\,Gyr after the first passage of the secondary
galaxy through the center of the primary for three different simulations.
The meaning of colors is the same as in Fig.~\ref{fig:halo-r-n-b}
(red: M0B0 -- without halo, green: M2B6 -- halo 20 times more massive
than the luminous component, blue: M6B10 -- 60 times more massive).
For the parameters of the mergers, see Table~\ref{tab:param-halo}.
\label{fig:halo-hist} }}
\end{figure}

Snapshots for three different times are shown in Fig.~\ref{fig:halo-film}
and radial histograms for time 3\,Gyr in Fig.~\ref{fig:halo-hist}.
For the simulation with a heavy halo (run M6B10) the particles cover
the largest span of energies (apocenters) and in both simulations
with a halo (runs M2B6 and M6B10), new shells on lower radii are created
in further passages of the secondary through the center of the primary
galaxy and many particles end up being caught in the center of the
primary. In the simulation without a halo (run M0B0) the secondary
decays in the first passage, but the particles have mostly sizable
energies at that time and thus have apocenters at larger galactocentric
radii or outright escape the system. The positions of the shells in
a given time are obviously different for different potentials of the
primary galaxy.

The main effect of the halo on the shell system is probably in that
its presence (through the increased mass of the primary galaxy) allows
for a faster development of shells at larger radii, despite the secondary
releasing in our case only a small part of its stars during its first
passage through the center of the host galaxy. Meanwhile, there are
additional shells created in the following passages, creating the
high radial range of shells observed in some galaxies which has continuously
proven difficult to reproduce in simulations.

The increased total mass of the host galaxy is apparently more important
than the difference in the dynamical friction caused by the differences
in local density and velocity dispersion for different halo configurations.
The more massive halo accelerates the secondary galaxy more, reducing
the loss of its energy via the dynamical friction and increasing the
time before a subsequent return of the secondary galaxy. But the higher
energy/velocity of the secondary galaxy allows the existence of shells
at larger radii - while it is important to note that in our simulations,
we see shells at 200 to 300\,kpc from the center of the host galaxy,
which is a distance where noone ever observed (or even looked for)
shells in real galaxies.

\begin{figure}[!t]
\begin{centering}
\includegraphics[width=14.5cm]{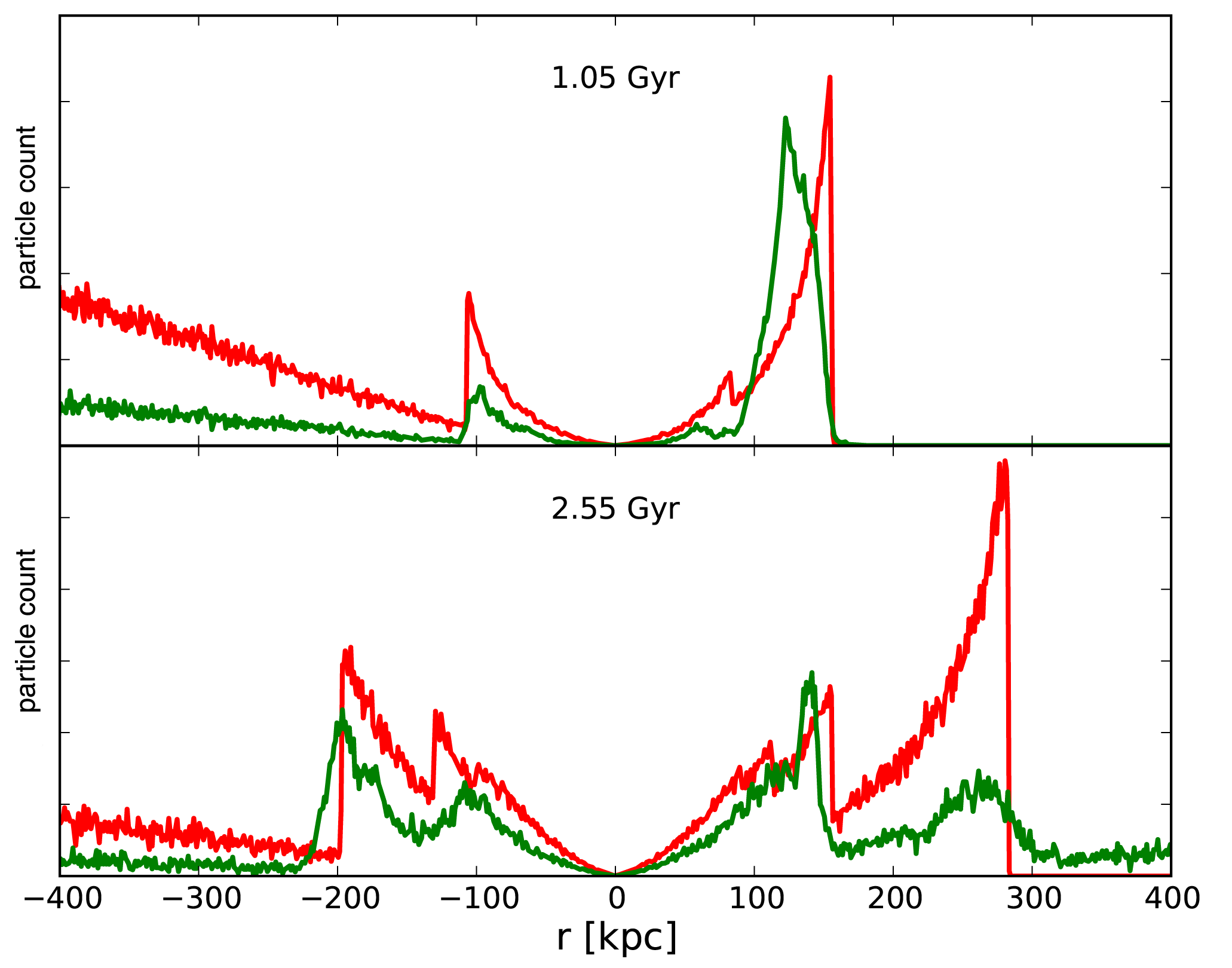}
\par\end{centering}

\caption{\textsf{\small Comparison of histograms of radial distances of shells\textquoteleft{}
particles in the self-consistent (green) and test-particle (red) simulations
at two different time steps. \label{fig:self-hist} }}
\end{figure}

\subsection{Self-consistent versus test-particle simulations \label{sub:Self-cons.}}

In this section, we compare two simulations with the same initial
conditions, one conducted in a self-consistent manner using GADGET-2
by Kate\v{r}ina Barto{\v s}kov{\'a}, the other one with test particles. Originally
we intended to keep the parameters of the primary galaxy, but a two-component
(luminous+dark matter) Plummer sphere is not a consistent system for
an arbitrary choice of parameters, particularly for those we have
used so far. The system is consistent when each physically distinct
component has a positive distribution function \citep{1996ApJ...471...68C}.
Thus we have chosen the following parameters for the merger:

The potential of the primary galaxy is a double Plummer sphere with
respective masses $M_{*}=2\times10^{11}$\,M\suns and $M_{\mathrm{DM}}=8\times10^{12}$\,M\suns,
and Plummer radii $\varepsilon_{*}=8$\,kpc and $\varepsilon_{\mathrm{DM}}=20$\,kpc
for the luminous component and the dark halo, respectively. The potential
of the secondary galaxy is chosen to be a single Plummer sphere with
the total mass $M=2\times10^{10}$\,M\suns and Plummer radius $\varepsilon_{*}=2$\,kpc.
The cannibalized galaxy is released from the distance of 200\,kpc
from the center of the host galaxy with the initial velocity 102\,km$/$s
in the radial direction (as always). 

\begin{figure}[H]
\begin{centering}
\includegraphics[width=10cm]{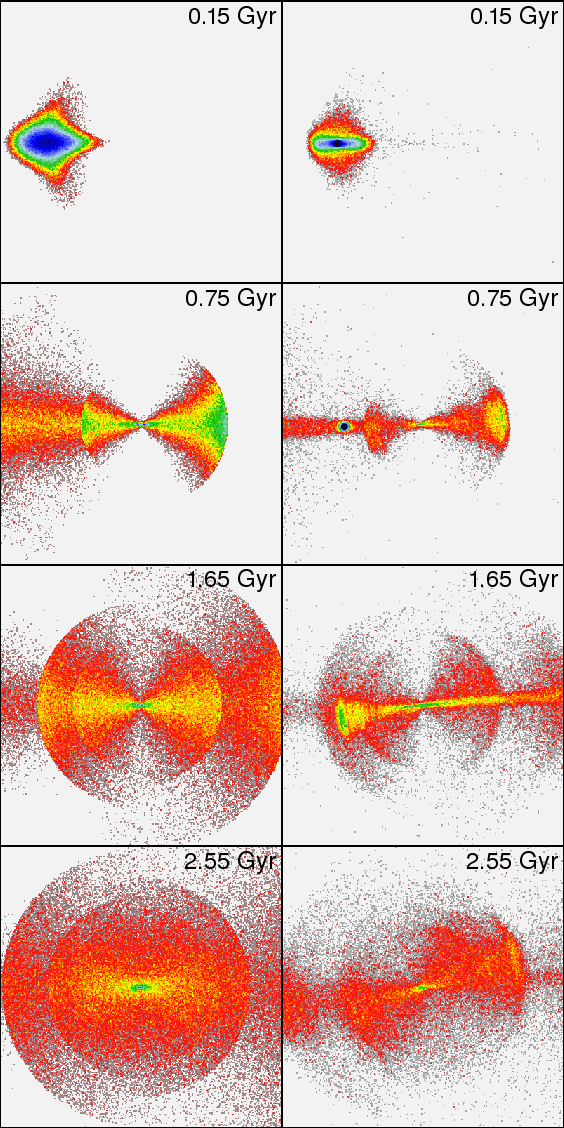}
\par\end{centering}

\caption{\textsf{\small Snapshots from a test-particle simulation (left) and
from the corresponding self-consistent simulation (right). Time equal
zero corresponds to the passage of the secondary galaxy through the
center of the primary galaxy. Each panel covers 400$\times$400\,kpc
and is centered on the host galaxy. Only the surface density of particles
originally belonging to the satellite galaxy is displayed. The density
scale varies between frames, so that the respective range of densities
is optimally covered. \label{fig:self-film}}}
\end{figure}

Snapshots from several times for both of the simulations are shown
in Fig.~\ref{fig:self-film}, radial histograms for the chosen times
in Fig.~\ref{fig:self-hist}. Video from the self-consistent simulation
is part of the electronic attachment. For the description, see Appendix~\ref{Apx:Videos}
point~\ref{enu:video5-self}. 

Unfortunately it turns out that for this choice of parameters, our
method of including the gradual decay of the secondary galaxy (Sect.~\ref{sub:TD-Implement.})
does not lead to a very gradual decay at all. In the case of simulation
with test particles, the secondary galaxy loses all its particles
near its first passage through the center of the primary galaxy. Thus
we use the model with instant disruption of the secondary instead.
To make the comparison even worse, the self-consistent simulations
behaves in yet another way: the core of the secondary galaxy survives
the first two passages through the center of the primary galaxy and
for some reason dissolves close to its apocenter. 

However, despite these significant differences, the results are surprisingly
similar. Most importantly, the radii of the outermost shells differ
by less than 10\%. In comparison with our enhanced test-particle simulations
(e.g., Sect.~\ref{sec:Sim-DF&TD} -- Runs\,2\,\&\,3), the self-consistent
simulation does not show a significant transport of particles of the
secondary galaxy to the area around the center of the host galaxy,
neither does it produce shells at low radii. Where on the other hand
the self-consistent simulation resembles more the enhanced test-particle
simulations than the simple test-particle simulation (with instantaneous
disruption of the secondary galaxy and no dynamical friction) is the
dramatic decrease of the brightness of the outermost shell on large
radii (compare with Sect.~\ref{sec:Sim-DF&TD} -- Fig.~\ref{fig:DF-hist}).

%\clearpage

\section{Discussion \label{sec:Discussion-Part-III}}

Our goal in this Part of the work was to include the dynamical friction
and the gradual decay of the secondary galaxy in the test-particle
simulations. It has been previously pointed out that coupling of
these phenomena is a key effect in the shell structure formation but
it was never modeled in much detail so far. Using these simulations,
we aimed to asses the plausibility of timing the shell-creating merger
using the outermost observed shell in a shell galaxy. 

For the dynamical friction we used our own modification of the Chandrasekhar
formula for radial trajectories, Appendix~\ref{Apx:OurDF}, which
is more faithful to the true stellar distribution function of the
host galaxy. The dynamical friction calculated in this way is fully
determined by the distribution function of the host galaxy and the
mass and velocity of the secondary, thus is contains no free parameters.
Comparison between our modification and the commonly used form of
the Chandrasekhar formula, Sect.~\ref{sApx:compare-to-Chf}, shows
that the use of a constant Coulomb logarithm is completely inappropriate
for radial mergers. But when compared with the self-consistent simulations,
our method is found to significantly overestimate the friction, Sect.~\ref{sub:Comparison-of-methods}.
In reality, the dynamical friction in a radial merger depends on the
whole merger history and thus can be hardly reproduced by any modification
of the Chandrasekhar formula, Sect.~\ref{sub:MTBAprinc}. Our simulations
thus have to be understood as the upper estimate on the true effect
of the dynamical friction on the shell formation.

Including the tidal disruption of the secondary galaxy in test-particle
simulation is even more complicated. We have tried several methods,
Sect.~\ref{sub:TD-Implement.}, and none of them is a priori better
than any other. Moreover we tried to reflect on the change of the
shape of the gravitational potential of the cannibalized galaxy during
the merger using a variable Plummer radius, Sect.~\ref{sub:b}. We
have carried out several simulations using different methods for the
decay of the secondary galaxy, focusing on qualitative effects in
which these simulations differ from simple simulations that assume
instantaneous breakdown of the secondary galaxy and no dynamical friction.
We can believe that effects that are independent of the method used
are more likely to participate in shell forming process in reality.

One such effect is that while the position of the outermost shells
of the first generation is not much affected by the inclusion of the
gradual decay and dynamical friction in the simulations, its brightness
is drastically lowered. The same effect is observed in our self-consistent
simulation, Sect.~\ref{sub:Self-cons.}. Even easily inferring the
age of the collision is rendered impossible (as already pointed out
by \citealp{1987A&A...185L...1D}). The shell systems in Fig.~\ref{fig:DF-hist}
(Sect.~\ref{sec:Sim-DF&TD}), all having the outermost shell at $+150$\,kpc,
are seen 5\,Gyr after the first passage of the cannibalized galaxy
through the center of the host galaxy. If we observationally identify
the leftmost shell (around $-80$\,kpc in Fig.~\ref{fig:DF-hist})
as being the outermost one, we would mistakenly estimate the merger
age to be only $\sim2.5$\,Gyr. We would also wrongly determine the
direction from which the secondary galaxy came: assuming the classical
picture (based on simulations without friction and with instantaneous
disruption), the outermost shell would be located on the side from
which the satellite came, so we would conclude it went from the left
while the opposite is true.

Furthermore, with respect to the simple simulation, in the simulations
with gradual decay of the secondary, we observe the creation of new
generations of shell during every passage of the remnant of the secondary
galaxy through the center of the primary. In the consecutive generations,
shells are created at lower radii and with higher brightness. It is
important to see that, in our simulations, the subsequent passages
of the secondary galaxy do not significantly disturb the existing
shell structure of the previous generation and thus a shell system
with a large range of radii is created. The radial range of shells
observed in some real shell galaxies is truly impressive and it is
impossible to reproduce in a simple simulation. It is also worth noting
that in simulations with the gradual decay of the secondary, a large
part of the mass of the secondary ends up in the proximity of the
center of the primary galaxy.

Presence of a dark matter halo in the primary galaxy, Sect.~\ref{sec:Sim-DF&TD&DM},
changes not only the dependence of the period of radial oscillations
on radius (Sect.~\ref{sec:rad_osc}), but also the range of stellar
energies through the change of the velocity of the accreted satellite.
The halo allows for a faster development of shells at larger radii.
A more massive halo creates a larger range of shell radii in our simulations
than a less massive one. The increased total mass of the host galaxy
is more important than the difference in the dynamical friction caused
by the differences in local density and velocity dispersion for different
halo configurations. The more massive halo accelerates the secondary
galaxy more, reducing the loss of its energy via the dynamical friction
and increasing the time before a subsequent return of the secondary
galaxy. The higher velocity of the secondary galaxy also means that
the primary galaxy catches only very few particles in the first passage
and a significant growth of the shell structure is observed only in
later phases of the merger. 

In general, it seems that test-particle simulations are not suitable
for a quantitative reproduction of observed shell systems. There is
no reliable (semi-)analytical method to calculate the dynamical friction
in radial and close-to-radial minor mergers. Apparently even more
importantly, there is no universal method to model the tidal decay
of the cannibalized galaxy in test-particle simulations. Unfortunately,
it turns out that it is exactly the details of the decay of the secondary
galaxy that affect significantly the overall shell structure. In two
simulations, with apparently small differences in the loss of mass
and energy of the secondary galaxy during the first passage and the
time of the second passage, shells of the second generations were
created at different radii with respect to the shells from the first
generation (which are otherwise very similar between the simulations).
Moreover, the brightness of these shells differs and with each farther
passage of the secondary galaxy, the difference in the appearance
of the shell system increases and the observability of shells in the
host galaxy changes by whole gigayears. Overall, an accurate reproduction
of a shell galaxy is a very delicate matter, as in practice we do
not know an exact distribution of mass in the host galaxy, the original
trajectory of the secondary galaxy, nor its own mass distribution
and our simulations suggest that the shell structure is very sensitive
even to small details in these quantities.

Nevertheless even despite the simplicity of the models we used, it
turned out that our test-particle simulations with gradual disruption
and dynamical friction of the secondary galaxy do better than the
simple simulations in reproducing observed features in real shell
galaxies. We thus conclude that also in real galaxies, these features
are the result of combined effects of the gradual decay and dynamical
friction.

At the end, we shall stress that while all these details have a large
effect on the overall appearance of the shell system, they are not
very important for the application of the method to measure the host
galaxy potential from kinematical data that we have introduced in
Part~\ref{PART II-S.kin}. This method relies only on the assumption
that the stars that form one particular shell are moving along radial
trajectories and were released in the center of the primary galaxy
together at some moment in the past. Within the framework the radial-minor-merger
model, neither the gradual decay of the secondary galaxy nor the dynamical
friction do not in principle have a large influence on the radiality
of the stellar trajectories. Also, even when these effects are present,
stars are being released in short time intervals when the secondary
galaxy passes through the center of the primary galaxy, however these
intervals are slightly larger than zero, which would be the case for
the instantaneous decay of the secondary galaxy. This fact causes
the shells to be slightly more diffuse and can interfere with an effort
to determine the positions of the spectral peaks and the shell edge.
Nevertheless, in principle the measurement of the potential should
be still possible.

\clearpage

\newpage{}

\part{Conclusions \label{PART IV-zaver}}

In Part~\ref{PART I} we have summarized observational and theoretical
knowledge about the shell galaxies according to the available literature.
Shell galaxies are mostly elliptical galaxies containing fine structures
which are made of stars and form open, concentric arcs that do not
cross each other. The most prominent observational characteristics
of shells are summarized in 22 points in Sect.~\ref{sec:characterictics}.
In Sect.~\ref{sec:Scenarios}, we introduce all proposed scenarios
of origin of shell galaxies. The most widely accepted theory, supported
by a multitude of observational evidence, is the close-to-radial minor
merger of galaxies introduced by \citet{1984ApJ...279..596Q}. In
the framework of this model, \citet{mk98} suggested using shell kinematics
to measure the potential of the host galaxy. The issue of the determination
of the overall potential and distribution of the dark matter in galaxies
is among the most prominent in galactic astrophysics since the most
successful theory of the evolution of the Universe so far seems to
be the theory of the hierarchical formation based on the assumption
of the existence of cold dark matter, significantly dominating the
baryonic one. Thus, independent measurement of the dark matter content
in galaxies is highly desirable. Measurement of galactic potential
is particularly difficult in elliptical galaxies at large distances
from the center of the galaxy. Incidentally, shells are found mainly
in elliptical galaxies and they do occur in distances up to 100\,kpc
from the center.

The method of \citet{mk98} is based on the approximation of a stationary
shell. Using positions of peaks in the line-of-sight velocity distribution
(LOSVD), it allows the calculation of the gradient of the potential
near the shell edge. We have developed this method further in Part~\ref{PART II-S.kin}
assuming validity of the radial-minor-merger model and spherical symmetry
of the host galaxy. Using both analytical calculations and test-particle
simulations, we have shown that the LOSVD has a quadruple shape in
this situation. Assuming a constant shell phase velocity and a constant
radial acceleration in the host galaxy potential for each shell, we
have developed three different analytical and semi-analytical approaches
(Sect.~\ref{sec:Compars}) for obtaining the circular velocity in
the host galaxy and the current shell phase velocity from the positions
of the peaks of the maxima of the LOSVD.

The applicability of our different approaches varies with the character
of measured data. As obtaining suitable data is at the very limit
of current observational tools and thus no such data is yet available
for analysis, we have applied our methods to results of a simulation
of a radial minor merger. We were able to reproduce the circular velocity
at shell radii to within $\sim1$\,\% from the actual value. Applying
the method of \citet{mk98} to the simulated data, we have derived
a circular velocity larger by 40--50\% than the true value.

All our approaches, however, derive the shell phase velocity systematically
larger, 7--30\%, than the real velocity is. That can be caused by
nonradial trajectories of the stars of the cannibalized galaxy or
by poor definition of the shell radius in the simulation. The method
of \citet{mk98} does not allow to derive the shell phase velocity
at all since it is based on the approximation of a stationary shell.

In the case of spherical symmetry, the value of the circular velocity
directly determines the amount of mass enclosed under the given radius,
thus determining the dark matter content of the galaxy. On the other
hand, the shell velocity depends on the serial number of the shell
and on the whole potential from the center of the galaxy up to the
shell radius and thus its interpretation is less straightforward.
A comparison of its measured velocity to theoretical predictions is
possible only for a given model of the potential of the host galaxy
and the presumed serial number of the observed shells. In such a case,
however, it can be used to exclude some parameters or models of the
potential that would otherwise fit the observed circular velocity.
Moreover, the measurement of shell velocities can theoretically decide
whether the outermost observed shell is the first one created; determine
the time from the merger and the impact direction of the cannibalized
galaxy; and reveal the shells from different generations, which can
be present in a shell galaxy \citep{katka11}.

In Part~\ref{PART II-S.kin} we have examined effects of the gradual
decay and dynamical friction of the cannibalized (secondary) galaxy
on the appearance of the shell structure. Our goal was to asses the
plausibility of timing the shell-creating merger using the outermost
observed shell in a shell galaxy. Attempts to date a merger from observed
positions of shells, using simple test-particle simulations, have
been made in previous work of \citet{canalizo07} supporting a potential
causal connection between the merger, the post-starburst ages in nuclear
stellar populations, and the quasar.

We have searched for a method to include the gradual decay and dynamical
friction of the secondary galaxy into the test-particle simulations.
While these effects are (along with many other physical processes)
naturally included in self-consistent simulations, using these has
also some serious drawbacks when compared to test-particle simulations.
For example, some effects seen in self-consistent simulations are
difficult or outright impossible to reproduce by analytical or semi-analytical
methods. At the same time, their manifestation in self-consistent
simulations is difficult to separate and sometimes they may even be
confused with non-physical outcomes of used methods. Moreover, self-consistent
simulations with high resolution necessary to analyze delicate tidal
structures such as the shells are demanding on computation time. This
demand is even larger if we want to explore a significant part of
the parameter space. 

For the dynamical friction we used our own modification of the Chandrasekhar
formula for radial trajectories, Appendix~\ref{Apx:OurDF}. The dynamical
friction calculated in this way is fully determined by the distribution
function of the host galaxy and the mass and velocity of the secondary,
thus is contains no free parameters. But when compared with the self-consistent
simulations, our method is found to significantly overestimate the
friction, Sect.~\ref{sub:Comparison-of-methods}. Our simulations
thus have to be understood as the upper estimate on the true effect
of the dynamical friction on the shell formation.

We have tried several methods for including the tidal disruption and
deformation of the secondary galaxy, Sect.~\ref{sec:Tidal-disruption},
and none of them is a priori better than any other. In our simulations
it turns out that the resulting shell system is very sensitive to
small differences during the decay of the cannibalized galaxy and
thus the test-particle simulations are not suitable for a quantitative
reproduction of observed shell systems. We have thus focused on qualitative
effects in which our enhanced simulations differ from simple simulations
that assume instantaneous breakdown of the secondary galaxy and no
dynamical friction. It turned out that these enhanced test-particle
simulations do better than the simple simulations in reproducing observed
features in real galaxies, including features that the simple simulations
cannot show at all. We thus conclude that also in real galaxies, these
features are the result of combined effects of the gradual decay and
dynamical friction.

One effect found commonly in all the enhanced test-particle simulations
is that while the position of the outermost shells of the first generation
is not much affected by the inclusion of the gradual decay and dynamical
friction in the simulations, its brightness is drastically lowered.
The same effect is observed in our self-consistent simulation, Sect.~\ref{sub:Self-cons.}.
Even just inferring the age of the collision is thus tricky: if we
observationally miss the weakened outermost shell, which should be
clearly visible according to simple simulations, we would underestimate
the merger age by a factor of 2. At the same time, we would also wrongly
determine the direction from which the secondary galaxy came.

Ideally, for systems with multiple shells we would like to combine
measurements of shell kinematics and their radial distribution, possibly
also with measurements of surface brightness profile (Sect.~\ref{sub:Projected-surface-brightness}).
The kinematical measurements supply us with the magnitude of acceleration
at the shell edge and an estimate of the phase shell velocity, which
allows us to separate the shells in different generations, if these
are present. Simulations with the dynamical friction and gradual decay
of the secondary galaxies that reproduce the kinematic and photometric
data will then constrain other parameters of the merger such as its
age and the trajectory and nature of the satellite galaxy. A similar
result has been obtained for M31, \citet{fardal07,fardal08,fardal12},
whereas for the other shell galaxies, obtaining the kinematical data
is a great challenge for the next generation of astronomical instruments.

\newpage{}

\appendix

\part{Appendix}

\section{Units and conversions \label{Apx:units}}

When dealing with galaxies, we need to describe objects and time spans
incommensurable with our daily experience that defines the standard
sets of units, such as SI. Throughout the text we thus use a set of
units adapted for this task -- we measure the mass in M\suns the
length in kpc and the time in Myr. Although their meaning is clear,
they sometimes give rise to rather awkward derived units. We will
briefly list the most prominent of them (together with the basic ones)
and give their relation to the SI and cgs units.
\begin{lyxlist}{00.00.0000.0000}
\begin{onehalfspace}
\item [{\textbf{Time:}}] \noindent 1\,Myr = $10^{6}$\,yr = $3.156\times10^{13}$\,s
\item [{\textbf{Distance:}}] \noindent 1\,kpc = 3\,262\,ly = $3.086\times10^{19}$\,m
= $3.086\times10^{21}$\,cm
\item [{\textbf{Mass:}}] \noindent 1\,M\suns = $1.989\times10^{30}$\,kg
= $1.989\times10^{33}$\,g
\item [{\textbf{Velocity:}}] \noindent 1\,kpc$/$Myr = 977.8\,km$/$s
= $9.778\times10^{7}$\,cm$/$s (the roundness of this value allows
for an easy conversion for most of our plots)
\item [{\textbf{Acceleration:}}] \noindent 1\,kpc$/$Myr$^{2}$= $3.098\times10^{-8}$\,m$/$s$^{2}$
= $3.098\times10^{-6}$\,cm$/$s$^{2}$
\item [{\textbf{Density:}}] \noindent 1\,M\sun$/$kpc$^{3}$= $6.768\times10^{-29}$\,kg$/$m$^{3}$
= $6.768\times10^{-32}$\,g$/$cm$^{3}$
\item [{\textbf{Grav.~unit:}}] \noindent 1\,kpc$^{3}$$/$Myr$^{2}$$/$M\suns
= 14.83\,m$^{3}$$/$s$^{2}$$/$kg = 14\,830\,cm$^{3}$$/$s$^{2}$$/$g
--\\
 thus G = $6.674\times10^{-11}$\,m$^{3}$$/$s$^{2}$$/$kg = $4.500\times10^{-12}$\,kpc$^{3}$$/$Myr$^{2}$$/$M\suns \end{onehalfspace}

\end{lyxlist}
\newpage{}

\section{List of abbreviations \label{apx:List-of-Abbreviations}}
\begin{description}
\item [{AU}] arbitrary unit, a relative placeholder unit for when the actual
value of a measurement is unknown or unimportant
\item [{DM}] dark matter
\item [{FWHM}] full width at half maximum, parameter of Gaussian function
\item [{GADGET-2}] free software used for self-consistent simulations,
see Sect.~\ref{sec:GADGET}
\item [{KDC}] kinematically distinct/decoupled cores of galaxies, see Sect.~\ref{sub:Other-features}
\item [{LOS}] line-of-sight
\item [{LOSVD}] line-of-sight velocity distribution
\item [{MK98}] paper about measuring gravitational potential using shell
kinematics \citet{mk98}
\item [{MTBA}] Multiple Three-Body Algorithm, a method used by \citet{1994A&A...290..709S}
to study dynamical friction in head-on galaxy collisions, see Sect.~\ref{sec:MTBA}
\item [{S$/$N}] signal-to-noise ratio
\item [{WIM}] Weak Interaction Model of origin of shell galaxies by \citet{1990MNRAS.247..122T},
see Sect.~\ref{sub:WIM}
\end{description}
\newpage{}

\section{Initial velocity distribution\label{Apx:IVD}}

The shell-edge density distribution, $\sigma_{\mathrm{sph}}\left(r_{\mathbf{s}}\right)$,
is defined by Eq.~(\ref{eq:sigma}). Note that, since, in the model
of the radial oscillations, all stars at the shell edge have the same
energy, the function $N\left(r_{\mathbf{s}}\right)$ determines the
distribution of stellar apocenters, the radial dependence of which
differs just slightly from $\sigma_{\mathrm{sph}}\left(r_{\mathbf{s}}\right)$.

Let $f(r_{\mathrm{ac}})$ and $g(v_{0})$ be the distribution function
of the stellar apocenters and the initial velocities, respectively,
then 
\begin{equation}
g(v_{0})=f(r_{\mathrm{ac}})\frac{\mathrm{d}r_{\mathrm{ac}}}{\mathrm{d}v_{0}}.\label{eq:DF-1}
\end{equation}
In almost all cases in the thesis $\sigma_{\mathrm{sph}}\left(r_{\mathbf{s}}\right)\propto1/r_{\mathbf{s}}^{2}$,
so the distribution function of the stellar apocenters is a constant
function $f(r_{\mathrm{ac}})=A$.

Initially, all stars are at the center of the host galaxy, so 
\begin{equation}
v_{0}=\sqrt{-2\left[\phi(r_{\mathrm{ac}})-\phi(0)\right]},\label{eq:v0-1}
\end{equation}
 where $\phi(r)$ is the spherically symmetric potential of the host
galaxy. Then 
\begin{equation}
g(v_{0})=A\left.\frac{\mathrm{d}\phi(v)}{\mathrm{d}v}\right|_{v=v_{0}}v_{0},
\end{equation}
 where $\phi(v)$ is inverse function to $v_{0}\left(\phi\right)$
given by Eq.~(\ref{eq:v0-1}).

The correspondence between the shell-edge density distribution and
initial velocity distribution is one-to-one, unlike for example the
one between the spatial (or projected) density and the shell-edge
density distribution, Eq.~(\ref{eq:F-rho}), as the density at one
radius receives contributions from particles with two distinct velocities.

\clearpage

\section{Introduction to dynamical friction \label{Apx:Intro-DF} \label{Apx:ChFvelka}}

Appendix~\ref{Apx:Intro-DF} was, with some adjustments, adapted
from the master thesis \citet{EbrovaMAT}.

\subsection{A thermodynamic meditation}

The dynamical friction is a braking force of gravitational origin,
caused by the sole fact that the area, through which the secondary
galaxy (or, in general, any object passing through a galaxy or another
extended object) flies is not an empty space filled with a smooth
potential, but a large sea of individual stars.

Thinking deeper, we can easily see that some slowdown of the secondary
galaxy is inevitable. Every system, where energy transfer is possible
tends to temperature equilibrium. In a system of at least three gravitating
bodies such a transfer is indeed possible and frequently happens.
The relatively fast and heavy secondary galaxy possesses a decent
amount of kinetic energy and as such it is just a hot piece thrown
into a colder sea of the stars of the primary galaxy. The slowdown
of the intruder that cannot be accounted for in the fixed-potential
model, is the way of leveling the temperatures. The kinetic energy
transfers to the primary's stars -- the same effect causes the heating
of the cold disk population in the week interaction model, as mentioned
in Sect.~\ref{sub:WIM}.

The reality of this process can be grasped from a different point
of view. The relatively massive secondary galaxy attracts the primary's
stars and thus creates an area of a higher density of stars behind
itself. The passing galaxy is attracted backwards by this condensation,
lowering its speed towards the primary.

\subsection{Chandrasekhar formula\label{sApx:ChF} }

An analytical derivation of such a braking force is based on the following
thought: In a distant encounter with just one star, the velocity of
an object cannot be changed, instead it is only deflected from the
original direction and thus enriched with a component of speed perpendicular
to the original direction. For a very massive body, as our secondary
galaxy is, the magnitude of this perpendicular component will not
be large, neither will be the loss of the velocity in the original
direction. But when it undergoes many such encounters, the contributions
add. The contributions in the perpendicular directions will have randomly
scattered azimuthal angles and thus add to zero (except for the overall
action of the smooth potential). On the other hand, the contributions
to the original direction of the velocity will always be opposite
to it, resulting in the braking of the galaxy.

The Chandrasekhar formula was originally derived by \citet{1943ApJ....97..255C}.
Here we present a short version of the presentation of the chapter~7.1
in the bible of the galactic astronomy, {}``Galactic Dynamics''
by \citet{1987gady.book.....B}.

To start, let us imagine the encounter of our object of interest with
a single star. When two bodies meet, energy is not transferred, but
the direction of velocity of our object changes. It is a matter of
a simple mechanics and as a result, the change of the component of
velocity parallel to its original direction, $\mid\mathbf{\bigtriangleup v}_{M\parallel}\mid$
between the times $t=-\infty$ and $t=\infty$ is given by (Eq.~7-10b
in \citealp{1987gady.book.....B}; see its derivation there):

\begin{equation}
\mid\bigtriangleup\mathbf{v}_{M\parallel}\mid=\frac{2\, m\, V_{0}}{M+m}\left[1+\frac{b^{2}V_{0}^{4}}{G^{2}(M+m)^{2}}\right]^{-1},\label{eq:dletaV}
\end{equation}
 where \textit{$M$} is our object's (the secondary galaxy) mass,
\textit{$m$} is the mass of the star, $b$ the impact parameter (the
length of $\mathbf{b}$, the vector indicating the position of the
star in a plane perpendicular to the original velocity of the galaxy)
and $\mathbf{V}_{0}$ is the difference between the original velocity
of the star $\mathbf{v}_{m}$ and velocity of our object $\mathbf{v}_{M}$,
so $\mathbf{V}_{0}=\mathbf{v}_{m}-\mathbf{v}_{M}$. The bold typeface
indicates vectors, and their length is indicated by the same symbol
in normal type.

For an object flying through a field of stars with the phase-space
number density of stars $f(\mathbf{v}_{m},\mathbf{b})$, the change
in the parallel component of velocity $\mathrm{d}\mathbf{v}_{M\parallel}$
in an infinitesimal time $\mathbf{\mathrm{d}}t$ will be given by
the integration of Eq.~(\ref{eq:dletaV}) multiplied by the density
$f(\mathbf{v}_{m},\mathbf{b})$ over the plane of $\mathbf{b}$ and
the space $\mathbf{v}_{m}$. For $\mathbf{b}$ is measured from a
given point in a plane, it is advantageous to use the polar coordinates
$(b,\varphi)$:

\begin{equation}
\frac{\mathrm{d}\mathbf{v}_{M\parallel}}{\mathrm{d}t}=\intop\intop\intop f(\mathbf{v}_{m},b,\varphi)\frac{2m\, V_{0}(\mathbf{v}_{m})\,\mathbf{V}_{0}(\mathbf{v}_{m})}{(M+m)\left[1+\frac{b^{2}V_{0}^{4}(\mathbf{v}_{m})}{G^{2}(M+m)^{2}}\right]}\mathrm{d}^{3}\mathbf{v}_{m}\, b\mathrm{d}b\,\mathrm{d}\varphi.\label{eq:iPol}
\end{equation}
To derive the Chandrasekhar formula we further assume the homogeneity
of the field of stars, so as the distribution function of the stars
does not depend on $\mathbf{b}$. The remaining $\mathbf{b}$-dependent
part is of the following form a can be easily integrated from 0 to
some $b_{\mathrm{max}}$:

\begin{equation}
\intop_{0}^{b_{\mathrm{max}}}\frac{b\mathrm{d}b}{1+c^{2}b^{2}}=\left[\frac{\ln(1+c^{2}b^{2})}{2\, c^{2}}\right]_{b=0}^{b=b_{\mathrm{max}}},
\end{equation}
where in our case $c=V_{0}^{2}/[G(M+m)]$. It is conventional to introduce
the notation

\begin{equation}
\Lambda=\frac{b_{\mathrm{max}}V_{0}^{2}}{G(M+m)}.\label{eq:L}
\end{equation}
A typical value of $\Lambda$ would be of the order of $10^{3}$,
thus we can neglect the one and put $\frac{1}{2}\ln(1+\Lambda^{2})\cong\ln(\Lambda)$.
This factor is often called the Coulomb logarithm. Furthermore we
assume that we do not err too much when replacing $V_{0}$ in $\Lambda$
by $v_{\mathrm{typ}}$, a typical speed. Then the Coulomb logarithm
does not depend on $\mathbf{v}_{m}$, and still $V_{0}=\mid\mathbf{v}_{m}-\mathbf{v}_{M}\mid$
and the whole Eq.~(\ref{eq:iPol}) goes to

\begin{equation}
\frac{\mathrm{d}\mathbf{v}_{M\parallel}}{\mathrm{d}t}=4\pi\ln(\Lambda)G^{2}m(M+m)\intop f(\mathbf{v}_{m})\frac{\mathbf{v}_{m}-\mathbf{v}_{M}}{\mid\mathbf{v}_{m}-\mathbf{v}_{M}\mid^{3}}\mathrm{d}^{3}\mathbf{v}_{m}.
\end{equation}
The integral is of exactly the same form as in the Newton's law of
gravity and if the stars move isotropically, the density distribution
is spherical and by Newton`s first theorem (see \citealp{1987gady.book.....B};
chapter~2), the total acceleration of our object by dynamical friction
is:

\begin{equation}
\frac{\mathrm{d}\mathbf{v}_{M\parallel}}{\mathrm{d}t}=-16\pi^{2}\ln(\Lambda)G^{2}m(M+m)\frac{\intop_{0}^{v_{M}}f(v_{m})v_{m}\mathrm{d}v_{m}}{v_{M}^{3}}\mathbf{v}_{M}\label{eq:ChF}
\end{equation}
 i.e., only stars moving slower then our object contribute to the
force and this force always opposes the motion. Eq.~(\ref{eq:ChF})
is usually called the \textit{Chandrasekhar dynamical friction formula}.

If $f(v_{m})$ is Maxwellian with dispersion $\sigma$

\begin{equation}
f=\frac{n_{0}}{(2\pi\sigma^{2})^{3/2}}\exp(-\frac{1}{2}v^{2}/\sigma^{2}),\label{eq:Maxw}
\end{equation}
we can integrate Eq.~(\ref{eq:ChF}). The density of the stars is
$\rho_{0}=n_{0}\, m$ and for $M\gg m$, what happens to be our case,
we can put $(M+m)\cong M$, and then Eq.~(\ref{eq:ChF}) reads:

\begin{equation}
\frac{\mathrm{d}\mathbf{v}_{M\parallel}}{\mathrm{d}t}=-\frac{4\pi\ln(\Lambda)G^{2}\rho_{0}M}{v_{M}^{3}}\left[\mathrm{erf}(X)-\frac{2\, X}{\sqrt{\pi}}\mathrm{e}^{-X^{2}}\right]\mathbf{v}_{M},\label{eq:DF}
\end{equation}
where $\Lambda$ is given by Eq.~(\ref{eq:L}), $X\equiv v_{M}/(\sigma\sqrt{2})$
and erf(\textit{$X$}) is the error function given by

\begin{equation}
\mathrm{erf}(X)\equiv\frac{2}{\sqrt{\pi}}\intop_{0}^{X}\mathrm{e}^{-t^{2}}\mathrm{d}t
\end{equation}
for which we can obtain tabulated values, or we can pre-generate them
numerically with an arbitrary precision.

\subsection{What a wonderful universe}

Giving it a deeper thought, one can consider the validity of the \textit{Chandrasekhar
formula} almost a miracle. We have by the way disclosed that it works,
at least approximately -- the confrontation with numerical simulations
of flybys through a galaxy or a cluster has been carried out by e.g.
\citet{1983ApJ...264..364L,1987MNRAS.224..349B}, who proved that
the analytical solution (given by the Chandrasekhar formula) is in
a good agreement with the simulations in a relatively wide range of
situations. The analytical solutions has some freedom in the Coulomb
logarithm which is not completely well-defined. Its correct choice
can help to better reproduce the numerical results and compensate
other drawbacks of the formula -- anyway, the freedom is small when
we demand the Coulomb logarithm to stay constant.

\begin{figure}[!h]
\begin{centering}
\includegraphics[width=0.6\columnwidth]{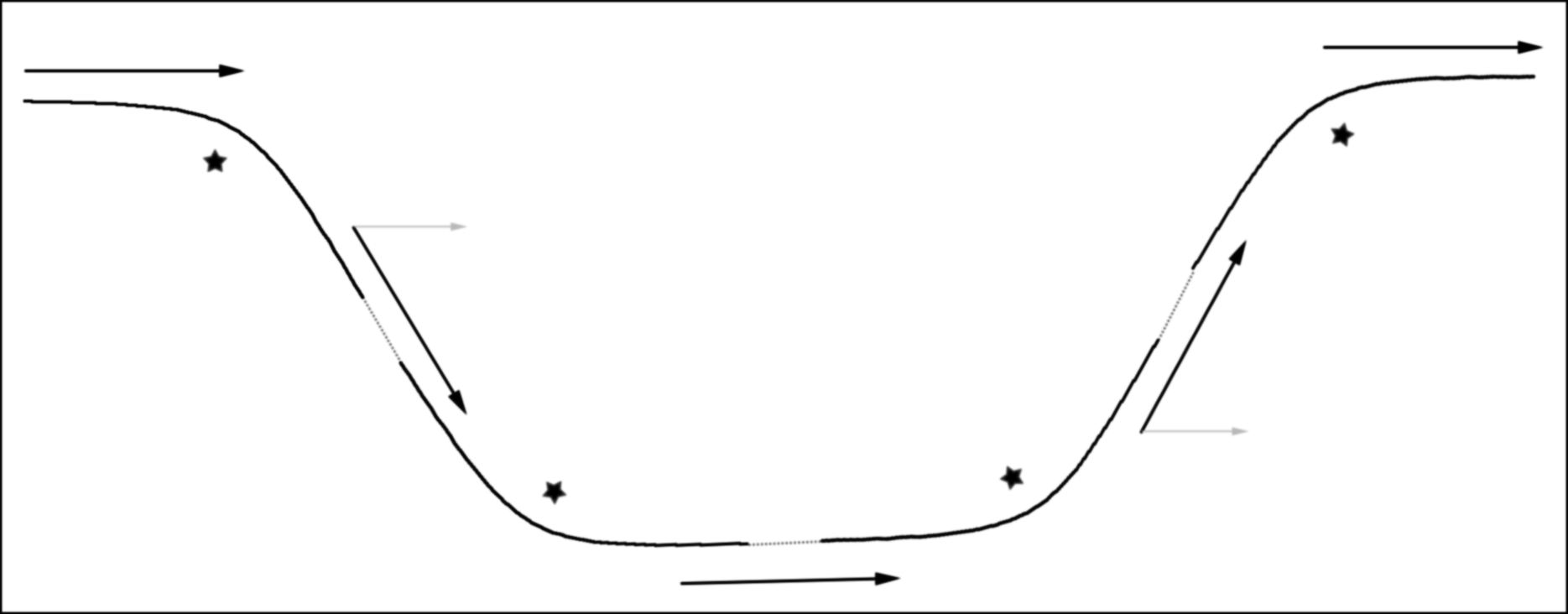}
\par\end{centering}

\caption{\textsf{\small The path and velocity changes of the objects undergoing
encounters with individual stars. The absolute value of the velocity
remains unchanged. \label{obr.DTzly}}}
\end{figure}

But back to our astonishment. When the secondary galaxy deviates from
its course, its speed in the original direction is reduced. But after
meeting another star that compensates the deviation, it also gets
back the original velocity in this direction, as is shown in Fig.~\ref{obr.DTzly}.

\begin{figure}[!h]
\begin{centering}
\includegraphics[width=0.6\columnwidth]{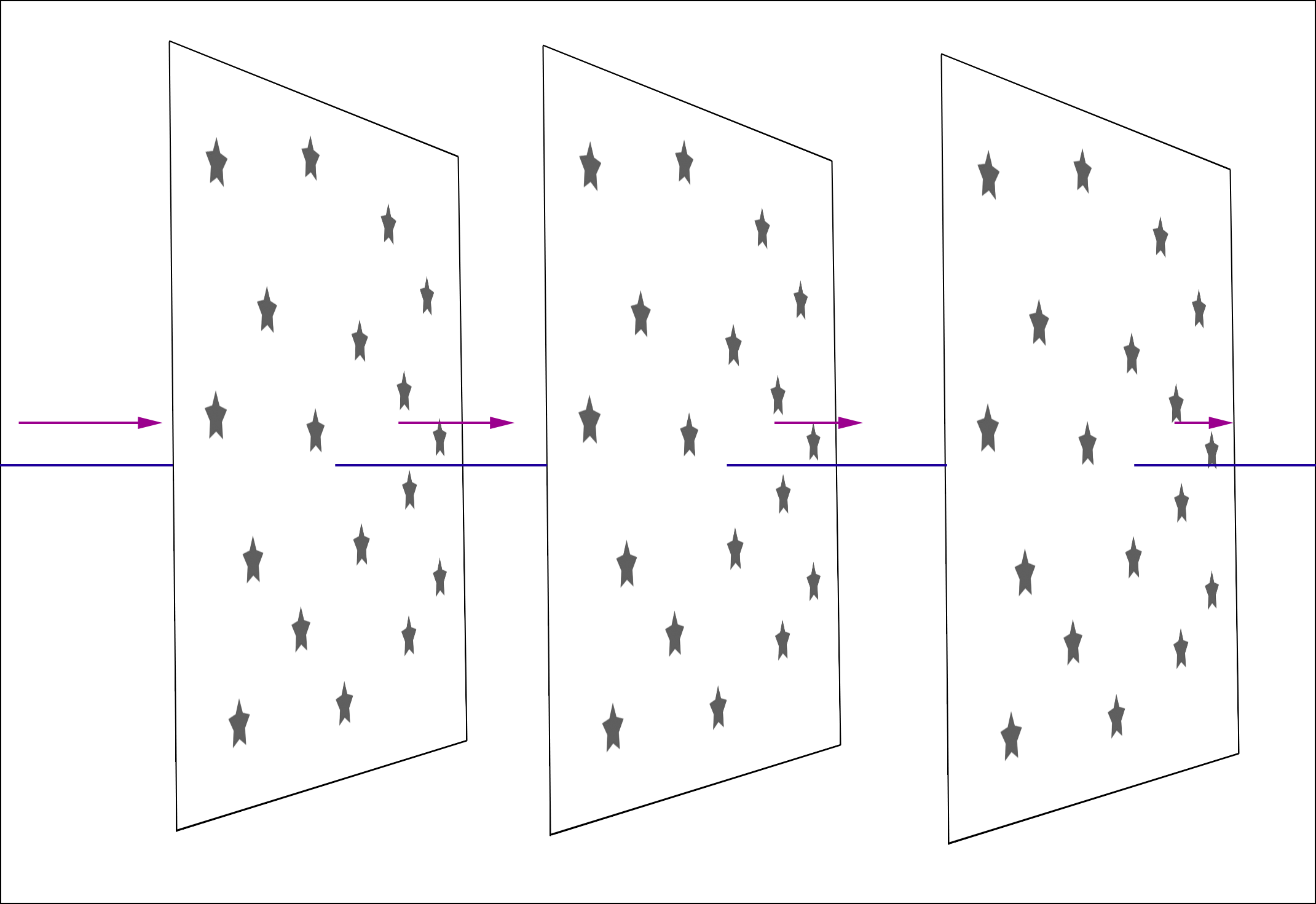}
\par\end{centering}

\caption{\textsf{\small A schematic depiction of the change of the velocity
of the secondary galaxy after three steps. In every moment, only the
influence of the stars lying in one plane perpendicular to the motion
of the galaxy is taken into account. \label{obr.DT3d}}}
\end{figure}

The point is that the \textit{Chandrasekhar formula} evaluates the
change of the parallel component of the velocity after the flyby from
infinity to infinity for every single star with the same initial conditions
and then adds these changes and applies them to the secondary galaxy
in one moment, the moment of the closest approach with these stars,
see Fig.~\ref{obr.DT3d}. The change of the parallel component of
the velocity and the compensation of the changes in the perpendicular
direction then happen somehow at the same time, although the magnitude
of their effect is calculated as if they happen consecutively -- and
by some wonder, it works.

Let us just remark that the fact that we account for the influence
of the stars in the moment of the closest approach is not so strong
neglection. During an encounter of two bodies, roughly one half of
the velocity change takes place around the point of the closest approach
on the scale of the impact parameter. For the encounter of the galaxy
with two stars, it is confirmed in the right panel of Fig.~\ref{obr.DTsim}.

\subsection{Why does it work?\label{sApx:Why?}}

\begin{figure}[!h]
\begin{centering}
\includegraphics[width=1\columnwidth,keepaspectratio]{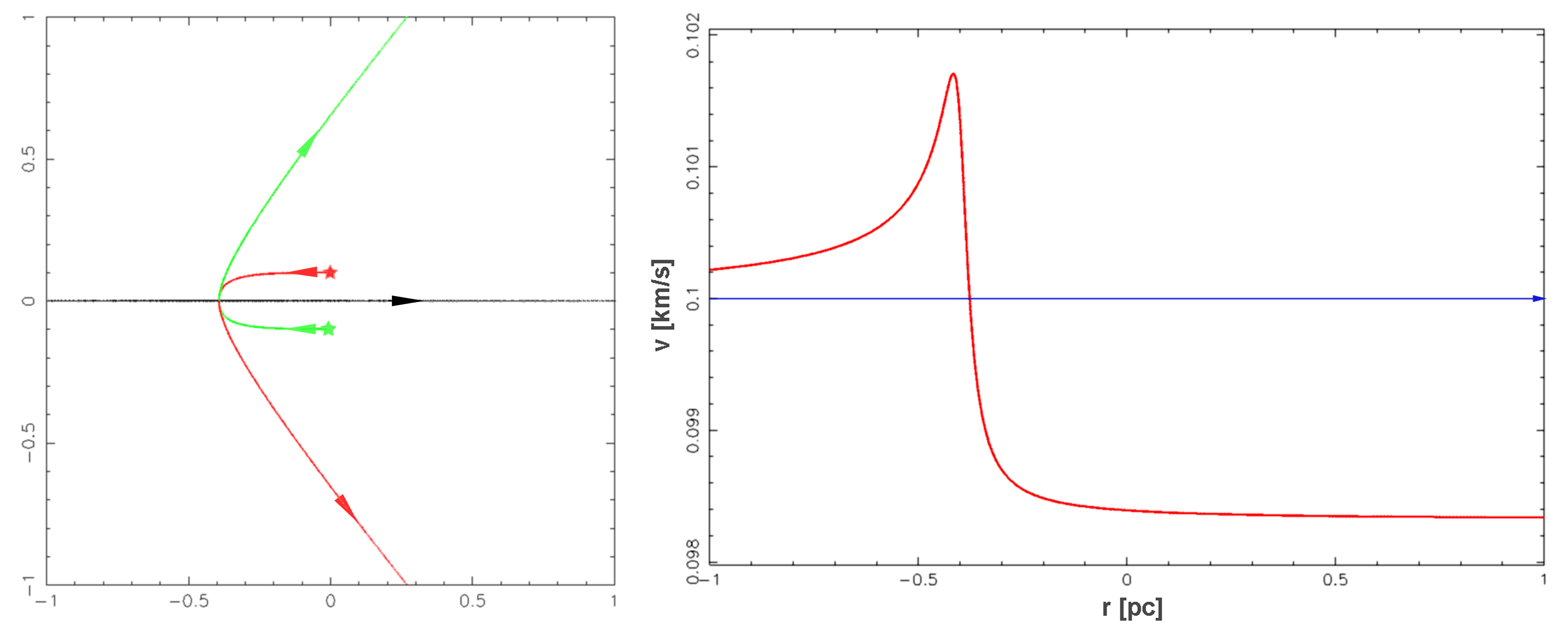}
\par\end{centering}

\caption{\textsf{\small The result of the simulation. A large body (with the
mass of 200\,M\sun, straight black line) moves in the direction
of the $x$-axis (with the velocity of only 100 m$/$s -- this and
the other unrealistic values have been chosen only to make the picture
more illustrative in a linear and uniform scale) and encounters a
pair of stars (2\,M\suns each) that are initially place symmetrically
with respect to its track (0.1\,pc from the track). The mutual gravitational
attraction of the stars is neglected. The right panel shows the development
of the velocity of the large body during the closest approach. The
blue line represents its original velocity, thus its path if the stars
were not present. \label{obr.DTsim}}}
\end{figure}

We can see the mechanism of the dynamical friction in action even
in a simple model of a {}``galaxy'' interacting with two {}``stars'',
results of which are seen in Fig.~\ref{obr.DTsim}. Although the
model is a very simple one, it allows us to see in practice that yet
in the system of three bodies (in contrary to two) the permanent energy
and momentum transfer is possible. The symmetry of the configuration
ensures that the galaxy will keep a straight line and thus any change
of velocity it undergoes will be a change in the magnitude of the
velocity. According to the idea of an infinite sea of stars, we take
into account only the interaction between the stars and the galaxy,
not mutually between the stars.

It is clear that due to the galaxy's gravity, the stars begin to move
towards its track (meanwhile also moving towards the galaxy along
the track, but let us not care for a moment). While the stars move
towards the track, the attraction accumulates and they gather speed.
When they cross the galaxy's track, the galaxy starts pulling them
back (at least when we speak about the perpendicular component of
the velocity) and they slow down. Anyway, thanks to the fact that
they cross the track \emph{after} the galaxy's passage, they spend
more time in the phase where their perpendicular velocity component
is increased than otherwise and finally they retain some speed in
this direction. But it means they have gathered kinetic energy, what
must be at the expense of the galaxy's kinetic energy and so the speed
of the galaxy must have decreased (that is the dynamical friction)
-- even though it has moved much faster than before during the closest
approach of the encounter. In reality, the situation is a little more
complex, because apart from the energy, the momentum has to be also
conserved -- the momentum of the galaxy has decreased and so the stars
must have also a non-zero parallel component of the velocity, to maintain
this component of momentum.

In accordance with the derivation of the Chandrasekhar formula, we
use Eq.~(\ref{eq:dletaV}) just multiplied by two to derive the analytical
formula for the change of the galaxy's velocity. For the impact parameter
\textit{$b$} we obviously put the original distance between the stars
and the galaxy's track. Our numerical tests for various values of
parameters (masses, initial velocity of the galaxy, impact parameter)
show that the analytical results obtained this way tend to overestimate
the decrease in the velocity, typically by about 15 per cent.

It could be anticipated that the numerical and analytical results
will differ, as the analytical formula counts with two separated encounters
from infinity to infinity. In such a case the galaxy follows a curved
trajectory and thus its interaction with the star is slightly different
than when both encounters happen at the same time and the galaxy is
forced to stay on a straight line. Let us remark that we have tested
the model by removing one of the stars and then the results for the
change of the parallel component of the velocity differ from the prediction
in fractions of per mille.

In reality, the situation is even more complex, there are many stars
in the game and they also mutually interact and undergo the influence
of all the surrounding stars that do not take part in the dynamical
friction directly.

\clearpage

\section{Our method \label{Apx:OurDF}}

In \citet{EbrovaMAT} we have introduced our method to calculate the
dynamical friction in restricted $N$-body simulations during the
radial merger. In this section we remind the reader of its characteristics
and derivation as introduced in the master thesis.

\subsection{Avoiding some approximations\label{sApx:Maple-sekce}}

The Chandrasekhar formula contains two kinds of inaccuracies. The
first of them is the principal one, namely the fact that the change
in the parallel component of the velocity from any individual star
is added instantaneously at the point of the closest approach (of
the secondary galaxy) to it. We have already shown that it is not
too wrong, but what is worse, the influence of the star is taken to
be such as if the galaxy passed it from infinity to infinity and there
was nothing in the universe but the star and the galaxy. Sects.~\ref{sApx:ChF}--\ref{sApx:Why?}
for details. 

The second source of inaccuracy lies in all the approximation that
have been done when passing from Eq.~(\ref{eq:iPol}) to Eq.~(\ref{eq:ChF}).
These will concern us in this section, leaving aside the assumptions
of the Maxwellian velocity distribution and the negligence of the
masses of the stars compared to that of the secondary galaxy, that
led us from Eq.~(\ref{eq:ChF}) to Eq.~(\ref{eq:DF}), which we
use in the simulations and keeping the {}``principal inaccuracy''
mentioned above.

The first approximations that allowed us to integrate Eq.~(\ref{eq:iPol})
over the plane of the impact parameter was the assumed homogeneity
of the star field, i.e. that the distribution function does not depend
on position. Then we have taken the Coulomb logarithm to be independent
of velocity of the stars $\mathbf{v}_{m}$ (it obviously isn't, but
it varies slowly) and this has allowed us to simplify the $\mathbf{v}_{m}$-integral
and given a suitable choice of the distribution function we could
even carry out the integration (see Sect.~\ref{sApx:ChF}). Both
steps are only approximate even in the simple case of the spherical
galaxy with the Plummer profile, as both the density -- Eq.~(\ref{eq:hust})
and the velocity dispersion -- Eq.~(\ref{eq:VD}) of the Plummer
sphere do depend on the radius.

If we wish to avoid these simplification, we have to turn back to
Eq.~(\ref{eq:iPol}) and put in e.g. the Maxwellian distribution,
Eq.~(\ref{eq:Maxw}), for $f(\mathbf{v}_{m},b,\varphi)$, together
with putting $n_{0}m=\rho$, where $\rho$ is the density of the primary
at a given point -- keeping in mind that the radius \emph{$r$} (the
distance of a point from the center of the primary galaxy) on which
the formulae depend is a function of \textit{$b$}\textit{\emph{,}}
$\varphi$ and in fact also of the direction of motion of the braked
body (the secondary galaxy). When dealing with the radial mergers,
this direction points towards the center of the primary galaxy and
\textit{$r$} becomes a particularly simple function of \textit{$b$}:

\begin{equation}
r=\sqrt{d^{2}+b^{2}},
\end{equation}
\textit{\emph{where}} \textit{$d$} is (also in the following) the
distance between the centers of the primary and the secondary galaxy.
There is no $\varphi$-dependence in the radial case and the integration
gives a trivial factor of $2\pi$. For simplicity, we put the Eq.~(\ref{eq:VD})
for the velocity dispersion, as the friction is essentially negligible
for both the simple and the cut-off dispersion in the areas where
they significantly differ (see Fig.~\ref{obr.disp}). Furthermore,
during the multiple passages that occur in the simulations (where
the friction becomes significant) the secondary galaxy does not reach
these areas at all. Using Eq.~(\ref{eq:VD}) for the cut-off velocity
dispersion would thus unnecessarily complicate the already complex
formulae.

Putting all this together, we get

\begin{eqnarray}
\frac{\mathrm{d}\mathbf{v}_{M\parallel}}{\mathrm{d}t} & = & \frac{3^{5/2}\varepsilon_{\mathrm{p}}^{2}}{(\pi G)^{3/2}M_{\mathrm{p}}^{1/2}M_{\mathrm{s}}}\intop\intop\frac{\mid\mathbf{v}_{m}-\mathbf{v}_{M}\mid(\mathbf{v}_{m}-\mathbf{v}_{M})}{(b^{2}+d^{2}+\varepsilon_{\mathrm{p}}^{2})^{7/4}}\times
\end{eqnarray}

\[
\times\left[1+\frac{b^{2}(\mathbf{v}_{m}-\mathbf{v}_{M})^{4}}{G^{2}M_{\mathrm{s}}^{2}}\right]^{-1}\mathrm{\exp\left[-\frac{3\mathbf{v}_{\mathit{m}}}{\mathit{GM_{\mathrm{p}}}}\sqrt{\mathit{b}^{2}+\mathit{d}^{2}+\varepsilon_{\mathrm{p}}^{2}}\right]}b\mathrm{d}b\,\mathbf{\mathrm{d^{3}}v}_{m},
\]
where the meaning of the variables is the same as when we derived
the Chandrasekhar formula in Sect.~\ref{sApx:ChF}. The indexes \textit{$p$}
and \textit{$s$} again stand for the parameters of the primary and
the secondary galaxy, respectively.

First, we shift the integration variable to $\mathbf{v}_{m}^{\prime}=\mathbf{v}_{m}-\mathbf{v}_{M}$
and immediately rename it back $\mathbf{v}_{m}^{\prime}\rightarrow\mathbf{v}_{m}$.
We then perform the scalar product with the unit vector $\mathbf{v}_{M}/v_{M}$
on both sides, getting the projection of the friction acceleration
to the direction of the velocity of the secondary galaxy. This is
by symmetry its only nonzero component in the radial case and it will
be advantageous to deal with a scalar-valued integral. The negative
value means that the friction acts in the direction opposite to the
motion of the braked body, what is the only feasible situation in
any setup with an isotropic velocity distribution in the primary galaxy.

Transforming to the spherical coordinates (taking the $z$-axis parallel
with the velocity of the secondary galaxy), we have $\mathbf{v}_{m}\cdot\mathbf{v}_{M}=v_{m}v_{M}\cos\theta$
and again no dependence on the azimuthal angle, leaving us with the
obligatory factor of $2\pi$. The $\theta$-integral then can be carried
out in the form that could be with some effort put on mere three lines:

\begin{equation}
\frac{\mathrm{d}\mathbf{v}_{M\parallel}}{\mathrm{d}t}\cdot\frac{\mathbf{v}_{M}}{v_{M}}=\frac{\sqrt{3\, M_{\mathrm{p}}}G^{3/2}M_{\mathrm{s}}\varepsilon_{\mathrm{p}}^{2}}{2\sqrt{\pi}v_{M}^{2}}\intop_{0}^{\sqrt{R^{2}-d^{2}}}\intop_{0}^{\infty}\frac{b\mathrm{d}b\, v_{m}^{2}\mathrm{d}v_{m}}{\left(\varepsilon_{\mathrm{p}}^{2}+d^{2}+b^{2}\right)^{11/4}\left(G^{2}M_{\mathrm{s}}^{2}+b^{2}v_{m}^{4}\right)}\times\label{eq:TENint}
\end{equation}

\begin{center}
$\times${\Huge {[}}$\mathrm{e}^{-3\,\frac{\sqrt{\varepsilon_{\mathrm{p}}^{2}+d^{2}+b^{2}}\left(v_{m}-v_{M}\right)^{2}}{GM_{\mathrm{p}}}}\left(\mathrm{G}M_{\mathrm{p}}-6\, v_{m}v_{M}\sqrt{\varepsilon_{\mathrm{p}}^{2}+d^{2}+b^{2}}\right)-$
\par\end{center}

\begin{center}
$-\mathrm{e}^{-3\,\frac{\sqrt{\varepsilon_{\mathrm{p}}^{2}+d^{2}+b^{2}}\left(v_{M}+v_{m}\right)^{2}}{GM_{\mathrm{p}}}}\left(\mathrm{G}M_{\mathrm{p}}+6\, v_{m}v_{M}\sqrt{\varepsilon_{\mathrm{p}}^{2}+d^{2}+b^{2}}\right)${\Huge {]}}
,
\par\end{center}

\noindent where \textit{$R$} is the considered cut-off of the primary
galaxy. We cannot proceed analytically with the integration (not even
in one of the variables), instead we have solved it numerically in
Maple for chosen values of the parameters.

We have come to a formula for the dynamical friction Eq.~(\ref{eq:TENint})
that is physically more accurate than the Chandrasekhar formula, but
it is valid only for a radially moving body in the Plummer sphere.
It is also only more accurate in the sense of avoiding the approximation
used between Eq.~(\ref{eq:ChF}) and Eq.~(\ref{eq:DF}) but it is
still built atop the {}``principal inaccuracies'' described above.

\begin{figure}[!h]
\begin{centering}
\includegraphics[width=0.75\columnwidth]{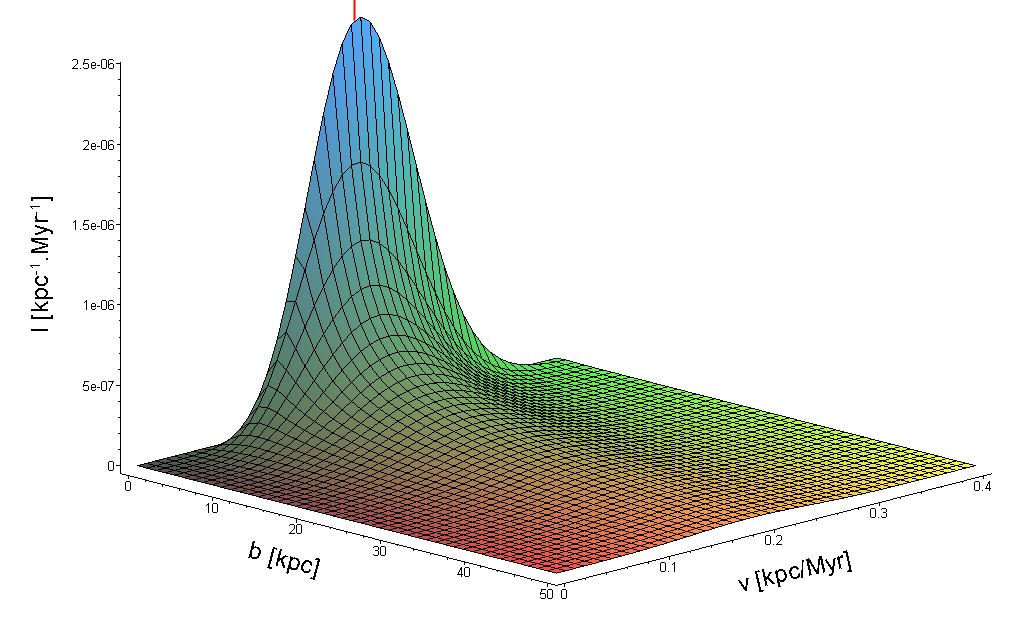}
\par\end{centering}

\caption{\textsf{\small The value of the integrand (including all the constants)
from Eq.~(\ref{eq:TENint}) in the dependence on the integration
variables (the impact parameter and the relative velocity between
the secondary galaxy and the stars) for the standard set of parameters
(see}\textsf{\emph{\small{} }}\textsf{\small Sect.~}\textsf{\emph{\small \ref{sub:SSoP}}}\textsf{\small )
and the distance of the braked body (the secondary galaxy) from the
center of the primary of 70\,kpc. The velocity of the body is taken
to be 0.2\,kpc$/$Myr (1\,kpc$/$Myr $\doteq$ 1000\,km$/$s, see
Appendix~\ref{Apx:units}). This value is also indicated in the graph
by a red marker -- it not surprising to find it near the peak, because
there is a strong contribution from the stars that are in rest with
respect to the center of the primary galaxy, as the Maxwellian distribution
peaks in zero. \label{obr.3Dint}}}
\end{figure}

The reader who considers a formula to be the best figure can enjoy
Eq.~(\ref{eq:TENint}) and who considers a figure to be the best
formula can explore Fig.~\ref{obr.3Dint}, where the integrand of
Eq.~(\ref{eq:TENint}) is shown in dependence of both integration
variables for a chosen set of parameters. It is clear that far most
of the acceleration comes from a close neighborhood of the braked
body both in the plane of the impact parameter and the velocity space.
However, the maximum of the integrand does not exactly coincide with
the actual speed of the body, as there is no reason for it to be so,
but it is very close.

For a primary galaxy made of two Plummer spheres -- one for the luminous
component and one for the dark matter halo -- the equivalent of Eq.~(\ref{eq:TENint})
becomes much more complicated. It can be obtain in much the same manner
as described in this chapter, only using Eq.~(\ref{eq:dispPP}) instead
of Eq.~(\ref{eq:VD}) for the velocity dispersion. But the angular
integration is not possible to analytically, and the resulting three-dimensional
integral cannot be written in a couple of lines' worth of space. A
numerical solution is necessary for specific values of parameters.

\subsection{Back to Chandrasekhar formula \label{sApx:compare-to-Chf}}

We have examined how the braking force according to Eq.~(\ref{eq:TENint})
differs from that calculated using the Chandrasekhar formula. The
Coulomb logarithm is in some sense a free parameter of the formula,
thus we have adjusted it to maximize the agreement between the two
methods of calculation of the friction. For further details, see \citet{EbrovaMAT}.

\begin{figure}[!h]
\begin{centering}
\includegraphics[width=0.95\columnwidth]{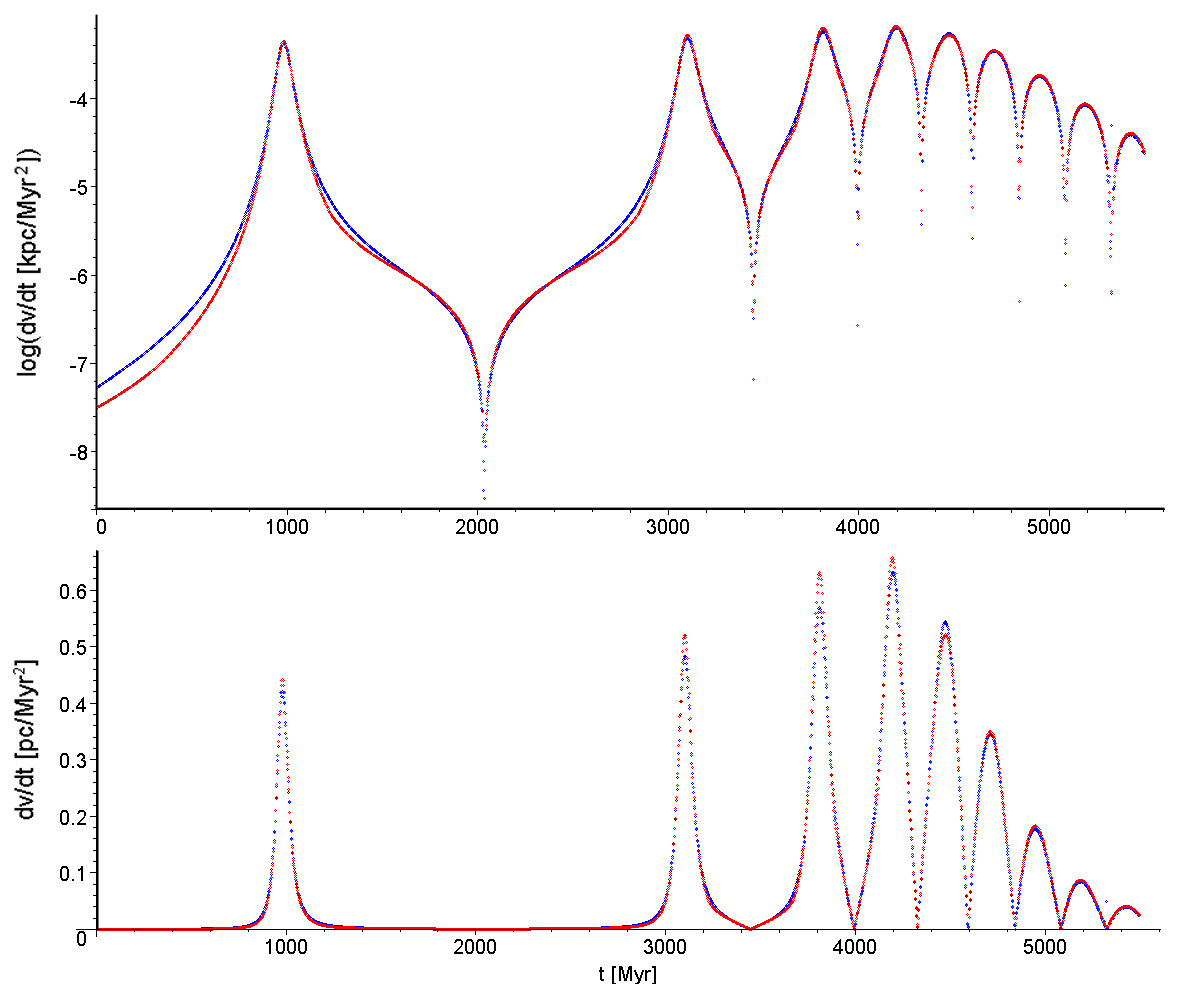}
\par\end{centering}

\caption{\textsf{\small The logarithmic and linear plots of the time dependence
of the dynamical friction for multiple passages of the secondary galaxy
for the standard set of parameters (Sect.~\ref{sub:SSoP}), using
$b_{\mathrm{max}}=10$\,kpc together with the lower limit of the
Coulomb logarithm }\foreignlanguage{british}{\textsf{\small $\ln\Lambda_{\mathrm{crit}}=2$}}\textsf{\small .
Red values are computed in the model, blue values are numerical solution
of Eq.~(\ref{eq:TENint}). \label{obr.bmax10ln2} }}
\end{figure}

Using a constant value of the Coulomb logarithm we did not obtain
a good agreement between the friction calculated using the Chandrasekhar
formula and using our method. The best option seems to be to calculate
the value of the Coulomb logarithm in every step from the actual value
of the velocity of the secondary galaxy. The $V_{0}$ in the definition
Eq.~(\ref{eq:L}) for $\Lambda$ is the difference between the velocities
of the stars and the secondary galaxy. As the stellar velocities are
isotropic, the average value is just the velocity of the secondary
galaxy with respect to the center of the primary. 

There is a uncertainty in the parameter $b_{\mathrm{max}}$ in the
same equation -- it should be theoretically equal to the distance
between the center of the secondary and the outer boundary of the
primary measured in the plane perpendicular to the motion of the secondary.
But Eq.~(\ref{eq:DF}) assumes a homogeneous field of stars across
all this distance, what is obviously not true. As the plane of the
impact parameter is the plane perpendicular to the radial motion of
the secondary galaxy, the density of the primary galaxy is always
the highest in its center and decreases outwards. Thus it may seem
that the $b_{\mathrm{max}}$ should be smaller than the normal distance
to the edge of the primary galaxy, but the approximation of the $V_{0}$
with the velocity of the galaxy and other circumstances make the situation
more complex. The value of $b_{\mathrm{max}}$ must be chosen in a
trial-and-error method for the chosen parameters of collision so that
the magnitude of the friction agrees best with the numerical solution
of the integral Eq.~(\ref{eq:TENint}).

The adaptive version of the Coulomb logarithm with a suitable chosen
$b_{\mathrm{max}}$ fits nicely in the high velocity regime. The problem
appears when the satellite gets close to its apocenter and also mainly
in the late parts of the merger when the velocity of the satellite
is much lower than during its first passage through the center of
the primary galaxy. Here the adaptive version of the Coulomb logarithm
with the chosen $b_{\mathrm{max}}$ significantly underestimates the
friction when compared with the numerical solution of the integral
Eq.~(\ref{eq:TENint}). So we use the adaptive Coulomb logarithm
until its value drops under a certain limit $\ln\Lambda_{\mathrm{crit}}$,
then we put this limit for the Coulomb logarithm instead. With this
modification of the Chandrasekhar formula, we can achieve a reasonable
agreement, see Fig.~\ref{obr.bmax10ln2}. $b_{\mathrm{max}}$ and
the lower limit for the Coulomb logarithm are free parameters and
they depend on the parameters of the radial merger -- the initial
mutual velocity of the galaxies, their masses and Plummer radii.

\subsection{Incorporation of the friction in the simulation\label{sApx:Zacleneni}}

The question of incorporation of the dynamical friction in the simulations
of the shell formation is tricky. In a fully self-consistent simulation,
the dynamical friction would be automatically included, but such a
simulation would be extremely demanding on resources -- for the friction
to be really well simulated, the number of particles of primary galaxy
should not be several orders of magnitude smaller than the true amount
of stars in the galaxies. Joining the stars in a smaller amount of
more massive objects systematically overcounts the friction. \citet{2004MNRAS.349..747P}
remarked that \citet{1992A&A...259...25P,1996A&A...312..431W} have
indeed shown that the dynamical friction is artificially increased
if the particle number is small. Using the analytical formula for
the friction is not devoid of problems, but in some respects it could
be more accurate than some of the self-consistent simulations. 

On the other hand, the number of the particles of the secondary is
an important quantity for the visibility of the shells in the simulations.
And for the large number of required test particles $(\sim10^{6})$
that represent just the secondary galaxy, even our {}``simple''
simulations take hours of computation on a contemporary desktop computer.
Furthermore, to explore the parameter space we have to run a lot of
simulations, so we can really use a handy (semi-)analytical formula.
We can easily add the acceleration calculated by Eq.~(\ref{eq:TENint})
into the equation of motion of the galaxies. 

It is worth mentioning that we departed in two aspects from the potential
that we chose to model the merging galaxies. We assumed Maxwell velocity
distribution, Eq.~(\ref{eq:Maxw}). This is not exactly true for
the Plummer sphere, but the difference is small and the true velocity
distribution in real galaxies is not known, so we cannot do much better,
or say exactly how big mistake do we make.

The secondary galaxy is here treated as a point mass what artificially
increases the friction, because the extended character of the galaxy
softens the force (Sect.~\ref{sub:Force}). Specially the stars with
a small impact parameter with respect to the center of the secondary
galaxy fly straight through it and their effect is significantly reduced
compared to the Chandrasekhar formula for the point mass. The overestimation
of the dynamical friction is not a crucial problem as we want to estimate
how much the shell system is influenced by  it -- we can assume that
the reality is not worse than our results and we get the upper bound
on the effect.

\clearpage

\section{Tidal radius \label{Apx:Tidal-radius}}

For starters, let us remind the reader of the derivation of the tidal
radius, as presented in \citet{EbrovaMAT}. The tidal forces acting
on an object are often derived using the following picture: A massive
body (secondary galaxy) as a whole follows the force acting on it
in its center of mass. But the force acting on outer parts of the
body is different, as it is at different distances of the source (the
primary galaxy). If this difference is larger than the binding force
with the secondary for a given star, it is stripped off.

The tidal radius $r_{\mathrm{tidal}}$ is then defined as the distance
(from the center of the secondary), where the difference of the force
of the primary from its force in the center of mass of the secondary
is just equal to the force from the secondary:

\begin{equation}
F_{\mathrm{p}}(d-r_{\mathrm{tidal}})-F_{\mathrm{p}}(d)=F_{\mathrm{s}}(r_{\mathrm{tidal}}),\label{eq:tidalDef}
\end{equation}
where \emph{$d$} is the separation between the centers of the galaxies
and $F_{\mathrm{p}}(r)$ and $F_{\mathrm{s}}(r)$ is the force from
the primary and the secondary for a given test particle (its mass
is immediately canceled out from the equation).

For two point-like bodies (with masses $M_{\mathrm{p}}$ and $M_{\mathrm{s}}$),
we can write Eq.~(\ref{eq:tidalDef}) as:

\begin{equation}
\frac{G\, M_{\mathrm{p}}}{d^{2}(1-r_{\mathrm{tidal}}/d)^{2}}-\frac{G\, M_{\mathrm{p}}}{d^{2}}=\frac{G\, M_{\mathrm{s}}}{r_{\mathrm{tidal}}^{2}}.\label{eq:tid2}
\end{equation}

\noindent Assuming further $r_{\mathrm{tidal}}\ll d$ we can use the
Taylor expansion $(1-x)^{-2}\cong1+2x$ for $x=r_{\mathrm{tidal}}/d$
as it is then a small quantity. under this assumption we get a simple
formula for the tidal radius:

\begin{equation}
r_{\mathrm{tidal}}=d\sqrt[3]{\frac{M_{\mathrm{s}}}{2\, M_{\mathrm{p}}}}.\label{eq:t-aprox}
\end{equation}

However, for two point masses we can get an exact result for the tidal
radius. Not making any approximation in Eq.~(\ref{eq:tid2}) we can
cast it as a fourth-order polynomial

\begin{equation}
X^{4}-2\, X^{3}+q\, X^{2}-2\, q\, X+q=0,
\end{equation}
where $X=r_{\mathrm{tidal}}/d$ and $q=M_{\mathrm{s}}/M_{\mathrm{p}}$.
A polynomial with an order less than five can be always solved. In
our case, where \emph{$q$} is positive, there are two real roots,
from which we take the one that gives $r_{\mathrm{tidal}}<d$ and
thus $X<1$. The second real root corresponds to a point of the other
side of the primary galaxy that is not of interest for us. The expression
for this root does not give much insight, but an interested reader
can find it in Appendix \ref{Apx:tidal-hnus}.

\begin{figure}[!h]
\begin{centering}
\includegraphics[width=15cm]{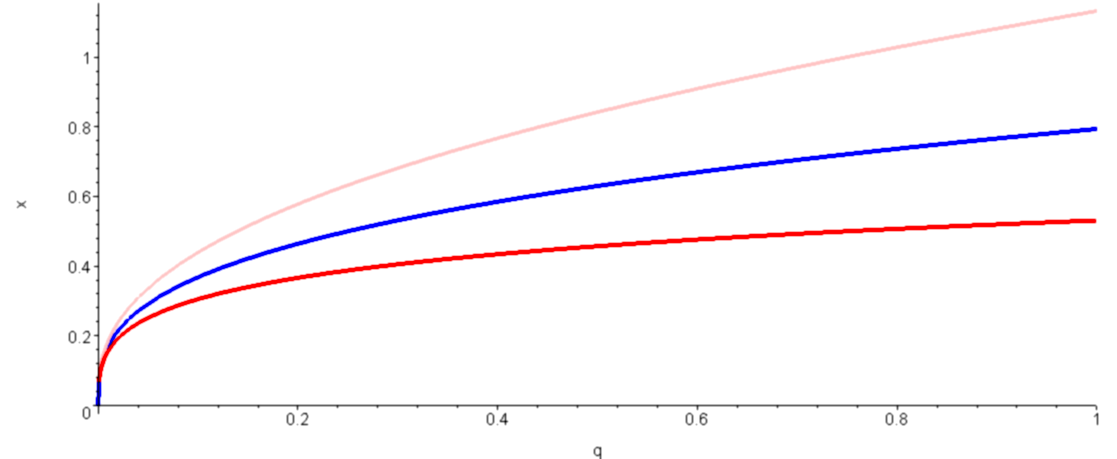}
\par\end{centering}

\caption{\textsf{\small Tidal radius for two point masses: the approximate
solution, Eq.~(\ref{eq:t-aprox}), is shown in blue, the exact solutions
in red (the outer one in light red, the inner in dark red). The shows
y-axis $X=r_{\mathrm{tidal}}/d$, the $x$-axis shows the secondary-to-primary
mass ratio. \label{obr.t-point} }}
\end{figure}

Eq.~(\ref{eq:tidalDef}) gives the tidal radius for the particles
on the line connecting the centers of the two bodies -- we call it
the inner tidal radius. Similarly we can write an equation for the
particles on the other side of the secondary than the center of primary
lies: 
\begin{equation}
F_{\mathrm{p}}(d)-F_{\mathrm{p}}(d+r_{\mathrm{tidal}})=F_{\mathrm{s}}(r_{\mathrm{tidal}}).\label{eq:vnejsi}
\end{equation}
It again leads to a fourth-order polynomial for which we can obtain
the root that we call the outer tidal radius. The approximate solution
Eq.~(\ref{eq:t-aprox}) is the same for both equations, Eq.~(\ref{eq:tidalDef})
and Eq.~(\ref{eq:vnejsi}). Let us remark that the tidal radius is
in any case just proportional to \emph{$d$} as there is no other
scale in the problem. Fig.~\ref{obr.t-point} shows the dependence
of the three radii on the mass ratio of the bodies. We can see that
for all relevant ratios the approximate formula is just between the
inner and the outer tidal radius.

\begin{figure}[H]
\begin{centering}
\includegraphics[width=12cm]{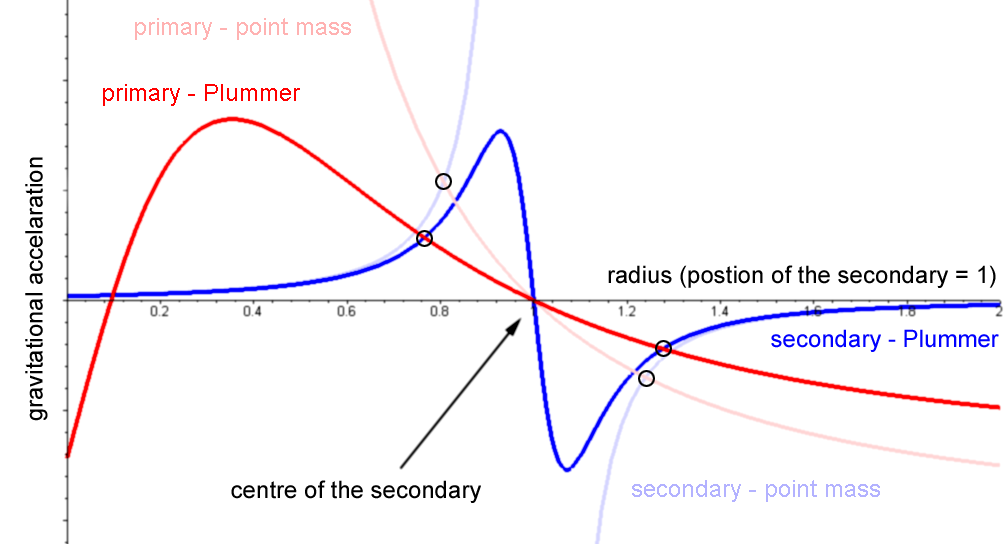}
\par\end{centering}

\caption{\textsf{\small The outer and inner tidal radii (marked with circles)
for the point masses and Plummer spheres with the secondary-to-primary
mass ratio of 0.02. In the Plummer case, the Plummer radius of the
primary is 0.5 of the distance between the bodies and the Plummer
radius of the secondary is 0.1 of the same quantity. Blue lines (light
blue for the point mass, dark blue for the Plummer sphere) show the
gravitational force of the primary in arbitrary units, red lines (light
red for the point mass, dark red for the Plummer sphere) show the
difference between the gravitational force of the primary in a given
point and its value in 1, where the center of the secondary is. The
tidal radii are the points of intersection of corresponding curves.
\label{obr.t-schem} }}
\end{figure}

The tidal radius for a point mass is in some sense an oxymoron, as
these objects have zero proportions by definition. For spherically
symmetric bodies we can write Eq.~(\ref{eq:tid2}) as

\begin{equation}
\frac{G\, M_{\mathrm{p}}(d-r_{\mathrm{tidal}})}{(d-r_{\mathrm{tidal}})^{2}}-\frac{G\, M_{\mathrm{p}}(d)}{d^{2}}=\frac{G\, M_{\mathrm{s}}(r_{\mathrm{tidal}})}{r_{\mathrm{tidal}}^{2}},
\end{equation}
where $M(r)$ is the mass enclosed in the radius \emph{$r$.} Particularly
for the Plummer sphere we get this value integrating Eq.~(\ref{eq:hust})
over the sphere with the radius \emph{$r$}:

\begin{equation}
M(r)=\frac{M}{(1+\varepsilon^{2}/r^{2})^{3/2}},\label{eq:Mr}
\end{equation}
where \emph{$M$} is the overall mass of the body and $\varepsilon$
is the Plummer radius. Unfortunately this makes the equation too complex
to be easily solved. Let us compare graphically the tidal radii for
point masses and Plummer spheres of the same overall masses just for
one particular case -- Fig.~\ref{obr.t-schem}.

The figure (or a simple thought) shows that the notion of the tidal
radius in a general potential makes sense only when the force grows
with the distance. Otherwise the tidal force acts in the same direction
as the gravitation of the secondary and thus cannot strip off any
mass. In the Plummer potential the force reaches its maximum in $\sqrt{2}\,\varepsilon/2$,
so the tidal radius is not defined under this radius, whereas for
the point masses it is defined everywhere. 

The idea of the tidal radius is just an approximation to the complex
processes during encounters of two extended bodies. It also does not
define a sphere around the center of the secondary galaxy, but as
we have seen, it is different for various locations, with the lowest
value towards the center of the primary galaxy and the highest on
the opposite side. For these reasons it is not really useful to improve
its evaluation and so we have used the approximate Eq.~(\ref{eq:t-aprox})
that as we have seen gives the values somewhere in the middle between
the two extreme values of the tidal radius. 

\clearpage

\section{Expressions for the tidal radius\label{Apx:tidal-hnus}}

Here we give the analytical formulae for the tidal radii in the system
of two point masses as discussed in Appendix~\ref{Apx:Tidal-radius}.
For the inner tidal radius we have:
\begin{equation}
\frac{r}{d}=\frac{1}{2}+\frac{\sqrt{3}}{6}\left(\frac{\sqrt{y}}{\sqrt[6]{z}}-\sqrt{6-4\, q-\sqrt[3]{qz}-\sqrt[3]{\frac{q^{5}}{z}}+6\,\sqrt[6]{z}\sqrt{\frac{3}{y}}(q+1)}\,\right),
\end{equation}
where
\begin{equation}
y=(3-2\, q)\sqrt[3]{z}+\sqrt[3]{qz^{2}}+q^{5/3}
\end{equation}

\begin{equation}
z=54+q^{2}+6\,\sqrt{81+3\, q^{2}}
\end{equation}
and for the outer tidal radius we get similar expressions:

\begin{equation}
\frac{r}{d}=\frac{1}{2}+\frac{\sqrt{3}}{6}\left(\frac{\sqrt{u}}{\sqrt[6]{v}}+\sqrt{6+4\, q-\sqrt[3]{qv}-\sqrt[3]{\frac{q^{5}}{v}}+6\,\sqrt[6]{v}\sqrt{\frac{3}{u}}(q-1)}\,\right),
\end{equation}
where

\begin{equation}
u=(3+2\, q)\sqrt[3]{v}+\sqrt[3]{qv^{2}}+q^{5/3}
\end{equation}

\begin{equation}
v=-54-q^{2}+6\,\sqrt{81+3\, q^{2}}
\end{equation}
and in all the expressions we use

\begin{equation}
q=\frac{M_{\mathrm{s}}}{M_{\mathrm{p}}}.
\end{equation}

\newpage{}

\section{Videos \label{Apx:Videos}}

Several videos are also part of the electronic attachment of the thesis.
Here we present their description. Information on details of the simulation
process can be found in Sect.~\ref{sec:Description-of-simulation}.
The videos can be also downloaded at: pc048b.fzu.cz/$\sim$ivana/shells/phd
\begin{enumerate}
\item 1-shells.avi -- Video from a simulation of a shell-producing radial
minor merger from a perspective perpendicular to the axis of the merger.
The bottom three panels show an area of $60\times60$\,kpc centered
on the primary which is the zoomed part of the upper panels of size
$300\times300$\,kpc. The first column shows the surface density
of both the primary and the secondary galaxy, the second only the
surface density of the particles originally belonging to the secondary
galaxy (corresponding to the host galaxy subtraction, a technique
used in processing real galaxy images). The third column shows the
surface density of particles originally belonging to the secondary
galaxy divided by the surface density of the primary galaxy (also
corresponding to an observational technique). The parameters of the
merger are the following: the mass of the primary is $3\times10^{11}$\,M\suns,
the secondary-to-primary mass ratio is 0.02, the Plummer radius of
the primary is 7.6\,kpc, of the secondary 0.76\,kpc. The initial
relative velocity of the galaxies was equal to the escape velocity
of the secondary and the separation of their centers was 90\,kpc.
When the centers of the galaxies pass through each other, the potential
of the secondary is suddenly switched off. \label{enu:video1}
\item 2-shells.mpg -- Video from a simulation of a shell-producing radial
minor merger used in Sect.~\ref{sec:N-Simulations}. The top panel
($300\times300$\,kpc centered on primary) shows the surface density
of the particles originally belonging to the secondary galaxy from
a perspective perpendicular to the axis of the merger; the bottom
panel shows the density of the particles originally belonging to the
secondary in the space of radial velocity (vertical axis) versus galactocentric
distance (horizontal axis). The potential of the host galaxy is the
same as the one described in Sect.~\ref{sec:param}. Primary is modeled
as a double Plummer sphere with respective masses $M_{*}=2\times10^{11}$\,M\suns
and $M_{\mathrm{DM}}=1.2\times10^{13}$\,M\suns, and Plummer radii
$\varepsilon_{*}=5$\,kpc and $\varepsilon_{\mathrm{DM}}=100$\,kpc
for the luminous component and the dark halo, respectively. The potential
of the cannibalized galaxy is chosen to be a single Plummer sphere
with the total mass $M=2\times10^{10}$\,M\suns and Plummer radius
$\varepsilon_{*}=2$\,kpc. The cannibalized galaxy is released from
rest at a distance of 100\,kpc from the center of the host galaxy.
When it reaches the center of the host galaxy in 306.4\,Myr, its
potential is switched off and its particles begin to oscillate freely
in the host galaxy. \label{enu:video2-kinem}
\item 3-projection.mpg -- Video shoes the simulation from point \ref{enu:video2-kinem}
(used in Sect.~\ref{sec:N-Simulations}) at the time 2.2\,Gyr after
the decay of the cannibalized galaxy (2.5\,Gyr of the simulation
time) from different perspectives. Angle of 0 degrees corresponds
to the perspective perpendicular to the axis of the merger. \label{enu:video3-proj}
\item 4-friction.avi -- Surface density of the particles originally belonging
to the secondary galaxy from two simulation of a radial minor merger
from Sect.~\ref{sec:Sim-DF&TD} (run\,1 -- right panels and run\,2
-- left panels). The first column corresponds to the simulation with
dynamical friction and gradual decay of the secondary; the other corresponds
to the simulation without friction and with the instant disruption
of the secondary near the center of the primary galaxy. The bottom
panels show an area of $60\times60$\,kpc centered on the primary
which is the zoomed part of the upper panels of size $300\times300$\,kpc.
The video covers 8\,Gyr since the release of the secondary galaxy
from distance of 180\,kpc from the center of the primary with the
escape velocity. Both simulations were executed for the the standard
set of parameters (Sect.~\ref{sub:SSoP}): the mass of the primary
is $3.2\times10^{11}$\,M\suns, the secondary-to-primary mass ratio
is 0.02, the Plummer radius of the primary is 20\,kpc, of the secondary
2\,kpc. \label{enu:video4-fri}
\item 5-selfconsistent.avi -- Video from self-consistent simulation of a
radial minor merger from Sect.~\ref{sub:Self-cons.}. The bottom
panel ($400\times400$\,kpc centered on primary) shows the surface
density of the particles originally belonging to the secondary galaxy
from a perspective perpendicular to the axis of the merger; the top
panel shows the density of the particles originally belonging to the
secondary in the space of radial velocity (vertical axis) versus galactocentric
distance (horizontal axis). The potential of the primary galaxy is
a double Plummer sphere with respective masses $M_{*}=2\times10^{11}$\,M\suns
and $M_{\mathrm{DM}}=8\times10^{12}$\,M\suns, and Plummer radii
$\varepsilon_{*}=8$\,kpc and $\varepsilon_{\mathrm{DM}}=20$\,kpc
for the luminous component and the dark halo, respectively. The potential
of the secondary galaxy is chosen to be a single Plummer sphere with
the total mass $M=2\times10^{10}$\,M\suns and Plummer radius $\varepsilon_{*}=2$\,kpc.
The cannibalized galaxy is released from the distance of 200\,kpc
from the center of the host galaxy with the initial velocity 102\,km$/$s.
\label{enu:video5-self}
\end{enumerate}
Videos 2--4 were made from simulated data by Miroslav K{\v r}{\'{\i}}{\v z}ek.

\newpage{}

\clearpage
\setcounter{page}{146}\bibliographystyle{klunamed}
\addcontentsline{toc}{section}{\refname}\bibliography{shells-GACR,shells-cl,shells-DP}

\end{document}